\documentclass[12pt]{article}
\usepackage{epsf, cite, amsmath, amssymb}
\usepackage{epsfig}

\makeatletter
\renewcommand\section{\@startsection {section}{1}{\z@}%
                                   {-3.5ex \@plus -1ex \@minus -.2ex}%
                                   {2.3ex \@plus.2ex}%
                                   {\normalfont\large\bfseries}}
\renewcommand\subsection{\@startsection{subsection}{2}{\z@}%
                                     {-3.25ex\@plus -1ex \@minus -.2ex}%
                                     {1.5ex \@plus .2ex}%
                                     {\normalfont\bfseries}}
\makeatother

\let\non\nonumber

\newcommand{\bea}{\begin{eqnarray}}
\newcommand{\eea}{\end{eqnarray}}

\newcommand{\hlf}{\frac{1}{2}}
\newcommand{\qrt}{\frac{1}{4}}
\newcommand{\Z}{{\mathbb Z}}
\newcommand{\R}{{\mathbb R}}

\newcommand{\Q}{{\mathbb Q}}

\newcommand{\mO}{\Omega}

\renewcommand{\P}{{\mathbb P}}
\newcommand{\C}{{\mathbb C}}

\newcommand{\half}{\hlf}

\def\a{\alpha} 
 
\def\g{\gamma} 
\def\G{\Gamma}
\def\e{\epsilon}
\def\h{\eta}
\def\th{\theta}

\def\v{\varphi} 

\def\Tor{{\rm Tor}\,}
\def\ov{\overline}
\def\Pee{\mathbb{P}}
\def\Cee{\mathbb{C}}
\def\Zee{\mathbb{Z}}

\newcommand{\be}{\begin{equation}}
\newcommand{\ee}{\end{equation}}

\begin{document}

\begin{titlepage}

\begin{center}
Revised April 13, 2001
\hfill                  hep-th/0103170 

\hfill ITFA-2001-11, HUTP-01/A009, EFI-01-06, DUKE-CGTP-01-05

\hfill PAR-LPTHE-01-10, LPTENS 01/07, NSF-ITP-01-12

\vskip 1.5 cm
{\Large \bf Triples, Fluxes, and Strings}\\
\vskip 1 cm 
{J. de Boer$^1$, R. Dijkgraaf$^{1,2}$, K. Hori$^3$, A. Keurentjes$^4$, }\\
\vskip 0.1cm
{J. Morgan$^5$, D. R. Morrison$^6$, and S. Sethi$^7$}\\
\vskip 0.5cm
{\sl \footnotesize
     {}$^1$ Institute for Theoretical Physics, University of Amsterdam,\\
Valckenierstraat 65, 1018 XE Amsterdam, The Netherlands\\
\vskip 0.2cm
     {}$^2$ Korteweg-de Vries Institute for Mathematics, University of
Amsterdam,\\ Plantage Muidergracht 24, 1018 TV Amsterdam, The Netherlands\\
\vskip 0.2cm
     {}$^3$ Department of Physics, Harvard University, Cambridge, MA
     02138, USA \\
\vskip 0.2cm {}$^4$ LPTHE, Universit\'e Pierre et Marie Curie, Paris
VI, \\Tour 16, 4 place Jussieu, F-75252 Paris Cedex 05, France

 Laboratoire de Physique Th\'eorique de l'Ecole
Normale Sup\'erieure,\\ 24 rue Lhomond, F-75231 Paris Cedex 05, France \\
\vskip 0.2cm
{}$^5$ Department of Mathematics, Columbia University, New York, NY
10027, USA \\
\vskip 0.2cm
{}$^6$ Department of Mathematics, Duke University, Durham, NC 27708, USA \\
\vskip 0.2cm
{}$^7$ Enrico Fermi Institute, University of Chicago, Chicago, IL
60637, USA\\}

\end{center}

\vskip 0.5 cm

\begin{abstract}

We study string compactifications with sixteen supersymmetries. The
moduli space for these compactifications becomes quite intricate in
lower dimensions, partly because there are many different irreducible
components. We focus primarily, but not exclusively, on
compactifications to seven or more dimensions.  These vacua can be
realized in a number ways: the perturbative constructions we study
include toroidal compactifications of the heterotic/type I strings,
asymmetric orbifolds, and orientifolds. In addition, we describe less
conventional M and F theory compactifications on smooth spaces. The
last class of vacua considered are compactifications on singular
spaces with non-trivial discrete fluxes.

We find a number of new components in the string moduli space. 
Contained in some of these components are M theory compactifications
with novel kinds of ``frozen'' singularities. We are naturally 
led to conjecture the existence of new dualities 
relating spaces with different singular geometries and fluxes.
As our study of these vacua unfolds, we also learn about 
additional topics including: F theory on spaces without section, 
automorphisms of del Pezzo surfaces, and novel physics (and puzzles) from 
equivariant K-theory. Lastly, we comment on how the data we gain 
about the M theory three-form might be interpreted.   

\end{abstract}

\end{titlepage}

\pagestyle{plain}
\baselineskip=18pt
\tableofcontents
\begin{center}
-----------------------------------------------------------------------------
\end{center}
\section{Introduction and Summary}

The moduli space of supersymmetric string compactifications is an
immensely complicated object. One of the aspects that we might hope to
understand are the discrete choices that characterize disconnected
components of the moduli space.  We shall focus on string
compactifications with sixteen supersymmetries. Familiar examples of
such compactifications are the heterotic string on a torus and M
theory on a $K3$ surface. With this much supersymmetry, the moduli
space cannot be lifted by space-time superpotentials. The number of
distinct components in the string moduli space can, however, change as
we compactify to lower dimensions. For example, when compactified on a
circle, there is a new component in the moduli space of the 
heterotic string which contains the CHL string \cite{CHL,
Chaudhuri95b}.

We shall primarily, but not exclusively, focus on compactifications to
7 or more dimensions. Our goal is to describe the different components
of the string moduli space in each dimension.  We also describe the
various dual ways in which a given space-time theory can be realized
in string theory or in M theory, its non-perturbative, mysterious
completion.  Down to 7 dimensions, it seems quite likely that our
study of the string moduli space captures all the distinct
components. However, a proof must await a deeper understanding of M
theory. As we analyze the various components of the string moduli
space, we will learn about new phenomena in string theory and some
interesting mathematical relations.  Many of the results we describe
call for a deeper analysis, or suggest natural paths for further
study. There are also tantalizing hints of how we might correctly
treat the M theory 3-form. Some of these hints suggest relations
between the 3-form and $E_8$ gauge bundles which are somewhat
reminiscent of \cite{Fabinger:2000jd, moore2}. 
In the remainder of the introduction, we
shall outline our results.

In the following section, we begin by describing the classification of
flat bundles on a torus. While this might seem trivial at first sight,
there are actually interesting new components in the moduli space of
flat connections on $T^3$, and on higher-dimensional tori. On $T^3$,
these new components correspond to ``triples'' of commuting flat
connections which are not connected to the trivial connection---the
case with no Wilson line.  If we pick a connection in a given
component of the moduli space, we can compute its Chern--Simons
invariant. This is constant over a given component of the moduli space
and actually uniquely characterizes each component of the moduli
space.  These new components in the gauge theory moduli space are the
basis for a new set of components in the moduli space of the heterotic
string on $T^3$. Our discussion of these heterotic/type I toroidal
compactifications begins with a discussion of anomaly cancellation
conditions, and continues with a study of asymmetric orbifold
realizations together with the structure of the moduli space for these
new components.

Let us summarize our findings: in 9 dimensions, we find only the 2
known components in the heterotic/type I string moduli space. The
``standard'' component unifies the conventional $E_8\times E_8$ string
together with the $Spin(32)/\Z_2$ heterotic/type I string. It is 
important for us to note that the gauge group for the $E_8\times E_8$
string 
is actually $(E_8\times E_8) \rtimes \Z_2$, as explained in section
\ref{circle}.\footnote{Nevertheless, throughout this paper we 
use the common nomenclature, ``$E_8\times E_8$ string.''}
The other
component contains the CHL string. In 8 dimensions, we still have the
standard component. There is still only one other component which now
contains both the CHL string and the compactification of the type I
string with no vector structure. The interesting new physics appears
in 7 dimensions where we find 6 components in the moduli space. These
components can be labeled by a cyclic group $\Z_m$ for
$m=1,2,3,4,5,6$. This cyclic group appears naturally in the
construction of the $\Z_m$-triple for an $E_8$ gauge bundle.  Like the
CHL string, these new components have reduced rank and interesting
space-time gauge groups like $F_4$ and $G_2$. The $m=1$ case is just
the standard compactification of the heterotic string, while the
$\Z_2$-triple contains the CHL string in its moduli space.

We then proceed in section \ref{hetduality} to describe duality chains
relating heterotic compactifications with different gauge
bundles. These chains generalize the usual T-duality relating the
$Spin(32)/\Z_2$ and $E_8\times E_8$ string on $S^1$. In some cases, we
follow these chains all the way down to 5 dimensions finding new
relations as we descend.  The relations involve ``quadruples'' and
``quintuples'' which are analogues of triples for $T^4$ and
$T^5$. Unfortunately, the classification of gauge bundles on tori of
dimension greater than 3 is unknown. This is an outstanding open
question. In section \ref{strongcouplingtriple}, we conclude our
discussion of the heterotic string by describing an intriguing
connection between the $\Z_2$-triple and a Ho{\accent20r}ava--Witten style
construction of the $E_8\times E_8$ string with background 3-form
flux.

In section \ref{orient}, we turn to orientifold string
vacua. Proceeding again dimension by dimension, we find only the two
previously known components in 9 dimensions: the first is the standard
component containing the type I string.  The second is the $(+,-)$
orientifold which contains no D-branes and has no enhanced gauge
symmetry. We use $+$ to refer to an $O^+$ plane, $-$ to refer to an
$O^-$ plane, and $-'$ to refer to an $O^-$ plane with a single stuck
D-brane. This notation and our conventions are explained more fully in
section \ref{orient}.  This is a new component in the string moduli
space beyond those with a dual heterotic description. In 8 dimensions,
we find three components: the standard one, the orientifold
realization of type I with no vector structure and the
compactification of the $(+,-)$ orientifold which is the $(+,+,-,-)$
orientifold. Again, there is only one new component beyond those
already described.

In 7 dimensions, we again find new physics. The compactifications of
the 8-dimensional constructions give three components. However, there
is now an interesting subtlety with the case of $(+^4, -^4)$.  We can
imagine arranging the 8 orientifold planes on the vertices of a
cube. However, there are 2 distinct ways of arranging the orientifold
planes which are not diffeomorphic.  The first arrangement is the one
obtained by compactifying $(+,-)$ on $T^2$. The four $+$ planes lie on
a single face of the cube. The $-$ planes lie on the four vertices of
the opposite face. If we exchange one adjacent pair of $+$ and $-$
planes, we find an inequivalent configuration. As perturbative string
compactifications, we show that these two configurations are
inequivalent.\footnote{An interesting paper with a similar conclusion 
appeared shortly after our paper~\cite{Bergman:2001rp}.} 
Whether these orientifolds are distinct
non-perturbatively is more subtle to determine, and we comment on this
in section \ref{threeformphysics}.  This question of how we order the
orientifold planes continues to be important in lower-dimensional
compactifications. Therefore, there are two new components in the
moduli space of perturbative string compactifications to 7
dimensions. We also give evidence against the existence of an
$O6^{-'}$ plane---a conclusion arrived at independently using
different arguments in \cite{imamura}.

In dimensions below 7, our classification of orientifold
configurations is no longer complete. However, we find evidence for a
number of interesting relations including a 6-dimensional duality
between $({+'}^4, {-'}^{12})$ and a quadruple compactification of type I
with no vector structure. We also find evidence for a 5-dimensional
equivalence between $({-'}^{32})$ and $(+^{16}, -^{16})$. There are a
host of open questions concerning the complete
classification of orientifold configurations
below 7 dimensions, the action of S-duality etc.

In section \ref{geometry}, we turn to M and F theory
compactifications. Our starting point is 6-dimensional M theory
compactifications without flux.  The compactifications we study are on
spaces of the form $(Z\times S^1)/G$ where $Z=K3$ or $T^4$ and $G$ is
a discrete group acting freely. For $Z=K3$, the choice of groups $G$
has been classified by Nikulin.  In our M theory context, the possible
choices are $G=\Z_m$ with $m=1,\ldots,8$, while for $Z=T^4$, $G=\Z_n$
with $n=2,3,4,6$. We describe both the lattices for these
compactifications and the singularities of $Z/G$. Only some of these M
theory compactifications can be lifted to 7-dimensional F theory
compactifications. For $Z=K3$, the cases $m=1,\ldots,6$ lift to new
7-dimensional theories which are dual descriptions of the heterotic
triples constructed in section \ref{asymorbifold}. It seems worth
mentioning that studying D3-brane probes on these backgrounds, along
the lines of \cite{probes,probes2,probes3,probes4,probes5,probes6},
should be interesting.

All of the $Z=T^4$ theories lift to 7 dimensions.  The case $G=\Z_2$
is another description of the compactified $(+,-)$ orientifold while
the 3 remaining cases are new components in the string theory moduli
space.  We also point out the existence of a new F theory vacuum in 6
dimensions associated with $G=\Z_2\times\Z_2$. In studying these vacua
and their dual realizations, we arrive at a natural interpretation of
F theory compactifications {\sl without} section
\cite{Morrison:WilsonFtheory}: the type IIB circle which should
decompactify under M theory/F theory duality 
as the volume of the elliptic fiber on the M-theory side
goes to zero has a
non-trivial twist. On decompactifying the circle, the twist becomes
irrelevant and we gain additional degrees of freedom beyond those that
we might have expected. The F theory compactification then
``attaches'' to a larger moduli space.

We proceed in section \ref{delpezzo} to study del Pezzo surfaces with
automorphisms. We show that the list of possible automorphisms of del
Pezzo surfaces is classified by exactly the same data that classifies
commuting triples of $E_8$. This is naturally suggested by the
existence of F theory duals for the heterotic triples, and confirmed
by direct analysis. This also suggests a possible way of classifying
$E_8$ bundles on higher-dimensional tori using a purely geometric
analysis.  We also recover our heterotic anomaly matching condition
directly from the geometric analysis.

Compactifications with flux are the next topic of discussion. In
section \ref{RRone}, we start by describing type IIA compactifications
on quotient spaces $Z/G$ 
with RR 1-form flux. These arise by reducing M theory on $(Z\times
S^1)/G$ to type IIA on the $S^1$ fiber. These models generalize the
6-dimensional Schwarz--Sen model which is dual to the 6-dimensional CHL
string \cite{sens}. We begin by describing equivariant line bundles on
$T^4$ and the computation of the relevant equivariant cohomology
groups. This approach is naturally suggested from our geometric M
theory starting point. We then proceed to explain in what sense the
1-form flux is actually localized at the singularities of $T^4/G$ by
studying the local holonomies for these bundles. We then generalize
our discussion to the case of singular $K3$ surfaces. This gives us a
technique for finding the group of 1-form fluxes given the sublattice
of vanishing cycles of the singular $K3$ surface.

The description of RR charges and fields in type II string theory
seems to involve K-theory rather than cohomology, at least at zero
string coupling.  In section \ref{secKtheory}, we study torsion RR
1-form and 3-form fluxes on orbifolds from the perspective of
equivariant K-theory. Our analysis is for local singularities of the
form $\C^2/G$. As usual, to preserve supersymmetry, $G$ should lead to
singularities of $ADE$-type. A torsion 1-form RR flux can be measured
by a D0-brane, while a 3-form RR flux can be measured by a
D2-brane. In both cases, the D-brane acquires an additional phase
factor in the string theory path-integral. We describe how this phase
can be computed for a given flux in terms of a reduced eta-invariant
for the virtual bundle representing the flux.
  
The group of RR 1-form fluxes (modulo higher fluxes in a sense
explained in section \ref{secKtheory}) is given by $H^1(G, U(1))$
which agrees with the result from equivariant cohomology. This is
reassuring since we expect to be able to trust a straightforward
analysis of fluxes for type II backgrounds that descend from {\sl
purely} geometric M theory compactifications.  The case of RR 3-form
flux is more interesting: with vanishing 1-form flux, we find that the
group of 3-form fluxes is given by $H^3(G, U(1))$. However, the full
group of 1-form and 3-form fluxes exhibits an unusual additive
structure. The physical interpretation of this effect is that 3-brane
flux can be induced by the presence of 1-brane flux: the 3-brane flux
has a shifted quantization law. It might be possible to verify this
from a dual description, perhaps one involving branes along the lines
of \cite{New}.  This is quite critical because our later results
suggest that it is far from clear that equivariant K-theory is the
right framework even in string theory. For example, from equivariant
K-theory, we find $\Z_{120}$ as the group of RR 3-form fluxes
supported by an $E_8$ singularity. Are all of these fluxes actually
possible, or are some choices inconsistent?
 
In section \ref{alternate}, we present an alternate algebraic method
for computing the desired K-theory quotients. The groups arrived at
via this method confirm the results obtained from the reduced
eta-invariant approach.

In section \ref{Mthreeflux}, we turn to the issue of M theory
compactifications with flux. We are immediately met by the challenge
of not knowing the correct framework in which to study the M theory
3-form. This is a basic problem for smooth compactifications.  In our
case, the problem is only compounded by the fact that our
compactifications involve singular geometries. The only previously
known case is that of a $D_{4+n}$ singularity which comes in two
flavors: a conventional resolvable singularity with space-time gauge
group $SO(8+2n)$, and a partially frozen variety with gauge group
$Sp(n)$ \cite{Landsteiner:1998ei, wittoroid}.  The $D_4$ frozen
singularity appears in the M theory description of $O6^+$ planes. In
section \ref{threeformphysics}, we argue that our new 7-dimensional
components in the string moduli space imply the existence of frozen
variants of $E_6, E_7$ and $E_8$ singularities. Each of these
singularities can support a variety of fluxes with different
associated space-time gauge groups. For example, $E_8$ comes in 5
frozen, or partially frozen, variants.  This result is starkly
different from what we might expect, for example, from equivariant
K-theory.

We propose M theory duals for our new 7-dimensional heterotic models,
and for our new 7-dimensional F theory models. The M theory duals are
on singular $K3$ surfaces with various combinations of frozen $D$ and
$E$ singularities.  We then proceed to argue for the existence of
dualities that map type IIA compactifications on singular spaces with
RR 1-form flux to type IIA compactifications on spaces with completely
different sets of singularities and RR 3-form flux.

In section \ref{threeformequiv}, we turn to the possibility that
3-form flux could be described by equivariant cohomology---perhaps
with additional consistency conditions from equations of motion, or
anomalies.  We describe the computation of the relevant equivariant
cohomology group using $T^4/\Z_2$ as an example. Section
\ref{threeformhol} extends our discussion of 1-form holonomies to
torsion 3-form fluxes. Working under the premise that the physical
choices for 3-form flux form a subset of choices predicted by
equivariant cohomology, we study the global orbifold
$T^4/\widehat{\cal D}_4$ in section \ref{dualCHL}. This orbifold has 2
$D_4$ singularities, and we show that there is a choice of flux with
holonomies localized at those singularities. This is a natural
concrete proposal for the M theory dual of the 7-dimensional CHL
string.

We turn to some puzzles in matching M theory with type IIA in section
\ref{compareiia}.  These puzzles involve the spectrum of 2-branes
computed both in M theory and type IIA.  A generalization of the
Freed--Witten anomaly \cite{Freed:1999vc}\ for D2-branes resolves the
puzzle and leads us to speculate about a generalization of the anomaly
in the context of K-theory. In section \ref{threeformgeom}, we present
some comments on the framework in which the M theory 3-form should be
studied. Using a line of reasoning suggested by anomalies in wrapped
branes, we are actually able to reproduce our list of frozen
singularities.  This is quite exciting, although the arguments are
preliminary, and leave many (interesting) unresolved questions.
The final section concludes with a brief summary of known F theory
compactifications with flux. We find no new models beyond those
previously studied.

As a guide for the reader, we summarize our results on the 
moduli space of 7-dimensional string compactifications in
table~\ref{table:overview}. This includes a listing
of all (known) dual ways of realizing a given component
of the moduli space. 

\begin{table}
\begin{center}
\renewcommand{\arraystretch}{1.5}
\begin{tabular}{|c|c|c|c|} \hline
Heterotic  & Orientifold  & M theory on K3 with frozen &
F theory       \\ 
description & description & singularities of type & compactified on\\ \hline
``standard component'' & $(-^8)$ & smooth $K3$ & $K3 \times S^1$ \\ \hline
$\Z_2$ triple & $(-^6,+^2)$ & $D_{4}\oplus D_4$ & 
 $(K3 \times S^1)/\Z_2$ \\ 
CHL string & & & \\
no vector structure & & & \\ \hline
$\Z_3$ triple & &  $E_6 \oplus E_6 $ & $(K3 \times S^1)/\Z_3$ \\ \hline
$\Z_4$ triple & &  $E_7 \oplus E_7 $ & $(K3 \times S^1)/\Z_4$ \\ \hline
$\Z_5$ triple & &  $E_8 \oplus E_8 $ & $(K3 \times S^1)/\Z_5$ \\ \hline
$\Z_6$ triple & &  $E_8 \oplus E_8 $ & $(K3 \times S^1)/\Z_6$ \\ \hline
& $(-^4,+^4)_1$ & $(D_4)^4 $ & $(T^4 \times S^1)/\Z_2$ \\ \hline
& $(-^4,+^4)_2$ &    & \\ \hline
& &  $(E_6)^3 $ & $(T^4 \times S^1)/\Z_3$  \\ \hline
& &  $D_4 \oplus E_7 \oplus E_7 $ & $(T^4 \times S^1)/\Z_4$  \\ \hline
& &  $D_4 \oplus E_6 \oplus E_8 $ & $(T^4 \times S^1)/\Z_6$  \\ \hline
\end{tabular}
\renewcommand{\arraystretch}{1.0}
\caption{A summary of 7-dimensional string theories with 16
 supercharges.
 } \label{table:overview}
\end{center}
\end{table}

\section{The Heterotic/Type I String on a Torus}
\label{asymorbifold}
\subsection{Gauge bundles on a torus}

Let us begin by reviewing the choice of gauge bundles on tori.  While
we need specific results only for the case of an $E_8$ or
$Spin(32)/\Z_2$ bundle, we shall include some general comments
independent of the choice gauge group. For a more detailed review of
this topic as well as further references, see \cite{arjan}. We want
our gauge fields to have zero curvature. This ensures that when we
turn to string theory, they contribute nothing to the energy. A flat
connection of Yang--Mills theory with gauge group $G$ on $T^n$ is
specified by a set of $n$ commuting elements of $G$, denoted
$\Omega_i$. These Wilson lines, which specify the holonomies around
the $n$ non-trivial cycles of the torus, are not unique. The same
classical vacuum is also described by any other choice $\Omega_i'$
obtained by a global gauge transformation,
$$\Omega_i'= g \, \Omega_i \, g^{-1}. $$ Classifying all flat
connections on $T^n$ with gauge group $G$ therefore amounts to
classifying all sets of commuting elements in $G$ up to simultaneous
conjugation in $G$.

The simplest way to construct a set of commuting elements is as
follows: exponentiating the Cartan subalgebra of $G$ gives a maximal
torus $T_G$, which is an abelian subgroup of $G$. By choosing our
$\Omega_i \in T_G$, we obtain a flat connection on $T^n$. For
particular groups like $G=SU(N)$ or $G=Sp(N)$, all flat connections
are gauge equivalent to a flat connection with holonomies on a maximal
torus, for any $n$.  However, in general the moduli space of flat
connections contains additional components beyond the one containing
the trivial connection. This insight was crucial in resolving some
puzzles about counting vacua in four-dimensional gauge theory
\cite{wittoroid, witten00, bfm, kacsmil, arjansmil, arjanalone,
arjanalone2}.

How do we describe the component of the moduli space containing the
trivial connection?  With all $n$ holonomies on a maximal torus, we
can use a gauge transformation to set the corresponding gauge
potentials $A_i$ to constant elements of the Cartan subalgebra. The
centraliser of this connection---the subgroup of $G$ commuting with
each $\mO_i$---clearly contains the maximal torus as a
subgroup. Therefore the rank of the centraliser of this flat
connection equals the rank of $G$. We can characterize elements of the
Cartan subalgebra by vectors on the space $\R^r$ with $r$ the rank of
$G$. To represent our $n$ holonomies, we can therefore choose $n$
vectors $\mathbf{a}_i$ where we identify vectors that differ by
elements of the coroot lattice. This identification simply corresponds
to quotienting out periodic gauge transformations. The resulting
moduli space is then compact. Lastly, we can conjugate each $\mO_i$
simultaneously by elements of the normalizer of $T_G$. This
corresponds to further quotienting our moduli space by the action of
the Weyl group $\mathcal{W}$ on each vector $\mathbf{a}_i$,
simultaneously. In later applications to string theory, we shall deal
exclusively with simply-laced groups where we can normalize the roots
to have length $\sqrt{2}$. The roots and coroots can then be
identified.  As an example, let us take the familiar case of $SU(N)$
for which the moduli space is $(T^{N-1})^n/\mathcal{W}$.

For other components of the moduli space, we typically have a
reduction of the rank of the centraliser of a flat connection. It is
clear in this case that we cannot simultaneously conjugate all
holonomies into a maximal torus. However, it is possible to gauge
transform to a set where each holonomy $\mO_i$ can be written as the
product of two commuting elements. One element is on a maximal torus
while the second element implements a discrete transformation: either
an outer automorphism, or a Weyl reflection. Let us now consider the
possibilities for various choices of $n$.

\subsubsection{Bundles on $S^1$}
\label{circle}

A flat connection on a circle is specified by a single holonomy
$\Omega$.  The topological types of bundles over $S^1$ are in natural
one-to-one correspondence with $\pi_0(G)$.  If $\mO$ is in a component
$G_c$ of $G$ connected to the identity, then we can always choose a
maximal torus $T_G$ containing $\mO$. The rank of the centraliser of
$\Omega$ then equals the rank of $G$. To find something new, we
require a component of $G$ not connected to the identity. It is clear
that conjugation with $\Omega$ maps $G_c$ to itself. Therefore
$\Omega$ represents an automorphism of $G_c$ and because $\Omega
\notin G_c$, it is an outer automorphism.  In order to realize a
holonomy which acts as an outer automorphism of $G_c$, the gauge group
$G$ must be disconnected.  The gauge group $G$ typically takes the
form $G=G_c \rtimes \G$ where $\G$ is a finite group (acting by outer
automorphisms) and $\rtimes$ denotes semi-direct product.

The outer automorphisms of a compact, simple, connected, and
simply-connected Lie group are in correspondence with the symmetries
of its Dynkin diagram. The only compact, connected, and
simply-connected simple Lie groups with outer automorphisms are
$SU(N)$, $Spin(2N)$ and $E_6$. These outer automorphisms permute the
nodes of the Dynkin diagram.  Thus, gauge theories with gauge groups
$SU(N)\rtimes \Z_2$ for $N>2$, $Spin(8)\rtimes \mathfrak{S}_3$,
$Spin(2N)\rtimes \Z_2$ for $N>4$, and $E_6\rtimes \Z_2$ all admit
non-trivial bundles over $S^1$.

The abelian group $U(1) = SO(2)$ admits an outer automorphism. The
group manifold $U(1)$ is a circle and the outer automorphism acts by
reflection on the circle. It can be represented as complex conjugation
on $U(1)$, or as an element of $O(2)$ with $\det = -1$ when $G_c =
SO(2)$. For the gauge group we take $G=U(1)\rtimes\Z_2=O(2)$.

A group $G$ with a subgroup containing multiple isomorphic factors
gives another example. There are outer automorphisms which permute the
isomorphic factors.

Turning to the cases of interest to us, we note that $Spin(32)/\Z_2$
does not have an outer automorphism, although $Spin(32)$ does. The
group $Spin(32)$ has two isomorphic spin representations that are
interchanged by its outer automorphism.  Only one of these spin
representations is present in $Spin(32)/\Z_2$, and therefore the outer
automorphism of $Spin(32)$ does not descend to a symmetry of
$Spin(32)/\Z_2$.

Although $E_8$ itself does not admit any outer automorphisms, the
product $E_8 \times E_8$ has two isomorphic factors and therefore has
an outer automorphism exchanging the two $E_8$ factors.  Compactifying
$(E_8 \times E_8)\rtimes\Z_2$ gauge theory on a circle with holonomy
interchanging the two $E_8$ factors leads to a theory that has rank
reduced by $8$ because the group elements invariant under the holonomy
must be symmetric in the two $E_8$ factors. This construction is the
one employed in \cite{Chaudhuri95b} in their realisation of the
9-dimensional CHL theory \cite{CHL}.

\subsubsection{Bundles on $T^2$}

A flat connection on $T^2$ is specified by two commuting holonomies.
Let us first dispense with some simple extensions of our prior
discussion.  We can always pick two holonomies, $\mO_i$, on a maximal
torus of $G$.  A second possibility is to have an outer automorphism,
as in our $S^1$ discussion, as one holonomy and a group element left
invariant by this automorphism as a second holonomy.

The next question we should ask is whether gauge bundles can have any
non-trivial topological types on $T^2$. The first obstruction is
measured by an element of $H^1(T^2; \pi_0(G))$ as we discussed in
section \ref{circle}. Let us assume that $G$ is connected so this
obstruction is trivial.  A non-trivial topological type then requires
a non-simply-connected group. For $E_8$, there are therefore no
non-trivial choices. However, for $Spin(32)/\Z_2$ there is a
non-trivial choice. We begin with a general discussion about how this
topological choice comes about. Take $G$ to be connected but pick a
holonomy $\Omega_1$ for one cycle that has a disconnected centraliser
$Z(\Omega_1)$. Elements of the disconnected part of $Z(\Omega_1)$ map
the connected part to itself, and are allowed choices for the second
holonomy $\Omega_2$. For simplicity, take $T^2=S^1\times S^1$.  By
dimensional reduction on the first circle with holonomy $\mO_1$, we
may regard this as a theory with gauge group $Z(\Omega_1)$ on the
remaining $S^1$. Therefore this is to some extent the same as our
previous example. This is the situation that occurs for 't Hooft's
twisted boundary conditions \cite{Hooft}.

The group $G$ should be non-abelian since we require a disconnected
centraliser for $\Omega_1 \in G$. Let us assume that $G$ is simple. A
theorem by Bott, as quoted in \cite{kacsmil, bfm}, states that the
centraliser of any element from a simple and simply-connected group is
connected. Therefore $G$ should be a non-simply-connected group.
Examples of non-simply-connected groups include $SU(n)$, $Sp(n)$,
$Spin(n)$, $E_6$ and $E_7$ quotiented by a non-trivial subgroup $Z$ of
their centers. Let us denote the simply-connected cover of $G$ by
$\tilde{G}$.  The allowed representations of the gauge group are then
restricted to those which represent $Z$ by the identity element. We
can now choose holonomies which commute in $G$ but commute to a
non-trivial element in the kernel of $\tilde{G} \to G$ \cite{sch}. The
obstruction for lifting $G$ bundles to $\tilde{G}$ bundles is measured by a
characteristic class $\widetilde{w}_2 \in H^2(T^2,Z_{\tilde{G}}/Z_G)$,
where $Z_{\tilde{G}}$ and $Z_G$ are the centres of ${\tilde{G}}$ and
$G$, respectively.

More explicitly for the case of $Spin(32)/\Z_2$, there is one choice,
measured by a generalized second Stiefel--Whitney class, which
determines whether the compactification does or does not have ``vector
structure'' \cite{nvs, wittoroid}.  The case of no vector structure
corresponds to taking Wilson lines, $(\mO_1, \mO_2)$, which commute to
the non-trivial element in the kernel of the map $Spin(32) \to
Spin(32)/\Z_2$.  In the component without vector structure, the rank
is reduced by $8$.  This is our only discrete choice on $T^2$.
 
\subsubsection{Bundles on $T^3$}

The simplest way to construct commuting triples is to pick an element
of the maximal torus that commutes with two holonomies constructed
with the methods that we just described.  Again, this is essentially a
dimensional reduction of our prior discussion. However, we shall meet
new possibilities on $T^3$.

For the groups $E_8$ and $Spin(32)/\Z_2$, compactification on $T^3$
introduces no additional topological choice beyond the choice of the
generalized Stiefel--Whitney class in $H^2(T^3, \Z_2)$. Up to
automorphisms of $T^3$, there are two topological types for the
case of $Spin(32)/\Z_2$: Bundles of trivial class which are liftable
to $Spin(32)$, and non-liftable bundles. If we choose coordinates
$(x_1, x_2, x_3)$ for $T^3$, we can always choose these non-trivial
bundles to be unliftable on the $T^2$ parametrized by $(x_1, x_2)$ and
liftable on all other two-tori.  For $E_8$ bundles, there are no
non-trivial topological choices.

Even after fixing this topological choice, there is the possibility of
additional components in the moduli space of flat connections. For
example, in the case with trivial generalized Stiefel--Whitney class,
these additional components consist of connections with three
holonomies, $\mO_i$, which commute but which are not connected by a
path of flat connections to the trivial connection.

Let us again begin by framing our discussion in more general terms,
before turning to the special groups of interest to us. Let $G$ be
simply connected. Pick an element $\Omega_1$ with centraliser
$Z(\Omega_1)$ so that $Z(\Omega_1)$ contains a semisimple part
$Z_{ss}(\Omega_1)$ that is not simply-connected. We can then choose
holonomies $\Omega_2$ and $\Omega_3$ from $Z_{ss}(\Omega_1)$ that obey
twisted boundary conditions: they commute in $Z_{ss}(\Omega_1)$ but
their lifts $\tilde{\Omega}_2$ and $\tilde{\Omega}_3$ to the
simply-connected cover $\tilde{Z}_{ss}(\Omega_1)$ do not commute. In
this way, we achieve rank reduction even in a connected and
simply-connected group.  The groups $Spin(n>7)$ and all exceptional
groups have non-trivial triples of this kind \cite{bfm, kacsmil,
arjansmil, arjanalone, arjanalone2, wittoroid}.

If $G$ is not simply-connected (but still semisimple), it is also
possible in specific cases to choose an element $\Omega_1$ with
centraliser $Z(\Omega_1)$ such that the semisimple part
$Z_{ss}(\Omega_1)$ has a fundamental group strictly larger than the
fundamental group of $G$. Then we can pick elements $\Omega_2$ and
$\Omega_3$ from $Z_{ss}(\Omega_1)$ so that their lifts
$\tilde{\Omega}_2$ and $\tilde{\Omega}_3$ to the simply-connected
cover $\tilde{Z}_{ss}(\Omega_1)$ commute up to an element that is not
contained in the fundamental group of $G$.  Compactifying with
$\Omega_1$, $\Omega_2$ and $\Omega_3$ as holonomies leads to rank
reduced theories, but the rank reduction can be larger than would be
the case for a compactification with twisted boundary conditions on a
two-torus and a third holonomy from the maximal torus.

The various possible flat connections are characterised by two sets of
data: first, the topological type of the bundle measured by the
generalized Stiefel--Whitney class. Second, for a fixed topological
choice, there can be different components of the moduli space.  An
important characteristic of a connection $A$ in some component of the
moduli space is its Chern--Simons invariant which is defined by,
\begin{equation}
\label{CS}
 \int_{T^3} CS(A) = \frac{1}{16 \pi^2 h} \int_{T^3} {\rm tr} \left( A dA + \frac{2}{3} A^3 \right),
\end{equation}
where $h$ is the dual Coxeter number.  The Chern--Simons ($CS$)
invariant is well-defined in $\R/\Z$ and is constant over a connected
component of the moduli space. These invariants, which have been
computed for all simple groups in \cite{bfm}, are typically rational
for non-trivial components\footnote{Those that do not contain the
trivial connection.}. A key result is that these components can be
distinguished by their $CS$ invariant \cite{bfm}.  In fact, fixing the
generalized Stiefel--Whitney class, $CS$ embeds the set of components
of the moduli space of a fixed type into $\Q/\Z$. By the order of a
component $k$, we mean the order of its $CS$ invariant in $\Q/\Z$.

Tables \ref{table:E_8} and \ref{table:SO(32)} summarize the structure
of the moduli space for $E_8$ and $Spin(32)/\Z_2$. In the latter case,
we include both bundles with and without vector structure.  Note that
there are $12$ distinct components for $E_8$ and $6$ for
$Spin(32)/\Z_2$. The Chern--Simons invariants of a component of order
$k$ are of the form $n/k$ with $1\le n \le k$ and $n$ relatively prime
to $k$. There is exactly one component of order $k$ for each such
$n$. For example in the $E_8$ case, there are $2$ components with
$k=6$. We can distinguish these two components by their $CS$
invariants which are $1/6$ and $5/6$ mod $\Z$, respectively.  Denoting
the moduli space of flat connections on $T^3$ of a given topological
type by $\mathcal{M}$ and letting $X$ be a component of $\mathcal{M}$,
we note that \cite{bfm}
$$ \sum_{X\in\mathcal{M} } (\frac{1}{3} \, {\rm dim}(X)+1) = h.$$
Since there are no topological choices for $E_8 \times E_8$, all
solutions are characterized by the $CS$ invariant for each $E_8$
factor. The case of integer $CS$ invariant corresponds to no rank
reduction. For $CS = \hlf$, the rank is reduced by $4$, $CS=
\left(\frac{1}{3},\frac{2}{3}\right)$ give a rank reduction of $6$,
$CS= \left(\frac{1}{4},\frac{3}{4}\right)$ give a rank reduction of
$7$, while $CS= \left(\frac{1}{5},\frac{2}{5},\frac{3}{5},\frac{4}{5},
\frac{1}{6}, \frac{5}{6}\right)$ give a rank reduction of $8$.  For
$E_8 \times E_8$ we therefore have $144$ different combinations with
possible rank reductions of $0,4,6,7,8,10,11,12,13,14,15$ and $16$.
Note that for an $E_8\times E_8$ bundle where both factors have
identical triples, we can further impose the CHL outer automorphism on
one of the holonomies leading to rank reductions of $12,14,15$ and
$16$. The moduli space of $Spin(32)/\Z_2$ is slightly more intricate,
but as we shall see in section \ref{anomaly}, we do not require a
detailed description of the non-trivial components in the moduli space
beyond the usual no vector structure compactification with $CS=0$.
 
\begin{table}
\begin{center}
\begin{tabular}{|c|c|c|c|} \hline
Order of               & Maximal                   &                &    \\
the Component    & Unbroken Gauge Groups     &     Degeneracy & Dimension \\ \hline
  1            &   $E_8$                   &       1        &  24 \\
  2            &    $F_4$, $C_4$                  &       1        &  12 \\
 3            &    $G_2$                   &       2        &   6 \\
 4            &    $ A_1 $                 &       2        &   3 \\
 5            &    $\{e\}$                 &       4        &   0 \\
 6            &    $\{e\}$                 &       2        &   0 \\ \hline
\end{tabular}
\caption{The structure of the moduli space for $E_8$.} \label{table:E_8}
\end{center}
\end{table}

\begin{table}
\begin{center}
\begin{tabular}{|c|c|c|c|c|} \hline
Order of       & Maximal                  &                &    & (No) \\
Component  & Unbroken Gauge Groups    &     Degeneracy & Dimension & Vector Structure\\ 
\hline
  1            &   $ D_{16}$              &       1        & 48 & VS  \\
  2            &   $ B_{12}$              &      1        & 36 & VS  \\ 
  2            &   $ D_n \times C_m, \, n+m=8$    &       2        & 24 & NVS \\   
  4            &   $ B_n \times C_m, \, n+m=5$ &      2        & 15 & NVS \\ \hline
\end{tabular}
\caption{The structure of the moduli space for $Spin(32)/\Z_2$.} \label{table:SO(32)}
\end{center}
\end{table}

\subsubsection{Bundles on $T^4$}
\label{four-torus}
As usual, we can extend our prior discussion to the case of $T^4$ in a
simple way. We add a circle to $T^3$ and choose the holonomy around
the circle to lie in the maximal torus of $G$, and commute with the
other holonomies. However, there are again new possibilities that
cannot be obtained this way. Beyond $T^3$, there is no complete
analysis for the general case so we shall restrict ourselves to
examples which naturally arise in string theory.

Let us begin our discussion with $Spin(32)$. We note that the group
$Spin(32)$ admits holonomies that break the group to a subgroup of the
form $Spin(2N) \times Spin(2N') \times G$, where $N, N' \geq 4$ and
$G$ is some product of $U(n)$-factors, possibly not semisimple, of
rank $16-N-N'$.  In fact the semisimple part of this subgroup is
two-fold connected so we should include a quotient by some $\Z_2$;
however, we will ignore this subtlety since it plays no role in our
current considerations. The point is that \emph{both} $Spin(2N)$ and
$Spin(2N')$ have non-trivial triples. We may therefore construct
$Spin(32)$ holonomies that implement a triple in each subgroup. This
leads to a non-trivial quadruple. It results in a rank reduction of
$8$ which is twice the reduction of a triple. We also note that the
$CS$ invariants defined for any sub-three-torus of $T^4$ are always
integer.

A similar construction can be applied to the case of $Spin(32)/\Z_2$
without vector structure.  Pick a holonomy that breaks the group to a
subgroup $Spin(2N) \times Spin(2N') \times G$, where $N, N' \geq 6$
and $G$ is some (possibly not semisimple) group of rank $16-N-N'$. We
have again ignored the topology of the subgroup. We imposed the
restriction $N,N' \geq 6$ because $Spin(2N \geq 12)$ admits triples
without vector structure that lead to rank reduction beyond the
reduction that follows from no vector structure \cite{bfm,
Keurentjes00}.  Constructing such a triple in each of the factors of
$Spin(2N) \times Spin(2N')$ leads to two possibilities with rank
reduction 14.  One of these has a half-integer $CS$ invariant on a
three-torus, but the other has integer $CS$ on all sub-three-tori. We
shall meet these gauge bundles when we discuss orientifold
constructions in section \ref{orient}.

A new topological possibility that occurs when compactifying
$Spin(32)/\Z_2$ without vector structure on $T^4$ was briefly
described in \cite{wittoroid}. Recall that the class $\widetilde{w}_2$
is an element of $H^2(T^4,\Z_2)=\Z_2^6$. The new possibility
appears when $\widetilde{w}_2$, viewed as an antisymmetric
$4 \times 4$ matrix, has maximal rank. This happens precisely
when $\tilde{w}_2^2\neq 0$. Such bundles have no vector structure
over two complementary two-tori. 
The orientifold realization of this type of
 bundle was discussed in~\cite{wittoroid}.  
 
 Thus for $Spin(32)/\Z_2$ bundles on $T^4$, we encounter 
altogether three new
types of bundles. On the other hand, 
for $E_8$ or $E_8\times E_8$ no new bundles
appear, since neither $E_8$ nor
$E_8 \times E_8$ admit non-trivial quadruples.

\subsubsection{Bundles on $T^5$}
\label{five-torus}

A $T^5$ compactification can be achieved in the usual trivial way: add
a circle to $T^4$ with a holonomy from the maximal torus chosen to
commute with the other holonomies.  For the groups of interest to us,
there are some interesting new possibilities to which we now turn.

The group $Spin(32)$ has an element with centraliser $(Spin(16) \times
Spin(16))/\Z_2$.  The group $Spin(16)$ admits a non-trivial quadruple
\cite{kacsmil}. The argument proceeds along the lines sketched for
$Spin(32)$: the group $Spin(16)$ has elements that have as centraliser
$(Spin(8) \times Spin(8))/\Z_2$. Since $Spin(8)$ is among the groups that
have a non-trivial triple, we construct one in each factor of $Spin(8)
\times Spin(8)$. This results in a non-trivial quadruple of
$Spin(16)$.  Constructing a non-trivial quadruple in each factor of
the $Spin(16) \times Spin(16) \supset Spin(32)$ leads in turn to a
non-trivial quintuple for $Spin(32)$. For this case, the rank
reduction is complete and equals $16$.

On the other hand, the group $E_8$ has an element that has as its
centraliser $Spin(16)/\Z_2$.  Constructing a non-trivial quadruple in
$Spin(16)$ leads to a non-trivial quintuple in $E_8$, with a complete
rank reduction of $8$.  For an $E_8 \times E_8$ bundle on $T^5$, we
can embed a quintuple in one or both $E_8$-factors. This leads to rank
reductions of $8$ and $16$, respectively.

\subsection{Anomaly cancellation}
\label{anomaly}
\subsubsection{A perturbative argument}

Our primary interest is in constructing consistent string
compactifications. We need to know which of the many possible gauge
bundle configurations actually give anomaly-free compactifications.
The issue can be addressed from multiple perspectives. Let us begin
with the familiar heterotic/type I anomaly cancellation
conditions. Let us phrase our discussion in the language of the
heterotic string. Up to irrelevant coefficients, 
the NS-NS $H$-field of the heterotic string
satisfies,
\begin{equation}
\label{defineH}
H = dB + CS(\omega) - CS(A), 
\end{equation} 
where $\omega$ is the spin connection, and $A$ is the connection for
either an $E_8\times E_8$ or $Spin(32)/\Z_2$ bundle. For a toroidal
compactification, $CS(\omega)$ vanishes. For a flat geometry to remain
a solution of string theory, the $H$-field cannot have energy which
would warp the background geometry. This requirement can only be
satisfied if the $H$-field is torsion or trivial in cohomology. On
$T^3$, there is no possibility for torsion so the $H$-field must be
trivial. Integrating equation (\ref{defineH}) over $T^3$ leads to the
requirement that the total $CS$ invariant for $A$ vanish.

Anomaly cancellation therefore rules out all the components in table
\ref{table:SO(32)} with non-vanishing $CS$ invariant. We are left with
two $Spin(32)/\Z_2$ compactifications: the component of order one with
vector structure, and the component of order two without vector
structure but with vanishing $CS$ invariant. Of the $144$ possible
choices of $E_8\times E_8$ gauge bundle, only $12$ survive. These
choices correspond to taking two $E_8$ gauge bundles with opposite
$CS$ invariants.

\subsubsection{An M theory argument}

For the case of the $E_8\times E_8$ string, we can revisit anomaly
cancellation from the perspective of the strong coupling
Ho{\accent20r}ava--Witten  
description of M theory on $S^1/\Z_2$ \cite{HW}.  Let us sketch the
argument without worrying about overall constants that are not needed
for this argument.  In the presence of boundaries at $x^{11}=0$ and at
$x^{11}=\pi$, the definition of the M theory four-form $G$ is
modified.  The component of $G$ with legs on the torus and a leg on
$x^{11}$ satisfies \cite{HW},
\begin{equation}
\label{hwanomaly}
G_{11 x y z} \sim \delta(x^{11}) CS(A_1) +\delta(x^{11}-\pi ) CS(A_2)
+ \ldots,
\end{equation} 
where $A_1$ and $A_2$ are connections for the $E_8\times E_8$ bundle.
The terms omitted involve the $C$-field which is constant on
$T^3$ (see section \ref{strongcouplingtriple}) . For a
 flat geometry like
$T^3$, we require that $G/2\pi$ be an integral
 cohomology class
\cite{fluxquant}. Integrating eq.~(\ref{hwanomaly})
 over $T^3 \times S^1/\Z_2$
then implies that the total $CS$ invariant
 must cancel between the two $E_8$
bundles. This is just the strong
 coupling version of perturbative anomaly
cancellation.

\subsection{Gauge bundles in string theory}

We now turn to a detailed discussion of toroidal compactifications of
the heterotic string. While our discussion is in the context of the
weakly coupled string, supersymmetry should guarantee that results
about moduli spaces remain uncorrected at strong coupling.  Since the
data specifying perturbative type I compactifications is identical to
the data specifying $Spin(32)/\Z_2$ heterotic compactifications, our
results also apply to the type I string.

{}From our prior discussion, we saw that all gauge bundles on a torus 
can be characterised by commuting holonomies $\Omega_i$ which we can write in the
form, 
\begin{equation} \label{holsplit}
\Omega_i = \exp \left( 2 \pi i \mathbf{a}_i \right) \ \Theta_i,
\end{equation}
where $\mathbf{a}_i$ an element of the Cartan subalgebra.  The second
factor $\Theta_i$ implements a discrete transformation (but may be set
to the identity on the group). For the decomposition of the holonomy
given in equation (\ref{holsplit}) to be unambiguous, we demand that
the two factors commute with each other. The $\Theta_i$ implement
automorphisms of the group lattice. These can be either inner
automorphisms which constitute the Weyl group, or outer
automorphisms. Let us denote the automorphism implemented by
$\Theta_i$ by $\theta_i$, and note that our requirement of
commutativity allows us to choose,
\begin{equation}
\Theta_i^{-1} \, \exp \left( 2 \pi i \mathbf{a}_i \right) \, \Theta_i
\ = \ \exp \left( 2 \pi i \theta_i(\mathbf{a}_i )\right) \
\Leftrightarrow \ \mathbf{a}_i = \theta_i(\mathbf{a}_i).
\end{equation} 
In string theory, this decomposition of the holonomy into a component
on the maximal torus and a discrete part is convenient since the two
factors are treated differently in the world-sheet conformal field
theory.

The factor representing the maximal torus contribution can studied
within the usual framework of Narain compactification \cite{Narain,
Narain87a}. The discrete factor $\Theta_i$ can be implemented by the
asymmetric orbifold construction \cite{Narain87b, Narain91}. Let us
first turn to Narain compactifications, postponing the asymmetric
orbifold discussion until later in this section.

\subsubsection{Holonomy in string theory I: Narain compactification} 

For simplicity, we will consider the heterotic string theory on a
rectangular torus. We therefore set the metric on the torus $g_{ij} =
\delta_{ij}$, and display the radii $R_i$ explicitly. The heterotic
NS-NS two-form field $B$ will not enter our considerations so we will
set it to zero. The Regge slope $\alpha'$ will eventually enter our
discussion so we shall keep it explicit. With these conventions, the
momenta of the heterotic string are denoted by:
\begin{eqnarray} 
{\mathbf{k}} & = & ({\mathbf{q}} + \sum_{i} w_i
{\mathbf{a}}_i)\sqrt{\frac{2}{\alpha'}}  \label{hmom1}, \\  k_{iL,R} & =
& \frac{n_i - {\mathbf{q} \cdot \mathbf{a}}_i - \sum_j \frac{w_j}{2}
\mathbf{a}_i \cdot \mathbf{a}_j}{R_i} \pm \frac{w_i R_i}{\alpha'}.
\label{hmom2}   
\end{eqnarray} 
In these formulae, no summation is implied unless explicitly stated.
The vector $\mathbf{q}$ takes values on the lattice $\Gamma_8 \oplus
\Gamma_8$ or $\Gamma_{16}$ for the $E_8 \times E_8$ or the
$Spin(32)/\Z_2$ heterotic string, respectively. The $n_i$ and $w_i$
are integers corresponding to the momentum and winding numbers,
respectively. When rescaled by $\sqrt{\alpha'/2}$, these momenta are
dimensionless and take values on an even self-dual lattice
$\Gamma_{16+d,d}$.

Restricting to states with $w_i$ equal to zero, the spectrum exhibited
in eqs (\ref{hmom1}) and (\ref{hmom2}) is that of compactified gauge
theory with holonomies parametrised by $\mathbf{a}_i$. We may take the
$\mathbf{a}_i$ used for a specific compactification of Yang--Mills
theory as a starting point for discussing a similar compactification
of heterotic string theory. However, the string theory spectrum is
richer because states with non-zero winding have to be added to the
spectrum.

In the decomposition of eqs (\ref{hmom1}) and (\ref{hmom2}), the
vectors $\mathbf{k}$ correspond to quantum numbers reflecting the
group lattice. In gauge and string theory, the group is broken to
subgroups by expectation values for the Wilson lines
$\mathbf{a}_i$. In gauge theory, the weight lattices for these
subgroups correspond to sublattices of the group lattice---an
observation which is crucial to the analysis of \cite{kacsmil, bfm}.
This essential point however is not true for string theory!  Winding
heterotic strings have bosons on their world-sheet that are charged
with respect to the holonomies. Therefore the momenta of these bosons
receive corrections reflected in (\ref{hmom1}). This implies the
existence of representations of the gauge group that would not be
present in pure gauge theory, where winding modes are absent. In
particular, the topology of the unbroken subgroup will be different
from the topology that would be deduced from an analysis of the low
energy gauge theory. With respect to the use of the phrase ``low
energy theory,'' note that the extra representations can be
arbitrarily heavy because they arise from strings that may wrap
arbitrarily large circles.

Which compactifications feel this difference between gauge and string
theory?  Those compactifications with holonomies implementing outer
automorphisms are not affected. By construction all $\mathbf{a}_i$
should commute with the outer automorphism.  This implies that
although there will be extra representations, these representations
are invariant under the outer automorphism and therefore cannot
obstruct the compactification. In particular, the construction of the
CHL string in \cite{Chaudhuri95b} completely parallels the
construction in gauge theory described earlier.

For compactifications with twisted boundary conditions, at least one
of the $\mathbf{a}_i$ is not invariant under the action implemented by
one of the $\Theta_j$ \cite{sch} (with $i \neq j$). Instead, we have
\begin{equation} \label{thetashift}
\theta_j (\mathbf{a}_i) = \mathbf{a}_i + \mathbf{z}_{ij} \quad
\Leftrightarrow  \quad \Theta_j^{-1} \, \exp \left( 2 \pi i
\mathbf{a}_i \right) \, \Theta_j \, \exp \left( -2 \pi i
\mathbf{a}_i \right) = \exp \left( 2 \pi i \mathbf{z}_{ij} \right).
\end{equation} 
This equation implies that $\mathbf{z}_{ij}$ is a vector on the
(co)weight\footnote{Actually it is the coweight lattice \cite{sch},
but since we will discuss string theory with simply-laced groups only,
we can identify the coweight lattice with the weight lattice.}
lattice of the original group, since the commutator on the left hand
side should be equal to the identity. The group element $\Theta_j$
implements a length preserving (orthogonal) automorphism, and hence
$\mathbf{a}_i$ has the same length as $\theta_j(\mathbf{a}_i )$.
Putting these facts together, we see that the insertion of winding
states not only leads to the introduction of states with group quantum
numbers $\mathbf{a}_i$, but also automatically generates images for
these states. Therefore $\Theta_j$ is also a symmetry of the string
theory.  Although the explicit representation content of string and
gauge theory are distinct, the constructions are completely parallel
so the arguments used in \cite{wittoroid} and \cite{Lerche98} are not
affected.

The construction of triples \emph{is} affected by the extra winding
states.  Recall that to construct a non-trivial triple, we have to
turn on holonomies that leave a non-simply-connected subgroup
unbroken, or a subgroup with fundamental group larger than the
fundamental group of the original group. The topology of the unbroken
subgroup can be deduced from the representation content of the theory,
and this is modified by the presence of winding states. We will return
to this point momentarily.

For constructions of quadruples and quintuples, it is hard to make a
general statement. The construction of quadruples and quintuples may
be viewed as an inductive process, where triples are constructed in an
intermediate step. These triple constructions may be restricted by
winding states, but one has to check the topology of the gauge
(sub)group case by case.

\subsubsection{The topology of subgroups in string theory} 

We have seen that some gauge theory vacuum configurations cannot be
reproduced in string theory. We will now investigate which
configurations remain in $E_8 \times E_8$ and $Spin(32)/\Z_2$ string
theory for the specific case of triples. This will provide an
alternate derivation of the constraints obtained in section
\ref{anomaly}.  Recall that we should look for group elements, to be
used as holonomies, with a non-simply-connected centraliser.

Our analysis is based on theorem 1 of \cite{kacsmil}. Equivalent
results can be found in \cite{bfm}. This theorem states that any
element of a simple group can be conjugated into the form,
\begin{equation} \label{wl}  
\exp \left( 2 \pi i \sum_{j=1}^r s_j \omega_j \right),
\end{equation} 
where $r$ is the rank of the group, and the $\omega_j$ are the
fundamental coweights of the group. The $s_j$ are a set of
non-negative numbers satisfying
$$\sum_{j=0} s_j g_j = 1$$ with $g_j$ the root integers. This last
relation determines the number $s_0$.  The theorem further states that
the centraliser of this element is obtained by erasing all nodes $i$
for which $s_i \neq 0$ from the extended Dynkin diagram, and adding
$U(1)$ factors to complete the rank of the group. The fundamental
group $\pi_1$ of the centraliser contains $\Z$ factors for the added
$U(1)$ factors.  In addition, there is a $\Z_n$ where $n$ is the
greatest common divisor (gcd) of the coroot integers of the erased
roots.

For the non-simple group $E_8 \times E_8$ any element can be
conjugated to an element that is a product of two elements of the form
(\ref{wl}). Since the group is simply-laced, we will drop the
distinction between root and coroot, weight and coweight. Let us first
consider one of the $E_8$-factors, setting the group element for the
other factor to the identity.

\begin{figure}[ht] 
\begin{center}
\includegraphics[height=2cm]{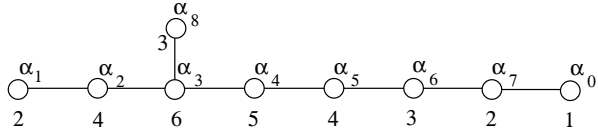}
\caption{The extended Dynkin diagram of $E_{8}$. The integers are the
(co)root integers associated to the respective nodes.} \end{center}
\end{figure}  
\noindent 
If we desire that the centraliser of an element contains an $m$-fold
connected factor with $m \neq 1$, we cannot erase the extended root
$\alpha_0$ which has root integer $1$. Therefore, we have to set $s_0
= 0$. The extended root $\alpha_0$ will now survive as a root of the
subgroup.  The simple factor of which $\alpha_0$ is a root is either
$SU(n)$ with $2 \leq n \leq 9$ or $Spin(16)$.

In heterotic string theory, there are winding states on the weight
lattice, 
$$\sum_{j=1}^8 s_j \omega_j + {\rm roots},$$ 
of the unbroken gauge group. We easily find that,
\begin{equation}  
<\sum_{j=1}^8 s_j \omega_j, \alpha_0 > =
s_0 -1 = -1,
\end{equation}
and therefore $\sum_{j=1}^8 s_j \omega_j$ projected onto the subgroup
containing $\alpha_0$ is minus the weight corresponding to the simple
root $\alpha_0$ in the unbroken gauge group. Therefore, there is at
least one state, with winding number $1$, which transforms in the
$\mathbf{\bar{n}}$ irreducible representation (the anti-fundamental
representation) of an $SU(n)$ factor, or the $\mathbf{16}$ (the vector
representation) of $Spin(16)$. This state is a singlet with respect to
other simple factors in the centraliser because $<\sum_{j=1}^8 s_j
\omega_j, \alpha_i > = s_i=0 $ when $\alpha_i$ is a root of a
surviving subgroup.  Note that this then
implies that the relevant state transforms in a
\emph{simply-connected} representation, unless it transforms in the
$\mathbf{16}$ of $Spin(16)$. In this single exceptional case, the
vector lattice of $Spin(16)$ has to be added to a group lattice that
already contains a spin weight lattice and the root lattice. Again,
the result is the lattice of a simply-connected group.

The conclusion then is that in $E_8 \times E_8$ string theory the
holonomies that are trivial in one $E_8$ factor do not give
non-simply-connected subgroups at all. Closer inspection shows that
the only way to get non-simply-connected subgroups in string theory,
is to use holonomies that in gauge theory would break \emph{each}
$E_8$ factor to a group containing a semisimple, non-simply-connected
factor. In gauge theory, this would result in a semisimple part with
fundamental group $\Z_{n_1} \times \Z_{n_2}$. An analysis of the group
elements that give rise to such a centraliser shows that in string
theory, the semisimple part has fundamental group $\Z_n$ with $n =
\gcd (n_1, n_2)$.

It is therefore possible to construct triples in the $E_8 \times E_8$
theory, but of the 144 components of the $E_8 \times E_8$ gauge
theory only a set of 12 ``diagonal'' constructions can be realized in
string theory. This is in complete accord with our earlier anomaly
cancellation results.  A similar analysis can be performed for the
$Spin(32)/\Z_2$ string. As the techniques involved are the same as for
the $E_8 \times E_8$ heterotic string, we will give fewer details.

\begin{figure}[ht] 
\begin{center}
\includegraphics[height=2cm]{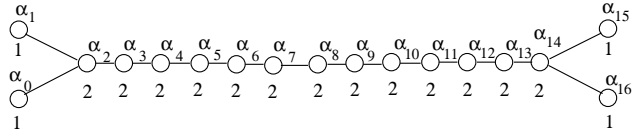}
\caption{The extended Dynkin diagram of $D_{16}$ ($Spin(32)$). The
integers are the (co)root integers associated to the respective nodes.}
\end{center} 
\end{figure}   

As remarked before, the standard compactifications with and without
vector structure are not obstructed in any way. Let us first consider
the gauge theory triple in $Spin(32)$ with vector structure. It is not
hard to show that it is impossible to construct this triple in string
theory because there is no element that gives a centraliser with
non-simply-connected semisimple part in $Spin(32)$. There are
elements that have a centraliser with non-simply-connected semisimple
part in $Spin(32)/\Z_2$, but this is because the group itself is not
simply-connected, and therefore results in compactifications without
vector structure rather than a triple with vector structure.
 
Moving to compactifications without vector structure, let us first
note that $Spin(32)/\Z_2$ has a non-trivial center which is isomorphic
to $\Z_2$. We need two holonomies to encode the absence of vector
structure. The third holonomy has to commute with the other two, but
is otherwise unrestricted. In particular, if $\Omega$ is an allowed
choice, then so is $z \Omega$, $z$ being the non-trivial centre
element in the centre of $Spin(32)/\Z_2$. These two choices are
represented as points in two disconnected components which have an
identical structure.

The element $z$ is represented by the identity in the vector
representation of $Spin(32)$. In particular, the two components
mentioned above cannot be distinguished by their holo\-nomies in
$SO(32)$, and have an identical orientifold description
\cite{Keurentjes00}. This degeneracy should not persist in a
consistent string theory, and indeed it does not. As an example, set
$\Omega$ to the identity. Then $\Omega$ has $Spin(32)/\Z_2$ as its
centraliser. On the other hand, for $\Omega = z = \exp \left( 2 \pi i
\mathbf{a}_i \right)$, with $\mathbf{a}_i$ on the vector weight
lattice, the centraliser in gauge theory would also be
$Spin(32)/\Z_2$; but in string theory it turns out to be\footnote{We
are ignoring the Kaluza--Klein gauge bosons here.}  $Spin(32)$---a
simply connected group! It therefore destroys the possibility for
absence of vector structure. By continuation, we see that the
degeneracy is completely lifted, and only one of the degenerate
components survives.

For gauge theory on $T^3$ with $Spin(32)/\Z_2$ gauge group, and
absence of vector structure, two more components exist. One may
construct holonomies parametrising these components by the methods of
\cite{bfm} as mentioned earlier, or with an alternative approach
\cite{Keurentjes00}.  Attempting to construct these holonomies in
heterotic string theory does not lead to flat bundles. These choices
therefore are not realized in toroidally compactified string theory.

\subsubsection{Holonomy in string theory II: asymmetric orbifolds} 
\label{asymm}

Having dealt with the intricacies of Narain compactification, we
should now implement the discrete transformations $\Theta_j$ appearing
in equation (\ref{holsplit}). This can be done by means of the
asymmetric orbifold construction \cite{Narain87b, Narain91}.  We will
briefly review the general formalism and then apply it to our problem.

Essentially by definition, an asymmetric orbifold uses the fact that
the left and right-moving degrees of freedom on the world-sheet are
largely independent. The orbifold group can therefore have a different
action on the left and right-movers. For heterotic strings, where left
and right-movers live on different spaces, this possibility is quite
natural.  Let us use $P_L$ ($P_R$) to denote the left-moving
(right-moving) momenta of the heterotic string. The group elements $g$
of the orbifold group act separately on the left and right-moving
momenta since mixing left and right-movers typically leads to
inconsistencies. The action of $g$ consists of a combination of a
rotation $\theta_L$ and a translation $a_L$ acting on the left-moving
sector. Similarly, a rotation $\theta_R$ and translation $a_R$ acting
on the right-movers. The action of $g$ on states in the Hilbert space
takes the form,
\begin{equation}
g|P_L,P_R\rangle = e^{2\pi i (P_L\cdot a_L - P_R \cdot a_R)}
 |\theta_L P_L ,\theta_R P_R\rangle.
\end{equation}                  
The orbifold construction leads to untwisted and twisted sectors in
the theory.  In the untwisted sector, describing states invariant
under the orbifold action, we encounter (in the partition function) a
sum over a lattice $I$ which is the sublattice of $\Gamma_{16+d,d}$
invariant under rotations by $\theta$. In the twisted sector, we find
a lattice $I^{\ast} + a^{\ast}$, where $I^{\ast}$ is the lattice dual
to $I$ and $a^{\ast}$ is the orthogonal projection of $a = (a_L, a_R)$
onto $I$ \cite{Narain87b}.

The left- and right-moving sectors of the closed string are not
completely independent. The constraint of level matching connects both
sectors.  This constraint leads to consistency conditions on the
asymmetric orbifold. Let the group element $g$ have finite order
$n$. The eigenvalues of $\theta_L$ are then of the form $\exp
\left(2\pi i r_i/n \right)$, $i=1,\ldots,19$, while $\exp \left( 2\pi
i s_i/n \right)$, $i=1,2,3$ are the eigenvalues of $\theta_R$. The
consistency conditions for $n$ odd are:
\begin{equation} \label{codd} 
\sum_i r_i^2 = (na^{\ast})^2 \, \mod n.
\end{equation} 
For even $n$, this condition is replaced by a more stringent one. 
There are supplementary conditions, 
\bea
\sum_i r_i^2  & = & (na^{\ast})^2  \mod 2n \label{ceven1}, \\
\sum_i s_i = 0 \mod 2, &\quad & v \, \theta_B^{n/2} \, v = 0 \mod 2. 
\label{ceven2}
\eea
The last condition should hold for any $v \in \Gamma_{16+d,d}$, where
$\theta_B$ is a block diagonal matrix with $\theta_L$ and
$\theta_R$ on the diagonal. 

In our applications, the asymmetric orbifold construction will be used
to implement outer automorphisms, or Weyl reflections on the gauge
group. Since the gauge group comes from left-moving excitations, we
set $\theta_R = 1$ and drop the subscript on $\theta_L = \theta$.
This conforms with notation used in previous sections. Notice that the
first condition in (\ref{ceven2}) is trivialized.

The shifts $a_{L,R}$ will be interpreted as physical translations. We
therefore take a minimal lightlike vector in a $\Gamma_{1,1}$
sublattice, divide it by $n$, and identify the shifts $(a_L,a_R)$ with
the components of that resulting vector. There is only one ambiguity
in this prescription: there is the possibility of a relative sign
between the components $a_L$ and $a_R$ (there is another overall sign
corresponding to a parity transformation). The difference in sign
comes from the choice of fractionalizing either the winding numbers,
or fractionalizing the momenta. Both choices are related by a
T-duality.  We shall return to this point later. In the following
discussion, we fractionalise the winding numbers because this has an
obvious space-time interpretation.

After quotienting string theory by $g$, we obtain a theory with a
clear geometric interpretation. Traversing the cycle on the spatial
torus in the direction of $(a_L, a_R)$ gives a holonomy implementing a
Weyl reflection or an outer automorphism (see also \cite{Chaudhuri95b,
Lerche98}). We have now gathered all the elements needed to construct
the orbifolds.

\subsubsection{Triples in string theory}

In this section, we present an analysis of some non-trivial
heterotic compactifications on a $3$-torus. For
non-trivial compactifications on lower-dimensional tori, we refer the
reader to \cite{Chaudhuri95b, Lerche98}.

Since we earlier ruled out the existence of triples in $Spin(32)/\Z_2$
string theory, we deal exclusively with the $E_8 \times E_8$ theory in
this section. Our previous analysis combined with results in $E_8$
gauge theory \cite{arjan, arjanalone, arjanalone2, kacsmil, bfm} lead
us to expect non-trivial triples for which one holonomy implements a
Weyl reflection generating a cyclic group $\Z_m$ with $m=2,3,4,5$ or
$6$.

In this section, we will construct asymmetric orbifolds for special
choices of holonomy. The extension to the general case will be delayed
until section \ref{moduli}. Our choice will be to embed the holonomies
for the $\Z_m$-triples in ``minimal'' subgroups of $E_8$.  These are
the smallest simply-laced subgroups that contain a $\Z_m$-triple. This
choice can be motivated along the lines sketched in
\cite{kacsmil}. The maximal torus of the group has a subtorus $T$ that
commutes with the triple. The centraliser of $T$ is a product of $T$
with a semisimple group $C$.  This semisimple group $C$ is the
``minimal'' subgroup which we require.  For $E_8$ gauge theory, $C=
D_4, E_6, E_7, E_8$ or $E_8$ for the $\Z_m$-triple with $m=2,3,4,5$ or
$6$, respectively. For $E_8 \times E_8$ string theory, we find that $C
= D_4 \times D_4$, etc. The subtorus $T$ corresponds to surviving
moduli, so we can interpret the group $C$ as representing eliminated
moduli. Our choice is dictated by the desire to make $C$ manifest in
the construction. The entries from the list of possible $C$'s will
reappear later in our paper.

The holonomies are essentially fixed by the decision to embed them in
a ``minimal'' subgroup. The only remaining freedom corresponds to
global gauge transformations, or equivalently, lattice symmetries of
the heterotic string.  Triples embedded in these subgroups have a
number of convenient properties. One is that all three holonomies are
conjugate to each other \cite{bfm}, which implies that they have the
same set of eigenvalues in every representation of the gauge
group. Further, from the minimality property, it follows that these
eigenvalues are of the form $\exp \left( 2 \pi i n/m \right)$ where $n
\in \Z$ \cite{arjanalone, arjanalone2}. This is convenient for
checking the asymmetric orbifold consistency conditions (\ref{codd}),
(\ref{ceven1}) and (\ref{ceven2}). Additional properties will emerge
in the construction.

Let us now specify values for some of the quantities appearing in the
formulae for the momenta, (\ref{hmom1}) and (\ref{hmom2}). We work on
a $3$-torus so $i=1,2,3$. We will turn on holonomies in the $1$ and
$2$-directions, and use direction $3$ for the shift accompanying the
orbifold projection. The holonomy in this direction, ${\mathbf{a}}_3$,
will be set to zero for the moment.  In equation (\ref{hmom2}), the
inner products $\mathbf{a}_i \cdot \mathbf{a}_j$ appear. Since the
holonomies at the relevant point in moduli space are conjugate to each
other, we see that
$${\mathbf{a}}_1^2= {\mathbf{a}}_2^2. $$ Let us introduce the notation
${\mathbf{a}}_i = (\tilde{a}_i^{I}, {\tilde{a}}_i^{II})$ to display
the Wilson lines in the `first' (I), and `second' (II) $E_8$
factor. It is convenient to set ${\mathbf{a}}_1 =
(\tilde{a}_1,\tilde{a}_1)$ and ${\mathbf{a}}_2 =
(\tilde{a}_2,-\tilde{a}_2)$. This eliminates the inner product
${\mathbf{a}}_1 \cdot {\mathbf{a}}_2$ from our formulae, leaving only
diagonal terms in the spatial momenta. We will use an orbifold
projection that is symmetric in both $E_8$ factors. Notice that in
this way, we implement the prescription that the contributions of each
$E_8$ factor to the Chern--Simons invariant cancel each other. There
are other ways of implementing this constraint; for example, by
choosing ${\mathbf{a}}_1$ and ${\mathbf{a}}_2$ symmetric in both
factors, and choosing opposite orbifold projections in the two
$E_8$'s. This would leave us with off-diagonal terms in the momenta,
and we consider this less convenient. Nevertheless, it should provide
equivalent results.

The value of ${\mathbf{a}}_1^2 = {\mathbf{a}}_2^2$ can be found in
various ways. In the setup we have chosen, the holonomy parametrized
by ${\mathbf{a}}_1$ eliminates only one node from each $E_8$ extended
Dynkin diagram, and from the discussion around eq.~(\ref{wl}), it
follows that ${\mathbf{a}}_1$ is of the form $(\omega_j h_j^{-1}, \,
\omega_j h_j^{-1})$. Here $\omega_j$ is the (co)weight and $h_j$ the
(co)root integer associated to the node.\footnote{For $m<5$, there
seem to be more options but only one corresponds to a minimal triple.}
We now easily find ${\mathbf{a}}_i^2$ by noting that the weight can be
expanded in the simple roots,
$$\omega_i =\sum_k p^k_i \alpha_k.$$ 
It is then trivial to show that
${\mathbf{a}}_i^2 =2 p^i_i / (h_i)^2$ (no summation implied), where $p^i_i$
is a diagonal element of the inverse Cartan matrix. For the cases
under consideration, we find that ${\mathbf{a}}_i^2 = 2 (m-1)/m$.\footnote{It can be
proven that this is the minimal value that ${\mathbf{a}}_i^2$ can have
for an $m$-triple. This provides another invariant way of characterising
these holonomies.} 

Combining these conventions and results, we find the momenta for the
compactified heterotic string \emph{before the orbifold projection}:
\begin{eqnarray} 
{\mathbf{k}} & = & ({\mathbf{q}} + \sum_{i=1,2} w_i
{\mathbf{a}}_i) \sqrt{\frac{2}{\alpha'}},  \label{mom1} \\ 
k_{iL,R} & = & \frac{m n_i - m {\mathbf{q} \cdot \mathbf{a}}_i - w_i
(m-1)}{mR_i} \pm \frac{w_i R_i}{\alpha'}, \qquad i=1,2  \label{mom2}\\
k_{3L,R} & = &  \frac{n_3}{R_3} \pm \frac{w_3 R_3}{\alpha'}.
\label{mom3}  
\end{eqnarray} 
Because of the eigenvalues of the holonomies, ${\mathbf{q}} \cdot
{\mathbf{a}}_i$ is always a multiple of $1/m$. Therefore, the
combination $m n_i - m{ \mathbf{q} \cdot\mathbf{a}}_i - w_i (m-1)$ is
always an integer, and actually can take any integer value.

We have now arrived at the point where we want to perform the orbifold
construction. From the gauge theory interpretation of the theory, we
should be confident that orbifolding will lead to a consistent
theory. Nevertheless, we shall verify that the theory given by
equations (\ref{mom1}), (\ref{mom2}) and (\ref{mom3}) has the right
symmetries, and that the orbifold operation obeys the consistency
conditions (\ref{codd}), (\ref{ceven1}) and (\ref{ceven2}).

The orbifolding operation consists of a shift $a$, and a
transformation $\theta$ acting on the gauge part of the lattice. The
transformation $\theta$ is an element of the Weyl group of $E_8 \times
E_8$. Since the rank of $E_8 \times E_8$ is $16$, the Weyl group is a
discrete subgroup of the orthogonal group $O(16)$. As discussed
before, the requirement of commuting holonomies does not necessarily
mean that $\theta({\mathbf{a}}_i)$ is equal to ${\mathbf{a}}_i$, but
rather implies the weaker condition (\ref{thetashift}) where
$\mathbf{z}_{ij}$ is some lattice vector. There is some ambiguity in
the choice of $\mathbf{z}_{ij}$, but in the cases of non-trivial
commuting triples, the lattice vector cannot be set equal zero. The
choice ${\mathbf{z}_{ij}}=0$ corresponds to a trivial triple, which
should be equivalent to a conventional Narain compactified theory.

To see that the lattice has the right symmetry, we construct the image
of a vector with labels $({\mathbf{q}}, n_i, w_i, n_3, w_3)$. There
should exist a vector labelled by $({\mathbf{q}}', n'_i, w'_i, n'_3,
w'_3)$ with ${\mathbf{q}}' = \theta({\mathbf{q}} + w_1 {\mathbf{a}}_1
+ w_2 {\mathbf{a}}_2)$. We expect existence for generic radii of the
spacial torus which implies $w_i = w'_i$, $w_3=w_3'$ and $n_3 =
n'_3$. We are therefore led to the equations,
\begin{eqnarray} {\mathbf{q}}' & = &
\theta({\mathbf{q}}) + w_1 (\theta({\mathbf{a}}_1) - {\mathbf{a}}_1)+ w_2
(\theta({\mathbf{a}}_2) - {\mathbf{a}}_2)  \label{pos1}, \\ n'_i - {\mathbf{q}}'
\cdot {\mathbf{a}}_i & = & n_i - {\mathbf{q}} \cdot {\mathbf{a}}_i
\qquad i=1,2. \label{pos2}
\end{eqnarray}
Equation (\ref{pos1}) is consistent by construction since both the
left and right hand side contain lattice vectors only. We still have
to verify that $({\mathbf{q}}'- {\mathbf{q}}) \cdot {\mathbf{a}}_i$ is
an integer for $i=1,2$. We will show that both $(\theta({\mathbf{q}})-
{\mathbf{q}})\cdot {{\mathbf{a}}_i}$ and $(\theta({\mathbf{a}}_i)-
{\mathbf{a}}_i)\cdot {\mathbf{a}}_j$ are integers, and hence
(\ref{pos2}) always has a solution.

The quantity $(\theta({\mathbf{a}}_i)- {\mathbf{a}}_i) \cdot
{\mathbf{a}}_j$ is actually zero for $i \neq j$ because of the
specific choice of ${\mathbf{a}}_1, {\mathbf{a}}_2$, and because
$\theta$ is symmetric in both $E_8$ factors. We remarked previously
that $\theta({\mathbf{a}}_i)- {\mathbf{a}}_i = {\mathbf{z}_i}$ for
some lattice vector $\mathbf{z}_i$. Then $(\theta({\mathbf{a}}_i))^2 =
{\mathbf a}_i^2 = {\mathbf z}_i^2 + 2 {\mathbf{a}}_i \cdot
{\mathbf{z}}_i + {\mathbf{a}}_i^2$, where use was made of the fact
that $\theta \in O(16)$. Since $\mathbf{z}_i$ is on an even lattice,
it immediately follows that ${\mathbf{a}}_i \cdot {\mathbf z}_i$ is an
integer. Finally, rewriting $(\theta({\mathbf{q}})- \mathbf{q})\cdot
{\mathbf{a}_i}$ as $(\theta^{-1}({\mathbf{a_i}})- \mathbf{a_i})\cdot
{\mathbf{q}}$, we notice that this is an inner product between two
lattice vectors, and hence also integer. Therefore an image point
always exists.

\begin{table}[ht]
\begin{center}
\renewcommand{\arraystretch}{1.5}
\begin{tabular}{|c|c|c|c|c|c|c|c|c|c|c|c|c|c|c|c|} \hline
$m$ & 2 & \multicolumn{2}{c|}{3} & \multicolumn{3}{c|}{4} &
\multicolumn{4}{c|}{5}& \multicolumn{5}{c|}{6}\\ \hline
$r_i$ & 1 & 1 & 2 & 1 & 2 & 3 & 1 & 2 & 3 & 4 & 1 & 2 & 3 & 4 & 5 \\
\hline multiplicity  & 8 & 6 & 6 & 4 & 6 & 4 & 4 & 4 & 4 & 4& 2 & 4 &
4 & 4 & 2 \\
\hline 
\end{tabular}
\renewcommand{\arraystretch}{1.0}
\caption{Eigenvalues $r_i \neq 0$ for the $\Z_m$ orbifolds}
\label{table:r_i} \end{center}
\end{table}
For the orbifold consistency conditions (\ref{codd}) and
(\ref{ceven1}), we need the eigenvalues $r_i$ of $\theta$. These can
be obtained from group theory\cite{arjanalone, arjanalone2, kacsmil,
bfm} and are listed in table \ref{table:r_i}.  We set the shift $a$ to
a light-like vector so $(a^{\ast})^2$ is always zero.  In all cases,
the orbifold consistency conditions are satisfied. For future use, we
remark that the eigenvalues and multiplicities appearing in table
\ref{table:r_i} are identical to the eigenvalues and multiplicities of
automorphisms K3 \cite{Nikulin79}.

The last condition that we need to check is the second condition of
eq.~(\ref{ceven2}). It can be shown that this leads to the same
condition for $n=2,4,6$ (compare with table \ref{table:r_i}).  In all
cases $\theta_B^{n/2}$ is a matrix that reflects $8$ orthogonal roots.
It is then easily checked that (\ref{ceven2}) is satisfied.

With the table of eigenvalues, it is also easy to calculate the
zero-point energies for the twisted sector(s). An eigenvalue $r_i$
contributes,
\begin{equation}
\frac{1}{48} - \frac{1}{16} \left( 2 \, \frac{r_i}{m} -1 \right)^2
\end{equation}
to the zero-point energy. A periodic boson has $r_i=0$, and hence
contributes $-\frac{1}{24}$. Summing all contributions leads to a
remarkably simple result: the zero point energies are $-1/m$ for the
twisted sector(s) of the $\Z_m$-orbifold. With all the requirements
checked, we found---as expected---that the asymmetric orbifolds are
consistent.

\subsubsection{Anomaly cancellation and winding states} 

In the previous section we barely mentioned the Chern--Simons
invariant, which provides another way to decide which orbifolds are
consistent.  Nevertheless, both orbifold analysis and the Chern--Simons
analysis lead to identical results. Let us examine the relation
between the approaches in more detail.

According to our analysis of the topology of subgroups in string
theory, it is the presence of winding states that rules out particular
gauge theory compactifications in string theory.  Let us consider such
a state with $w_1 =1$ and $w_2 = w_3 =0, \mathbf{q} = 0$. We have seen
that such a state carries a gauge group representation vector equal to
$\mathbf{a}_1$. We denote this state by $| \mathbf{a}_1
\rangle$. Consider parallel transport of this state along the
following path: we start by going around a closed cycle in the $2$
direction then a closed cycle in the $3$ direction, around the $2$
direction with the opposite orientation, then around the $3$ direction
with the opposite orientation. Because of the background gauge fields,
the state transforms in the following way, 
\bea | 
\mathbf{a}_1 \rangle
\quad \rightarrow & e^{2 \pi i (\mathbf{a}_1 \cdot \mathbf{a}_2 )} \,
| \mathbf{a}_1 \rangle & \rightarrow \quad e^{ 2 \pi i (\mathbf{a}_1
\cdot \mathbf{a}_2 )} \, | \theta(\mathbf{a}_1) \rangle \non \\
\rightarrow & e^{ 2 \pi i ((\mathbf{a}_1 - \theta(\mathbf{a}_1)) \cdot
\mathbf{a}_2 )} \, | \theta(\mathbf{a}_1) \rangle & \rightarrow \quad
e^{ 2 \pi i ((\mathbf{a}_1 -\theta(\mathbf{a}_1))\cdot \mathbf{a}_2 )}
\, | \mathbf{a}_1 \rangle.  \label{anwin} 
\eea 
With the results from
\cite{bfm}, the Chern--Simons invariant can be expressed in terms of
the gauge fields as $(\theta(\mathbf{a}_1)-\mathbf{a}_1)\cdot
\mathbf{a}_2$. On the other hand, we transported a state around a
contractible curve in a flat background, and consistency requires that
the final phase factor appearing in (\ref{anwin}) equal unity. The
conclusion is that the Chern--Simons invariant must be integer,
precisely as was argued in section \ref{anomaly}.  Analogous arguments
apply to other winding states.

To complete the connection, we remark that (\ref{anwin}) only
expresses the change in phase caused by the gauge fields. The state
$|\mathbf{a}_1 \rangle $ is due to a winding string, and in the
transport process sketched above, it sweeps out a two-dimensional
world sheet. It therefore also picks up another contribution $\exp({2
\pi i \int B})$ to the phase, where the integral is over the world
sheet area sweeped out.  The surface integral can be converted to a
volume integral giving the total phase change:
\begin{displaymath}
\exp \left( 2 \pi i \int \{ dB - CS(A) \} \right) = \exp \left( 2 \pi i \int
H \right).
\end{displaymath} 
The right hand side states that the total phase change should be
attributed to the gauge field strength to which the string couples.
This equation is just a global version of anomaly cancellation (\ref{defineH}).

\subsection{Moduli spaces} \label{moduli}

Our previous discussion focused on orbifold descriptions appearing at
specific points in the orbifold moduli space. In this section, we
extend the discussion to cover the whole moduli space of asymmetric
orbifold theories.

\subsubsection{Lattices for the orbifolds}

In the standard toroidal or Narain compactification of the heterotic
string, a central role is played by the even self-dual lattice
$\Gamma_ {d+16,d}$ \cite{Narain, Narain87a}. The momenta lie on this
lattice.  In the construction of the moduli space for a Narain
compactification, we further divide out by a discrete subgroup
corresponding to the symmetries of $\Gamma_ {d+16,d}$. In an attempt
to set up a similar structure for the CHL string and its
compactifications, Mikhailov introduced lattices for these theories
\cite{Mikhailov98}. In a somewhat more laborious construction, the
same can be done for the asymmetric orbifolds of the previous section.

Recall that for all orbifolds corresponding to triples, we had a
transformation $\theta$ which has order $m$. For each $\theta$, we can
define a projection $P_{\theta} $ acting on $\R_{3,19}$ by, 
\begin{equation} 
P_{\theta} = \frac{1}{m} \sum_{n=0}^{m-1} \theta^n.  
\end{equation}
{}From $\theta P_{\theta} =  P_{\theta} \,\theta= P_{\theta}$,  we see
that $P_{\theta}$ projects all lattice vectors in $\Gamma_{3,19}$ onto
the space invariant under $\theta$. In particular, for the holonomies
introduced in the previous section we have $P_{\theta}(\mathbf{a}_i) =
0$.

As our starting point, we return to the momenta (\ref{mom1}),
(\ref{mom2}) and (\ref{mom3}) of the heterotic string prior to
orbifolding.  In the orbifolded theory, the untwisted sector consists
of those states that are left invariant under the projection. These
are of the form
\begin{equation}
N \sum_{n=0}^{m-1} \exp \left( 2 \pi i n  a \cdot p \right) |
\theta^n(\mathbf{k}), p \rangle. 
\end{equation}
Here, we symbolically denote the spatial momenta by $p$, the group
quantum numbers by $\mathbf{k}$, while $a$ and $\theta$ are the shift
and rotation of the orbifold symmetry. There is also a normalization
constant $N$.

To these states, we associate a lattice in the following way: 
first we define 
$$\mathbf{q}_{inv} = \sqrt{m} P_{\theta}({\mathbf q}), $$
where the reason for the factor of $\sqrt{m}$ will soon become clear. We also
set 
$$\tilde{n}_i = m n_i - m{ \mathbf{q} \cdot \mathbf{a}}_i - w_i (m-1),
$$ for $i=1,2$. We will rescale the radii for the $1$ and $2$
directions by defining $R'_i = m R_i$. We also define $\alpha'' = m
\alpha'$. Note that the invariant radii $R_i/\sqrt{\alpha'}$ are only
rescaled by a factor $\sqrt{m}$.

Now define a lattice by projecting the Narain lattice (\ref{mom1}),
(\ref{mom2}) and (\ref{mom3}) onto the invariant subspace of
$\theta$. We call this lattice $\tilde{I}$. With the
reparametrisations introduced above, its vectors are given by:
\begin{equation} \label{latuntw} {\mathbf{v}} = ({\mathbf{q}}_{inv})
\sqrt{\frac{2}{\alpha''}}, \qquad v_{iL,R} = \frac{\tilde{n}_i}{R'_i}
\pm \frac{w_i R'_i}{\alpha''}, \quad i=1,2 \qquad v_{3L,R} =
\frac{n_3}{R_3} \pm \frac{m w_3 R_3}{\alpha''}. \end{equation} The
vectors $v_{iL,R}$ and $v_{3L,R}$ form a lattice, which when rescaled
by $\sqrt{\alpha''/2}$ may be called $\Gamma_{2,2} \oplus
\Gamma_{1,1}(m)$. Here we follow the notation of Mikhailov
\cite{Mikhailov98}, defining the lattice $\Gamma_{1,1}(m)$ to be a
lattice of signature $(1,1)$ generated by 2 vectors $e$ and $f$ with
scalar product $(e \cdot f) =m$. For a summary of our lattice
conventions, see Appendix \ref{app:lattice}.  This lattice arises as
an intermediate step because we have not yet included the twisted
sectors.

The vectors $\mathbf{v}$ are the vectors of $\Gamma_8 \oplus \Gamma_8$
projected onto the subspace invariant under $\theta$ and suitably
rescaled. This defines a different lattice for every $m$, which can be
deduced from group theory. The lattices are $D_4 \oplus D_4$, $A_2
\oplus A_2$, $A_1 \oplus A_1$ for $m=2,3,4$, respectively. The lattice
for the cases $m=5$ and $m=6$ is the empty lattice. These lattices are
usually defined so their roots are normalized with length $\sqrt{2}$.
In the symmetry groups which arise in gauge theory, these form the
\emph{short} roots of non-simply-laced algebras at level 1. For
example, at the point in moduli space constructed here the gauge group
is $F_4 \times F_4$ for the $m=2$ case, and $G_2 \times G_2$ for the
$m=3$ case, with long roots which have length $2$ and $\sqrt{6}$,
respectively. The gauge group $SU(2)$ in the $m=4$ case has roots of
length $\sqrt{8}$ (it is at level 4). The vectors with length
$\sqrt{2}= \sqrt{8}/2$ are on the weight lattice of $SU(2)$. Although
there is no simple 4-laced algebra, there is a 4-laced affine Dynkin
diagram that plays a role in the description of the group theory
\cite{bfm}. Interestingly, the lattices of $F_4$, $G_2$ and $A_1$ at
level 4 all satisfy a `generalised self-duality' in the sense that
their weight lattice is identical to the original lattice. For the
simply-laced $E_8$ lattice, this notion of `generalised self-duality'
coincides with self-duality. There is an interesting connection
between this observation and S-duality of four-dimensional theories,
which we shall discuss later.

In the twisted sectors, the momenta lie on the lattice $I^{\ast}$,
which is dual to the lattice $I$ of invariant vectors. As in the case
of the untwisted sector, we will treat the parts of the lattices that
represent the group quantum numbers, and the part that represents the
space quantum numbers separately.

Obviously the lattice $I$ of invariant vectors is a sublattice of the
lattice $\tilde{I}$. It can be verified that the group part of the
lattice $I$ of invariant vectors is the lattice which we will
denote $\sqrt{2}(D_4^* \oplus D_4^*)$, $\sqrt{3}(A_2^* \oplus
A_2^*)$, $2(A_1^* \oplus A_1^*)$ for $m=2,3,4$, respectively. As usual, 
the star denote the dual lattice which is, of course, the (co)weight lattice.
The stars arise because we define the lattices $I$ relative to the
lattices $\tilde{I}$, which forces us to keep track of relative
orientations. It is now trivial to construct the group part of the
lattices $I^{\ast}$: these are  $(D_4 \oplus D_4)/\sqrt{2}$,
$(A_2 \oplus A_2)/\sqrt{3}$, $(A_1 \oplus A_1)/2$, for $m=2,3,4$.

We now construct the spatial part of the invariant lattice $I$.
First note that for invariant vectors,
\begin{displaymath}
P_{\theta} (\mathbf{q}+ \sum_i w_i \mathbf{a}_i) = P_{\theta}
(\mathbf{q}) = \mathbf{q}+ \sum_i w_i \mathbf{a}_i. 
\end{displaymath}
Since $P_{\theta} (\mathbf{q})$ is a sum of elements of the root
lattice, it again lies on the root lattice. It then follows that $w_i
\mathbf{a}_i$ is on the root lattice. Because $\mathbf{a}_i$ is on the
weight lattice, we deduce that for invariant vectors, $w_i$ has to be
a multiple of $m$, say $l_i m$. Another way to see this is from the
value of ${\mathbf a}_i^2 = 2(m-1)/m$. Also, if ${\mathbf q} + \sum
w_i {\mathbf{a}}_i$ is on the invariant lattice then its dot product
with either ${\mathbf{a}}_j$ has to vanish:
\begin{displaymath}
({\mathbf q} + \sum_i
w_i {\mathbf{a}}_i) \cdot \mathbf{a}_j = P_{\theta}({\mathbf q} + \sum_i
w_i {\mathbf{a}}_i) \cdot \mathbf{a}_j = ({\mathbf q} + \sum_i
w_i {\mathbf{a}}_i) \cdot P_{\theta}(\mathbf{a}_j) = 0.
\end{displaymath} 
This leads immediately to ${\mathbf q}
\cdot {\mathbf a}_i = -2l_i(m-1)$. The spatial momenta on the invariant
lattice are thus given by, 
\begin{equation}
\frac{n_i + l_i (m-1)}{R_i} \pm \frac{m l_i R_i}{\alpha'}, \quad i=1,2
\qquad \frac{n_3}{R_3} \pm \frac{w_3 R_3}{\alpha'}. 
\end{equation}
Note that $n_i + l_i(m-1)$ can take any integer value, while $l_i m$ is
always a multiple of $m$. The momenta on the dual to the spatial
part of the invariant lattice are then given by the vectors,
\begin{equation}
\frac{n'_i}{mR_i} \pm \frac{w'_i R_i}{\alpha'}, \quad
i=1,2 \qquad \frac{n'_3}{R_3} \pm \frac{w'_3 R_3}{\alpha'}.
\end{equation}
We complete the construction of $I^{\ast}$ with the same
reparametrisations as in the untwisted sector: multiply the group
parts of the lattices by $\sqrt{m}$, define $R_i' = mR_i$ for $i=1,2$
and set $\alpha''= m \alpha'$. Note that in all cases, we have the
simple result $\tilde{I} = I^{\ast}$, confirming a result from
Appendix A of \cite{Narain87b} for our specific case.

To construct the twisted sectors, we still need the shift $a^{\ast}$. It
is given by multiples of 
\begin{equation}
\Delta v_{3L,R} = \pm \frac{R_3}{m\alpha'} = \pm \frac{R_3}{\alpha''}. 
\end{equation}
We remind the reader that the sign choice between the left and
right-moving parts of the shift reflects our choice of fractionalizing
the winding numbers. Because the lattices $I^{\ast}$ are identical to
$\tilde{I}$, and because the momenta in the $n^{th}$ twisted sector
are given by $I^{\ast} + n \Delta v_{3L,R}$ with $n = 1, \ldots,
(m-1)$, we can assemble the lattices into a single lattice
$\Lambda$. In the process of assembly, the spatial part of the lattice
is completed to $\Gamma_{3,3}$.

We only computed the lattices for very specific orbifolds with special
values of the holonomies and other background fields.  To extend to
the general case, first note that the metric and antisymmetric tensor
field did not play any role so far, and the moduli corresponding to
these fields survive the orbifold projection.  For the holonomies, we
took special values that had
\begin{displaymath} 
P^{\perp}_{\theta} (\mathbf{a}_i) = (\mathbf{1} - P_{\theta})
(\mathbf{a}_i) = \mathbf{a}_i. 
\end{displaymath}
For the general case, we take holonomies parametrized by $\mathbf{a}_i'$, subject to
\begin{equation}
P^{\perp}_{\theta} (\mathbf{a}_i') = \mathbf{a}_i. 
\end{equation}
The possible moduli for varying the holonomies are then given by
$P_{\theta}(\mathbf{a}_i')$. We may use general formulae from
\cite{Narain, Narain87a, ginsparg} to show that this results in a
moduli space that locally has the form,
$$O(19-\Delta r,3)/ \left(O(19-\Delta r) \times O(3)\right), $$ 
where $\Delta r$ is the rank reduction for the $\Z_m$ orbifold. 

As usual, we should also divide on the left by a discrete group of
lattice symmetries. Following Mikhailov \cite{Mikhailov98}, we propose
that this discrete group is formed by the symmetries of the lattice
$\Lambda$ constructed above. This is a non-trivial statement with
regard to those symmetries in $\Lambda$ that connect different twisted
and untwisted sectors.  Mikhailov demonstrates this explicitly for his
lattice.  For our cases, we will not attempt to prove this. However,
we note that from a gauge theory point of view, all holonomies are on
equal footing (and in special situations, even conjugate to each
other).  We stress that the asymmetry in our treatment of the various
holonomies is purely technical, caused by the fact that we are working
at the level of the algebra rather than the group.  There is therefore
every reason to expect symmetry transformations that connect the
various sectors, and lift the apparent asymmetry between the
holonomies.

We therefore propose that the moduli space of these asymmetric orbifolds is
given by,
\begin{equation}
O(\Lambda)\, \backslash \,O(19-\Delta r,3) \, / \, \left(O(19-\Delta r)
\times O(3) \right).
\end{equation}
The lattices $\Lambda$ and the rank reduction $\Delta r$ are collected
in table \ref{table:lattices}, where other relevant results are
summarized. We have included the standard Narain compactification,
denoted by $\Z_1$, which fits perfectly in the picture when we take
trivial holonomies and $m=1$. In a separate column, we list the
lattices $\Lambda^{\perp}$ which are the lattices of vectors
orthogonal to the $\Lambda$ sublattice of $\Gamma_{19,3}$. Not
surprisingly, the lattices are those of the `minimal subgroups' $C$
for $m$-triples.

\begin{table}[ht]
\begin{center}
\renewcommand{\arraystretch}{1.5}
\begin{tabular}{|c || c | c | c | c |}
\hline
        & $\Lambda$ & $\Lambda^{\perp}$ & $\Delta r$ & $E_t$ \\
\hline
\hline
$\Z_1$ & $\Gamma_{3,3} \oplus E_8 \oplus E_8$ & $\emptyset$      & $0$  &  $-1$ \\
$\Z_2$ & $\Gamma_{3,3} \oplus D_4 \oplus D_4$ & $D_4 \oplus D_4$ & $8$  &  $-1/2$ \\ 
$\Z_3$ & $\Gamma_{3,3} \oplus A_2 \oplus A_2$ & $E_6 \oplus E_6$ & $12$ &  $-1/3$ \\
$\Z_4$ & $\Gamma_{3,3} \oplus A_1 \oplus A_1$ & $E_7 \oplus E_7$ & $14$ &  $-1/4$ \\
$\Z_5$ & $\Gamma_{3,3}$                       & $E_8 \oplus E_8$ & $16$ &  $-1/5$ \\
$\Z_6$ & $\Gamma_{3,3}$                       & $E_8 \oplus E_8$ & $16$ &  $-1/6$ \\
\hline
\end{tabular}
\renewcommand{\arraystretch}{1.0}
\caption{Lattices $\Lambda$, complements $\Lambda^{\perp}$, rank
reduction $\Delta r$ and zero-point energies in the twisted sector
$E_t$ for the $Z_m$ asymmetric orbifolds corresponding to triples.}
\label{table:lattices} 
\end{center} 
\end{table}

We note that the entries in the $\Z_2$ row are identical to those for
the CHL string: the same Mikhailov lattice, the same rank reduction
and the same zero-point energy. It can indeed be proven that the
$\Z_2$-triple and the CHL string are equivalent. We will encounter
these and many other dualities in the next section. To conclude this
discussion of the moduli spaces for these compactifications, let us
make a preliminary count of the number of distinct $7$-dimensional
heterotic compactifications that we have constructed. We begin by
noting an obvious discrete symmetry: for example in the $\Z_3$ case,
we can embed a bundle with $CS=1/3$ in one $E_8$ factor and a bundle
with $CS=2/3$ in the other $E_8$. Flipping the choice of embedding
does not generate a new theory. There is a single theory associated to
the $\Z_3$ orbifold. This counting gives us a single component for
$\Z_1, \Z_2, \Z_3, \Z_4$ and $\Z_6$. At first sight, the case of
$\Z_5$ seems to give two distinct embeddings $( \frac{1}{5},
\frac{4}{5})$ and $( \frac{2}{5}, \frac{3}{5})$. However, the two
theories are actually equivalent.  This can be argued directly using
discrete gauge symmetries, but is more easily seen in the dual
description which we shall meet in section \ref{geometry}. Therefore,
we find $6$ distinct components in the moduli space.

\subsubsection{Dualities}
\label{hetduality}
It is well known that the heterotic $E_8 \times E_8$ string
compactified on a circle is equivalent to the heterotic
$Spin(32)/\Z_2$ string compactified on a circle. This may be deduced
from the form of their moduli spaces \cite{Narain}, and can be made
explicit by constructing a map between the two theories at a
particular point in the moduli space. The rest of the moduli space is
covered by continuation \cite{ginsparg}. This duality can be shown to
imply a duality between the CHL string, and a compactification of the
$Spin(32)/\Z_2$ string without vector structure \cite{Lerche98, Bianchi:1997rf,
wittoroid}, which are $\Z_2$-asymmetric orbifolds of the $E_8 \times
E_8$ string and the $Spin(32)/\Z_2$ string, respectively.

We now present a list of dualities between heterotic theories
with various bundles. The previously mentioned duality between the
CHL string and the $Spin(32)/\Z_2$ compactification without vector
structure is included as part of a much larger chain. We also find
new dualities for theories with rank reduction bigger than
$8$. These dualities should be expected on general grounds, such as the
structure of the orbifold groups and the moduli spaces of various
heterotic asymmetric orbifolds.  A more detailed study of T-duality
for the heterotic string, performed in appendix \ref{app:hetdual}, gives
us tools that allow us to make these statements more precise.

\subsubsection*{The $\Z_{2}$-chain}

Our first example starts with the CHL string and will feature toroidal
compactifications down to $5$ dimensions.  We take the heterotic $E_8
\times E_8$ string compactified on a circle of radius $R_1$. We will
use standard coordinates for the $E_8 \times E_8$ lattice, giving 16
numbers $u_i$ $i = 1, \ldots, 16$ of which the first 8 denote the
first $E_8$ factor, and the second group of 8 denotes the other $E_8$
factor. The construction from \cite{Chaudhuri95b} involves modding out
this theory by a shift over $\pi R_1$ combined with the following
transformation on the group lattice:
\begin{equation} \label{theta1}
{\theta}(u_{1}, \ldots, u_{16}) = -(u_{16},\ldots, u_1).
\end{equation} 
The transformation $\theta$ interchanges the two $E_8$ factors so we
end up with a theory with gauge group $(E_8)_2$ on a circle with
radius $R_1/2$.  The extra subscript denotes that the gauge group is at level
$2$.
 
We compactify this theory on a second circle with radius $R_2$. We may
turn on a holonomy provided it is invariant under $\theta$.
Therefore, we can smoothly deform the theory and introduce a holonomy
parametrised by,
\begin{equation}
\mathbf{a}_2 = ( 1, 0^{14}, -1).
\end{equation}
The notation $0^{14}$ denotes 14 subsequent entries of zero.
Introducing this holonomy breaks the gauge group to $Spin(16)_2$.

This theory is interpreted as the CHL string on a torus with radii
$(R_1/2, R_2)$. We can now find an element of the T-duality group that inverts $R_2$ (for details, see appendix~B)
to obtain a compactification of the $Spin(32)/\Z_2$ theory on a torus
with radii $(R_1/2, R'_2= \alpha'/2 R_2)$ with holonomy given by,
\begin{equation} \label{a2prime}
\mathbf{a}_2' = \left( \left(-\hlf \right)^8, 0^8 \right).
\end{equation}
Of course, the gauge group is still $Spin(16)_2$. The vector
$\mathbf{a}_2'$ is no longer invariant under $\theta$. However,
we note that
\begin{displaymath}
\theta(\mathbf{a}_2') - \mathbf{a}_2' =  \left( \hlf^{16} \right),
\end{displaymath}
which is allowed since $\left(\hlf^{16} \right)$ is a lattice vector
of the $Spin(32)/\Z_2$ lattice. That it lies on the spin-weight
lattice indicates that we are dealing here with a compactification
without vector structure. We have rederived the result of
\cite{Lerche98}, which was also discussed in \cite{Bianchi:1997rf, wittoroid}.

We compactify this theory on a third circle with radius $R_3$, and
turn on a holonomy parametrised by
\begin{equation}
\mathbf{a}_3 = \left( \hlf^2, \left(-\hlf \right)^2, 0^8, \hlf^2,
\left(-\hlf \right)^2 \right). \end{equation}
This breaks the group to $(Spin(8)_2)^2$. We have chosen $\mathbf{a}_3$
to be invariant under $\theta$,  and $\mathbf{a}_2' \cdot
\mathbf{a}_3 = 0$. We can dualize again  in the $3$ direction
to obtain an $E_8 \times E_8$ theory on a 3-torus with radii
$(R_1/2, R_2', R_3')$ (with $R_i' = \alpha' /2R_i$), and holonomies
given by $\mathbf{a}_2'$, (\ref{a2prime}), and  
\begin{equation}
\mathbf{a}_3' = \left(-1, 0^3, 1, 0^{11} \right). 
\end{equation}
Along the first circle,  there is still the action of $\theta$ given
by (\ref{theta1}). However, in the present $E_8 \times E_8$ theory, the root
lattice is organized differently. Here one of the $E_8$ root lattices
is denoted by the 8 coordinates $u_i$ with $i = 5,\ldots, 12$, while the
second $E_8$ resides in the remaining 8 positions. The transformation
$\theta$ therefore acts within each $E_8$ factor and on both factors
simultaneously. This is a particular instance of the $\Z_2$-triple
construction described previously. Note that $\theta(\mathbf{a}_i')
- \mathbf{a}_i' \neq 0$ for either $i=2$ or $i=3$, but that in both
cases the difference is a root vector of $E_8 \times E_8$.

We see that the CHL string, $Spin(32)/\Z_2$ compactification without
vector structure and the $\Z_2$-triple are indeed equivalent, as
claimed. Our duality chain does not end here, but by proceeding
straightforwardly, we would end up with non-standard coordinates on
group-lattices for the theories that we encounter. To avoid possible
confusion, let us perform a coordinate transformation on the group
lattice of $E_8 \times E_8$ so that the first $E_8$ ends up in the
first $8$ positions again, and the second in the second
$8$. Furthermore, the coordinates are chosen such that, 
\bea 
\theta (u_{i}) & = & -u_{i}, \quad 
i = 5,\ldots, 12, \qquad 
\theta (u_{i}) = u_{i}, \quad i \neq 5,\ldots, 12 \\ 
\mathbf{a}_2' & = & \left( \hlf^2,0^4,\hlf^4,0^4,\hlf^2\right), \label{a2} \\ 
\mathbf{a}_3' & = & \left( \left(0, \hlf \right)^4,\left(-\hlf,0 \right)^4 
\right ).
\label{a3} 
\eea 
We stress that we have not changed anything in the
theory. It is easy to check that the coordinate transformation can be
chosen so that it corresponds to a lattice symmetry, or equivalently,
a gauge transformation. To emphasize this, we will continue using the
symbols $\theta$ and $\mathbf{a}_i'$ since we are working with the
same theory as before.

Let us continue our study of the $\Z_2$-triple in $E_8 \times E_8$ on a
$3$-torus with radii $(R_1/2, R_2', R_3')$ and gauge group
$(Spin(8)_2)^2$. We compactify the theory on a fourth circle with
radius $R_4$ and turn on a holonomy parametrised by,
\begin{equation}
\mathbf{a}_4 = (1, 0^{14},-1).
\end{equation}
This breaks the gauge group to $(Spin(4)_2)^4 = (SU(2)_2)^{8}$. We
dualise along the 4-direction to a $Spin(32)/\Z_2$ theory on a
fourth circle with radius $R_4' = \alpha'/2R_4$ and holonomy
para\-metrised by,  
\begin{equation} \label{a4}
\mathbf{a}_4' = \left( \left(0,-\hlf \right)^2,
\left(0,\hlf \right)^2, \left(0,-\hlf \right)^2, \left(0,\hlf \right)^2
\right).
\end{equation}
We have discussed the physical interpretation of this theory briefly
in the section on gauge bundles over the $4$-torus. The transformation
$\theta$ does not leave any of the $\mathbf{a}_i'$ invariant, but now
$\theta(\mathbf{a}_i') - \mathbf{a}_i'$ are roots for all $i$. This
theory is therefore not a compactification without vector structure.
Instead, we should interpret this theory as a particular case of a
quadruple of $Spin(32)/\Z_2$ on a $4$-torus.

To arrive at the final theory on this chain, we compactify on a fifth
circle with radius $R_5$, and turn on a holonomy parametrised by,
\begin{equation} \label{b5}
\mathbf{a}_5 = \left( \hlf^2, \left(-\hlf \right)^2, 0^8, \hlf^2,
\left(-\hlf \right)^2 \right).
\end{equation}
The resulting gauge group is $U(1)^8$. We dualize along the
$5$ direction to obtain an $E_8 \times E_8$ theory where the fifth circle
has radius $R'_5 = \alpha'/2R_5$ and holonomy,   
\begin{equation} \label{a5}
\mathbf{a}_5' = (0^2,-1,0,1, 0^{11}).
\end{equation}
The first $E_8$ factor is in the positions $i =5, \ldots, 12$, while the
second occupies the positions $i=1,\ldots, 4, 13, \ldots,16$. None of the
$\mathbf{a}_i$ are invariant. The transformation $\theta$ inverts all
coordinates of one of the $E_8$'s; this compactification should
therefore be interpreted as a quintuple embedded in one of the
$E_8$ gauge factors.

This chain of $5$ theories shows again that the list of toroidal
compactifications in string theory is much shorter than the list of
gauge theory compactifications. Among the gauge theory
compactifications that can be implemented consistently in string
theory, there are many equivalences. Various choices of bundles just
correspond to different limits in the string moduli space. It also
shows that the topology of the gauge bundle, although entering
crucially in our analysis, is a not a sharp notion to a string. For
example, note that according to our preceeding analysis, certain
compactifications of $Spin(32)/\Z_2$ without vector structure are
connected to compactifications with vector structure and another
mechanism of rank reduction. As we will see later, this provides
connections between seemingly different dual theories.  In particular,
between choices of bundles in configurations of D-branes on
orientifolds.

In the $5$ dual descriptions obtained above, $\mathbf{a}_i$ and
$\mathbf{a}_i'$ are always of length $\sqrt{2}$. This is not the
minimum value. As we remarked in the previous section, the minimal
value for the $\Z_2$-theory is $1$. To reach this minimal value, we
can deform the $\mathbf{a}_i'$ by adding vectors in $\R_{16}$ which
are left invariant by $\theta$. By doing so, we can flow to a
`canonical' point in the moduli space. At such a point, there is
non-abelian gauge symmetry. We have listed the gauge symmetries
appearing at these `canonical points' in table \ref{table:canpoint},
where we have also included the Kaluza--Klein gauge bosons.  We have
also listed the Mikhailov lattices $\Gamma_{(10-n)}$ derived in
\cite{Mikhailov98}. The expression for $\Gamma_{(5)}$ is not found in
\cite{Mikhailov98}, but is easily derived with our methods.

\begin{table}[ht]
\begin{center}
\renewcommand{\arraystretch}{1.5}
\begin{tabular}{|c | c | c || c | c |}
\hline
Theory & bundle & $n$ & Symmetry group & Mikhailov lattice \\
\hline
$E_8 \times E_8$ & CHL &  1 & $(E_8)_2 \times U(1) \times U(1)$ & $E_8
\oplus \Gamma_{1,1}$ \\ 
$Spin(32)/\Z_2$ & NVS & 2  & $(Sp(8)/\Z_2) \times U(1)^2 \times
U(1)^2$ & $D_8 \oplus \Gamma_{2,2}$  \\
$E_8 \times E_8$ & triple & 3 & $F_4 \times F_4 \times U(1)^3 \times
U(1)^3$ & $D_4 \oplus D_4 \oplus \Gamma_{3,3} $\\
$Spin(32)/\Z_2$ & quadruple & 4 & $Spin(17) \times U(1)^4 \times
U(1)^4$ & $D_8^{\ast}(2) \oplus \Gamma_{4,4}$ \\
$E_8 \times E_8$ & quintuple & 5 & $E_8 \times U(1)^5 \times
U(1)^5$ & $E_8(2) \oplus \Gamma_{5,5}$ \\
\hline 
\end{tabular} 
\renewcommand{\arraystretch}{1.0}
\caption{Asymmetric $\Z_2$-orbifolds of heterotic
theories on an $n$-torus. } \label{table:canpoint} 
\end{center}  \end{table} 

On the canonical points in the moduli space, the connection
between the groups and in particular their lattices, and the lattices of
Mikhailov is clear:
The lattices are related to the short roots of the symmetry groups.
This provides a physical link to aspects of
Mikhailov's mathematical constructions. Mikhailov argues that his
lattices are also useful for understanding aspects of dual theories.
This point will be revisited later in our paper.

\subsubsection*{More quintuples}

Let us now set aside compactifications of the CHL string, and turn to
other theories that admit dual realizations.  The endpoint of our
$\Z_2$-chain was given by the $E_8 \times E_8$ theory with a quintuple
in one $E_8$. Actually, with an $E_8$ theory on a $5$-torus and
holonomies $\mathbf{a}_2'$, $\mathbf{a}_3'$, $\mathbf{a}_4'$ and
$\mathbf{a}_5'$ given by eqs.  (\ref{a2}), (\ref{a3}), (\ref{a4}) and
(\ref{a5}), it is possible to have quintuples in both $E_8$
factors. For this purpose, we construct the asymmetric orbifold
obtained by shifting over $\pi R_1$ combined with the complete
reflection on the gauge group lattice:
\begin{equation}
\theta(u_i) = -u_i.
\end{equation}
Only discrete gauge symmetries survive in this construction. All
continuous gauge symmetry is broken.

This $E_8 \times E_8$ theory on a 5-torus with radii $(R_1/2, R_2,
R_3, R_4,R_5)$ with quintuples in each $E_8$ may be easily dualized in
either the $2,3,4$ or $5$ directions, since these are all equivalent. For ease
of notation, we will dualize in the $5$ direction with,
\begin{equation} R_5 \rightarrow \frac{\alpha'}{2R_5}
\qquad \mathbf{a}_5' \rightarrow \mathbf{a}_5,
\end{equation}
and where $\mathbf{a}_5$ is given in eq.~(\ref{b5}). We have obtained a
$Spin(32)/\Z_2$ theory on the $5$-torus with completely broken gauge
group. Checking the action of $\theta$ on the $\mathbf{a}_i'$ and
$\mathbf{a}_5$ reveals that this is a case of a quintuple in
$Spin(32)/\Z_2$. Notice that this duality connects two intrinsically
five dimensional theories: decompactification of dimensions on either
side of the duality would restore some gauge symmetry.

\subsubsection*{The $\Z_4$ chain}

Take the heterotic $E_8 \times E_8$ theory with a $\Z_4$-triple on a
$3$-torus. We use standard coordinates on $E_8 \times E_8$ so that the
first $E_8$ is in the first $8$ positions, and the second $E_8$ in the
remaining $8$. We turn on holonomies $\mathbf{a}_2$ and
$\mathbf{a}_3$.  To obtain the $\Z_4$-triple, we divide by a shift
over $\pi R_1/2$ combined with a $\theta$ of order $4$. In a special
case, the expressions for $\theta$ and $\mathbf{a}_i$ can be taken to
be, 
\bea 
\theta (u_i, u_{i+1}, u_{i+2}) & = & (-u_{i+2}, -u_{i+1},
u_{i} ), \quad i = 1,9\\ \theta (u_{i},u_{i+1}, u_{i+2}, u_{i+3},
u_{i+4}) & = & (-u_{i+4}, -u_{i+3}, -u_{i+2}, -u_{i+1}, u_{i} ), \quad
i = 4, 13 \\ \mathbf{a}_2 & = & \left( \hlf^3, 0^5, \left( -\hlf
\right)^3, 0^5 \right), \\ \mathbf{a}_3 & = & \left(\hlf, \qrt, 0,
\hlf, \left( \qrt \right)^3 ,0, \hlf, \qrt, 0, \hlf, \left( \qrt
\right)^3,0 \right).  
\eea 
To clarify these formulae a little, let us
elaborate somewhat on the steps in the triple construction. Starting
with a torus without holonomies, turning on $\mathbf{a}_2$ will break
$E_8 \times E_8$ to $(Spin(6)^2 \times Spin(10)^2)/\Z_4$. The $\Z_4$
is generated by the products of the generators of the $\Z_4$ centers
of $Spin(6) = SU(4)$ and $Spin(10)$. Therefore, it acts diagonally on
all factors. The reader may anticipate that $(Spin(6)^2 \times
Spin(10)^2)/\Z_4$ is a group that can also be obtained by
compactifying $Spin(32)/\Z_2$ with \emph{two} holonomies. The
remaining holonomies embed twisted boundary conditions: $\mathbf{a}_3$
breaks $Spin(6)$ to $U(1)^3$ and $Spin(10)$ to $SU(3) \times
U(1)^3$. Finally, the $\theta$ quotients eliminate all $U(1)$'s and
mod $SU(3)$ by its outer automorphism to $SU(2)$ which is the
surviving gauge group. The holonomies defined above are equivalent to
the ones described abstractly in the previous section. The theory is
at what we have called a canonical point in the moduli space.

We will not dualize this theory directly, but instead make a slight
detour to stress some subtle points. We start by compactifying on a
fourth circle, with radius $R_4$, \emph{without} turning on a holonomy
on this circle. Instead we will perform an $SL(4,\Z)$ transformation
on the $4$-torus:
\begin{equation}
x_4 \rightarrow x_4 - 2x_2  \qquad x_i \rightarrow x_i \quad i \neq 4.
\end{equation} 
In this way, we obtain a theory with identical spectrum but with
holonomies parametrised by $\theta$, $\mathbf{a}_2$, $\mathbf{a}_3$
and a holonomy around the fourth circle given by $\mathbf{a}_4 = 2
\mathbf{a}_2$. Although none of the $\mathbf{a}_i$ is invariant under
$\theta$, this theory on the $4$-torus can be decompactified to one on
the $3$-torus by decompactifying in the $x_4 - 2x_2$ direction.

We now turn on a $B$-field, which has as its only non-zero components
$B_{24} = -B_{42} = 3/2$. The purpose of this, as may be verified with
the aid of the formulae given in appendix \ref{app:hetdual}, is so we
can use lattice symmetries to re-express this as an equivalent theory
with $B_{ij}= 0$ and holonomies parametrised by $\theta$,
$\mathbf{a}_2$, $\mathbf{a}_3$. The holonomy around the fourth circle
is now given by,
\begin{equation}
\tilde{\mathbf{a}}_4 = \left( 0^7, 1,0^7, -1 \right).
\end{equation}

We have arrived at the theory we wish to dualize. Doing so gives a
theory where the fourth radius is $R'_4 = \alpha'/2R_4$.  There is a
holonomy parametrised by $\mathbf{a}'_4$ around the $4$ direction, and
there is a non-zero $B$-field. The non-zero components are given by,
\begin{equation} \mathbf{a}'_4 = \left( 0^8,
-\hlf, \hlf^2, -\hlf, \hlf^4 \right) \qquad B_{24} = - B_{42} = \qrt. 
\end{equation}
The remaining moduli are given by $\mathbf{a}_2$ and $\mathbf{a}_3$.

We may ignore the $B$-field, which does play a role in the dualities,
but not in the gauge field interpretation.\footnote{Besides, we can
always deform the $B$-field away.} We have found a $Spin(32)/\Z_2$
theory. It is not too hard to verify that $\theta(\mathbf{a}_3)
-\mathbf{a}_3$ is on the spin lattice of $Spin(32)/\Z_2$, indicating
that we are dealing with a compactification without vector
structure. The rank reduction is, however, not equal to $8$ but to
$14$. This is a realization of a quadruple without vector
structure. All the Chern--Simons invariants that can be defined over
sub-three-tori of the four-torus are integer.

The existence of a $Spin(32)/\Z_2$ description of this theory implies
the existence of a type I theory on the $4$-torus with the same
bundle.  By T-dualities, this translates into orientifolds of type II
theories to be described in section \ref{orient}.

\subsubsection*{$\Z_2 \times \Z_2$ asymmetric orbifolds}

Consider an $E_8 \times E_8$ theory on a $4$-torus with holonomies
parametrised by,
\bea  
\mathbf{a}_3 & = & \left( 1,0^{14},-1 \right), \\ 
\mathbf{a}_4 & = & \left(0^4,\hlf^4, \left(-\hlf \right)^4,
0^4 \right).   
\eea
We construct  a $\Z_2$-triple in this theory by dividing this theory
by a $\pi R_1$ shift over the first circle combined with    
\begin{displaymath}  
\theta_1: \quad (u_{1}, \ldots, u_8,u_9, \ldots
u_{16}) \rightarrow -(u_{8}, \ldots, u_1, u_{16}, \ldots u_9). 
\end{displaymath}
We have called this Weyl reflection $\theta_1$, because there is a
second $\Z_2$-symmetry that we wish to use in an orbifold construction.
Consider
\begin{equation}  
\theta_2: \quad (u_{1}, \ldots, u_{16}) \rightarrow -(u_{16},
\ldots, u_1).  
\end{equation}
Note that $\mathbf{a}_3$ and $\mathbf{a}_4$ are invariant under
$\theta_2$, which is an outer automorphism. Hence it is consistent to
divide this theory by a $\Z_2$-shift over $\pi R_2$ combined with
$\theta_2$. The resulting theory combines the CHL construction with the
$\Z_2$-triple. The rank of the gauge group is reduced by $12$. 

We may dualize, for example, in the $3$ direction to obtain a
$Spin(32)/\Z_2$ theory with $\mathbf{a}_3$ replaced by:
\begin{equation}
\mathbf{a}'_3= \left( \left( - \hlf \right)^2, \hlf^2, \left( - \hlf
\right)^2, \hlf^2, 0^8 \right).
\end{equation}
Note that $\mathbf{a}'_3$ is invariant under $\theta_1$, while
$\theta_2(\mathbf{a}'_3)-\mathbf{a}'_3$ lies on the spin weight
lattice. On the other hand $\mathbf{a}_4$ is invariant under
$\theta_2$, while $\theta_1(\mathbf{a}_4)-\mathbf{a}_4$ lies on the
spin weight lattice. The resulting theory is interpreted as a theory
without vector structure, which has $\tilde{w}^2_2 \neq 0$.
Notice that we have now encountered all three types of bundles
for $Spin(32)/\Z_2$ on $T^4$ that we met in section~2.1.4. 
They appear for respectively for $G=\Z_2$, $G=\Z_4$, and $G=\Z_2
\times \Z_2$. 

\subsubsection{Degeneration limits: connections to other models}
\label{hetdeg}

We have constructed moduli spaces for a number of asymmetric orbifold
theories. These moduli space are non-compact and the infinities
correspond to decompactification limits.
The moduli space of the $D=9$ CHL string is \cite{Mikhailov98},
\begin{equation}
O(\Gamma_{1,1} \oplus E_8)\, \backslash \,O(9,1) \, / \,
\left(O(9) \times O(1) \right).
\end{equation}
We have omitted a factor $\R^+$ corresponding to the expectation
value of the dilaton $\phi$.  Moving to the end points of $\R^+$
corresponds to the weak and strong-coupling limits of the theory.
Our preceeding discussion has been limited to weak coupling, where
$e^{\phi} \rightarrow 0$. There are various strong coupling limits
which depend on how the string scale is treated as
$e^{\phi} \rightarrow \infty$. At least in certain regions of the 
moduli space, there are M and F theory dual descriptions of the 
strongly-coupled heterotic string. We shall discuss some of these
duals in section four.

The remaining infinities correspond to decompactification limits. In
\cite{Mikhailov98}, it is shown that there exists only one light-like
vector in $\Gamma_{1,1} \oplus E_8$, modulo the symmetry group. One of
the elements of the symmetry groups inverts this vector.  Therefore,
there are two directions in which one can decompactify---roughly by
taking $R \rightarrow \infty$ or $R \rightarrow 0$. These two limits
appear to correspond to physically distinct theories, because the
orbifold projection involves a shift over the compactification circle,
and therefore explicitly involves $R$.

In the standard picture \cite{Chaudhuri95b} that we have also used so far, 
the winding numbers are fractionalized.
The masses of excitations from the twisted sector are then
offset by a contribution that is linear in $R$ for large $R$. As $R
\rightarrow \infty$, the states in the twisted sector become infinitely
heavy, and therefore decouple. In this limit, the shift
becomes irrelvant and so we expect to recover the conventional
ten-dimensional $E_8\times E_8$ heterotic string. 

On the other hand, in the
limit $R \rightarrow 0$, there are states in the twisted sector that
become massless. These massless states from the twisted sector
transform in the adjoint of $E_8$, and as $R$ goes to zero, these
enhance the gauge group to $E_8 \times E_8$ with one $E_8$ coming from
the twisted sector and one from the untwisted sector. This limit is
therefore also smooth. 

A perhaps more apropriate way to describe this is by T-dualizing,
taking $R \rightarrow \alpha'/R$. This also exchanges momenta and
winding numbers so that now we have a theory with fractionalized
momenta and integer winding numbers.  Fractionalized momenta are
common in theories with background gauge fields. This may seem unusual
for the CHL theory, but is an appropriate way to think about
compactifications without vector structure, triples, quadruples and
quintuples. Indeed, since the CHL outer automorphism is an orbifold
action, it is a discrete gauge symmetry. Associated to this discrete
gauge symmetry is a discrete gauge-field modulus which is
fractionalizing the momenta.

At least in one limit, the asymmetric orbifold, which is often deemed
non-geometric, has a natural interpretation in terms of a gauge
bundle, which is a geometric concept. It is an interesting observation
that asymmetric orbifolds can be interpreted in terms of bundles.
Combining such bundles with symmetric orbifolds may offer geometric
interpretations for at least a particular class of asymmetric
orbifolds.

Having settled this subtlety,  the rest of our discussion is parallel to
the last section of \cite{Mikhailov98}. Large volume limits can be
identified from our discussion of dualities. The $8$-dimensional
CHL string has various decompactification limits in which a two-torus
becomes large. One of these corresponds to the
$Spin(32)/\Z_2$ compactification without vector structure, and one to
the CHL string.  The $7$-dimensional CHL string has various limits which
we have identified as the CHL string, $Spin(32)/\Z_2$  without vector
structure and the $E_8 \times E_8$ $\Z_2$-triple.
This is in disagreement with \cite{Mikhailov98} where two $Spin(32)$
and one $E_8 \times E_8$ degenerations were found. That analysis, however, 
is far less concerned with identifying the theories that appear in these
limits.
The $6$-dimensional CHL string brings us one new limit where a
$4$-torus becomes large, which is identified as a quadruple, which has
vector structure. Another new limit is found in $5$-dimensions in terms
of a quintuple in one $E_8$. Also, in the cases of $\Z_4$ and $\Z_2
\times \Z_2$, we have identified the various limits.

The duality chains of this section connect a diverse set of
theories encountered later in this paper. Note in particular the
$Spin(32)/\Z_2$ theories in various chains. By S and T-dualities these
will translate into type I theories and type II orientifolds, which
make their appearance in section \ref{orient}.

We will make a brief excursion to four dimensions where it is believed
that the heterotic string theory is S-dual. In \cite{Mikhailov98},
this translates into a property of the Mikhailov lattice: namely, that
it is isomorphic to its dual lattice up to a rescaling by a factor of
$\sqrt{2}$. One way to heuristically understand the factor of
$\sqrt{2}$ is that it is related to the possible appearance of
non-simply-laced groups in CHL theories. For non-simply-laced groups
$G$, the gauge group in the S-dual theory is given by the dual group
$G_D$. The roots of $G_D$ are the coroots of $G$ up to a suitable
rescaling \cite{Goddard77}, while the weight lattice of $G_D$ is the
coweight lattice of $G$, also rescaled. The rescaling is precisely the
factor of $\sqrt{2}$ for the $2$-laced theories appearing in CHL
compactifications.

By analogy the asymmetric $\Z_m$-orbifolds with $m=3,4,5,6$ should
have lattices for their $4$-dimensional theories that are isomorphic
to their dual lattices up to rescaling by a factor of $\sqrt{m}$. The
reader may verify that using the lattices $\Lambda$ from table
\ref{table:lattices}, the lattices $\Gamma_{3,3}(m) \oplus \Lambda$
indeed have this property. This is again closely related to the fact
that the $F_4$, $G_2$ and $A_1$ at level $4$ are their own dual groups
up to rotations \cite{Goddard77}. It is amusing to see structure in a
theory compactified on a $6$-torus reflected in a similar theory
compactified on the $3$-torus.

Also the table \ref{table:canpoint} suggests that the $7$-dimensional
theory has special status because the groups and lattices listed there
for $(7+d)$ are the duals of those in $(7-d)$.  This is somehow
related to the $\Z_2$ nature of the duality operation, and to the fact
that $3 =6/2$, but a more precise understanding of this relation is in
order.

We also wish to briefly comment on other heterotic theories appearing
in the literature. In particular we note that $\Z_2$, $\Z_3$, $\Z_4$
$\Z_5$, $\Z_6$ and $\Z_2 \times \Z_2$ asymmetric orbifolds in
dimensions less than $7$ do appear in \cite{chaudlowe,
AspinwallFMA}. The models in these papers are similar to the
construction of \cite{sens}, and originate in exploiting symmetries of
K3. That heterotic duals should exist is obvious from the
constructions, but an explanation for their existence has been
lacking. The gauge theory based analysis presented here fills that
gap. It also makes clear that these models can be traced back to
constructions in higher dimensions. This again presents new challenges
for finding dual descriptions which we take up in the following
sections.

\subsubsection{A strong coupling description of the $\Z_2$ triple}
\label{strongcouplingtriple}

Let us conclude our discussion of heterotic string theory by pointing
out an intriguing relation between the $\Z_2$ triple and a discrete
$3$-form flux appearing in its strong coupling description.  

The type I string on $T^2$ without vector structure can be viewed 
as an orientifold of type IIB with a half-integral NS-NS $B$-field flux 
through the $T^2$ \cite{sagnotti, sensethi}. It is natural to ask whether 
the strong coupling Ho{\accent20r}ava--Witten description of the $E_8\times E_8$ 
heterotic string might permit a similar discrete flux. Consider M theory
on $S^1/\Z_2 \times T^3$. The $\Z_2$ action projects out the M theory
3-form $C$-field. After all, there are no membranes in heterotic string
theory. The component of the $C$-field with a leg on $S^1/\Z_2$, however,
survives projection and couples to the perturbative heterotic string. 

It is natural to ask whether we can turn on a half-integral $C$-field
on $T^3$ and then quotient by $\Z_2$. It is not clear that such a
compactification is consistent but the following chain of dualities
suggests that it exists and is an alternate description of the
$\Z_2$-triple. Let us perform a $9-11$ flip and reduce from M theory
to string theory on a circle of the $T^3$. This gives an orientifold
of IIA on $S^1/\Z_2 \times T^2$ with a half-integral $B$-field through
$T^2$.  A further T-duality on $S^1/\Z_2$ turns this compactification
into type I on $S^1\times T^2$ with a half-integral $B$-field. As we
recalled above, this is just type I with no vector structure which is
in the same moduli space as the $\Z_2$-triple. This suggests an
intimate connection between background 3-form fluxes and non-trivial
Chern--Simons invariants for the $E_8\times E_8$ gauge bundle.

\section{Orientifolds}
\label{orient}

\newcommand{\RP}{{\R \P}}

\subsection{Background and definitions}

We begin our discussion of orientifolds with some background and
some words on our notation.
We use ${\cal I}_{9-p}$ to denote the sign flip of the last $9-p$ spatial
coordinates of $\R^{10}$,
\bea
(x_0,\ldots,x_p,x_{p+1},\ldots,x_{9})\mapsto
(x_0,\ldots,x_p,-x_{p+1},\ldots,-x_{9}).
\eea
Type II string theory on $\R^{10}$ is invariant under the action of 
${\cal I}_{9-p}$ when combined with world-sheet orientation reversal $\Omega$,
where $p$ is even for type IIA and odd for type IIB.
As is standard, we use $(-1)^{F_L}$ to denote the symmetry that flips the sign of 
all R-NS and RR states.
It is not hard to check that, 
\bea
{\cal I}_{9-p}\Omega, 
\eea
is an involution, i.e.,  squares to the identity, for $p=0,1,4,5,8,9$
while,
\bea
{\cal I}_{9-p}(-1)^{F_L}\Omega,
\eea
is an involution for $p=7,6,3,2$.  We can then consider an orbifold of
Type II string theory by the $\Z_2$ symmetry group generated by this
involution.  This is called a Type II orientifold on $\R^{10}/{\cal
I}_{9-p}$, or just $\R^{10}/\Z_2$ when there is no room for confusion.
The $\Z_2$ fixed plane $x_{p+1}=\cdots=x_9=0$ is called an orientifold
$p$-plane, or $Op$ plane for short.  We can also extend this
construction to the case where $\R^{10}$ is replaced by a non-trivial
ten-manifold $M^{10}$ with an involution ${\cal I}$, as we will do in
the case of $T^{9-p}\times \R^{p+1}$ where ${\cal I}$ acts by
inversion on the $(9-p)$ coordinates of $T^{9-p}$.

There are two kinds of orientifold planes, $Op^-$ and $Op^+$, which
are distinguished by the sign of the closed string $\RP^2$ diagram
surrounding the plane---$Op^+$ has an extra $(-1)$ factor when
compared to $Op^-$.  When $N$ D$p$-branes are placed on top of $Op^-$
(resp.~$Op^+$), they support an $SO(N)$ (resp.~$USp(N)=Sp(N/2)$) gauge
group, where we count the number of D-branes on the double cover.  An
$Op$ plane carries D$p$-brane charge, and the two kinds of $Op$ planes
are also distinguished by the sign of this charge---$Op^{\pm}$
carries D$p$-brane charge $\pm 2^{p-5}$ when counted on the double
cover. The superscript $^+$ or $^-$ in the name of the plane has its
origin here.

An $Op^-$ plane together with an even number of D$p$-branes is quite
different in character from an $Op^-$ plane together with an odd
number of D$p$-branes. We distinguish these two cases by using the
notation $Op^{-'}$ for an $Op^{-}$ plane with a single D$p$-brane
stuck to it.  It has been shown that $Op^{-'}$ has a non-trivial
$\Z_2$ flux associated with the RR $(6-p)$-form $G_{6-p}$, while the
flux is trivial for $Op^-$.  It has also been shown that there are two
kinds of $Op^+$ planes distinguished by the same flux; we denote the
trivial one by $Op^+$ and the non-trivial one by $Op^{+'}$.  In total,
there are four kinds of orientifold planes:
\bea
Op^-,~Op^{-'},~Op^+,~Op^{+'}.
\eea
These four kinds of orientifold planes---especially the planes
$Op^{-'}$ and $Op^{+'}$ with non-trivial $\Z_2$ flux---have been
identified for the cases $p=5,4,3,2,1,0$ \cite{New,kh,baryon,ss,hkr,
berkkapustin, gimon}.  We refer to these references for a fuller
discussion of the $\Z_2$ flux associated with $G_{6-p}$.

\subsubsection{The closed string perturbation expansion}

As mentioned above, $Op^+$ and $Op^{+'}$ have an extra $(-1)$ sign
when compared to $Op^-$ and $Op^{-'}$ for the fundamental string
$\RP^2$ diagram.  The meaning of an `$\RP^2$ surrounding the plane' is
clear only if the transverse dimension $(9-p)$ is greater or equal to
$3$. We can actually make this statement more precise so that it also
applies to the cases $(9-p)=1,2$.  We note that the phase of the
closed string diagram is given by the $B$-field, and therefore it is
natural to use equivariant cohomology to determine the allowed
$B$-field configurations.  Let us consider the closed string
perturbation expansion for a type II orientifold on $M/{\cal I}$
following \cite{wittoroid}.  The world-sheet is an orientable Riemann
surface $\Sigma$ with orientation reversing freely-acting involution
${\cal I}_{\Sigma}$.  Note that the quotient $\Sigma/{\cal
I}_{\Sigma}$ is a smooth unorientable surface if $\Sigma$ is
connected. However, it is smooth and orientable but not oriented when
$\Sigma$ consists of two identical components exchanged by ${\cal
I}_{\Sigma}$.  The world-sheet path-integral is over maps
$\phi:\Sigma\to M$ which commute with the involution,
\bea
{\cal I}\circ \phi=\phi\circ {\cal I}_{\Sigma}.
\label{equivar}
\eea
We would like to assign a phase factor in a way that is
consistent with all physical consistency conditions, such as
factorization etc.
This can be accomplished in the following way.
Let us fix an element of the equivariant cohomology group,
\bea
y\in H_{\Z_2}^2(M,\Z_2).
\eea
This means choosing an element of
$H^2(M_{\Z_2},\Z_2)=H^2(S^N\times_{\Z_2} M,\Z_2)$ for sufficiently
large $N$, where $\Z_2$ acts on $S^N$ via the antipodal map and on
$M$ by ${\cal I}$.\footnote{ For a discussion of equivariant
cohomology, see appendix \ref{app:equiv}, and section \ref{RRone}.
Here, we provide a brief summary of some basic definitions.  The
equivariant cohomology $H^*_G(X,R)$, with $G$ a group acting on a
space $X$ and $R$ a coefficient ring, is defined to be the ordinary
cohomology $H^*(M_G,R)$ where $M_G$ is the fibre product
$$
M_G=EG\times_GM:=(EG\times M)/G.
$$
Here $EG$ is the universal $G$-bundle over the classifying space $BG$.
In the context of our orientifold discussion, 
$G=\Z_2=\{1,{\cal I}\}$ for $X=M$ and $R=\Z_2$. 

For $G=\Z_2$, we can take $EG=S^N$ so that $BG=\RP^N$ with
$N\to\infty$ strictly speaking.  However, it is enough to take large
but finite $N$ for most of the practical purposes.  If $G$ acts on $X$
freely (as is the case for $X=\Sigma$ and $G=\{1,{\cal
I}_{\Sigma}\}$), then $H_G^*(X,R)=H^*(X/G,R)$ since $EG$ is
contractible.}  The map $\phi$, obeying the condition (\ref{equivar}),
induces a map $\hat{\phi}:\Sigma_{\Z_2}\to M_{\Z_2}$. Therefore the
element $y\in H^2(M_{\Z_2},\Z_2)$ can be pulled back to an element
$\hat{\phi}^*y\in H^2(\Sigma_{\Z_2},\Z_2)$.  We denote this element
simply by
$$\phi^*y\in H_{\Z_2}^2(\Sigma,\Z_2).$$ Since ${\cal I}_{\Sigma}$ acts
on $\Sigma$ freely, $\phi^*y$ can be viewed as an element of
$H^2(\Sigma/{\cal I}_{\Sigma},\Z_2)$ which can be integrated over
$\Sigma/{\cal I}_{\Sigma}$:
\bea
\langle \Sigma/{\cal I}_{\Sigma},\phi^*y\rangle:=
\int\limits_{\Sigma/{\cal I}_{\Sigma}}\phi^*y~\in \Z_2.
\eea
We can then assign the following sign factor to the world-sheet path-integral, 
\bea
(-1)^{\langle \Sigma/{\cal I}_{\Sigma}, \phi^*y\rangle}.
\eea
Now, we can make precise the meaning of the sign of the $\RP^2$
diagram.  Suppose ${\cal I}$ acting on $M$ has a fixed point $P$.  The
constant map $\phi_0$ from $S^2 \to P$ trivially obeys condition
(\ref{equivar}).  The map $\hat{\phi}_0:S^2_{\Z_2}\to M_{\Z_2}$ then
collapses to the identity map $\RP^N\to \RP^N$ and the integral
$\langle \RP^2,\phi_0^*y\rangle$ is simply evaluating $y$ on the
$2$-cycle $S^2\times_{\Z_2} \{P\}\cong \RP^2$ of
$M_{\Z_2}=S^N\times_{\Z_2}M$.  This definition agrees with the
standard one when ${\cal I}$ acts around $P$ by the action ${\cal
I}_{9-p}$ with $(9-p)\geq 3$.  To see this, first consider the space
$M^{\circ}$ obtained from $M$ by deleting all the $\Z_2$ fixed points.
By restriction, $y\in H^2_{\Z_2}(M,\Z_2)$ yields an element
$y^{\circ}\in H^2_{\Z_2}(M^{\circ},\Z_2)$. Since $\Z_2$ acts on
$M^{\circ}$ freely, $y^{\circ}$ can be viewed as an element of
$H^2(M^{\circ}/{\cal I},\Z_2)$.  We now note that the map
$\hat{\phi}_0:S^2_{\Z_2}\to M_{\Z_2}$ can be continuously deformed to
a map $\hat{\phi}_{\epsilon}$ where $\phi_{\epsilon}$ is a map of
$S^2$ to a small $\Z_2$ invariant $2$-sphere surrounding $P$ and lying
in $M^{\circ}$.  Or equivalently, a map from $\RP^2$ to a small
$\RP^2$ surrounding $P$ and lying in $M^{\circ}/{\cal I}$.  It is then
clear that,
\bea
\langle \RP^2, \phi_0^*y\rangle=
\langle \RP^2, \phi_{\epsilon}^*y\rangle
=\int\limits_{\RP^2} \phi_{\epsilon}^*y^{\circ},
\eea
which is the standard meaning of the sign of the $\RP^2$ diagram surrounding $P$.

To summarize, {\it the possible configurations of plus ($+$ or $+'$)
versus minus ($-$ or $-'$) orientifolds can be classified from the
closed string perspective by the group $H_{\Z_2}^2(M,\Z_2)$.}  As we
shall see below, there can be further constraints.

\subsubsection{Some pathologies}

We note that both $-'$ and $+'$ orientifolds are inconsistent for
$p=9,8,7,6$ if we insist, as we shall, on backgrounds preserving
sixteen supersymmetries.  There are a number of ways to see this.  For
example, for $p=6$ we can probe the $O6^{-'}$ with a D$2$-brane. The
theory on the probe is an $N=4$ $Sp(1)$ gauge theory with a
half-hypermultiplet in the fundamental representation.  This
three-dimensional gauge theory has a $\Z_2$ anomaly, and is therefore
inconsistent. The only way to cancel such an anomaly is by including a
Chern--Simons term which makes the gauge-field massive.  However,
Chern--Simons terms can only be written down for $N\leq 3$
supersymmetry in $d=3$ \cite{klee}; $N=4$ supersymmetry does not allow
a short massive vector multiplet.  Rather, there are arguments given
in \cite{imamura} which suggest that $O6^{-'}$ planes require a
non-zero cosmological constant, and so can only be found in massive
type IIA supergravity.  Similarly, if $O6^{+'}$ is distinguished from
$O6^+$ by a flux associated with $G_0$, it is expected to be in a
theory with a cosmological constant.

In our subsequent discussion, we shall classify $-$ and $-'$
configurations by using the fact that we can T-dualize these
configurations to type I. The type I requirement that the $O(32)$
gauge bundle lift to a $Spin(32)/\Z_2$ bundle with vanishing
Chern--Simons invariant constrains the possible $-$ and $-'$
configurations.  This actually gives an alternative and stronger
argument for excluding $O6^{-'}$ in a $T^3/\Z_2$ orientifold.  The
details of the derivation are found in appendix \ref{app:orient}.
This anomaly cancellation argument and the D$2$-brane probe brane
argument are not completely independent. Suppose that we na\"{\i}vely
construct the $T^3/\Z_2$ orientifold with 8 $O6^{-'}$
planes. Somewhere, we should encounter an inconsistency.  {}From the
chain of dualities connecting this type IIA orientifold to the
heterotic string on $T^3$, we find a hint of where to look.  In the
heterotic theory, we argued that obstructions to certain
compactifications arise from states associated to winding strings.
After S-duality, these correspond to winding D-strings in a type I
compactification. The states on these winding strings transform
non-trivially under a $\Z_2$ that should be divided out in the
construction of the bundle. As a consequence, the phases of these
states have a $\Z_2$ ambiguity, and the theory is not
well-defined. Nevertheless, proceeding with this inconsistent theory
and T-dualizing $3$ directions leads to the $T^3/\Z_2$ orientifold
with 8 $O6^{-'}$ planes. In the process, the D-string becomes a
D$2$-brane, and the $\Z_2$ ambiguity in the phases of winding states
becomes the $\Z_2$ anomaly in the gauge theory on the probe brane. The
anomaly therefore has multiple dual realizations.

\subsection{The classification}

Our discussion below focuses on type II orientifolds of $T^{9-p}\times
\R^{D=p+1}$.  We note that there are $2^{9-p}$ $\Z_2$ fixed points,
and the type of orientifold plane has to be specified at each fixed
point.  We use the notation
$$\left\{-,-',+,+'\right\}$$ for $Op^-$, $Op^{-'}$, $Op^+$ and
$Op^{+'}$, respectively.  For example, the notation $(-,+,+,-')$
specifies a (possible) configuration in eight dimensions, or
equivalently, a particular orientifold of $T^2$.  For low enough
dimensions, ordering is also required to completely specify the
configuration.

In what follows we provide the complete list of possible orientifolds
in dimensions $D:=p+1=10,9,8,7$. We also give the complete list of
orientifolds involving only $-$ and $-'$ in dimensions $D=6,5$.  In
$D=6,5$ we do not attempt to completely classify configurations
involving $+$ and $+'$ but we will comment on some of the new issues
that arise.  To see what kind of $\{+,+'\}$ configurations are
possible from the closed string point of view, we have included an
explicit computation of the equivariant cohomology
$H_{\Z_2}^2(T^n,\Z_2)$ for $n=1,2,3$ in appendix \ref{app:equiv}.  The
results are stated in the main text below.

\subsubsection{$D=10$}

The equivariant cohomology $H_{\Z_2}^2(\R^{10},\Z_2)$ is the
cohomology of the classifying space $B\Z_2=\RP^{\infty}$ which is
$H^2(\RP^{\infty},\Z_2)=\Z_2$.  The two choices correspond to $O9^-$
and $O9^+$.  $O9^-$ has D$9$-brane charge $-32$ and can be cancelled
by introducing $32$ D9-branes.  Of course, this is the standard type I
string on $\R^{10}$ which indeed preserves $16$
supersymmetries. However, $O9^+$ has positive D$9$-brane charge which
is cancelled by introducing anti-D$9$-branes.  Since $O9^+$ and the
anti-D$9$-branes preserve different supersymmetries, the system is not
supersymmetric.

\subsubsection{$D=9$}

The equivariant cohomology has rank $2$,
$H^2_{\Z_2}(T^1,\Z_2)=\Z_2\oplus \Z_2$.  The two generators are
non-zero on $\RP^2$ at the two fixed points of $S^1$ in the sense
described above.  The four elements correspond to
$$(-,-), \, (+,-), \, (-,+), \, (+,+).$$ The configuration $(+,+)$ is
not supersymmetric for the same reason that $O9^+$ is not
supersymmetric. The two configurations $(+,-)$ and $(-,+)$ are related
by a diffeomorphism.  Therefore in $9$ dimensions, there are
essentially $2$ possible orientifold configurations.  The first,
$(-,-)$, has $32$ D$8$-branes and is simply T-dual to type I
compactified on a circle.  This component also contains the $E_8\times
E_8$ string \cite{ginsparg}.

The second component, $(-,+)$, has no net D-brane charge. Therefore no
branes can be added while preserving supersymmetry so there is no
enhanced gauge symmetry. This compactification is T-dual to type IIB
on $S^1/\delta\Omega$ where $\delta$ is a half-shift along the circle
and $\Omega$ is world-sheet parity.  For a more detailed discussion,
see \cite{park, wittoroid}.  Together with the CHL string, this gives
a total of $3$ distinct components in the moduli space of perturbative
string compactifications.

\subsubsection{$D=8$}

The equivariant cohomology now has rank $4$,
$H_{\Z_2}^2(T^2,\Z_2)=(\Z_2)^{ 4}$.  The four generators are in
one-to-one correspondence with the four $\Z_2$ fixed points---each
has a non-trivial value on $\RP^2$ at the corrsponding fixed point and
is trivial at the other three points.  The possible orientifolds up to
diffeomorphisms are
$$(-,-,-,-), \quad (+,-,-,-), \quad (+,+,-,-).$$ 
The last is obtained from the sum of two generators. In addition, there
are non-super\-symmetric configurations.

The first component $(-,-,-,-)$ has $32$ D7-branes and is T-dual to
the usual type I string on $T^2$.  The second component $(-,-,-,+)$ is
more interesting. The orientifold planes have total of $-16$ units of
D$7$-brane charge and therefore require $16$ D$7$-branes. This
orientifold is T-dual to type IIB on $T^2$ modded out by the
world-sheet parity, $\Omega$, in the presence of a half-integral
background NS-NS $B$-field \cite{sagnotti, sensethi}.  Although
world-sheet parity sends $B\to -B$, a half-integral value of $B$ is
permitted since $B$ takes values in a torus. Orientifolds with
quantized $B$-field backgrounds have been explored, for example,
in~\cite{Angelantonj:1999jh, Angelantonj:1999xf}. This orientifold is
equivalent to a compactification without vector structure
\cite{sensethi}. By the T-duality argument of section
\ref{hetduality}, we know that this component containing type I
without vector structure also contains the $8$-dimensional CHL string.

The third and final component $(-,-,+,+)$ is T-dual to the
compactification of the $9$-dimensional $(-,+)$ orientifold to eight
dimensions. This has no D-branes and therefore no gauge symmetry.

\subsubsection{$D=7$}
\label{orientpuzzle}

The equivariant cohomology has rank $7$,
$H^2_{\Z_2}(T^3,\Z_2)=(\Z_2)^{ 7}$.  The seven generators
$y_1,\ldots,y_7$ having the following property: let us use
$P_1,\ldots,P_8$ to denote the eight $\Z_2$ fixed points of $T^3$.
Then $y_i$ has a non-trivial value on both the $\RP^2$ surrounding
$P_i$ and the one surrounding $P_8$, while it is trivial at the other
six points.  This shows that the allowed orientifolds are $(-^8)$,
$(-^6,+^2)$ and its permutations, and $(-^4,+^4)$ and its
permutations.  We exclude those cases with more $+$ than $-$ planes
since they require anti-D$6$-branes.  That an odd number of $O6^+$
planes is not allowed can also be understood by an elementary
argument.  On $T^3/\Z_2$, we can enclose each $O6$ plane with an
$\RP^2$.  The eight of them together correspond to a trivial cycle,
and therefore the product of the eight $\RP^2$ diagrams must have sign
$+1$.  Therefore, the number of $O6^+$ planes must be even.

The first case, $(-^8)$, has $32$ D$6$-branes.  It is T-dual to the
standard compactification of type I on $T^3$.  The second case,
$(-^6,+^2)$, requires $16$ D$6$-branes.  All permutations are
diffeomorphic to each other so there is essentially one configuration
of this type.  This is obtained from the $(-,-,-,+)$ orientifold on
$T^3/\Z_2$ by compactification along another circle and T-duality on
that circle.

The third case $(-^4,+^4)$ has no D$6$-branes and therefore no open
strings.  In this case, there is an interesting subtlety.  The various
permutations are not necessarily diffeomorphic to each other. There
are essentially $2$ distinct ways to place the orientifold planes at
the vertices of the cube.  The first can be characterized by placing
the four $-$ orientifold planes on a single face of the cube. The
remaining four $+$ planes are found at the remaining vertices. This is
the configuration that follows from a dimensional reduction of the
$(-,-,+,+)$ case. It is therefore automatically consistent.  However,
we could also consider the case where a single adjacent pair of $+$
and $-$ planes are interchanged. This gives a distinct configuration.
For a fixed ordering of the $\Z_2$ fixed plane, let us denote these
$2$ cases by $(-,-,-,-,+,+,+,+)$ and by $(-,-,-,+,+,+,+,-)$.  Both
possibilities are allowed from the closed string point of view.  Both
are realized as elements of $H^2_{\Z_2}(T^3,\Z_2)$. This is easy to
see because the equivariant cohomology is a group. Noting that all
permutations of $(-,-,-,-,-,-,+,+)$ are realized as the elements of
the group, we see that the first case is simply the sum of
$(-,-,-,-,+,+,-,-)$ and $(-,-,-,-,-,-,+,+)$.  The second possibility
is the sum of $(-,-,-,+,+,-,-,-)$ and $(-,-,-,-,-,+,+,-)$.  Thus, both
are consistent configurations in perturbative closed string theory and
they are distinct.  Although the two are distinct configurations
perturbatively, it is possible that they are equivalent
non-perturbatively (if they are both consistent).  We shall mention
how this possibility can be checked in section \ref{threeformphysics}
on M theory compactifications with flux.  This subtlety involving the
ordering of orientifold planes also appears for lower-dimensional
orientifold configurations.

Therefore, we have a total of $4$ distinct orientifold
compactifications, and an additional $4$ components in the $E_8\times
E_8$ string moduli space. The standard $E_8\times E_8$
compactification together with the CHL string/$\Z_2$-triple are
contained in the orientifold moduli spaces.

\subsubsection{$D=6$}

In this dimension, all four kinds of orientifold planes are possible,
and indeed each is realized in a particular $T^4/\Z_2$ orientifold.
Let us first restrict our attention to $O5^-$ and $O5^{-'}$ planes
only. With this restriction, there are only two possibilities: one is
$(-^{16})$ and the other is $({-'}^{16})$.  The latter case is quite
interesting. If we coalesce the $8$ D-brane pairs on one of the
$O^{-'}$ planes, we get an $SO(17)$ maximal gauge group. The rank
reduction is therefore $8$. As discussed in appendix \ref{app:orient},
the Chern--Simons invariant for any sub-three-torus is integer as required
by anomaly cancellation. However, this compactification clearly has
vector structure.  The type I dual of this orientifold
compactification therefore has a gauge bundle with vector structure
which is not connected to the trivial bundle. We met this bundle in
section \ref{four-torus}, it is a non-trivial quadruple. Here we have
found its orientifold realization.

In addition to these two cases, we have the dimensional reductions of
the higher-dimensional cases which are $( -^{12}, +^4)$ and $(-^8,
+^8)$. Recall that the first case includes the CHL string, the
compactification with no vector structure and the $\Z_2$ triple in its
moduli space. However, the duality chain of section \ref{hetduality}
showed that the quadruple found above is in the same moduli space as
the CHL string on $T^4$. Somewhat surprisingly, this implies that the
$( -^{12}, +^4)$ and $({-'}^{16})$ orientifolds are in the same moduli
space. We can give further motivation for this inference in a simple
way.  Compactify both configurations to four dimensions on $T^2$. This
gives $( -^{48}, +^{16})$ and $({-'}^{16}, -^{48})$. S-duality maps $
O^{-'} \leftrightarrow O^+$ and $O^{-}$ to itself.  These
configurations are therefore S-dual and their moduli spaces must
necessarily agree.

Unlike the prior cases, we will not attempt a complete analysis of
orientifold configurations in $D=6$. There are two other cases worth
mentioning, however.  The first is the case of $(-^{10}, +^{6})$ which
made an appearance in \cite{wittoroid}. This orientifold corresponds
to a gauge bundle with $\widetilde{w}_2^2$ non-zero. The last case is
$({+'}^{4}, {-'}^{12})$.  In this case, we need an additional $2$
pairs of D-branes. The maximal gauge group is then $Sp(2)$ and the
rank reduction is $14$. Based on the structure of its moduli space, it
is natural to conjecture that this orientifold is dual to the
quadruple with no vector structure heterotic/type I compactification
described in section \ref{four-torus}.  It would certainly be
interesting to analyze this case further.

\subsubsection{$D=5$}
\label{plusminusorient}

In this dimension, all four flavors of orientifold plane can again be
realized by $T^5/\Z_2$ orientifolds.  If we restrict our attention to
$O4^-$ and $O4^{-'}$, there are only three possibilities up to
diffeomorphisms: $(-^{32})$, $({-'}^{32})$ and $(-^{16},{-'}^{16})$.
We need a few additional words to actually describe these
orientifolds.  On a torus of sufficiently high dimension, the number
of $-$ and $-'$ planes does not in general completely specify the
configuration up to diffeomorphisms.  The actual pattern of the
distribution must be specified.  There is no room for such an
ambiguity for $(-^{32})$ and $({-'}^{32})$ but there are several
possibilities for $(-^{16},{-'}^{16})$.  In this case, the only
allowed configuration corresponds to the one where the $T^5$ can be
factorized into $S^1\times T^4$ so that all the $O4^-$ planes sit at
the $16$ fixed points at the ``origin'' ($\theta=0$ in natural
coordinates) of the $S^1$. All the $O4^{-'}$ planes reside at the $16$
fixed points in the ``middle'' of the $S^1$ ($\theta=\pi$).

We can find M theory duals for each of these cases.  Recall that a
single $O4^-$ plane at the origin of $\R^5/\Z_2$ is dual to M theory
on $\R^5/\Z_2\times S^1$. Note that the $\Z_2$ action acts on the
3-form $C$ of M theory by inversion.  On the other hand, $O4^{-'}$ is
dual to M theory on $(\R^5\times S^1)/\Z_2$ where the $\Z_2$ acts on
the last circle by a shift of a half period \cite{kh}.  The dual of
$(-^{32})$ is then M theory on $T^5/\Z_2\times S^1$ as we would
naturally expect.  The dual of $({-'}^{32})$ is M theory on
$(T^5\times S^1)/\Z_2$.  The dual of $(-^{16},{-'}^{16})$ is given by
M theory on $(T^5\times S^1)/\Z_2$. The $\Z_2$ now acts on the
coordinates $x_1,x_2,x_3,x_4,x_5,x_{11}$ of $T^5\times S^1$, where
$x_i\equiv x_i+2\pi$, by
\begin{equation}
(x_1,x_2,x_3,x_4,x_5,x_{11})
\longrightarrow
(-x_1,-x_2,-x_3,-x_4,-x_5,x_{11}+x_1).
\end{equation}
In the neighborhood of the ``origin'' $x_1=0$ of the first circle, the
$\Z_2$ action is of the type $\R^5/\Z_2\times S^1$ and the O4-planes
are all $-$.  In the neighborhood of the ``midpoint,'' $x_1=\pi$, the
$\Z_2$ action is of the type $(\R^5\times S^1)/\Z_2$ and indeed the
O4-planes are all $-'$.  We note that we would not be able to
construct an M theory dual if other distributions of 16 $-$ and 16
$-'$ planes were permitted.

As in the $D=6$ case, we shall only discuss select additional
examples.  From dimensional reduction, we obtain $(-^{24}, +^8)$ and
$(-^{16}, {-'}^{16})$. However, because of the duality explained in
$D=6$, these should be part of the same moduli space.  In addition, we
have $(+^{16}, -^{16})$ and $(-^{20}, +^{12})$. Note that
$({-'}^{32})$ has no enhanced gauge symmetry so the rank reduction is
$16$. Further, it does admit vector particles.  This suggests that it
is dual to type I with a non-trivial quintuple. This particular bundle
made an appearance in section \ref{five-torus}. By the chain of
dualities in section \ref{hetduality}, we see that this orientifold is
further equivalent to a compactification of the $E_8$ string with a
quintuple in both $E_8$ factors.

At first sight, it also seems plausible that $({-'}^{32})$ could be
identified with $(+^{16}, -^{16})$. By $(+^{16}, -^{16})$, we mean the
configuration obtained by toroidally compactifying $(+,-)$ in nine
dimensions. As support for this conjecture, note that on
compactification to four dimensions, we find two configurations
$(-^{32}, {-'}^{32})$ and $(-^{32}, {+}^{32})$ which are S-dual. Of
course, this alone does not demonstrate the equivalence. We can also
study the M theory description of $(+^{16}, -^{16})$.  Recall that the
$D=9$ $(+,-)$ orientifold is described, after T-duality, by IIB on
$S^1/\delta \Omega$ where $\delta$ is a half-shift along the circle
\cite{wittoroid, park}. Let us compactify this configuration on a
further $T^4$. We want to determine the corresponding M theory
description.  It is convenient to first compactify on one additional
circle $S^1$. We can then T-dualize five times on the $T^5$ which
leaves us in type IIA sending \bea \delta \, \Omega \quad \rightarrow
\quad \delta \, \Omega \Z_2., \eea where the $\Z_2$ acts by inversion
on the $T^5$.  The operation $\Omega \Z_2$ lifts in M theory to
inversion of the $T^5$ and the 3-form $C$ \cite{park}.  This leaves us
with M theory on $(T^5 \times S^1)/\Z_2 \times S^1$ where the $\Z_2$
acts as $\delta$ on the $S^1$ factor and by inversion on $T^5$.  It
also inverts the 3-form $C$.  This suggests that the M theory
description of $( +^{16}, -^{16})$ is the same as the description of
$({-'}^{32})$ which is further evidence in favor of their equivalence.

\section{Compactifications of M and F Theory}
\label{geometry}
\subsection{Some preliminary comments}

In prior sections, we have discussed aspects of perturbative string
compactifications: either heterotic/type I or type II
orientifolds. These descriptions are valid when the string coupling
constant is small, regardless of the size of the compactification
space. It is natural to ask what kind of description is valid when the
string coupling constant is large. The answer to this question depends
on how we treat the string scale $\alpha'$ and the volume of the
compactification space as $ g_s \rightarrow \infty.$ As an example,
let us take the CHL string in $9$ dimensions. If we wish to hold the
11-dimensional Planck scale fixed then $\alpha' g_s^{2/3}$ must be
held constant.  In this limit, the strong coupling description will
involve M theory compactified on a space which has been argued to be
the M\"obius strip \cite{Park:1998it}. This is analogous to the
Ho{\accent20r}ava--Witten description of the strongly-coupled $E_8\times E_8$
heterotic string.

On the other hand, another strong coupling description of the
$E_8\times E_8$ string on $T^2$ is given by F theory on $K3$. When is
this a valid description? Like M theory, F theory generically has no
perturbative expansion and the condition for validity is that the base
$B$ of the elliptic fibration (in this case $K3$) be large in string
units.  It is convenient to analyze this relation at the point in the
moduli space where the gauge group is broken to $(Spin(8)^4\times
U(1)^4)/\Z_2^3$. Let the 
torus be square with volume $V$, and let $g_H$ denote the heterotic
string coupling. T-duality along one cycle of the torus takes us to
the $Spin(32)/\Z_2$ heterotic string with ten-dimensional coupling
$(g_{H}')^2 = g_{H}^2 \alpha ' /V $ on a torus of volume $\alpha'.$
S-duality then takes us to the type I string with, 
\bea 
g_{I}^2 = \frac{V}{g_H^2 \alpha'}, \qquad 
V_{I} = \frac{V}{g_H^2}.  
\eea 
Two further T-dualities on the resulting torus take us to the type IIB
on $T^2/\Omega (-1)^{F_L} \Z_2$, with couplings: 
\bea 
g_B^2 = \frac{\alpha' g_H^2}{V}, \qquad V_{B} = \frac{(\alpha')^2
g_H^2}{V}.  
\eea
This is an orientifold limit of F theory on $K3$ \cite{Sen:1996vd}. We
see that F theory is a good description when $g_H$ becomes large with
the volume $V$ fixed in string units. In this regime of the moduli
space, we can use F theory to describe the physics.

A more general statement goes as follows. With sixteen
supersymmetries, each component of the moduli space is highly
constrained; for example, the moduli space metric does not receive
quantum corrections.  If the effective theory is formulated in $d$
space-time dimensions, then a given component of the moduli space can
be described in terms of an even lattice $L$ of signature
\[(s+10-d,10-d).\]
(We use the convention in which there are more positive than negative
eigenvalues in the lattice.)  When $d > 4$, this space takes the form
\begin{equation}\label{eq:component}
\mathcal M_L:=O(L)\backslash \mathcal{D}_L \times \mathbb{R}^+
\end{equation}
where $\mathcal{D}_L=O(s+10-d,10-d)/(O(s+10-d)\times O(10-d))$ is the
symmetric space associated to the lattice, and $O(L)$ is the
orthogonal group.

This must be modified somewhat in low dimension: when $d=4$, the
universal cover of the given component of the moduli space is
$\mathcal{D}_L\times \mathfrak{h}$ where $\mathfrak{h}$ is the upper
half plane, and when $d=3$, the universal cover of the given component
is $\mathcal{D}_{\widetilde{L}}$, where
$\widetilde{L}$ is the direct sum of $L$ and a lattice of signature
$(1,1)$. 

A given moduli space has boundaries that correspond to the various
ways in which a theory can degenerate. A different physical
description is typically valid as we approach a boundary of the moduli
space.  Our goal in this section is to study a class of M and F theory
compactifications which naturally include dual descriptions for the
perturbative compactifications described in earlier sections.  These
include purely geometric models and also models with background
fluxes, as in the case studied by Schwarz and Sen \cite{sens}.

We now require a more general discussion of the boundaries of moduli spaces 
than appeared in our initial discussion of section \ref{hetdeg}.
The possible boundary components of 
$\mathcal M_L$ are determined as follows (setting $d\ge5$ for
simplicity).  One type of boundary is given by approaching one end or
the other of the $\mathbb R^+$ factor.  These include non-stringy
limits: for example, in a conventional heterotic or type II
compactification one of these limits is the zero-coupling limit which
yields a conformal field theory rather than a string theory. This is
the kind of description that we studied in prior sections. As
discussed above, the strong coupling limit will typically have an M or
F theory description. Let us here instead focus on the other class of
limits, given by boundary components of the $O(L)\backslash
\mathcal{D}_L$ factor.  These boundary components typically correspond
to a limiting stringy theory whose effective dimension is greater than $d$.

A boundary component of $O(L)\backslash \mathcal{D}_L$ is determined
by an {\it isotropic sublattice}\/ $M\subset L$, that is, every $x\in M$
satisfies $q(x)=0$.  The lattice associated to the boundary component is then
given by $L_M:=M^\perp/M$,
and the boundary component takes the form
\[O(L_M)\backslash \mathcal{D}_{L_M}
,\] 
where we suppress the $\mathbb R^+$ factor.  To determine all
boundary components, all isotropic sublattices $M$ must be
found, modulo the action of the orthogonal group $O(L)$.
If the sublattice $M$ has rank $m$, the limiting theory will have effective
dimension $d+m$.

In general, $L_M$ only determines {\it part of}\/ a component of the
moduli space of the limiting theory in effective dimension $d+m$.
That is (suppressing the $\mathbb R^+$ factors), the space
$O(L)\backslash\mathcal D_L$ is glued to a space $O(L')\backslash
\mathcal D_{L'}$ along the boundary component $O(L_M)\backslash
\mathcal D_{L_M}$, where $L'$ is a lattice of signature
$(s'+10-(d+m),10-(d+m))$ which represents the limiting effective
theory.  The gluing is specified by an inclusion $L_M\subset L'$ with
$L'/L_M$ a positive definite lattice of rank $s'-s\ge0$.  The lattice
$L'$ must be determined by analyzing the physics of the limiting
process; it agrees with $L_M$ for some boundary components but is
larger than $L_M$ for others.

For example, the CHL string in nine dimensions has lattice
\cite{Mikhailov98} $L\cong \Gamma_{1,1}\oplus E_8$.  There is a unique
boundary component, corresponding to $L_x\cong E_8$.  However, as we
discussed in section \ref{hetdeg}, in the decompactification limit we
actually obtain the heterotic string in ten dimensions with lattice
$L'\cong E_8\oplus E_8$.

Decompactification limits of components corresponding to non-trivial
discrete choices of Wilson lines exhibit a similar phenomenon: viewed
from the perspective of the higher-dimensional theory, the non-trivial
types of Wilson lines can only be turned on for special values of the
moduli.  Thus, it is along a subspace $O(L_M)\backslash \mathcal
D_{L_M}$ of $O(L')\backslash\mathcal D_{L'}$ that the moduli space of
the lower dimensional theory ``attaches,'' and one recovers additional
degrees of freedom in the decompactification limit.

A similar phenomenon occurs in F theory \cite{Morrison:WilsonFtheory},
where compactification along an additional circle gives theories which
are dual to M theory on elliptically fibered manifolds.  The discrete
Wilson line degrees of freedom correspond to the possibility of
compactifying M theory on an elliptically fibered manifold without a
section, which typically is only possible for special values of the
moduli.

\subsection{Six-dimensional M theory compactifications without fluxes}

Our starting point is M theory compactified to $6$ dimensions, but we
will also include remarks about lower-dimensional
compactifications.\footnote{The corresponding type IIA string 
compactifications were studied in detail in \cite{chaudlowe,AspinwallFMA}.} 
Let us begin by excluding any background flux so that this is a purely
geometric compactification. We shall also restrict to supersymmetric
compactifications which take the form of a compact Riemannian manifold
times Minkowski space.  The metric on the compact part must then admit
a covariantly constant spinor, which leads to restrictions on the holonomy.
In fact, the list of possibilities can be determined by examining
the holonomy classification of Riemannian metrics, which (when
formulated carefully \cite{Besse}\footnote{We thank B. McInnes for
helpful correspondence on this point.}) implies that every compact Riemannian
manifold $Y$ admitting a covariantly constant spinor takes the form
\begin{equation}\label{eq:decomp}
Y= (T^k \times (X_1 \times \cdots \times X_m)/\Gamma)/G ,
\end{equation}
where $T^k$ is a torus of dimension $k\ge0$, each $X_i$ is a compact
simply-connected
Riemannian manifold whose holonomy is either $SU(n_i)$, 
$Sp(n_i)$, $G_2$, or $Spin(7)$, 
and $\Gamma$ and $G$ are finite groups which act without fixed points.  The
effective dimension of the physical theory is $d=11-\dim Y$.

In order to guarantee at least $16$ supercharges in the effective
theory, there must be at least half as many holonomy-invariant spinors
on this manifold as there are on flat space.  Because each $X_i$ in
eq.~\eqref{eq:decomp} reduces the set of holonomy-invariant spinors by
at least a factor of two, and the factor is greater than two except in
the case of holonomy $SU(2)=Sp(1)$, there are two cases: either (1) there
is a single $X_i$ with holonomy $SU(2)=Sp(1)$ (i.e., a K3 surface),
the group $\Gamma$ is trivial,
and the group $G$ preserves all of the spinors on $T^k\times X$, or
(2) there is no $X_i$ at all (and hence no $\Gamma$)
and the group $G$ preserves one-half of the
spinors on $T^k$.  In the second case, possibly after replacing the
torus by a finite cover or a finite quotient, we can assume that
\cite{fujiki} $T^k = T^4 \times T^{k-4}$ with the group action
preserving the $1$-forms on $T^{k-4}$ and the holomorphic $2$-form on
$T^4$, but leaving no invariant $1$-forms on $T^4$.  To get the
correct holonomy, the image of $G$ in $SO(k)$ must lie in an $SU(2)$
subgroup corresponding to a complex structure on the $T^4$ factor.  In
both cases, then, we can write $Y=(T^{\ell}\times Z)/G$ where $Z$ is
either a K3 surface or a complex $2$-torus (that is, a real $4$-torus
on which a complex structure has been specified).

The lattice $L$ can be directly determined from the cohomology of $Y$.
When $d>4$, the possible gauge charges for the theory are described by
\begin{equation}\label{eq:charges}
H^1(Y,\mathbb{Z}) \oplus H^2(Y,\mathbb{Z}) \oplus H^5(Y,\mathbb{Z}),
\end{equation}
(with the first factor coming from Kaluza--Klein modes, and the latter
two coming from the M theory three-form and its dual six-form).  This
cohomology group comes equipped with a natural quadratic form, to be
described below.  Bearing in mind the sign conventions, we can
identify the free part of eq.~\eqref{eq:charges} with the lattice $L(-1)$
if $d>5$. (There is also the possibility of torsion in
eq.~\eqref{eq:charges}, which we will not explore in any detail.)  When
$d=5$, the free part of the gauge  
lattice in eq.~\eqref{eq:charges} takes the form $L(-1)\oplus \langle
x\rangle$
with $q(x)=0$; the element $x$ is unique up to $\pm1$, so $L(-1)$ can
be recovered by modding out the free part of the gauge lattice by the span
of $x$. 
When $d=4$, the free part of the gauge lattice is simply
\begin{equation}\label{eq:fourD}
H^1(Y,\mathbb{Z})\oplus H^2(Y,\mathbb{Z})
\end{equation}
due to the self-duality of gauge fields in this dimension, and this
coincides with $L(-1)$.

The description of the quadratic form on $L$ depends on the dimension
$d$ of the effective theory.  If $d=6$, both $H^1(Y)$ and $H^5(Y)$ are
$1$-dimensional and this part of the lattice is isomorphic to
$\Gamma_{1,1}$.  The quadratic form on $H^2(Y,\mathbb
Z)/\text{torsion}$ is inherited from the intersection form on the
resolution $\widetilde{Z/G}$ of $Z/G$ via the isomorphism
\begin{equation}
\begin{aligned}
H^2(Y,\mathbb Z)/\text{torsion}&\cong H^2(Z/G,\mathbb Z)/\text{torsion}\\
&\cong \text{ the orthogonal complement of the }\\
&\hphantom{\cong} \quad
\text{exceptional divisors in }
H^2(\widetilde{Z/G},\mathbb Z).
\end{aligned}
\end{equation}
Thus, although the action of $G$ on $S^1\times Z$ has no fixed points
and a smooth quotient, keeping track of the singular points on $Z/G$
provides a convenient bookkeeping device for analyzing the lattice
associated to $(S^1\times Z)/G$.  There is one subtlety associated to
this, however.  If $E\subset H^2(\widetilde{Z/G},\mathbb Z)$ denotes
the lattice spanned by the exceptional divisors, then
$(E^\perp)^\perp$ will be larger than $E$: there are $\mathbb
Q$-linear combinations of exceptional divisors which belong to
$H^2(\widetilde{Z/G},\mathbb Z)$.  In fact, the finite group
$(E^\perp)^\perp/E$ provides another important invariant in this
situation.

\begin{table}[t]
\begin{center}
\renewcommand{\arraystretch}{1.5}
\begin{tabular}{|c|l|c|} \hline
&&Maximum dimension\\[-2mm]
$k$&$G$& of effective theory\\ \hline
$0$&$\{e\}$&$7$\\
$1$&$\mathbb{Z}_m$, $m=2, 3, 4, 5, 6, 7, 8$ & $6$\\
$2$ & $\mathbb{Z}_2\times \mathbb{Z}_m$, $m=2, 4, 6$
or & \\ 
    & $ \mathbb{Z}_m \times \mathbb{Z}_m$, $m=3, 4$ & $5$ \\
$3$ & $(\mathbb{Z}_2)^3$ & $4$ \\
$4$ & $(\mathbb{Z}_2)^4$ & $3$ \\ \hline
\end{tabular}
\renewcommand{\arraystretch}{1.0}
\caption{Automorphisms of K3 surfaces and the resulting M theory vacua.}\label{tab:k3aut}
\end{center}
\end{table}

Consider first the case in which $Y=(T^\ell\times Z)/G$ where $Z$ is a
K3 surface.  The action of $G$ preserves the two factors $T^\ell$ and
$Z$.  In order to preserve the invariant spinors on $T^\ell$ it must
act by translations on that factor, and in order to preserve the
invariant spinors on $Z$ it must preserve the holomorphic $2$-form on
$Z$.  Abelian group actions which preserve the holomorphic $2$-form on
$Z$ were classified by Nikulin \cite{Nikulin79}; there are $15$ cases,
including the trivial group.  The resulting vacua are displayed in
table~\ref{tab:k3aut}, where $k$ denotes the number of generators in
the group, and hence the minimal dimension of a torus factor in $Y$.

\begin{table}[t]
\begin{center}
\renewcommand{\arraystretch}{1.5}
\begin{tabular}{|c|c|c|c|} \hline
$G$                   &   Lattice $L$ &  Singularities on K3$/G$    &$(E^\perp)^\perp/E$ 
\\ \hline
$\Z_1$ & $\Gamma_{4,4}\oplus E_8\oplus E_8$ & none &$\{e\}$\\
$\Z_2$ & $\Gamma_{1,1}(2) \oplus \Gamma_{3,3} \oplus D_4 \oplus D_4 $ & $8 A_1$ &$
{\mathbb Z}_2$\\
$\Z_3$ & $\Gamma_{1,1}(3) \oplus \Gamma_{3,3} \oplus A_2 \oplus A_2 $ & $6A_2$ & $\Z_3$\\ 
$\Z_4$ & $\Gamma_{1,1}(4) \oplus \Gamma_{3,3} \oplus A_1 \oplus A_1 $ & $4A_3+2A_1 $
& $\Z_4$\\
$\Z_5$ & $\Gamma_{1,1}(5) \oplus \Gamma_{3,3} $ & $4A_4$&$\Z_5$\\[2mm]
$\Z_6$ & $\Gamma_{1,1}(6) \oplus \Gamma_{3,3} $ & $2A_5+2A_2+2A_1$&$\Z_6$ \\[2mm] 
$\Z_7$ & $ \Gamma_{2,2} \oplus \begin{pmatrix} 28 & 7 \\ 7 & 2 \end{pmatrix}   $ 
& $3A_6$&$\Z_7$\\
$\Z_8$ & $ \Gamma_{2,2} \oplus \begin{pmatrix} 4 & 0 \\ 0 & 2 \end{pmatrix} $ 
& $2A_7+ A_3 + A_1$&$\Z_8$ \\ \hline
\end{tabular}
\renewcommand{\arraystretch}{1.0}
\caption{Choices for $G$ together with their associated lattices for $Y=(S^1\times K3)/G$.} 
\label{table:K3}
\end{center}
\end{table}

The lattices $H^2(Z/G)$ in these cases are also known
\cite{Nikulin79}.  In table \ref{table:K3} we exhibit the lattices $L$
associated to six-dimensional effective theories built from M theory
on $Y=(S^1\times K3)/G$ when $G$ is trivial or cyclic;\footnote{The
descriptions we give of the lattices can be inferred from the
descriptions in \cite{Nikulin79}\ using techniques from
\cite{Nikulin80}. } we also describe the singularities which
are found on $Z/G$ itself, using the $ADE$ notation for rational
double points, and the finite group $(E^\perp)^\perp/E$ where $E$ is
the sublattice of $H^2(\widetilde{Z/G},\mathbb Z)$ spanned by the
exceptional divisors.  The corresponding facts about non-cyclic groups
$G$ (where the effective theory has lower dimension) are given in
table~\ref{table:lowerK3}.  (The ``discriminant group'' which appears in
that table is discussed in appendix~\ref{app:lattice}.)

\begin{table}[t]
\begin{center}
\renewcommand{\arraystretch}{1.5}
\begin{tabular}{|c|c|c|c|} \hline
 Compactification  & $G$ & Rank of & Discriminant \\[-2mm] 
dimension & & $H^2(K3/G)$ & group \\ \hline
5 & $\Z_2 \times \Z_2 $ & $10$ & $(\Z_2)^8$ \\
5 & $\Z_2 \times \Z_4 $ & $6$ & $(\Z_2)^2 \times (\Z_4)^2$ \\ 
5 & $\Z_2 \times \Z_6$ & $4$  & $(\Z_2) \times (\Z_6)$\\
5 & $\Z_3\times \Z_3$ & $6$ & $(\Z_3)^4$\\ 
5 & $ \Z_4 \times \Z_4$ & $4$  & $(\Z_4)^2$\\
4 & $\Z_2 \times \Z_2 \times \Z_2$ & $8$ & $(\Z_2)^8$ \\
3 & $\Z_2 \times \Z_2 \times \Z_2 \times \Z_2 $ & $7$ &  $(\Z_2)^7$ \\ \hline
\end{tabular}
\renewcommand{\arraystretch}{1.0}
\caption{Choices for $G$ for lower-dimensional compactifications.} 
\label{table:lowerK3}
\end{center}
\end{table}

Turning to the case where $Y=(T^\ell\times Z)/G$ with $Z$ a complex
$2$-torus, we need an abelian group $G$ acting on $Z$ in such a way
that the holomorphic $2$-form is preserved by the $G$-action.
Moreover, the group action must not leave any holomorphic $1$-form
invariant, so $G$ cannot act entirely by translations.  Such group
actions were classified in modern language by Fujiki \cite{fujiki}
(although the classification was essentially done more than ninety
years ago by Enriques and Severi \cite{enriques-severi}).  The
computations of the corresponding lattices in the case of abelian
group actions were made in
\cite{nikulin:kummer,morrison:k3picard,bertin} and were applied in the
physics literature in \cite{Wendland}.  The only possibilities are
cyclic groups $G=\Z_m$ with $m=2$, $3$, $4$, or $6$, and all of these
lead to theories of effective dimension six.  The associated lattices
are described in table~\ref{table:torus},\footnote{Again, matching the
descriptions in table~\ref{table:torus} with those in
\cite{nikulin:kummer,morrison:k3picard,bertin} requires techniques
from \cite{Nikulin80}.} where we also give the singularities on $Z/G$
and the finite group $(E^\perp)^\perp/E$.

\begin{table}[t]
\begin{center}
\renewcommand{\arraystretch}{1.5}
\begin{tabular}{|c|c|c|c|} \hline
$G$                   &   Lattice $L$  &  Singularities  on $T^4/G$  &$(E^\perp)^\perp/E$\\ \hline
$\Z_2$ & $\Gamma_{3,3}(2)\oplus \Gamma_{1,1}$ & $ 16A_1$ &$(\Z_2)^5$\\
$\Z_3$ & $\Gamma_{1,1}(3)\oplus \Gamma_{1,1}\oplus \begin{pmatrix} 2 & 1 \\ 1 & 2 \end{pmatrix} $ & $9A_2$
&$(\Z_3)^3$\\
$\Z_4$ & $\Gamma_{1,1}(4)\oplus \Gamma_{1,1}\oplus \begin{pmatrix} 2 & 0 \\ 0 & 2 \end{pmatrix} $ & $4A_3+ 6A_1$
&$\Z_4 \times(\Z_2)^2$\\
$\Z_6$ & $\Gamma_{1,1}(6)\oplus \Gamma_{1,1}\oplus  \begin{pmatrix} 6 & 0 \\ 0 & 2 \end{pmatrix} 
$ & $A_5 +4 A_2 +5 A_1$ & $\Z_6$\\ \hline
\end{tabular}
\renewcommand{\arraystretch}{1.0}
\caption{Choices for $G$ together with their associated lattices for
$Y=(S^1\times T^4)/G$.}
\label{table:torus}
\end{center}
\end{table}

\subsection{F theory compactifications without flux}

It is common to describe an F theory vacuum in terms of a Ricci-flat
manifold $Y$ together with an elliptic fibration $\pi:Y\to B$.
However, to specify an F theory vacuum, we actually only need
\begin{enumerate}
\item the manifold $B$ with a subset $\Delta$
of real codimension $2$ (where $\Delta$ specifies the location of the singular fibers in
the fibration $\pi$), 
\item a monodromy representation $\pi_1(B-\Delta,p)\to SL(2,\mathbb
Z)$ and a ``$j$-function'' $j:B\to\mathbb{CP}^1$ compatible with the
monodromy (which are specified by the complex structure on the fibers
of $\pi$), and \item a metric on $B-\Delta$ whose asymptotics near
$\Delta$ are described by the Greene--Shapere--Vafa--Yau ansatz
\cite{GSVY} (which can be seen as a limit of metrics on $Y$ as the
area of the elliptic fiber approaches zero \cite{gross-wilson}).
\end{enumerate} 
The F theory vacuum is then described as type IIB string theory
compactified on $B$ with the given metric and with branes along
$\Delta$, using the S-duality of type IIB theory to compensate for the
$SL(2,\mathbb Z)$ monodromy.

If we begin from M theory compactified on $Y$, and take the limit as
the area of the fibers of $\pi$ approaches zero, then one dimension of
the effective theory decompactifies \cite{SchwarzFMA, AspinwallFMA}
and we obtain the F theory vacuum in the limit.  This is sometimes
referred to as ``F theory compactified on $Y$'' although the full data
of $Y$ is not needed.  Conversely, if the elliptic fibration on $Y$
has a section, then the standard M theory/F theory duality
\cite{Ftheory} asserts that F theory on $Y\times S^1$ (with a trivial
Wilson line) is dual to M theory on $Y$.

When the elliptic fibration on $Y$ does {\it not}\/ have a section,
there is always an associated manifold $\mathcal J(Y)$, the {\it
Jacobian of the fibration,}\/ which has an elliptic fibration with a
section that gives rise to the same monodromy and $j$-function data as
the elliptic fibration on the original manifold.  Thus, F theory
cannot distinguish between the compactification on $Y$ and the
compactification on $\mathcal J(Y)$.

The M theory/F theory duality can be extended to cover this case
\cite{Morrison:WilsonFtheory}, where it becomes the assertion that
when Wilson line data is included, F theory on $\mathcal J(Y)\times
S^1$ is dual to the union of the M theory moduli spaces on $Y_k$ for
all manifolds $Y_k$ with the same Jacobian fibration $\mathcal
J(Y_k)=\mathcal J(Y)$.\footnote{For other comments on F theory
compactifications without section, see~\cite{Berglund:2000rq}. }
Thus, discrete choices for Wilson lines in F
theory correspond in M theory to different elliptic fibrations with
the same Jacobian fibration.  Typically, such discrete choices are
only present for special values of moduli.

An elliptic fibration on a Ricci-flat manifold $Y$ always determines a
class $x\in H^2(Y,\mathbb Z)$ with $q(x)=0$.  Thus, in the case of
$16$ supercharges, the boundary lattice associated to taking the F
theory limit is $L_x=x^\perp/\langle x\rangle$.
If the elliptic fibration admits a section, then this lattice can be
used to describe the entire component of the F theory moduli space.
On the other hand, if the elliptic fibration does not admit a section,
then the lattice $L'$ for the component of F theory moduli space
associated to $\mathcal J(Y)$ is typically larger than $L_x$.

To find these components in detail, we need to examine possible
elliptic fibrations on the Ricci-flat manifolds $Y=(T^\ell\times Z)/G$
(with $Z$ either a K3 surface or a $T^4$).  An elliptic fibration
$\pi:(T^\ell\times Z)/G\to B$ will lift to an elliptic fibration
$\widetilde{\pi}:T^\ell\times Z\to\widetilde{B}$ which is
$G$-invariant.  If $\pi$ has a section then its inverse image will be
a $G$-invariant section of $\widetilde{\pi}$.  The group $G$ acts on
the base $\widetilde{B}$ and $B=\widetilde{B}/G$.  We let $G_0$ be the
subgroup of $G$ which acts trivially on the base $\widetilde{B}$.  It
is easy to see that if $G_0$ is non-trivial, there cannot be a section
for the fibration $\pi$: the group $G_0$ acts by translations on the
fibers, so cannot preserve a section of $\widetilde{\pi}$, but
$\widetilde{\pi}$ must have a $G$-invariant section when $\pi $ has a
section.  Thus, when $\pi$ has a section the action of $G$ on
$\widetilde{B}$ must be faithful.

Let us first analyze the cases in which $\pi$ does {\it not}\/ have a
section.  As indicated above, the lack of a section can be attributed
to a non-trivial group $G_0$ which acts trivially on the base
$\widetilde{B}$.  Let us begin with the case $Z=K3$ and return to the
case $Z=T^4$ in a little while.  In fact, it is easy to see that the
Jacobian of the elliptic fibration on $(T^\ell\times Z)/G_0$ is just
$T^\ell\times (Z/G_0)$.  The manifold $Z/G_0$ is a singular K3
surface, and this is in fact part of a larger family of manifolds of
the form $T^\ell \times K3$.  The possible $G_0$'s which can occur
here can be classified: we need to know which finite abelian groups
$G_0$ could act as a group of translations on the fibers of an
elliptic fibration on a K3 surface.  This classification was carried
out by Cox \cite{cox}, who found almost exactly the same list as
Nikulin's classification of abelian automorphism groups
\cite{Nikulin79}, except that $(\mathbb Z/2\mathbb Z)^3$ and $(\mathbb
Z/2\mathbb Z)^4$ cannot occur as translations on the fibers.  Thus, in
all of these cases, there is a limit of the M theory moduli space in
which the limiting F theory vacua gain additional degrees of freedom
which allow them to be part of the ``standard component'' of the F
theory moduli space on $S^1\times K3$.  Since the lattice of the
component of M theory moduli space takes the form
$L(-1)\cong\Gamma_{1,1}\oplus H^2(Z/G_0,\mathbb Z)$, we must have
$L_x(-1)=H^2(Z/G_0,\mathbb Z)$ in order to allow the lattice $L_x$ to
be embedded into the ``standard lattice'' $\Gamma_{3,3}\oplus E_8
\oplus E_8$.  This is exactly the type of lattice limit which is found
for the standard component.

There can be ``mixed'' cases as well, in which both $G_0$ and
$G_1:=G/G_0$ are non-trivial.  In such a case, the Jacobian of the
elliptic fibration on $(T^\ell\times Z)/G$ will be an elliptic
fibration on $(T^\ell \times (Z/G_0))/G_1$.  The limiting theory
``attaches'' to the moduli space $(T^\ell\times Z')/G_1$, and the
lattice decomposition $L\cong M_1\oplus L_x$ should use the rank two
lattice $M_1$ associated with $G_1$.  (We will determine those
lattices below.)

Turning now to the cases in which $\pi$ has a section, there will be a
component of the F theory moduli space for each case in which $\pi$
has a section.  When $Z$ is a K3 surface, both $B$ and $\widetilde{B}$
are isomorphic to $\mathbb{CP}^1$ so $G$ must have a faithful action
on $\mathbb{CP}^1$, that is, there must be an injective homomorphism
$G\to SO(3)$.  Since $G$ is abelian, the only possibilities are that
$G$ is trivial, or $G$ is cyclic, or $G\cong (\Z_2)^2$.

When $G$ is trivial, we can take $\ell=0$ and we get the ``standard''
F theory component in $8$ dimensions.  This leads to standard
components in lower dimensions as well, which can be treated as F
theory on $T^\ell\times K3$.

When $G\cong\Z_m $ is cyclic, there will be two fixed points for the
action of $G$ on $\widetilde{B}=\mathbb{CP}^1$.  All of the fixed
points for the action of $G$ on $Z$ must lie in one of the two
elliptic curves fixed by the $G$-action, and the quotient $Z/G$ will
have an elliptic fibration which degenerates to have a irreducible
fiber of multiplicty $m$ at each of the two fixed points.  Note that
$Z/G$ has singularities along these two fibers as well, and that the
resolved surface $\widetilde{Z/G}$ will have a more conventional
elliptic fibration, with section (since there is a section up on $Z$
by assumption).  Since the only multiplicities which can occur within
fibers in such a fibration are $m\le 6$, we learn that the only
possible cyclic group actions in this case are $\Z_m$ with $2\le
m\le6$.

\begin{figure}[t]
\begin{center}
\begin{tabular}{rrrrr}
\includegraphics[scale=.6]{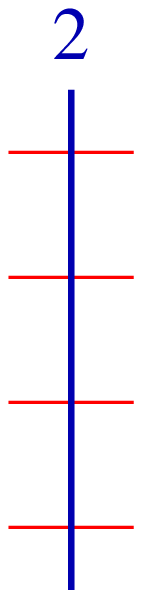} &
\includegraphics[scale=.6]{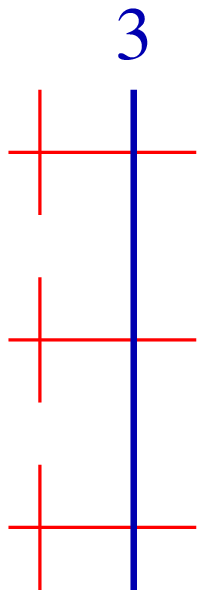}&
\includegraphics[scale=.6]{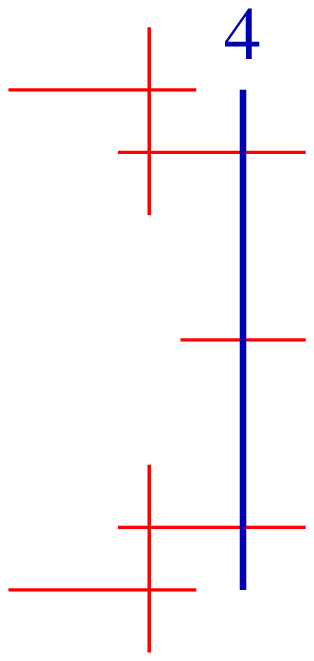}&
\includegraphics[scale=.6]{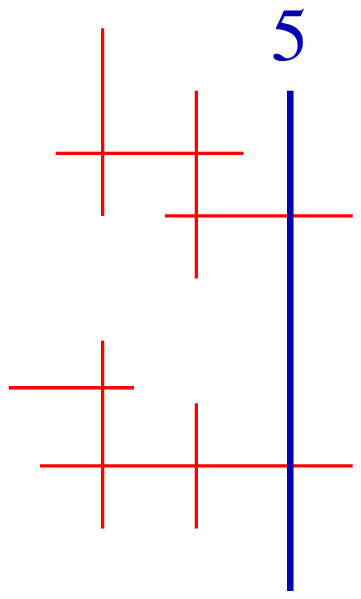}&
\includegraphics[scale=.6]{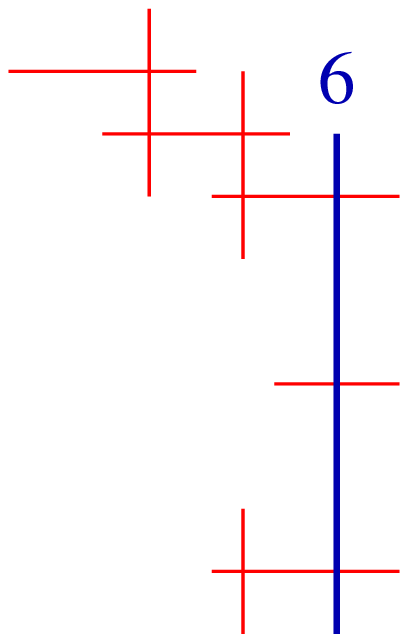}\\
\multicolumn{1}{c}{$4A_1$} &
\multicolumn{1}{c}{$3A_2$} &
\multicolumn{1}{c}{$2A_3+A_1$} &
\multicolumn{1}{c}{$2A_4$} &
\multicolumn{1}{c}{$A_5+A_2+A_1$} \\
\end{tabular}
\caption{The singular fibers on $Z/G$.}\label{fig:ellipmult}
\end{center}
\end{figure}

In fact, it is possible to see the geometry of these group actions
quite explicitly.  We have already enumerated the possible singular
points on $Z/G$ in table~\ref{table:K3}.  These singular points are
grouped together into elliptic fibers as indicated in
figure~\ref{fig:ellipmult}.  In each case, the elliptic curves on
$Z/G$ degenerate to an irreducible curve passing through $2$, $3$, or
$4$ singular points, and the irreducible curve has multiplicity $m$ in
the elliptic fibration on $Z/G$.  All of the values $2\le m\le6$ do
occur, as shown in figure~\ref{fig:ellipmult}.\footnote{A related 
geometric structure appears in \cite{katzmorrison}. It would be interesting
to understand how this is related to the frozen singularities that we will
later discuss.}  
On each fiber in the
figure, the thick (blue) line represents the irreducible curve (which
is labeled by its multiplicity $m$ in the fiber), and the thin (red)
lines represent the curves in the resolutions of the various
singularities.  We have labeled each fiber with the types of singular
points that occur on it.

\begin{figure}[t]
\begin{center}
\begin{tabular}{ccccc}
\includegraphics[scale=.6]{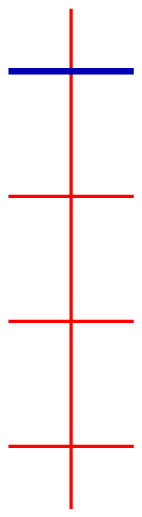} &
\includegraphics[scale=.6]{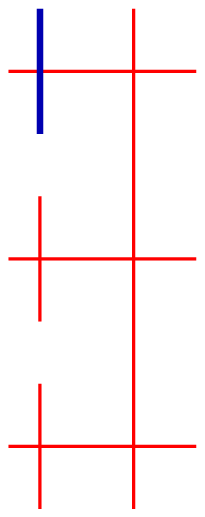} &
\includegraphics[scale=.6]{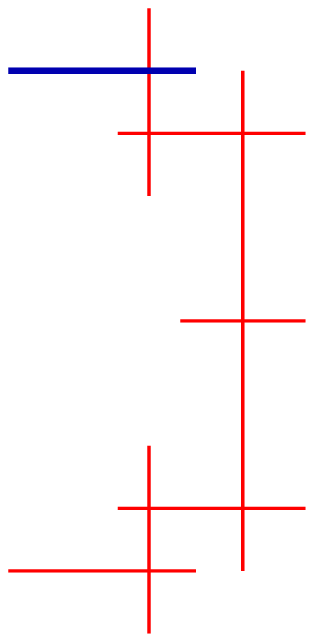} &
\includegraphics[scale=.6]{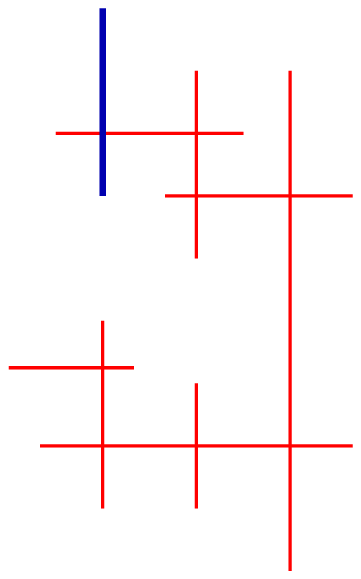} &
\includegraphics[scale=.6]{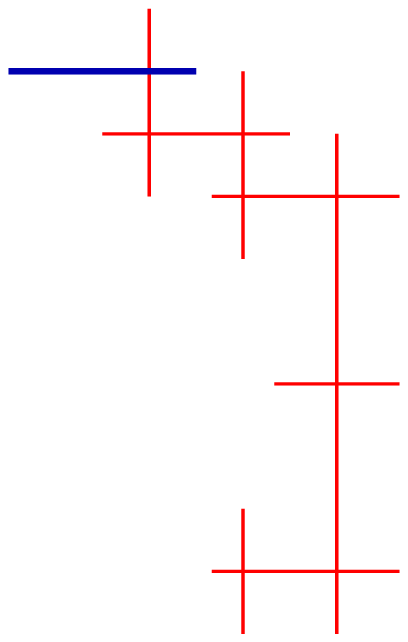} \\
$I_0^*$ & $IV^*$ & $III^*$ & $II^*$ & $II^*$ \\
\end{tabular}
\end{center}
\caption{The singular fibers for F theory on $\widetilde{Z/G}$.}\label{fig:ellipfib}
\end{figure}

We remind the reader that the singularities of $Z/G$ do not actually
occur in our M theory and F theory vacua, which are compactified on
$(S^1\times Z)/G$; the surface $Z/G$ is just a convenient device for
determining the lattice $L$ of the M theory compactification.  To
determine the lattice of the F theory compactification, we should use
a different birational model of $Z/G$: either the nonsingular surface
$\widetilde{Z/G}$, or the {\it Weierstrass model}\/ obtained from
$\widetilde{Z/G}$ by blowing down all components of fibers other than
the ones meeting a section.\footnote{
Again, the singularities on the Weierstrass model do not directly show up
in our F theory vacuum, but they will reappear as 
``frozen singularities'' in an M theory limit with $3$-form flux described 
in section 
\ref{threeformphysics}.}  Each singular fiber can then be labeled
by its {\it Kodaira type}; the labels for the fibers for different
values of $m$ are shown in figure~\ref{fig:ellipfib}.  The fibers in
that figure are in one-to-one correspondence with the fibers in
figure~\ref{fig:ellipmult}, and for each fiber in
figure~\ref{fig:ellipfib}, the thick (blue) line represents the fiber
component which meets the section, and the thin (red) lines represent
the components which are blown down to give the Weierstrass model.

\begin{table}[t]
\begin{center}
\renewcommand{\arraystretch}{1.5}
\begin{tabular}{|c|c|c|} \hline
$G$                   &   Lattice $L_x$ &  Singular fibers on $\widetilde{K3/G}$    \\ \hline
$\Z_1$ & $\Gamma_{3,3}\oplus E_8\oplus E_8$ & none \\
$\Z_2$ & $ \Gamma_{3,3} \oplus D_4 \oplus D_4 $ & $I_0^*+I_0^*$\\
$\Z_3$ & $ \Gamma_{3,3} \oplus A_2 \oplus A_2 $ & $IV^*+IV^*$\\ 
$\Z_4$ & $ \Gamma_{3,3} \oplus A_1 \oplus A_1 $ & $III^*+III^* $\\
$\Z_5$ & $ \Gamma_{3,3} $ & $II^*+II^*$\\
$\Z_6$ & $ \Gamma_{3,3} $ & $II^*+II^*$\\ \hline
\end{tabular}
\renewcommand{\arraystretch}{1.0}
\caption{F theory lattices and singular fibers for $Y=(S^1\times K3)/G$.} 
\label{table:FK3}
\end{center}
\end{table}

The configurations of singular points on $Z/G$ from
table~\ref{table:K3} are thus collected into Kodaira fibers, giving
the results in the right hand column of table~\ref{table:FK3}.\footnote{
Note that the free part of the lattices for the
$\Z_5$ and $\Z_6$ cases appearing table \ref{table:FK3} are
identical. It would be interesting to check whether the full cohomology
lattices 
differ by torsion classes.}  From
the Kodaira fibers, a lattice can easily be computed, as shown in the
middle column of table~\ref{table:FK3}, and we claim that this is the
lattice which describes the moduli space for the corresponding F
theory component.  In fact, this same lattice can be arrived at in two
ways: either directly in terms of the Kodaira fibers for the elliptic
fibration on $\widetilde{Z/G}$, or by using the $x^\perp/\langle
x\rangle$ construction, using the fact that $L=\Gamma_{1,1}(m)\oplus
L_x$ in the case that $Y=(S^1\times Z)/G$.  (This latter result
becomes obvious when comparing tables~\ref{table:K3}
and~\ref{table:FK3}.)  This gives six different F theory vacua. These
vacua are dual to the six heterotic asymmetric orbifolds in 7
dimensions that we constructed in section \ref{asymorbifold}.

Note that in the two remaining cyclic cases for M theory vacua, namely
$\Z_7$ and $\Z_8$, it is not possible to split off a factor of
$\Gamma_{1,1}(m)$ from the lattice $L$ as given in
table~\ref{table:K3}.  This gives further confirmation that there are
no F theory vacua associated with these cases.

\begin{table}[t]
\begin{center}
\renewcommand{\arraystretch}{1.5}
\begin{tabular}{|c|c|c|} \hline
$G$                   &   Lattice for $\widetilde{K3/G}$ &  Singular fibers on $\widetilde{K3/G}$    \\ \hline
$(\Z_2)^2$ & $ \Gamma_{2,2} \oplus \Gamma_{1,1}(2) \oplus D_4 $  & $I_0^*+I_0^*+I_0^*$\\ \hline
\end{tabular}
\renewcommand{\arraystretch}{1.0}
\caption{The additional F theory vacuum in dimension $6$.}
\label{table:FK3bis}
\end{center}
\end{table}

The one remaining F theory vacuum with $Z=K3$ is associated to the
group $G=\Z_2\times \Z_2$; see table~\ref{table:FK3bis}. The lattice
we list is for $K3/G$. To this lattice, we must add a lattice of
signature $(2,2)$ for the 5-dimensional M theory compactification and
a lattice of signature $(1,1)$ for the F theory compactification.  The
fixed points for the action of $G$ on $\widetilde{B}=\mathbb{CP}^1$
have stabilizer $\Z_2$ in each case, so on the quotient $Z/G$ we will
find fibers with multiplicity $2$ which become $I_0^*$ Kodaira fibers
on $\widetilde{Z/G}$.  Once again the lattices satisfy
$L=\Gamma_{1,1}(2)\oplus L_x$.

\begin{table}[t]
\begin{center}
\begin{tabular}{|c|c|c|} \hline
$G$                   &   Lattice $L_x$  &  Singular fibers  on $\widetilde{T^4/G}$  \\[2mm] \hline
$\Z_2$ & $\Gamma_{2,2}(2)\oplus \Gamma_{1,1}$ & $ I_0^*+I_0^*+I_0^*+I_0^*$\\[2mm]
$\Z_3$ & $ \begin{pmatrix} 2 & 1 \\ 1 & 2 \end{pmatrix} \oplus \Gamma_{1,1}$ & $IV^*+IV^*+IV^*$\\[2mm] 
$\Z_4$ & $ \begin{pmatrix} 2 & 0 \\ 0 & 2 \end{pmatrix} \oplus \Gamma_{1,1}$ & $III^*+III^*+I_0^*$\\[2mm]
$\Z_6$ & $  \begin{pmatrix} 6 & 0 \\ 0 & 2 \end{pmatrix} \oplus \Gamma_{1,1}
$ & $II^*+IV^* +I_0^*$\\[2mm] \hline
\end{tabular}
\caption{F theory lattices and singular fibers for $Y=(S^1\times T^4)/G$.} 
\label{table:Ftorus}
\end{center}
\end{table}

Turning to the case in which $Z=T^4$, we find a similar story.  The
base $B$ of the elliptic fibration on $Z/G$ is still $\mathbb{CP}^1$,
but the base $\widetilde{B}$ of the elliptic fibration on $Z$ is now
an elliptic curve.  The action of $G$ on $\widetilde{B}$ must be
faithful, and in fact it suffices to consider the case in which $G$
does not contain any translations (else we could mod out by the
translations first).  As an action on an elliptic curve with a fixed
point, the only possibilities are $G=\Z_m$ with $m=2$, $3$, $4$, or
$6$.  In fact, these are {\it exactly}\/ the group actions that we
have (in table~\ref{table:torus})!  As in the previous case, the
singular points on $Z/G$ can be collected into Kodaira fibers for
$\widetilde{Z/G}$; the results of this are displayed in
table~\ref{table:Ftorus}.  The lattices once again satisfy $L\cong
\Gamma_{1,1}(m)\oplus L_x$.

The entries in tables~\ref{table:FK3}, \ref{table:FK3bis},
and~\ref{table:Ftorus} thus describe the possible F theory vacua with
16 supercharges.  The first entry in table~\ref{table:FK3} gives the
``standard'' F theory vacuum in eight dimensions, the remaining five
entries in table~\ref{table:FK3} together with the four entries in
table~\ref{table:Ftorus} give the new F theory vacua in seven
dimensions, and table~\ref{table:FK3bis} shows the one new F theory
vacuum in six dimensions.  (Of course the higher dimensional theories
can be reduced to lower dimensional ones by compactification on
additional circles.)

\subsubsection{From F theory to type I$'$}

Decompactification limits of the components of the F theory moduli
space which have $D$-dimensional effective theories should lead to
$(D+1)$-dimensional effective theories (using an isotropic sublattice of
rank $1$).  It would be desirable to give
a description of these directly in terms of type I$'$ theory. However,
despite the interesting pictures presented in 
\cite{Morrison:1997xf ,Cachazo:2000ey}, at
present we do not have enough control over type I$'$ vacua to be able
to do this.  So our analysis will be somewhat indirect. In this
section, we describe the overall picture leaving a detailed discussion
for the following section.

The most common decompactification limit from these components involves a decomposition
of the lattice in the form 
\begin{equation}\label{eq:boundary}
L\cong \Gamma_{1,1}\oplus L_x.
\end{equation}
We saw in the previous section that when going from six dimensions up
to seven dimensions, a lattice decomposition of this form corresponded
to a boundary component which gained additional degrees of freedom in
the limit, and which ``attached'' to the standard boundary component.

To see this, we will give a geometric construction of these components
and their decompactification limits, and note that for the vast
majority of components in dimension seven, there is {\it only one}\/
decompactification limit, which must therefore be of this type.  We
thus argue by analogy that all limits of this type must have the
lattice decomposition given in eq.~\eqref{eq:boundary}; this leaves
only a few additional components in dimension eight for which we must
account.

Our geometric construction---to be described in section
\ref{delpezzo} for $Z=K3$---is designed for easy comparison with the
heterotic duals of these vacua, where this phenomenon is known: the
seven-dimensional components we have found are dual to non-trivial
triples of Wilson lines on the heterotic side, and their
eight-dimensional limits all attach to the standard eight-dimensional
component.  The exceptional case is $G=\Z_2$ which contains the CHL
string. This case does have a non-trivial 8 and 9-dimensional limit.

A similar discussion applies to the 8-dimensional limits of the F
theory moduli spaces coming from $Y=(S^1\times T^4)/G$. Since the
lattices for these models also have a $\Gamma_{1,1}$ summand,
8-dimensional limits exist. However, in each case, again with the
exception of $G=\Z_2$, these boundaries attach to a conventional
toroidal type IIB compactification. The case of $G=\Z_2$ has
non-trivial 8 and 9-dimensional limits as reflected in its lattice
given in table \ref{table:FK3}.  This is natural since this model has
a dual description as the compactification of the $(+,-)$ orientifold
to 7 dimensions. This duality can be seen immediately by studying the
geometric description of the compactified $(+,-)$ orientifold obtained
in section \ref{plusminusorient}.

The last case we need to discuss is the 6-dimensional F theory vacuum
associated to $G=\Z_2\times \Z_2$. From the lattice for $\widetilde{K3/G}$
given 
in table \ref{table:FK3bis}, we that there are two ways of
decompactifying to 8 dimensions. Peeling off a $\G_{1,1}$ factor gives
us a 7-dimensional theory that attaches to the standard component as
in our preceeding discussion. The other limit involves peeling off a
$\G_{1,1}(2)$ factor. This leads to a 7-dimensional theory that
attaches to the component of the moduli space containing the CHL
string.  Note that we also need to worry about the additional
signature $(1,1)$ summand in the lattice for F theory on $(K3\times
T^2)/G$.  The correct relative normalization for this summand is
tricky to determine, although preliminary computations suggest that it
is either $\G_{1,1}$ or $\G_{1,1}(2)$. This is certainly supported
from a study of the $\Z_2\times \Z_2$ asymmetric orbifold of the
heterotic string. From that approach, it seems clear that there is no
new 7-dimensional theory to which the 6-dimensional orbifold could
decompactify. From this 6-dimensional theory, we therefore arrive at
no new 7-dimensional theory.

\subsubsection{Automorphisms of del Pezzo surfaces}
\label{delpezzo}

Let us begin by explaining why del Pezzo surfaces play a natural role
in our F theory discussion. Consider an elliptically fibered $K3$
surface $Z$ whose base $B$ is large, and whose complex structure is
close to a degenerate one in which the $K3$ surface degenerates into a
pair of del Pezzo surfaces $Z_1$, $Z_2$ meeting on an elliptic curve
$E$.  This is the limit in which comparison with the $E_8\times E_8$
heterotic string becomes possible \cite{Friedman:1997yq}. 
(In fact, it corresponds to both large volume and weak string coupling on
the heterotic side of the duality.)  It is then natural to 
expect that automorphisms of del Pezzo surfaces will be classified by
the same data used to classify $E_8$ gauge bundles on $T^3$. Further,
the anomaly cancellation condition should have a purely geometric
realization as a constraint on whether we can ``glue'' two del Pezzo
surfaces with automorphisms into a $K3$.

So let us begin by considering a rational elliptic surface $X$ in Weierstrass
form.  This implies that $X$ has an elliptic fibration $\pi\colon X\to
\C\Pee^1$, and a section $\sigma\colon \C\Pee^1\to X$ contained in the
smooth points of $X$.  Since $X$ is rational, $\sigma^2=-1$, that is
to say $\sigma$ is an exceptional curve.  We fix a fiber $E\subset X$
of $\pi$, and we wish to study the group of automorphisms of $X$ which
are the identity on $E$ and which stabilize $\sigma$ (ie map the image
of $\sigma$ to itself).  We call these automorphisms of $(X,E,\sigma)$
Equivalently, we could consider a degree one del Pezzo surface $\ov X$
obtained by collapsing $\sigma$.  The elliptic fibration on $X$
becomes the anti-canonical pencil of elliptic curves (or more
generally Weierstrass curves) on this surface. This surface has at
worst rational double point singularities. The automorphism group
$(X,E,\sigma)$ acts naturally on $\ov X$ and is identified with the
subgroup of the automorphism group of $X$ fixing $E$ pointwise.

As a first step, let us argue that the automorphism group of
$(X,E,\sigma)$ is a finite cyclic group that preserves the elliptic
fibration structure of $X$.  Along the way, we shall also see that the
induced action of the group of automorphisms on the base $\C\Pee^1$ of
the elliptic fibration is faithful. The group acts on $\C\Pee^1$ fixing
two points, i.e., the automorphism group of $(X,E,\sigma)$ stabilizes
$E$ and exactly one other fiber.

To see this, let $f\subset X$ be a fiber of the elliptic fibration and
let $\alpha\colon X\to X$ be an automorphism of $(X,E,\sigma)$. Then
$\alpha(f)$ is a divisor in $X$ with zero algebraic intersection with
$E$. Hence, its projection to $\C\Pee^1$ must be a single point. That is
to say $\alpha(f)$ is contained in a fiber of the elliptic fibration.
By homological considerations, we see that it is exactly a fiber of
the elliptic fibration structure. This shows that $\alpha$ preserves
the elliptic fibration structure and hence induces an automorphism
$\ov \alpha$ of the base $\C\Pee^1$. If $\ov \alpha$ is trivial, then
$\alpha$ stablizes each fiber of the elliptic fibration. Since it acts
by the identity on $E$, it acts by the identity on the homology of
each smooth fiber and hence, it is a translation on each smooth
fiber. But, $\alpha$ also stablizes the section $\sigma$, and hence it
must be the identity on each smooth fiber.  Since the smooth fibers
are dense, it follows that $\alpha$ is the identity.  This proves that
the automorphism group of $(X,E,\sigma)$ acts faithfully on the base
$\C\Pee^1$.

The elliptic fibration structure on $X$ has at least two singular
fibers. This means that the automorphism group of $(X,E,\sigma)$ is
faithfully represented as a group of automorphisms of $\C\Pee^1$ fixing
a point of $\C\Pee^1$ and permuting a finite set of points of
cardinality at least two.  All such groups are finite cyclic and fix
two points of $\C\Pee^1$.

We use $E'$ to denote the fiber other than $E$ stabilized by the
automorphism group of $(X,E,\sigma)$. Suppose that an automorphism
$\alpha$ is the identity on $E'$.  It is also the identity on $E$.  We
consider the degree one del Pezzo model $\ov X$ where the section has
been collapsed. The image of $\sigma$ is a smooth point of this
surface, and $\alpha$ descends to an automorphism $\ov \alpha$ of $\ov
X$ fixing the point of intersection $x$ of $E$ and $E'$ and acting by
the identity on $E$ and on $E'$. If follows that the differential of
$\ov \alpha$ at $x$ is the identity, and hence that the restriction of
$\alpha$ to the exceptional curve in $X$ obtained from blowing up $x$
is the identity. It follows that $\alpha$ stabilizes each fiber of the
elliptic fibration.  But we have already seen that this implies that
$\alpha$ is the identity.  The action of the automorphism group of
$(X,E,\sigma)$ on $E'$ is therefore faithful.

Let us examine the automorphism groups of the various types of
Weierstrass curves fixing a given smooth point of the curve.  The
automorphism group of a generic elliptic curve fixing a point is
$\Zee_2$ acting by $-1$ and fixing $4$ points.  In the case of special
elliptic curves the automorphism group is either $\Zee_4$ acting with
two fixed points and two points with stabilizers $\Zee_2$, or $\Zee_6$
acting with one fixed point, $3$ points with stabilizer $\Zee_2$ and
two points with stabilizer $\Zee_3$. The automorphism group of an
ordinary double point fiber fixing a smooth point is $\Zee_2$ fixing
the singular point and the other point.  The automorphism group of a
Weierstrass cusp fixing a smooth point is $\Cee^*$ acting so as to fix
only the singular point and the given smooth point.

Let $Y$ be an elliptic surface and let $f\subset Y$ be a fiber. Let
$p\in \C\Pee^1$ be the image of $f$ under projection mapping. Let
$\widetilde Y\to Y$ be the minimal resolution of $Y$ and let
$\widetilde p\colon \widetilde Y\to \C\Pee^1$ be the induced projection
mapping.  Let $\widetilde f\subset \widetilde Y$ be the full preimage
of $f$. Fix a disk $\Delta\subset \C\Pee^1$ centered at $p$
sufficiently small so that the preimage of $\Delta-\{p\}$ contains the
preimage of no singular point of $Y$ or of the projection mapping. We
define {\sl the local contribution of $f$ to the Euler characteristic
of the smooth model of $Y$} to be the Euler characteristic of
$\widetilde p^{-1}(\Delta)$. It is easy to see that $f$ is a smooth
fiber if and only if its local contribution to the Euler
characteristic is zero and otherwise that the local contribution to
the Euler characteristic is positive. Also, the local contribution to
the Euler characteristic is $1$ if and only if $f$ misses the
singularities of $X$ and contains an ordinary double point. Lastly, if
the elliptic surface in question is rational, then the sum over all
fibers of the local contributions to the Euler characteristic is
$12$.

The quotient of $X$ by the automorphism group is a surface $Y$ with an
induced elliptic fibration and with rational double point
singularities.  In the singular model, the fiber which contains the
image of $E'$ has multiplicity equal to the order of the automorphism
group of $X$. On the other hand, in the minimal resolution, the strict
transform of this curve is one of the components of the set of
rational curves indexed by the nodes of an extended Dynkin diagram of
type $A$, $D$, or $E$ and which intersect each other as indicated by
the bonds of the extended Dynkin diagram. Furthermore, the
multiplicities of the various components in the fiber in the smooth
model are given by the coroot integers on the corresponding nodes of
this diagram.  These numbers are all at most $6$, and hence every
component has multiplicity at most $6$ in the fiber. It follows in the
singular model, that the fiber containing the image of $E'$ has
multiplicity at most $6$, and hence the order of the automorphism
group is at most $6$.  (Notice that the cases of smooth and nodal
fibers $E'$ follow directly from the classification of the
automorphism groups of these fibers given above.) The automorphism
group of $(X,E,\sigma)$ is therefore a cyclic group of order at most
$6$.
 
Now we simply list the possibilities: the automorphism group is
$\Zee_2$ and the fiber $E'$ stabilized by the action is smooth.  The
quotient surface has four $A_1$ singularities and the image of $E'$
passes through all of these. The local contribution of this fiber to
the Euler characteristic of the smooth model is $6$.  The local
contribution to the Euler characteristic of the smooth model of $X$
from the singular fibers of the elliptic fibration is $12$, and since
the automorphism group acts freely on these fibers, the local
contribution of the images of these fibers to the Euler characteristic
of the smooth model of the quotient is $12/2=6$.  In the minimal
resolution of the quotient surface, the preimage of this fiber is a
tree of rational curves intersecting according to the extended Dynkin
diagram of $D_4$. The strict transform of $E'$ has multiplicity $2$ in
the fiber, i.e., it corresponds to the central node in the extended
Dynkin diagram, the one with coefficient two in the dominant root.

The automorphism group is $\Zee_3$ and the fiber $E'$ stabilized by
the action is smooth. The quotient surface has three $A_2$
singularities and the image of $E'$ passes through all of them. In the
minimal resolution of the quotient surface, the preimage of this fiber
is a tree of rational curves intersecting according to the extended
Dynkin diagram of $E_6$ and this fiber contributes $8$ to the Euler
characteristic of the smooth model. The singular fibers contribute
$12$ to the Euler characteristic of the smooth model of $X$ and hence
their images in the quotient contribute $12/3=4$ to the Euler
characteristic of its smooth model.  The strict transform of $E'$ is
the curve of multiplicity three in the fiber, i.e., it corresponds to
the central node in the extended Dynkin diagram, the one with
coefficient three in the dominant root.  The local contribution of
this fiber to the Euler characteristic of the smooth model is $8$.

The automorphism group is $\Zee_4$ and the fiber $E'$ stabilized by
the action is smooth. The quotient surface has two $A_3$ singularities
and an $A_1$ singularity.  The image of $E'$ passes through all these
singularities. In the minimal resolution of the quotient surface the
preimage of this fiber is a tree of rational curves intersecting
according to the extended Dynkin diagram of $E_7$ and the local
contribution of this fiber to the Euler characteristic of the smooth
model is $9$. The singular fibers contribute $12$ to the Euler
characteristic of the smooth model of $X$, and hence the images in the
quotient of the singular fibers of $X$ contribute $12/4=3$ to the
Euler characteristic of the smooth model of the quotient.  The strict
transform of $E'$ is the curve of multiplicity $4$ in the fiber, i.e.,
corresponds to the node with coefficient $4$ in the dominant root.

The automorphism group is $\Zee_5$ and the fiber $E'$ is a Weierstrass
cusp. The quotient surface has two $A_4$ singularities and the image
of $E'$ passes through both of them. The preimage of this fiber in the
minimal resolution of the quotient is a tree of rational curves
intersecting according to the extended Dynkin diagram of $E_8$ and
contributes $10$ to the Euler characteristic of the smooth model of
the quotient. The other singular fibers of $X$ contribute $10$ to the
Euler characteristic of the smooth model of $X$ and their images in
the quotient contribute $10/5=2$ to the Euler characteristic of its
smooth model.  The strict transform of $E'$ is the curve in this
configuration with multiplicity $5$ in the divisor representing the
fiber. That is to say it corresponds to the node with coefficient $5$
in the dominant root.  The local contribution of this fiber to the
Euler characteristic of the smooth model is $10$. Notice that since
$E'$ is a cusp, the sum of the contributions of the all other fibers
to the Euler characteristic of the smooth model of $X$ is $10$. Thus,
in the quotient the sum of the local contributions of the fibers
besides the image of $E'$ to the Euler characteristic of the smooth
model is $2$, giving us a total of $12$ as required.

The automorphism group is $\Zee_6$ and the fiber $E'$ is smooth. The
quotient surface has three singularities---of types $A_5$, $A_2$ and
$A_1$, respectively, reflecting the three singular orbits of the
action of this cyclic group on a smooth elliptic curve.  The image of
$E'$ passes through all these singularities and its preimage in the
minimal resolution is a tree of rational curves intersecting according
to the extended Dynkin diagram of $E_8$. The strict transform of $E'$
has multiplicity $6$ in the fiber and hence corresponds to the
trivalent node in the extended Dynkin diagram, the one with
coefficient $6$ in the dominant root.  The local contribution of this
fiber to the Euler characteristic of the smooth model is $10$. The
singular fibers of $X$ contribute $12$ to the Euler characteristic of
its smooth model, and the images of these fibers in the quotient
contribute $12/6=2$ to the Euler characteristic of its smooth model.

This completes the list of possibilities. Notice how the extended
Dynkin diagram of $E_8$ predicts the automorphism groups and the
singularities of the quotient surfaces.  First of all there will be an
automorphism group of order $k$ if and only if one of the coefficients
on the extended Dynkin diagram of $E_8$ is divisible by $k$.  Given an
integer $k$ with this property, the singularity in the quotient
surface when the automorphism group is cyclic of order $k$ is
determined as follows: One takes the extended Dynkin diagram of $E_8$
with the usual coefficients and removes all nodes whose coefficients
are divisible by $k$. There remains a collection of $A_{n_i}$
diagrams.  These label the singularities of the quotient
surface. Furthermore, we can connect all of these diagrams to a
central node and form an extended Dynkin diagram of some subgroup of
$E_8$.  In that new diagram the coefficient of the central node that
we added will be exactly $k$.  This is the fiber in the minimal
resolution of the quotient.  The strict transform of the image of $E'$
is the component corresponding to the central node that we added.

Note that the list of possibilities matches perfectly with the gauge
theory picture of commuting pairs in $E_8$ and their centralisers. Of
course, this is not an accident: one can prove by abstract methods
that equivalence between the group theory and del Pezzo surfaces which
goes through the Looijenga space of $E\otimes \Lambda(E_8)/W(E_8)$ is
categorical and hence that the automorphism groups of the three
classes of objects---commuting pairs in $E_8$, $E\otimes
\Lambda(E_8)/W(E_8)$ (here, $\Lambda(E_8)$ is the coroot lattice of
$E_8$ and $W(E_8)$ is its Weyl group) and del Pezzo surfaces must be
the same.\footnote{We wish to thank Bob Friedman for pointing this out
to us.}

One last point is worth remarking on: given two rational elliptic
surfaces with smooth fibers and sections $(X_1,E_1,\sigma_1)$ and
$(X_2,E_2,\sigma_2)$ and given an isomorphism from $E_1$ to $E_2$
matching up the intersections with the sections, we can glue $X_1$ and
$X_2$ together to form a singular surface $X$ with a normal crossing
along $E=E_1=E_2$. This surface fibers over $\C\Pee^1\cup\C\Pee^1$ with a
section $\sigma$. Of course, it has a marked fiber $E$.  It is a
singular model of an elliptically fibered $K3$ surface, and in fact
represents a point in a divisor at infinity in a compactification of
period space for these surfaces.\footnote{On the heterotic side, this
represents the infinite-volume, zero-coupling limit.}  Given automorphisms
$\alpha_i$ of 
$(X_i,E_i,\sigma_i)$ we can glue them together to determine an
automorphism of $(X,E,\sigma)$.  To smooth the singular surface $X$ to
an elliptically fibered $K3$ we need a trivialization of the tensor
product of the normal bundles of $E$ in $X_1$ and $X_2$. To carry
along the group action requires then trivializing the action of this
tensor product. This is possible if and only if the actions of
$\alpha_i$ on the disks in the base $\C\Pee^1$'s centered at the image
points of $E_i$ are inverses of each other. In particular, the orders
of $\alpha_1$ and $\alpha_2$ must be the same and these two
automorphisms must be in inverse components.  This corresponds in the
gauge theory language to the fact that the Chern--Simons invariant of
the commuting triples must be inverses of each other. As promised, we
therefore recover our anomaly matching constraint from this gluing
condition.

\subsection{Type IIA compactifications with RR one-form flux}
\label{RRone}
\subsubsection{Equivariant flat line bundles on $T^4$}

We now take a different tack and consider compactifications with
flux. This is a quite different class of models from the purely
geometric compactifications just discussed. Among compactifications of
this kind, we shall find new dual descriptions for the perturbative
asymmetric orbifolds of section \ref{asymorbifold} both in 6 and 7
dimensions.

Let us first consider flat RR 1-form fields in a type IIA string
compactification on some (possibly singular) manifold $X$. Such a
1-form field $A$ can be seen as a connection in a principal $U(1)$
bundle $P \to X$. Of course the interpretation in M theory of such a
RR 1-form field configuration will be as a compactification on the
manifold $P$. In our preceeding discussion, $P$ took the form of $(X
\times S^1)/G$. As we shall see, it is natural from this purely
geometric M theory picture to treat $A$ via equivariant cohomology.
In section \ref{secKtheory}, we revisit this treatment from the
perspective of K-theory where we find a group of 1-form fluxes in
agreement with the results from equivariant cohomology.  K-theory,
however, will give us new physics for compactifications with more
general combinations of fluxes.

$U(1)$ principal bundles, or equivalently complex Hermitian line
bundles, over a smooth manifold $X$ are classified topologically by
their first Chern class $c_1$, which is an element of the cohomology
group $H^2(X,\Z)$. This Chern class equals $[F/2\pi]$ in real
cohomology, where $F=dA$ is the curvature two-form. Flat line bundles,
i.e., bundles that satisfies $F=0$, are classified by the cohomology
group $H^1(X,U(1))$. One can think of this group of homomorphisms of
$H_1(X,\Z)$ into $U(1)$ as the holonomies of the flat connection
around the non-trivial homology 1-cycles of $X$.  Note that because $F$
vanishes, in general such a flat bundle can have only have a torsion
first Chern class
\bea
c_1 \in \Tor H^2(X,\Z).
\eea
If there is no torsion in $H^2(X)$ then a flat line bundle is
necessarily topologically trivial.  It can still have non-trivial
holonomies. In fact the group $H^1(X,U(1))$ of flat line bundles fits
in the exact sequence
\bea
1 \to  H^1(X,\R)/ H^1(X,\Z) \to H^1(X,U(1)) \to \Tor H^2(X,\Z) \to 1
\eea
where the Jacobian torus
\bea
 H^1(X,\R)/ H^1(X,\Z)
\eea
gives the holonomies of flat connections on bundles that are
topologically trivial.

As explained in appendix \ref{app:equiv}, for the case of an 
orbifold $X/G$, we have to
consider equivariant line bundles on $X$. These are bundles over $X$
with a compatible $G$ action. Equivariant flat line bundles are
classified by the equivariant cohomology group
\bea
H^1_G(X,U(1)).
\eea
This will be the group of RR 1-form fluxes. 

We will be considering the case where $X$ is either a $T^4$ or a $K3$
manifold. In these cases, there is a well-known list of orbifold groups
$G$ such that $X/G$ is again Calabi--Yau, i.e., a K3 manifold.

For the case $X=T^4$, the groups are listed in \cite{fujiki}
\bea
G=\Z_2,\Z_3,\Z_4,\Z_6,\widehat{\cal D}_4,\widehat{\cal D}_5, {\mathbb T}.
\eea
Here $\widehat{\cal D}_N$ denotes the binary dihedral group of order
$4N$, a double cover of the ordinary dihedral group ${\cal D}_N$. In
the $ADE$-labeling of finite subgroups of $SU(2)$, this group corresponds
to the Dynkin diagram $D_{N+2}$, just as the cyclic group $\Z_N$
corresponds to $A_{N-1}$. The last case ${\mathbb T}$ is the binary
tetrahedral group, which corresponds to $E_6$. For a list of the
group actions along with the singularities of $T^4/G$, 
see table \ref{table:singquot} of appendix \ref{app:equiv}.

To be concrete let us first consider this group for the case of the
$\Z_2$ action $x \to -x$ on the torus $T^4$. Here the group
$H^*_{\Z_2}(T^4,U(1))$ can be computed by a spectral sequence
technique\footnote{We thank D. Freed and G. Segal for help with these
calculations.} which is given in appendix \ref{app:equiv}.

This calculation shows that it essentially suffices to compute the
equivariant cohomology with coefficients in the field $\Z_2$ which has
a rather simple description. Is this case there are no differentials
in the $E_2$-term of the spectral sequence and no extension problems
so that
\bea
H^k_{\Z_2}(T^4,\Z_2) = \bigoplus_{p+q=k} H^p(\R\P^\infty, H^q(T^4,\Z_2)).
\eea
The cohomology of $T^4$ with $\Z_2$ coefficients is given by
generators $\th^i$ of degree 1 satisfying $(\th^i)^2=0$. The $\th^i$
can be thought of as the mod 2 reductions of the one-forms
$dx^i$. Because we work with $\Z_2$ coefficients, they are invariant
under the orbifold group, since the group acts as $\th^i \to
-\th^i=\th^i$. In the equivariant cohomology of $T^4$, we have to add
an extra generator $\xi$ also of degree one.  The element $\xi$ is the
generator of the group cohomology $H^*(B\Z_2,\Z_2)$.

There is subtle point in that the relation among the generators
$\th^i$ is modified to
\bea
\th^i(\xi-\th^i)=0.
\label{rel}
\eea
This relation replaces the relation $(\th^i)^2=0$ in the cohomology of
$T^4$. It has a simple geometric interpretation that becomes already
clear if we consider the case of $S^1$.  Since we work with $\Z_2$
coefficients, let us consider a {\it real} line bundle over the
orbifold $S^1/\Z_2$. Upstairs, we have a bundle over $S^1$ and this
bundle is either trivial or the M\"obius bundle. The $\Z_2$ action on
$S^1$ has two fixed points at $x=0$ and $x=\pi$. We consider this
bundle locally around one of the fixed point, say $x=0$. Let $y$ be
the coordinate in the fiber. The group action will map $x \to -x$ and
for the fiber coordinate we have two possibilities: either the trivial
action $y \to y$ or the non-trivial action $y \to -y$. Since we have
two fixed points this gives a total of $2^2=4$ possibilities.

These 4 possibilities have a clear interpretation as equivariant
bundles on $S^1$. If the group action is trivial at both fixed points
we obtain the trivial bundle on $S^1$ with the trivial group
action. If the group action is only non-trivial at one fixed point, we
obtain the M\"obius bundle with the two possible $\Z_2$
actions. Finally if the group action is non-trivial at both fixed
points, the corresponding bundle on $S^1$ is trivial, but it has a
non-trivial group action that can globally be written as $y \to -y$. This
represents the equivariant bundle that we denoted as $\xi$.  So, if
$\th_0$ and $\th_1$ denote the classes that represent the non-trivial
bundles around the fixed points $x=0$ and $x=\pi$, then we clearly
have the relation
\bea
\th_0 + \th_1 = \xi.
\eea

Now the $\Z_2$-equivariant cohomology of $S^1$ is very simple. As
explained in appendix \ref{app:equiv}, it is the cohomology of the homotopy
quotient $S^1_{\Z_2}$ which is a circle bundle over the classifying
space $\R\P^\infty$. We can think of $S^1$ as the union of two line
intervals, glued together at $x=0$ and $x=\pi$. The corresponding
bundles of these intervals over $\R\P^\infty$ have contractible fibers
and so are homotopic to $\R\P^\infty$ itself.  The intersection of the
two intervals gives rise to a $\Z_2$ bundle over $\R\P^\infty$ with a
total space that, by definition, is contractibe to a point. Therefore
$S^1_{\Z_2}$ is homotopically a wedge $\R\P^\infty \vee \R\P^\infty$,
i.e., two copies of $\R\P^\infty$ glued together at a point. The
classes $\th_0$ and $\th_1$ are the generators of the two copies of
$H^*(\R\P^\infty,\Z_2)$, and clearly we have
\bea
\th_0 \cup \th_1 =0.
\eea
So the equivariant cohomology of $S^1$ is given by the ring
\bea
H^*_{\Z_2}(S^1,\Z_2) \cong \Z_2[\th_0,\th_1]/(\th_0 \th_1).
\eea
Both the elements $\th_0$ and $\th_1$ map to the invariant
element in $H^1(S^1,\Z_2)$ that represents the M\"obius bundle. We can
pick as generators $\th:=\th_1$ and $\xi$. But we should impose the
relation
\bea
\th (\xi-\th)=0.
\eea
Note that the translation $S^1$ over a half-period that interchanges
the fixed points is represented by $\th \to \th + \xi$ which indeed
interchanges $\th_0$ and $\th_1$.

Generalizing this construction to the $4$-torus gives the relation
(\ref{rel}) among the $\th^i$ and $\xi$. In that case a translation
$x^i \to x^i + \pi q^i$, $q^i \in \Z_2$ acts on the cohomology
generators as
\bea
\th^i \to \th^i + q^i \xi.
\label{trans}
\eea

A direct computation shows that all classes in $H^1_{\Z_2}(T^4,\Z_2)$
correspond to classes with $U(1)$ coefficients, so that we
find that
\bea
H^1_{\Z_2}(T^4,U(1)) \cong \Z_2^5.
\label{Z5}
\eea
We can pick a basis given by $\th^1,\th^2,\th^3,\th^4,\xi$ and
write a general equivariant flat bundle as
\bea
A = \sum_{i=1}^4 a_i\th^i+a_0\xi,\qquad a_i,a_0\in \Z_2.
\eea
Here the coefficients $a_i,a_0$ take value in the field of two
elements that we simply write as $\Z_2$, i.e., the set $\{0,1\}$ with
addition and multiplication modulo (which is often written as ${\mathbb
F}_2$). So altogether we have $2^5=32$ equivariant flat bundles on
$T^4$.

We can understand this result directly in terms of bundles on $T^4$.
We first pick a flat connection $A=A_i dx^i$ on $T^4$.  Such a connection
has holonomy
\bea
\exp \oint_{\g_i} A = \exp {2\pi i A_i}
\eea
around the one-cycle $\g_i$ along the $x^i$ axis. The holonomy takes
value in the torus 
$$H^1(T^4,U(1))\cong (T^4)^*.$$ 
The orbifold group
acts on the flat gauge field by $A_i \to -A_i$, where we identify $A_i
\cong A_i+1$. So the invariant connections are given by
\bea
A_i = 0,1/2 
\eea
which are the points of order two on the Jacobian. This gives a total
of $2^4=16$ invariant bundles.  Now we have to decide how the group
$\Z_2$ acts on the fiber of the line bundle. Here we can have either
the trivial action or the non-trivial action. This multiplies the total
number of bundles by two with a total of 32 bundles.

In terms of the equivariant cohomology classes that we described above
we have the following interpretation of these 32 bundles.  The element
$\xi$ clearly represents the trivial flat bundle on $T^4$ with the
non-trivial group action.  It corresponds to the non-trivial element in
the group cohomology $H^1(B\Z_2,U(1))$.

Let us consider the corresponding M theory compactifications. First we
remark that these 32 cases decompose into orbits of $SL(4,\Z_2)$. In
particular the 16 invariant flat connections $A=a_i\th^i$ behave very
much as spin structures on a two-torus. They form two orbits: a
singlet of trivial holonomy $A=0$ (the ``odd spin structure'') and a
remaining orbit of 15 non-trivial holonomies $A=a_i\th^i$ with not all
$a_i=0$ (the ``even spin structures''). (Here {\bf 15} is the adjoint
representation of $SL(4,\Z_2)$.) A similar decomposition of ${\bf 16}
= {\bf 1} +{\bf 15}$ holds for the bundles of type $A=a_i\th^i +\xi$.
Both $a_i\th^i$ and $a_i \th^i+\xi$ map to the flat connection
$A_i=\half a_i$. 

Note that a half shift along the torus can interchange the two {\bf
15} orbits. In fact the full symmetry group is
\bea
(\Z_2)^4 \ltimes SL(4,\Z_2)
\eea
which is the affine group acting on $(\Z_2)^4$. We can think of the
expression $(a_i,a_0)$ as defining the affine linear function $a_i x^i
+ a_0$. If not all $a_i$ vanish this defines one of the 30 affine
planes. This gives the orbit {\bf 30}. The two constant functions are
singlets. 

This split ${\bf 32}={\bf 1} + {\bf 1} + {\bf 30}$ also manifests
itself geometrically in the M theory interpretations.  The singlet
$A=0$, of course, corresponds to the trivial compactification $T^4\times
S^1 \cong T^5$. The other singlet $A=\xi$ gives a compactification
manifold
\bea
(T^4 \times S^1)/\Z_2
\eea
where the $\Z_2$ acts as a shift $x^{11} \to x^{11} + \pi$. The action
is fixed point free. This manifold breaks half the
supersymmetries. The moduli are simply the NS fields (metric and
$B$-field) on $T^5$ plus the string coupling constant (radius
$S^1$). There are no continuous RR field moduli. That is, this
compactification has exactly the same number of moduli as the CFT on
$T^4$.

All of the other bundles can also be obtained as quotients of $T^5$,
now considered as the trivial circle bundle over a double cover
$\widehat{T^4}$ of the original IIA 4-torus $T^4$. First, by an
$SL(4,\Z_2)$ transformation we can map $A$ to the form $A=\th^1$ or
$A=\th^1+\xi$. In the case $A=\th^1$ we define the cover
$\widehat{T^4}$ by making the coordinate $x^1$ periodic modulo
$4\pi$. We then obtain the bundle by the quotient
\bea
(\widehat{T^4}\times S^1)/\Z_2\times \Z_2.
\eea
Here the first $\Z_2$ (the deck transformation of the cover) acts as
\bea
g_1:\ x^1 \to x^1 + 2\pi,\quad x^{11}\to  x^{11} + \pi
\eea
while the second $\Z_2$ (the original orbifold group) acts as
\bea
g_2:\ x^i \to -x^i, \quad i=1,\ldots,4.
\eea
If we first divide out $g_2$, we obtain a description
as the quotient
\bea
(K3 \times S^1)/\Z_2
\eea
where $K3$ is the Kummer surfaces $\widehat{T^4}/\Z_2$ and the $\Z_2$ is
the involution that exchanges the 16 fixed points $x^i=0,2\pi$ in pairs.

Similar remarks hold for the case $A=\th^1+\xi$. Here we find a $\Z_2
\times \Z_2$ action generated by
\bea
g_1:\ x^1 \to x^1 + 2\pi,\quad x^{11}\to x^{11} + \pi
\eea
and
\bea
g_2:\ x^i \to -x^i,\qquad x^{11}\to x^{11} + \pi.
\eea
In this case it makes sense to first quotient by,
\bea
g_1g_2:\ x^i \to 2\pi-x^i.
\eea
This again produces a singular $K3$ manifold, now with fixed points at
$x^i=\pi,3\pi.$ This is no surprise because we already remarked that a
half shift interchanges these two types of compactification.

In summary, the M theory geometry of a type IIA orbifold $T^4/\Z_2$
with possible 1-form flux is one of the following three forms:
\bea
T^5,\ (T^4\times S^1)/\Z_2,\ \left(T^4/\Z_2 \times S^1 \right)/\Z_2
\eea
reflecting the ${\bf 32}={\bf 1} + {\bf 1} + {\bf 30}$ decomposition
of $(\Z_2)^5$. The extension to more general quotients is discussed
in appendix \ref{app:generalquotient}. 

\subsubsection{Local holonomies}
\label{oneformhol}

We now turn to the local description of these bundles on the orbifold
$T^4/\Z_2$ in terms of type IIA string theory. Note that the
quotient is a singular $K3$ manifold. On a smooth $K3$ manifold there
can be no non-trivial RR 1-form fluxes. The fundamental group is
trivial so that there can be no non-trivial flat $U(1)$ bundles. So the 1-form
flux is necessarily located at the singularities. This we want to make
more precise.

In the case of $T^4/\Z_2$, the $K3$ manifold has 16 $A_1$
singularities that look locally like $\C^2/\Z_2$. If we cut out a small
neighbourhood of the singularities, we obtain a manifold $X_0$ with a
boundary $Y$ that consists of 16 copies of $S^3/\Z_2
\cong \R\P^3$.  Since $\pi_1(\R\P^3) = \Z_2$ there is a non-contractible
curve $\g_p$ around each fixed point $p$. Since we are given a
(smooth) line bundle with a flat connection $A$ over $X_0$, we can
compute the holonomy of the connection around that curve for every
fixed point $p$
\bea
\exp i \oint_{\g_p}A= \exp i\pi \v(A;p),\qquad \v(A;p) \in \Z_2 = \{0,1\}.
\eea
The two possibilities $\v(A;p) =0,1$ correspond to the two possible
equivariant flat line bundles on $\C^2$ with either the trivial or the
non-trivial group action.

A clear interpretation of a Type IIA string theory compactification on
the orbifold $\C^2/\Z_2$ with a discrete flux for the RR 1-form
gauge field $A$ is given in \cite{New}.  This flux can be measured by
performing an Aharanov--Bohm scattering process by sending a D0-brane
around the loop $\g$. Because of this discrete gauge field, the
singular orbifold cannot be deformed to a smooth ALE hyperk\"ahler
manifold, as in the case without discrete RR flux. Such a
deformation would kill the fundamental group and the corresponding
non-trivial flat connection $A$.  Therefore the singularity is
``frozen.''  String loop corrections make this apparent singular
configuration non-singular.  In M theory, this compactification is
represented by the five-dimensional orbifold
\bea
(\C^2 \times S^1)/\Z_2
\eea
where the $\Z_2$ acts on the circle by $x^{11} \to x^{11} + \pi$.
This action is fixed point free so that we are dealing with a smooth
M theory compactification. In weak coupling, when the radius of the
$S^1$ shrinks to zero, the string theory becomes free everywhere
outside the core of the singularity, but an interacting string theory
remains at the singularity, which can be viewed as a dual represention of
a Neveu--Schwarz five-brane.

Since there are 16 fixed points, this seems to give a priori $2^{16}$
possibilities for putting discrete flux at the singularity. But in
this global situation not all of these possibilities are realized. 
A constraint comes about because the loops
$\g_p$ are not all independent in the homology group $H_1(X_0,\Z)$,
where we recall that $X_0$ is the orbifold with the singularities cut
out. There are relations among the 16 generators.  This means that not
any arbitrary combination of the 16 fluxes leads to a flat bundle that
can be extended over the interior of the $K3$ manifold.

In fact, it is a classical result in the theory of Kummer
surfaces that the first homology group of $X_0$ is given by
\bea
H_1(X_0,\Z) \cong (\Z_2)^5.
\eea
We will explain this result in more detail in the next section where
we discuss flat line bundles on singular $K3$ manifolds in
general. For the moment, we just observe that we recover the group
$(\Z_2)^5$ of the equivariant computation (\ref{Z5}).

We now want to connect the local and the global computation.  How do
we determine the holonomy of a given equivariant bundle on $T^4$ around
a particular fixed point $p$? This is a straightforward computation,
see also \cite{Kronheimer, Reid}.

Note that from an abstract point of view, we are doing the following:
consider a neighbourhood $U_p$ of the point $p$ on the cover
$T^4$. This we can identify as a ball in $\C^2$ equiped with the
non-trivial $\Z_2$ action.  The inclusion $i_p : U_p \to T^4$ is an
equivariant map. It therefore gives rise to a map
\bea
i^*_p:\ H^1_{\Z_2}(T^4,U(1)) \to H^1_{\Z_2}(U_p,U(1))\cong \Z_2.
\eea
This map describes the restriction of the bundle on $T^4$ to the
bundle in the neighbourhood of the fixed point.  Locally, there is only
the generator $\xi$. So under restriction the equivariant class $A$
will become some multiple of $\xi$:
\bea
i^*_p(A)=\v(A;p)\xi,
\eea
and this defines the holonomy $\v(A;p)$ around $p$.

Since the fixed points are the points of order two on $T^4$, we can
label them by $p=(p^1,p^2,p^3,p^4) \in (\Z_2)^4$, so that the fixed
point $p$ has coordinates $x^i=\pi p^i$. We now claim that the
holonomy of the bundle $A= a_i \th^i + a _0\xi$ is given by the linear
affine function
\bea
\v(A;p)= a_i p^i + a_0
\label{holphi}
\eea
{}From our abstract description this follows rather directly, because we
defined the generator $\th^i$ such that it restricts to zero at
$p^i=0$ and equals $\th^i=\xi$ at $p^i=1$. So we can write the
restriction simply as
\bea
i^*_p(\th^i) = p^i \xi.
\label{loc}
\eea

We display this function for the five generators, where we listed
the 16 fixed points $p$ lexicographically.

$$
\renewcommand{\arraystretch}{1.5}
\begin{tabular}{|c|cccccccccccccccc|}
\hline
$\th^1$  &  1 &  1 &  1 &  1 &  1 &  1 &  1 &  1 
       &  0 &  0 &  0 &  0 &  0 &  0 &  0 &  0 \\
$\th^2$  &  1 &  1 &  1 &  1 &  0 &  0 &  0 &  0 
       &  1 &  1 &  1 &  1 &  0 &  0 &  0 &  0 \\
$\th^3$  &  1 &  1 &  0 &  0 &  1 &  1 &  0 &  0 
       &  1 &  1 &  0 &  0 &  1 &  1 &  0 &  0 \\
$\th^4$  &  1 &  0 &  1 &  0 &  1 &  0 &  1 &  0 
       &  1 &  0 &  1 &  0 &  1 &  0 &  1 &  0 \\
$\xi$    &  1 &  1 &  1 &  1 &  1 &  1 &  1 &  1 
       &  1 &  1 &  1 &  1 &  1 &  1 &  1 &  1 \\
\hline
\end{tabular}
\renewcommand{\arraystretch}{1.0}
$$

\noindent Apart from the bundles $A=0$ and $A=\xi$ that have fluxes
at none or all of the 16 fixed points, every other bundle has fluxes
at 8 of the 16 fixed points. This is in line with the description of
the bundle as $(T^4/\Z_2\times S^1)/\Z_2$.

We can of course also do a straightforward computation of the holonomy
by integrating the gauge field over the closed path $\g_p$.  In terms
of the cover $T^4$, this loop can be represented as a path that starts
at the point $x_0=\pi p -\e$ and ends at $x_1=\pi p +\e$ with $\e$ a
small 4-vector. This is an open path on $T^4$, so to find the holonomy
along this path, we have to relate the bundle at the two end points.

The original connection on $T^4$ we picked to be the constant one-form
$A=\half a_i dx^i$. We can gauge this connection away at the cost of
introducing twisted boundary conditions. One describes these boundary
conditions by accompanying the identification $x^i \to x^i + 2\pi
n^i$, with $n^i\in \Z$, by a gauge transformation $\exp i\pi a_i n^i$.
Under the quotient map $x^i \to -x^i$, we can have a further gauge
transformation $\exp i\pi a_0$. With these boundary conditions, we can
simply compute the holonomy around the path $\g_p$.  Since the local
connection is trivial, the holonomy is complete expressed in the gauge
transformation that accompanies the identification of the two end
points. Since
\bea
x_1 = -x_0 + 2\pi p,
\eea
the gauge transformation gives the phase 
\bea
\exp i\pi (a_i p^i + a_0),
\eea
which gives our formula (\ref{holphi}).

\subsubsection{Singular $K3$ manifolds with one-form flux}

We will now turn to the case of a general $K3$ surface. As we
explained, smooth $K3$'s cannot support 1-form flux, so the $K3$
surface is necessarily singular. Locally these singularities will be
of type $\C^2/G$ with $G$ a finite subgroup of $SU(2)$ as given
by the $ADE$-classification. This implies
that we can consider $X$ to be a local orbifold, i.e., a manifold that can
be covered by coordinate patches that are orbifolds. In such a case, we
still have well-defined equivariant cohomology groups. However, in the
case of 1-form flux, there is no need to use these equivariant classes;
instead, we simply excise the singularities and work on the smooth
remainder.

This result is formulated as follows using the language of algebraic
topology. Let $X_*$ be the singuar $K3$, $U_*$ a neighbourhood of the
singularities, and
\bea
X_0= X_* - U_*,
\eea
the smooth manifold obtained by cutting out the singularities.  The
boundary $\partial X_0 = Y$ will consist of a union of smooth
three-manifolds of the type $S^3/G$. We clearly want to compute the group
\bea
{\cal Z} = H_1(X_0,\Z).
\eea

Now let $X$ be the {\it resolved} $K3$ surface, and let $U$ be the
neighbourhood of the {\it resolved} singularities. That is we replace
every component of $U_*$ by its corresponding smooth ALE-space. We
have $X-U=X_0$ and $\partial U=Y$. It is a standard result that
\bea
H_1(X_0,\Z) \cong H^3(X_0,Y,\Z) \cong H^3(X,U,\Z).
\eea
Here we use the relative cohomology groups $H^k(X,U)$.  These describe
pairs $(a,b) \in C^k(X)\times C^{k-1}(U)$ of a $k$-cochain $a$ on $X$
and a $(k-1)$-cochain $b$ on $U$ that satisfy $da=0,$ $i^*a=db$, modulo
the equivalence $(a,b) \sim (a+du,b+i^*u -dv)$ with $(u,v) \in
C^{k-1}(X)\times C^{k-2}(U)$. So the elements of $H^k(X,U)$ represent,
roughly, cocycles on $X$ that are trivial when restricted to $U$.
Lefschetz duality gives
\bea
H^k(X,U)  \cong H_{4-k}(X-U),\qquad 
H^k(X-U) \cong H_{4-k}(X,U).
\eea
Relative cohomology groups fit in a long exact sequence
\bea
\to H^k(X,U) \to H^k(X) \to H^k(U) \to H^{k+1}(X,U) \to
\eea
In our case, using the cohomology of $X=K3$ and $U$, this gives the
short exact sequence
\bea
0 \to H^2(X,U,\Z) \to H^2(X,\Z) \to H^2(U,\Z) \to H^3(X,U,\Z) \to 0.
\eea
Here $H^2(X,\Z) \cong \G_{3,19}$ is the familiar second cohomology of
$K3$. The group $H^2(U,\Z)$ is the dual of group of vanishing cycles $Q =
H_2(U,\Z)$. The lattice $Q$ is a sublattice of $H_2(X,\Z) \cong
\G_{3,19}$. It consist of the 2-cycles that are contracted to a point 
in the singularity. There is a second sublattice of $H_2(X,\Z)$, the
lattice $P = H_2(X_0,\Z)$ of 2-cycles in the smooth part
$X_0$. Duality gives us $P \cong H^2(X,U,\Z) \cong H^2(X_0,Y,\Z)$. So
$P$ represents the bundles that become trivial on the boundary $Y$. Using
these results the exact sequence gives
\bea
0 \to P \to \G_{3,19} \to Q^* \to {\cal Z} \to 0.
\eea
This equation has the following interpretation: the singularity gives
rise to a sublattice $Q \subset \G_{3,19}$ of vanishing cycles.  This
gives a dual map $ \G_{3,19} \to Q^*$. This map is not necessarily
onto. The quotient of $Q^*$ by the image of $\G_{3,19}$ gives us the
group of 1-form fluxes ${\cal Z}$. Equivalently, the lattice $Q \subset
\G_{3,19}$ is not necessarily primitively embedded, and this is measured
by ${\cal Z}$.

There are some other definitions of ${\cal Z}$ that are directly
equivalent to this by duality. Since we also have
\bea
{\cal Z} = \Tor H^2(X_0,\Z) \cong \Tor H_2(X,U,\Z),
\eea
we find for example
\bea
H^2(X_0,\Z)=H_2(X,U,\Z) \cong P^* \oplus {\cal Z}.
\eea

Note that the lattices $Q$ and $Q^*$ have an interpretation in terms
of line bundles over the resolved singularity $U$. The exact sequence
\bea
0 \to H^2(U,Y,\Z) \to H^2(U,\Z) \to H^2(Y,\Z) \to 0
\eea
becomes
\bea
0 \to Q \to Q^* \to {\rm disc}(Q) \to 0
\eea
where the discriminant of the lattice $Q$ also equals the first
homology of the three-manifold $Y$, which is a disjoint sum of quotients
$S^3/G$.  To be precise, ${\rm disc}(A_n)=\Z_{n+1}$, ${\rm disc}(D_n) =
\Z_2\times \Z_2$ or $\Z_4$ for $n$ even or odd, respectively,
and ${\rm disc}(E_n)=\Z_3,\Z_2,\{1\}$ for $n=6,7$ and $8$, respectively. 
Here $Q^*=H^2(U,\Z)$ are the line bundles on $U$, $Q=H^2(U,Y,\Z)$ are
the line bundles that become trivial when restricted to $Y$, and ${\rm
disc}(Q) = H^2(Y,\Z) \cong H_1(Y,\Z)$ represent the (necessarily
torsion) line bundles on $Y$.

\subsection{Fluxes and K-theory}
\label{secKtheory}

We now turn to the K-theory description of fluxes for the RR 1-form
and 3-form fields in type IIA string theory. 

\subsubsection{K-theory description of RR fluxes}

There is now considerable evidence that 
K-theory should play a role in classifying 
RR charges in type II string theory~\cite{Minasian:1997mm, Witten:1998cd}, at least 
for zero string coupling.
There are also arguments suggesting that
the RR fields themselves, and in particular
their flux quantization, should be formulated in terms of K-theory
\cite{Moore:2000gb,freed-hopkins}. To be more precise, there is a
local formulation of K-theory called differential K-theory that
provides a precise characterization of the RR fields in terms of
bundles with connections \cite{freed}. According to this point of view
the flat RR fluxes on a compactification $X$ are given by the groups
$K^i(X,U(1))$ where $i=1$ for the IIA theory and $i=0$ for the IIB
($i$ only matters modulo 2 by Bott periodicity). These groups have
been studied in detail in \cite{lott}.

The most important property
of the groups $K^i(X,U(1))$ is that they fit in the exact sequence
\begin{equation}
\renewcommand{\arraystretch}{1.5}
\begin{tabular}{cccccc}
$K^{1}(X)$ & $\to$ & $H^{odd}(X,\R)$ & $\to$ & $K^{1}(X,U(1))$\\
$\uparrow$ &       &                 &       &   $\downarrow$ \\
$K^{0}(X,U(1))$ & $\leftarrow$ & $H^{even}(X,\R)$ & $\leftarrow$ & $K^{0}(X)$\\
\end{tabular}
\renewcommand{\arraystretch}{1.0}
\label{sequence}
\end{equation}
Here the maps from K-theory to cohomology are the usual Chern character
maps, that are isomorphisms over the reals.

For a finite-dimensional smooth manifold $X$, we therefore find a description
of the RR phases as 
\begin{equation} 
0 \to \frac{H^{odd}(X,\R)}{K^1(X)/\Tor} \to K^1(X,U(1)) \to \Tor K^0(X) \to 0. 
\end{equation}
So the component group is given by the torsion classes in $K^0(X)$, and 
the trivial component consists of the torus $H^{odd}(X,\R)/(K^1(X)/\Tor)$. 

K-theory is not graded, so one cannot distinguish in a unique way
the contributions of the RR fields of fixed degree. Instead there is a 
filtration
\begin{equation}
K^0(X) \supseteq K^0_1 \supseteq K^0_2 \supseteq \cdots
\end{equation}
where the group $K^0_k(X)$ is defined in terms of vector bundles that
are trivial on the $(k-1)$-skeleton of $X$. This gives a description
of RR fluxes where the field strength is a form of degree $k$ or
higher. So, within the K-theory formalism fluxes always have
ambiguities up to higher degree terms. 

In this paper we want to restrict to fluxes of the 1-forms and 3-forms
in the IIA theory as measured by the world-volumes of D0-branes and D2-branes.
These are given by the quotient
\begin{equation}
K^1(X,U(1))/\left(K^1(X,U(1)\right)_5. 
\end{equation}

We will not give the general interpretation of the elements of the
group $K^1(X,U(1))$, but restrict our discussion here to the case that 
$H^{odd}(X,\R)$ vanishes, so that only torsion classes appear. In that
case the group of fluxes is simply
\begin{equation}
K^{1}(X,U(1)) \cong \Tor K^0(X).
\end{equation}
The corresponding RR flux can be represented by
a pair $(E_0,E_1)$ of {\it flat} vector bundles with ${\rm rk}(E_0)=
{\rm rk}(E_1)$, i.e., a virtual flat bundle $E_0-E_1$ of rank zero. 

A torsion RR flux will give an additional phase factor to a D-brane in
the string theory path-integral.  If we represent a brane by a
K-homology class, that is a map an odd spin-manifold $M$ (possibly
equipped with an additional Chan--Paton vector bundle $F$) into the
space-time $X$, the holonomy of the RR-fields over this brane can be
represented by the $\eta$-invariant of virtual bundle $E_0-E_1$ restricted
to $M$. 

More precisely, consider the Dirac operator $D$ on
$M$ coupled to the bundles $E_0$ and $E_1$. Such a Dirac operator 
is self-adjoint and has a $\eta$-invariant, which is the regularized 
sum of the signs of the eigenvalues $\lambda_i$ of $D$ \cite{APS} 
\begin{equation}
\eta = \eta(0),\qquad
\eta(s) = \sum_{\lambda_i  \not=0} {\rm sgn}(\lambda_i) 
|\lambda_i|^{-s}.
\end{equation}
Closely related is the so-called reduced $\eta$-invariant, defined by
\begin{equation}
\overline\eta = (\eta + {\rm dim\, ker}\, D)/2 \quad ({\rm mod}\ \Z).
\end{equation}
This invariant does not jump under variations of the parameters. It
appears in the index theorem on manifold with boundaries. If we can write
$M$ as the boundary of a manifold $B$, and extend the bundle $E_0$ to a 
bundle $E$ over $Z$, then with the appropriate APS boundary conditions,
the twisted Dirac operator $D_{E}$ on $B$ has index
\begin{equation}
{\rm index}\, D_{E} = \int_B ch(E) \widehat{A}(B) - \overline\eta_{E_0}
\end{equation}
The phase (modulo $\Z)$ associated to $M$ is now
\begin{equation}
\v(M) = \overline\eta_{E_0} - \overline\eta_{E_1}.
\end{equation}
In view of the APS index theorem, this can be written as
\begin{equation}
\v(M) = \int_{M\times I} ch(E) \widehat{A}(M) 
\end{equation}
where we have picked a bundle $E$ with a (no longer necessarily flat)
connection on the manifold $M \times I$
with $I=[0,1]$ that reduces to $E_{0}$ and $E_{1}$  on the two boundaries.

\subsubsection{K-theory on orbifolds}

For the case of an orbifold $X/G$, it seems reasonable to assume that
equivariant K-theory is the appropriate description. For the case of
the RR charges the relevance of equivariant K-theory has been
demonstrated in great detail \cite{Witten:1998cd, equiv-K, equiv-K2, BGK}. 
It leads to an elegant description of 
fractional branes pinned on the orbifold singularities. The boundary
state corresponding to these fractional branes has components in the
twisted sectors of the closed string.

Equivariant K-theory $K_G(X)$ classifies equivalence classes of
equivariant bundles on $X$, that is bundles with an action of $G$.
Just as in equivariant cohomology, we can make a model for $K_G(X)$ in
terms of the ordinary K-theory on the homotopy quotient
$X_G$. However, in the equivariant case there is a remarkable
difference in the prediction of fluxes coming from a K-theory or a
cohomology formulation.

As an example we can study the basic orbifold $\C^2/\Z_2$ again. In
this case we have 
\begin{equation} 
K^0_{\Z_2}(\C^2) \cong K^0_{\Z_2}(pt) \cong K^0(\R\P^\infty) = \Z \oplus
\Z.
\end{equation} 
This generalizes as follows.  For an arbitrary finite group $G$ the
ring $K_G^0(pt)$ is given by the (completion of the) representation
ring $R(G)$ of $G$. One has $R(G) = \Z^r$ with $r$ the number of
irreducible representations (or equivalently the number of conjugacy
classes). So as an example for $\Z_2$, the representation ring is generated 
by the trivial and the non-trivial representation.
 
The equivariant K-group should be contrasted by the equivariant 
cohomology that is given by
\begin{equation}
H^{even}_{\Z_2}(\C^2,\Z) = H^{even}(\R\P^\infty,\Z)= 
\Z \oplus \Z_2 \oplus  \Z_2 
\oplus  \cdots
\end{equation}
This difference is a consequence of a familiar effect:
if we consider a finite-dimensional
approximation by $\R\P^{2N+1}$ then the even 
cohomology and K-theory have the same associated graded, but 
differ in the extensions
\begin{equation}
K^0(\R\P^{2N+1}) = \Z \oplus \Z_{2^N},\qquad
H^{even}(\R\P^{2N+1},\Z) = \Z \oplus (\Z_2)^{N}.
\end{equation}
We see how in the large $N$ limit the K-group becomes free because
$\Z_{2^N}$ tends to $\Z$. This effect has been recently described in 
terms of fractional branes on the orbifold in \cite{BGK}.

According to the equivariant version of the sequence (\ref{sequence}),
the IIA RR fluxes for the orbifold $\C^2/G$ are given by the kernel of 
the map 
\begin{equation}
K_G^0(pt) \to H^{even}(G,\R) \cong \R.
\end{equation}
This is the so-called reduced K-theory  $\widetilde{K}_G^0(pt)$
that is represented by virtual bundles
of dimension zero. So we have the prediction that the group of fluxes
is given by
\begin{equation}
K_G^1(\C^2,U(1)) \cong \widetilde{K}_G^0(pt) \cong \widetilde{R}(G).
\end{equation}
Here $\widetilde{R}(G)$ is the reduced representation ring that consists
of virtual representations of dimension zero, i.e., pairs of representations
\begin{equation}
\rho_0,\rho_1:\ G \to U(N)
\end{equation}
with ${\rm dim}\,\rho_0 ={\rm dim}\,\rho_1$. Given such a representation, 
we can construct the corresponding equivariant flat bundles 
$E_0,E_1$ on $\C^2$, by taking trivial bundles $\C^N$ and implementing the
action of $G$ through the representations $\rho_{0}$, $\rho_{1}$.

Now this description raises some questions. It seems to predict an
infinite set of fluxes at a simple orbifold. If $G$ is of $ADE$-type,
then the lattice $\widetilde{K}_G^0(pt)$ can be identified with the
weight lattice $Q^*$ of the corresponding simple Lie group. However, as we
remarked before, in a K-theory description one cannot distinguish the
dimension of the corresponding flux. Since the equivariant fluxes are
defined in terms of an infinite-dimensional classifying space, this
would imply that the fluxes are of arbitrary high dimension and should be
measured by branes of arbitrary high dimension. This is
clearly not a physical description. In our case one would like to restrict to
1-form and 3-form fluxes that can be measured by the world-volumes of
zero-branes and two-branes. Therefore one would argue that the physical
fluxes are restricted to the quotient
\begin{equation}
\widetilde{K}^0_G(pt)/\left(\widetilde{K}^0_G(pt)\right)_5
\end{equation}
where we have taken the quotient by the subgroup that describes bundles
that vanish on the 4-skeleton of $BG$. 

\subsubsection{One-form fluxes in K-theory}

How do we measure such a flux in the orbifold $\C^2/G$? As we
explained above, a RR flux flux will give an extra phase to an
Euclidean D-brane instanton. Let us also restrict to branes with smooth
world-volumes that do not pass through the orbifold singularity. So we
consider the space $(\C^2- \{0\})/G$. This can be retracted to the
smooth 3-dimensional space $S^3/G$.

Let us first consider the RR 1-form field. Fluxes of this field are
described by the quotient
\begin{equation}
\widetilde{K}^0_G(pt)/\left(\widetilde{K}^0_G(pt)\right)_3.
\end{equation}
Given a closed one-manifold $\gamma$ in $S^3/G$ the holonomy of the
1-form field is described as follows. The K-theory class is given by
the pair of representations $(\rho_0,\rho_1)$ and the
corresponding flat equivariant bundles $(E_0,E_1)$. To such a virtual 
vector bundle we can associate a flat equivariant line bundle $L$, the
determinant bundle
\begin{equation}
L = \det E_0 \otimes (\det E_1)^*
\end{equation}
Now the phase of the RR 1-form field is simply the holonomy of the
flat connection $A$ of the line bundle $L$
\begin{equation}
\exp i \oint_\gamma A.
\end{equation}
The first Chern class of this line bundle is represented by an 
equivariant cohomology class
\begin{equation}
c_1(L) = c_1(\rho_0) - c_1(\rho_1) \in H^2(G,\Z) 
\end{equation}
Here we define the Chern classes of a representation $\rho:\  G \to U$
through the associated map of classifying spaces
\begin{equation}
\rho^*:\ H^*(BU) \to H^*(G,\Z).
\end{equation}
The cohomology of the classifying space $BU$ is generated by the Chern
classes $c_1,c_2,\ldots$ and the images $\rho^*c_i$ we denote by
$c_i(\rho)$. 

Since any element of $H^2(G,\Z)$ represent an equivariant line bundle,
we see that the equivariant K-theory description coincides with the
description in terms of equivariant cohomology classes given in section
\ref{RRone}. The fluxes are given by
\begin{equation}
H^2(G,\Z) \cong H^1(G,U(1)).
\end{equation}

\subsubsection{Three-form fluxes in K-theory}

According to the formalism explained in the preceeding discussion, 
a 3-form flux will
associate a phase to a Euclidean D2-brane world-volume $M$ given by
the eta-invariant of the virtual bundle $E_0-E_1$ restricted to $M$
\begin{equation}
\v(M) = \overline\eta_{E_0} - \overline\eta_{E_1}. 
\end{equation}
In our case there is an obvious choice for $M$, namely the manifold
$S^3/G$ that surrounds the singularity. (Note that any oriented
three-manifold is spin, although there could be in principle more than one
inequivalent spin structure.)  We can express this phase directly in
terms of the Chern--Simons invariant, since
\begin{equation}
\v(M) = \int_{M\times I} ch(E) = CS(E_0) - CS(E_1).
\end{equation}
Here we have defined the Chern--Simons invariant for a $U(N)$ flat gauge
field as
\begin{equation}
CS(E) = \int_B ch_2(E) 
\end{equation}
where $B$ is a four-manifold with $\partial B = M$ over which we have extended
the bundle $E$ and $ch_2$ is the second Chern character
\begin{equation}
ch_2 = \half c_1^2 - c_2= \frac{1}{8\pi^2} {\rm Tr}\, F \wedge F.
\end{equation}
For the $U(1)$ part, this is half the usual Chern--Simons term. As
explained in \cite{dijkgraaf-witten}, this is perfectly fine.  Because
the relevant cobordism groups vanish, for every three-manifold $M$ one can
find a four-manifold $B$ and an extension of the $U(N)$ bundle such that
$B$ is spin. The fact that the intersection form on $B$ is now even
allows one to define the $CS$ invariant using $\half c_1^2$. 

Given a representation $\rho$ of $G$, we can compute the Chern--Simons invariant
in terms of group cohomology as follows. We have the Chern classes
\begin{equation}
c_1(\rho) \in H^2(G,\Z),\qquad c_2(\rho) \in H^4(G,\Z).
\end{equation}
Now it is not difficult to see that for a group $G$ of $ADE$-type, we have
\begin{equation}
H^4(G,\Z) \cong \Z_{|G|}.
\end{equation}
In fact a spectral sequence argument \cite{mark} gives $H^4(G,\Z)$
as the cokernel of the composition
\begin{equation}
H^3(SU(2)/G,\Z) \to H^3(SU(2),\Z) \to H^4(BSU(2),\Z).
\end{equation}
The first map is clearly of order $|G|$ whereas the second
(transgression) is an isomorphism. So the generator of $H^4(G,\Z)$ is the
class $c_2(\rho_2)$ where $\rho_2$ is the defining 2-dimensional
representation of $G$.  

The $CS$ invariant for the manifold $S^3/G$ is defined by the map
\begin{equation}
CS:\ H^3(G,U(1)) \to H^3(S^3/G,U(1))\cong U(1)
\end{equation}
where we identify $S^3 \cong SU(2)$. So for the representation $\rho_2$,
this is simply given by
\begin{equation}
 CS(\rho_2)= \frac{1}{|G|} \quad ({\rm mod}\ \Z).
\end{equation}

For a general representation $\rho$, let $c_1$ be its first Chern
class considered as an element of $H^2(S^3/G,\Z)\cong H^2(G,\Z)$,
and let $\a$ be the representative in $H^1(S^3/G,U(1))$. 
Similarly, let $\beta$ be the representative of $c_2$ in $H^3(S^3/G,U(1))$.
Then the Chern--Simons invariant of $\rho$ is defined by 
\begin{equation}
CS(\rho) = \half \a \cup c_1 - \beta. 
\end{equation}
The phase associated to a D2-brane wrapped on $S^3/G$ in a background
described by the virtual flat bundle $\rho_0-\rho_1$ is given by
\begin{equation}
\v(S^3/G) = CS(\rho_0) - CS(\rho_1).
\end{equation}
Note that the first term in the Chern--Simons term (coming from $\half
c_1^2$) has a familiar description. There is an isomorphism 
\begin{equation}
H^2(G,\Z) \cong Q^*/Q,
\end{equation}
where $Q$ is the root lattice of type $G$. Since $Q$ is an even
lattice, there is a quadratic form (essentially the discriminant form
of the lattice $Q$---see appendix \ref{app:lattice}),
\begin{equation}
x \in Q^*/Q \ \mapsto \ q(x)=\half x\cdot x \in \Q/\Z \subset U(1).
\label{quad}
\end{equation}
This is the $CS$ invariant of the line bundle $x$. 

If the 1-form flux is zero, we can define the group of 3-form fluxes
as the quotient
\begin{equation}
(\widetilde{K}^0_G)_3/(\widetilde{K}^0_G)_5.
\end{equation}
In this case $c_1=0$ so the flux is measured by $c_2 \in H^3(G,U(1))$.
Since this group is generated by the 2-dimensional representation, we
see that this group of pure 3-form fluxes is simply given by 
\begin{equation}
H^4(G,\Z) = H^3(G,U(1) = Z_{|G|}.
\end{equation}

Indeed, in this case it is not difficult to compute the groups of fluxes
directly in K-theory.
For a finite subgroup $G \subset SU(2)$, the non-vanishing cohomology
occurs in even degree, where for $k\geq 1$
\begin{equation}
H^{4k-2}(G,\Z) = G_{ab} = G/[G,G],\qquad
H^{4k}(G,\Z) = \Z_{|G|}.
\end{equation}
This implies immediately that the Atiyah--Hirzebruch spectral sequence
degenerates. If
\begin{equation}
K_p = \left(\widetilde{K}^0_G(pt)\right)_{p}
\end{equation}
then we have the successive quotients 
\begin{equation}
K_p/K_{p+1}=H^p(G,\Z),\qquad p>0.
\end{equation}
This immediately gives that the group of 1-form fluxes modulo 3-form
fluxes is
\begin{equation}
K_0/K_3 = H^2(G,\Z) 
\end{equation}
as we have seen explicitly. The group of 3-form
fluxes with vanishing 1-form flux  is also easily computed to be
\begin{equation}
K_3/K_5 \cong H^4(G,\Z)
\end{equation}
confirming our computation via the $CS$ invariants.  Finally the full
group of 1-form and 3-form fluxes
\begin{equation}
M(G)=K_0/K_5
\end{equation}
is given by an extension
\begin{equation}
0 \to H^4(G,\Z)\cong \Z_{|G|} \to M(G) \to H^2(G,\Z) \to 0.
\end{equation}
The corresponding cocycle of this extension is given by the map
\begin{equation}
b:\ H^2(G:\Z) \times H^2(G,\Z) \to H^4(G,\Z),\qquad b(x,x') = x \cup x'.
\end{equation}
This cocycle is definitely exact as an extension by $U(1)$ since
\begin{equation}
b(x,x')=q(x+x') - q(x) - q(x'),\qquad q(x) = \half x \cdot x
\end{equation}
where we use the identification $H^2(G,\Z) \cong Q^*/Q$. So the
question is simply whether $q(x)$ is of the form $n/|G|$. Since $2
q(x)= x\cdot x$ is always in $H^4(G,\Z)$, the only question is whether
$x\cdot x$ is divisible by two in $H^4(G,\Z)$. 

This extension question changes the additive structure on the group of
fluxes. Physically this effect can be described as follows. In the
presence of a RR 1-form background $x\in H^2(G,\Z)$, there is an
induced 3-form flux given by the $CS$ invariant
\begin{equation}
CS(x) = \int_B \half c_1^2 = q(x).
\end{equation}
Now suppose $x$ has order $n$. That is to say, $x$ represents a line
bundle $L$ and $nx$ represents the trivial line bundle $L^{\otimes
n}$. The 1-form flux of $nx$ as measured by D0-branes vanishes, but
the 3-form flux will be given by $n q(x)$ and this is not necessarily
0 mod 1. As we argued above one can have $n q(x) = \half$ mod 1.

In terms of K-theory, such an element $x$ corresponds to a virtual
bundle $L-1$. But this element has infinite order using the group
structure in K-theory which is {\it addition} of bundles. If we take
it $n$ times, we get the class $nL -n$ which is a virtual vector
bundle. Of course the determinant line is trivial, but the bundle has
higher order secondary invariants, in particular the $CS$ invariant
measured by a D2-brane.

The simplest example is $G=\Z_2$. In that case
we have 
\begin{equation}
K_0= \widetilde{K}_{\Z_2}^0(pt) = \Z
\end{equation}
generated by the class $x=\rho-1$ where $\rho$ is the non-trivial
representation of $\Z_2$ (or the tautological line bundle over $\R\P^\infty$).
We can now compute the phases for a flux $nx$, $n\in \Z$. As we explained
we are only interested in the 1-form and 3-form fluxes. These are labeled
by
\begin{equation}
M(\Z_2) = K_0/K_5 = \lim_{N\to \infty} \Z_{2^N}/\Z_{2^{N-2}} \cong
\Z_4.
\end{equation}
This should be contrasted with the answer obtained from
equivariant cohomology\footnote{An explanation about how to compute
this equivariant cohomology group appears in appendix \ref{app:equiv}.} 
\begin{equation}
H^2(\Z_2,\Z) \oplus H^4(\Z_2,\Z)= \Z_2 \oplus \Z_2.
\end{equation}
Since $c_1(x)$ is the generator of $H^2(\Z_2,\Z)\cong \Z_2$, the
1-form flux is simply $n/2$ (mod 1). So the holonomy of the virtual
bundle $nx$ is  $(-1)^n$.

For the 3-form flux, we have to measure the phase of a membrane wrapped
over $S^3/\Z_2$. This equals the Chern--Simons invariant of the bundle
$\rho$. Since $c_2=0$ we obtain
\begin{equation}
CS(x) = CS(\rho) = \int_B \half c_1^2 = \frac{1}{4}.
\end{equation}
For the bundle $nx$ we have
\begin{equation}
CS(nx) = \frac{n}{4}  \quad ({\rm mod}\ \Z).
\end{equation}
We see that in the presence of a RR 1-form flux (that is, if $n$ is odd),
there an induced 3-form flux. More precisely, the 3-form flux has a shifted
quantization, where the phase factor satisfies
\begin{equation}
\v = \frac{1}{4} \quad ({\rm mod}\ \frac{1}{2}).  
\end{equation}

This computation can be repeated for general $G$.  We use the same
notation for the root lattice $Q$ and the corresponding finite
subgroup $G$ of $SU(2)$.  For $A_n$ we have an extension
\begin{equation}
0 \to \Z_{n+1} \to M \to \Z_{n+1} \to 0.
\end{equation}
Let $x$ be the generator of $H^2(A_n,\Z) = \Z_{n+1}$.  One computes
$q(x) = n/2(n+1)$ so that $(n+1)q(x)=n/2$. For $n$ odd this is $\half$
mod 1, so we find a non-trivial extension
\begin{equation}
M(A_{2k}) = \Z_{2k+1} \times \Z_{2k+1},\qquad
M(A_{2k-1}) = \Z_{4k} \times \Z_{k}.
\end{equation}
For $D_n$ ($n>3$) one finds $H^2(D_n,\Z)=\Z_2 \times\Z_2$ for $n$ even,
$H^2(D_n,\Z)=\Z_4$ for $n$ odd,  $H^4(D_n,\Z) = \Z_{4(n-2)}$ and
$q(x)=\frac{n}{8}$ mod 1 for the generator(s).  So, again for $n$ odd we have
a non-trivial extension since $4q(x)=n/2$
\begin{equation}
M(D_{2k}) = \Z_2\times \Z_2 \times \Z_{8(k-1)},\qquad
M(D_{2k-1}) = \Z_{8} \times \Z_{2(2k-3)}.
\end{equation}
For $E_{6,7,8}$ the induced fluxes $q(x)$ are respectively $\frac{2}{3}, 
\frac{3}{4},0$ and there are no extensions
\begin{equation}
M(E_6) = \Z_3 \times \Z_{24},\quad M(E_7) = \Z_2\times \Z_{48},\quad
M(E_8) = \Z_{120}.
\end{equation}
Note that $E_8$ does not support 1-form flux.

\subsubsection{An alternate method of computation}
\label{alternate}
In this section, let us make a brief digression and
describe a purely algebraic construction of the quotients
\begin{equation}
K^0_G(pt)/( K^0_G(pt) )_{2n+1}  
\end{equation}
which appeared in our preceeding discussion of fluxes. We shall
discuss the case $G=\Z_m$. 
Recall that $K^0_{2n}(X)$ involves bundles that are trivial on
the $(2n)$-skeleton of $X$. For the equivariant case, we need
$X=BG$, and a convenient set of spaces that approximate $X$
is given by the set $BG^{(n)}$ that appears in Milnor's construction
of $BG$, see, for example \cite{atiyahsegal}. Motivated by the results of 
\cite{atiyahsegal} and the calculations in appendix~\ref{app:equiv}, 
we propose the following characterization of
$ K^0_{\Z_m}(pt)/( K^0_{\Z_m}(pt) )_{2n+1}$. 
Let 
$\phi$ be the augmentation map 
\begin{equation} \phi: \quad R[G]\to\Z
\end{equation} 
that sends $\sum n_i
R_i$ to $\sum n_i d_i$. Here $d_i$ is the dimension of the 
representation $R_i$. The kernel of $\phi$ is the 
augmentation ideal $I$ of $R[G]$. Then
\begin{equation}
K^0_{\Z_m}(pt)/( K^0_{\Z_m}(pt) )_{2n-1}
\cong R[{\Z_m}]/I^n  .
\end{equation}
For example, for $G=\Z_2$ the augmentation ideal is
generated by $\rho-1$, where $\rho$ is the non-trivial representation
of $\Z_2$. Since $(\rho-1)^n=2^{n-1}(\rho-1)$, $I^n$ is generated
by $2^{n-1}(\rho-1)$. From this, we see that
\begin{equation}
 K^0_{\Z_2}(pt)/( K^0_{\Z_2}(pt) )_{2n-1}
\cong \Z \oplus \Z_{2^{n-1}} 
\end{equation}
in agreement with the result that we found before.
A few additional calculations are needed to show that
\bea
K^0_{\Z_m}(pt)/( K^0_{\Z_m}(pt) )_3 & = & 
\Z \oplus \Z_m \\
K^0_{\Z_m}(pt)/( K^0_{\Z_m}(pt) )_5 & = & 
\Z \oplus \Z_{d_1} \oplus \Z_{m^2/d_1} \\
K^0_{\Z_m}(pt)/( K^0_{\Z_m}(pt) )_7 & = & 
\Z \oplus \Z_{d_1} \oplus \Z_{d_2} \oplus \Z_{m^3/d_1 d_2}
\eea
where
$$ d_1={\rm gcd}\left(m, \frac{m(m-1)}{2}\right),\qquad
d_2={\rm gcd}\left(m, \frac{m(m-1)}{2}, \frac{m(m-1)(m-2)}{6}\right) . $$
These groups agree with the results of our earlier computations using
the reduced eta-invariant.

\subsection{M theory compactifications with three-form flux}
\label{Mthreeflux}

We now turn to M theory compactifications with 3-form flux. In
studying these compactifications, we face a difficult issue: namely,
we do not yet understand how to correctly treat the 3-form of M
theory. There are various suggestions involving gerbes \cite{sharpe},
or suggestive relations with $E_8$ gauge bundles
\cite{moore2, Moore:2000nn}. 
However, it seems likely that the correct way to treat
the 3-form involves a framework that is currently unknown. In the
following section, we describe how the data we have obtained point to
the existence of new ``frozen'' singularities which support 3-form
flux. We also conjecture the existence of a new class of dualities
between compactifications with different singular geometries and
fluxes.

In section \ref{threeformequiv}, we consider the possibility that
equivariant cohomology is the right framework for the 3-form. As an
example, we describe the relevant equivariant cohomology groups for
the global orbifold $T^4/\Z_2$.  Section \ref{threeformhol} extends
the discussion of 1-form holonomies which appeared in section
\ref{oneformhol} to the case of 3-form fluxes. This leads to some
quite beautiful results which are likely to be useful in other
contexts. In section \ref{dualCHL}, we compute the equivariant
cohomology of the global orbifold $T^4/\widehat{\cal D}_4$. We point
out that there is a choice of flux with the properties we expect for
the M theory compactification dual to the CHL string.

In section \ref{compareiia}, we resolve some puzzles in comparing M
theory with type IIA. The resolution suggests a natural generalization
of the Freed--Witten anomaly.  Section \ref{threeformgeom} contains
some thoughts on the the geometry of the M theory 3-form, and a number
of related topics. Finally, we wrap up the discussion with a brief
mention of F theory compactifications with flux.

\subsubsection{Frozen singularities and new dualities}
\label{threeformphysics}

So far, we have related our asymmetric orbifolds in $6$ dimensions to
M theory compactifications on $(Z \times S^1)/G$ where $Z=K3$. We have
also described cases with $Z=T^4$ without heterotic duals. As we have
seen, the smooth geometric M theory compactifications give type IIA
compactifications with torsion RR 1-form flux.

Some of these theories are obtained by compactifying a 7-dimensional
theory on a circle.  We studied three classes of these 7-dimensional
theories: namely, type IIA orientifolds, heterotic asymmetric
orbifolds and F theory compactifications. An example like the CHL
string can be realized by all three constructions.
 
However, it is natural to ask whether there are other strong coupling
descriptions which involve M theory compactified on a four-manifold.
Since the only bosonic fields of M theory are the metric and the
$3$-form field, we can only turn on $3$-form fluxes on the
four-manifold. Compactifying such a theory on a circle gives a 
6-dimensional IIA compactification on the same four-manifold with RR
$3$-form flux.

Let us begin by recalling what is known about $7$-dimensional M theory
compactifications. Apart from the standard compactification of M
theory on K3 without flux, only the strong coupling limit of the CHL
string has been discussed. The CHL string can be realized by a
$7$-dimensional orientifold with 2 $O6^{+}$ planes and 6 $O6^{-}$
planes. The strong coupling limit of $O6^{-}$ planes is smooth in M
theory \cite{Seiberg:1996bs, Seiberg:1996nz, Sen:1997kz}, while that
of an $O6^{+}$ plane appears to be a ``frozen'' $D_4$ singularity in M
theory \cite{Landsteiner:1998ei, wittoroid}.  The $D_4$ singularity
can be seen to arise in various ways. In type IIA near an $O6^+$ plane,
the geometry of a three-surface around the orientifold is
$S^2/\Z_2=\RP^2$. In addition, the $O6^{+}$ orientifold has RR charge
$+4$, which is measured by the field strength of the RR $1$-form gauge
field. This means that the circle on which we reduce from M theory to
type IIA forms a circle bundle with first Chern class $+2$ on
$\RP^{2}$. The total space of that circle bundle over $\RP^2$ is then
$S^3/D_4$. This explains why we expect a $D_4$ singularity to
appear. Alternatively, before modding out by $\Z_2$, we have an object
with charge $+4$ which is therefore described by an $A_3$ singularity
in M theory. Modding out by $\Z_2$ yields a $D_4=A_3/\Z_2$
singularity.

The same $D_4$ singularity can also be found in the geometry seen by a
D2-brane probe. On the probe, there is an $N=4$ $d=3$ Yang--Mills
theory with gauge group $O(2)$ and a hypermultiplet transforming in
the symmetric representation.  The Coulomb branch geometry contains a
$D_4$ singularity. The existence of this $D_4$ singularity is also
required to reproduce the Seiberg--Witten curves of $SO$ and $Sp$ gauge
theories from 5-brane geometry \cite{Landsteiner:1998ei}.

Since an $O6^{+}$ plane has no moduli, the $D_4$ singularity should
also have no moduli. In particular, the standard set of moduli
associated with resolving the singularity should be lifted.  If the
$D_4$ singularity were purely geometrical, we could certainly resolve
it. So the only possible means by which it could be frozen must
involve the $3$-form.  In $7$ dimensions, the two types of orientifold
plane $O6^-$ and $O6^+$ are distinguished by the sign of the
contribution of the $\RP^2$ world-sheets to the string path
integral. This sign is related to a discrete choice of NS-NS
$B$-field.  This discrete $B$-field should have its origin in a
discrete $3$-form field in M theory that gives different signs to
various contributions to the membrane path integral.  Unfortunately,
this path integral is not understood, and we do not have a precise
geometrical definition of the discrete allowed values of the $3$-form
field for a given singularity.  We shall return to a discussion of the
geometry of the M theory $3$-form later. For now, the primary message
is that M theory appears to have frozen singularities with discrete
$3$-form flux.  In particular, the $7$-dimensional CHL string is
described by M theory on $K3$ with two frozen $D_4$ singularities with
some discrete $3$-form flux \cite{wittoroid}.

As this point, let us return to an issue that appeared in our analysis
of 7-dimensional orientifolds of section \ref{orientpuzzle}.  We found
2 orientifold compactifications with equal numbers of $O6^+$ and
$O6^{-}$ planes. These configurations are clearly distinct as
pertubative string theories. Are they also distinct
non-perturbatively? With the strong coupling description of $O6^+$ in
hand, it seems worth mentioning that one check on the number of
non-perturbative configurations is to count the number of
non-isomorphic embeddings of the weight lattice for $(D_4)^4$ into the
$K3$ lattice, $\G_{3,19}$. If the orientifold configurations are
distinct non-perturbatively then it is possible that the difference
will be visible in the number of inequivalent ways of embedding
$(D_4)^4$. There are analogues of this embedding question for
lower-dimensional orientifold compactifications.  It might be possible
to do this counting using the techniques developed in \cite{Nikulin80,
daveold}.\footnote{A preliminary computation indeed suggests that
there are 2 non-isomorphic embeddings.}

\begin{table}
\begin{center}
\renewcommand{\arraystretch}{1.5}
\begin{tabular}{|c|c|c|} \hline
Singularity     & Discrete Fluxes & Enhanced Gauge Symmetry\\ \hline
 $D_{4+n} $ & $\Z_2$ & $C_n$ \\ \hline
 $E_6$ & $\Z_2, \Z_3$ & $C_2$, $\{ e\}$ \\ \hline
 $E_7$ & $\Z_2, \Z_3, \Z_4$ & $B_3$, $A_1$, $\{ e\}$ \\ \hline
 $E_8$ & $\Z_2, \Z_3, \Z_4, \Z_5, \Z_6$ & $F_4$, $G_2$, $A_1$, $\{ e\}$, $\{ e\}$ \\\hline
\end{tabular}
\renewcommand{\arraystretch}{1.0}
\caption{Singularities, their allowed $3$-form fluxes, and 
enhanced gauge symmetries.} \label{table:fluta}
\end{center}
\end{table}

Now, in order to find M theory compactifications dual to 
our new $7$-dimensional theories, for which no orientifold
descriptions exist, the lattices discussed in section \ref{moduli}
turn out to be very helpful. Recall that the lattice for the
CHL string in $d=7$ took the form,
\bea
\Lambda_{\rm CHL} = \Gamma_{3,3} \oplus D_4 \oplus D_4.
\eea
The orthogonal complement of $\Lambda_{\rm CHL}$
in the usual $7-d$ heterotic lattice $\Gamma_{3,19}$ is 
\bea
\Lambda^{\perp}_{\rm CHL} = D_4 \oplus D_4.
\eea
As was already observed in \cite{Mikhailov98}, $\Lambda^{\perp}_{\rm
CHL}$ seems to be directly related to the singularities appearing in
the dual M theory description. Indeed, on the heterotic side there are
no moduli for $\Lambda^{\perp}_{\rm CHL}$, and in M theory, the
corresponding singularities are frozen.  By looking at table
\ref{table:lattices}, we see that the generalizations of
$\Lambda^{\perp}_{\rm CHL}$ are $E_6 \oplus E_6$ for $\Z_3$, $E_7
\oplus E_7$ for $\Z_4$, $E_8 \oplus E_8$ for the finite groups $\Z_5$
and $\Z_6$. This leads us to conjecture that these lattices describe
frozen singularities in M theory with the corresponding discrete
fluxes.  By tuning moduli, we can enhance the unbroken gauge symmetry
in the asymmetric orbifold compactification. In the geometric picture,
this means that we can enhance certain singularities with flux. This
shows that we can also turn on a $\Z_3$ flux in an $E_8$
singularity. With such a $\Z_3$ flux we can resolve the $E_8$ into an
$E_6$, but no further resolution is possible.

\begin{table}
\begin{center}
\renewcommand{\arraystretch}{1.5}
\begin{tabular}{|c|c|} \hline
Frozen Singularities     & Dual Description \\ \hline
 $D_{4}\oplus D_4$ & $\Z_2$ triple \\ \hline
 $E_6 \oplus E_6 $ & $\Z_3$ triple \\ \hline
 $E_7\oplus E_7$ & $\Z_4$ triple \\ \hline
 $E_8 \oplus E_8$ & $\Z_5$ triple \\ \hline
 $E_8 \oplus E_8$ & $\Z_6$ triple \\ \hline
 $(D_4)^4 $ & $\Z_2$ F \\ \hline
 $(E_6)^3 $ & $\Z_3$ F \\ \hline
 $D_4 \oplus E_7 \oplus E_7 $ & $\Z_4$ F \\ \hline
 $D_4 \oplus E_6 \oplus E_8 $ & $\Z_6$ F \\ \hline
\end{tabular}
\renewcommand{\arraystretch}{1.0}
\caption{A list of frozen singularities for $7$-d M theory compactifications 
on $K3$ with flux. 
In labeling the dual theories, $G$ triple refers to 
the $G$ heterotic 
asymmetric orbifold, while $G$ F refers to F theory 
on $(T^4\times S^1)/G$. } \label{table:Mduals}
\end{center}
\end{table}

A summary of the $ADE$-singularities with their allowed discrete
$3$-form fluxes is given in table \ref{table:fluta}.  Notice that for
the $E_6$ singularity, we can turn on either a $\Z_2$ or a $\Z_3$
discrete flux, but not both. In particular, the set of allowed fluxes
do not form a group. This is a first sign that these fluxes are
classified by a rather intricate geometrical object.  In our heterotic
string analysis, we always found that the singularities came in
pairs. This is presumably related to the fact that on a compact
manifold like $K3$, the total flux must vanish. We therefore always
need two singularities with equal fluxes of opposite sign. This
appears to be similar to the cancellation of the two Chern--Simons
invariants in the $E_8 \times E_8$ heterotic string, and to the gluing
conditions for the two del Pezzo surfaces appearing in F theory
descriptions.

It is also natural to ask whether similar dual descriptions exist for
the $7$-dimensional F theory compactifications on $(T^4\times S^1)/G$.
Indeed, in essentially the same way as described above, we are led to
a conjectured dual description in terms of M theory on a $K3$ surface
with a particular set of frozen singularities. The list of frozen
singularities for the M theory models dual to asymmetric orbifolds and
to these F theory compactifications appears in table
\ref{table:Mduals}.

If we further compactify M theory on a singular $K3$ with discrete
$3$-form flux on a circle, we obtain type IIA on the same $K3$ with RR
$3$-form flux. However, we also had a different description of the
same theory in terms of another orbifold $K3$ with RR $1$-form
flux. This has to be a $K3$ with different geometry because the $K3$
surfaces that appear with $3$-form flux are in general {\it not}
orbifold $K3$ surfaces: they are not quotients of another $K3$ surface
by some discrete group. This picture can therefore only be consistent
if there is a duality between IIA on $K3$ with discrete RR $1$-form
flux and IIA on a different $K3$ surface with discrete $3$-form flux.

To argue that such a duality indeed exists, let us examine the charge
lattices of both theories obtained by wrapping D0, D2 and D4-branes on
the $K3$. For an orbifold $K3$ with discrete RR $1$-form flux, the
charge lattice is constructed as follows: the vanishing cycles in the
orbifold $K3$ generate a lattice $\Lambda^{(1)}_{\rm vanishing}$.
These lattices have been constructed by Nikulin \cite{Nikulin79}\ for
the $K3$ case, and appear in table \ref{table:K3}. For the case of
$X=T^4$, the lattice of vanishing cycles appears in table
\ref{table:torus} \cite{Wendland}.  We should not wrap D-branes over
any of these vanishing cycles.  This becomes particularly clear by
looking at M2-branes wrapping $2$-cycles in the M theory description
on $(X \times S^1)/G$ (we will elaborate on the connection
between the IIA and M theory description in section \ref{compareiia}) .
However, we can wrap D2-branes over any cycle
 orthogonal to $\Lambda^{(1)}_{\rm
vanishing}$, which yields the
 lattice $\Lambda^{(1)\perp}_{\rm vanishing}$,
the orthogonal
 complement of $\Lambda_{\rm vanishing}$ in $\Gamma_{3,19}$. 
In
 addition, we can also consider D0 and D4-branes on the orbifold $K3$,
leading to a total charge lattice
 \bea \label{lat1}
\Lambda_{\mbox{\rm D-branes}} 
= \Gamma_{1,1} \oplus \Lambda^{(1)\perp}_{\rm vanishing}.
\eea
For $K3$ surfaces with discrete RR $3$-form flux, the charge lattice
is constructed in a similar way. However, a new subtlety arises.
Namely, not all brane wrappings are allowed. This conclusion follows
by applying dualities to the Freed--Witten anomaly \cite{Freed:1999vc},
which states that the class of the pullback of $H$, the field
strength of the NS-NS $2$-form $B$, to the D-brane world-volume must
vanish.  In particular, if the class $[H]$ is torsion of order $k$,
only branes that come in stacks of $k$ are allowed since $k[H]=0$
\cite{Kapustin:2001di}. If we apply this constraint to a D3-brane and
further S and T-dualize, we see that the pullback of the $3$-form
field strength to the D4-brane world-volume has to vanish. In the
cases at hand, the field strength of the RR $3$-form will be torsion,
say of order $k$, which is directly related to the order of the finite
group appearing in the $1$-form discussion above.  Therefore, we can
only wrap $D4$ branes $nk$ times, with $n$ an integer. The
contribution to the charge lattice from D0 and D4-branes will
therefore not be $\Gamma_{1,1}$, but $\Gamma_{1,1}(k)$. In the
$3$-form case, we therefore conclude that the lattice is given by:
\bea \label{lat3}
\Lambda_{\mbox{\rm D-branes}} 
= \Gamma_{1,1}(k) \oplus \Lambda^{(3)\perp}_{\rm vanishing}.
\eea

If the theories with the $1$-form and the theories with the $3$-form
flux are to be dual to each other, the lattices (\ref{lat1}) and
(\ref{lat3}) should be equivalent to each other. Therefore,
$\Lambda^{(1)\perp}_{\rm vanishing}$ has to be an index $k$ sublattice
of $\Lambda^{(3)\perp}_{\rm vanishing}$. This explains how the
geometries of the two $K3$ surfaces can be so different from one
another: the lattices of vanishing cycles differ by an element of
order $k$. For example, in the case of the CHL string, the lattice
$\Lambda^{(1)\perp}_{\rm vanishing}$ is generated by eight vanishing
cycles, and in addition, half of the sum of these cycles is also an
element of integral homology.  On the other hand, the lattice
$\Lambda^{(3)\perp}_{\rm vanishing}$ is the lattice of two $D_4$
singularities. It is also generated by eight vanishing cycles, but
this time half the sum of the first four cycles and half the sum of
the second set of four cycles each are in integral homology. This
lattice is twice as large as $\Lambda^{(1)\perp}_{\rm vanishing}$, as
we expect from the preceeding discussion.

The duality between $K3$ surfaces with $1$-form and $3$-form fluxes
differs considerably from the usual T-dualities. Typically, T-duality
preserves the nature of singularities (since it preserves enhanced
gauge symmetries), while in our case, it changes the geometry of the
$K3$ in a much more dramatic way. We note, however, that the $K3$
surfaces are still birationally equivalent. That this is true can be
seen from the F theory discussion of the previous section. In the
construction involving two del Pezzo surfaces, we found singular
fibers of type $I_0^*$, $IV^*$, $III^*$ and $II^*$.  The corresponding
extended Dynkin diagrams are those of $D_4,E_6,E_7$ and
$E_8$. Birationally equivalent geometrical models of these
elliptically fibered $K3$ surfaces have singularities described by the
extended Dynkin diagram minus one node.  The singularities of the $K3$
with $1$-form flux are obtained by removing nodes with Dynkin label
$k$ from the singularities. The singularities of the $K3$ with
$3$-form flux are obtained by removing the extended nodes of the
Dynkin diagrams, leading to singularities of type $D_4,E_6,E_7$ and
$E_8$. One can show that the lattices of vanishing cycles obtained
this way indeed differ by an element of order $k$, and by construction
the $K3$ surfaces are birationally equivalent.

The novel duality described here raises all kinds of questions.  One
puzzle that arises is the following: we already described in section
\ref{RRone} that type IIA on $K3/\Z_2$ with a particular choice of
$1$-form flux is dual to the CHL string. Na\"{\i}vely, this string theory
appears to be quotient of IIA on $T^4$ with a smooth $\Z_2$ invariant
$1$-form i.e., a $\Z_2$ invariant element of $H^1(T^4,U(1))$.  It is
well-established that T-duality maps $T^4$ with a constant $1$-form to
the dual torus with a constant $3$-form.  In addition, if the $1$-form
is $\Z_2$ invariant then so is the $3$-form which is now an element of
$H^3(T^4,U(1))$. If we mod out the theory with the $3$-form by the
action of the T-dual $\Z_2$, we would expect to arrive at IIA on
$T^4/\Z_2$ with a $3$-form flux. This, however, is not consistent with
the picture given above.

The reason for this apparent discrepancy is that the $\Z_2$ quotient
makes sense as a perturbative orbifold only when the vanishing
$2$-cycles have a half-integral $B$-field flux
\cite{Aspinwall:1995zi}. In cases without this flux like geometric
orbifolds, we do not have perturbative control over the theory.
Therefore, modding out by $\Z_2$ yields a duality between IIA on
$T^4/\Z_2$ with $1$ and $2$-form flux and IIA on $T^4/\Z_2$ with $2$
and $3$-form flux.  Neither of these theories has a smooth
$7$-dimensional limit. On the other hand, we are interested in a
duality between IIA on $T^4/\Z_2$ with only $1$-form flux and IIA on
another $K3$ with only $3$-form flux. These theories are not
perturbative $\Z_2$ orbifolds of string theory, and so we cannot
na\"{\i}vely use T-duality to relate them. As we argued above, the duality
between these theories is considerably more involved.

However, this still leaves us with the following question: by starting
with a $T^4$ with $\Z_2$ invariant $3$-form, we can construct the
singular manifold $T^4/\Z_2$ with a non-trivial discrete $3$-form
flux. Does string theory compactified on this space make any 
sense\footnote{An example involving discrete zero and four form
fluxes is discussed in \cite{narain,narain2}.}?
We have not encountered this string theory anywhere in our prior
discussion, and we suspect it is inconsistent.
It would be nice to give a direct argument for this, perhaps using
supersymmetry or anomaly cancellation.  A related situation that is
worth mentioning arises for type IIA strings: consider an orientifold
$6$-plane with RR charge $q$. If there is no $B$-field, this plane is
supersymmetric for $q\geq -2$.  However, with a half-integral discrete
$B$-field, the orientifold plane is an $O6^+$ plane and is only
supersymmetric if $q\geq +2$. This can be seen from a tadpole
calculation in string theory, but it is not clear whether there is a
direct low-energy argument giving this conclusion.

To summarize our discussion so far: we found $7$-dimensional M theory
compactifications on $K3$ with frozen singularities and discrete
$3$-form fluxes. These theories are dual descriptions of either
7-dimensional asymmetric heterotic orbifolds or F theory
compactifications.  These 7-dimensional theories give 6-dimensional
IIA theories by compactifying on a circle, and the $6$ and
$7$-dimensional lattices differ by a factor of $\Gamma_{1,1}(k)$.
These models serve as data from which we obtain a list of new
frozen, or partially frozen, singularities in M theory.

\subsubsection{Three-form flux as equivariant cohomology}
\label{threeformequiv}

The most natural geometric object with which to identify the M theory
3-form is a generalized connection on a ``2-gerbe''; see, for
example,~\cite{Hitchin:1999fh}.  Although 2-form $B$-fields and
``1-gerbes'' have a relatively straightforward interpretation in terms
of bundles of algebras of operators on infinite-dimensional Hilbert
spaces, this description is less clear for the next case of 3-form
fields. For a smooth manifold $X$ we expect that flat 3-forms are
given by the cohomology group $H^3(X,U(1))$. There are some
indications that for an orbifold $X/G$ one should again consider the
corresponding equivariant group $H^3_G(X,U(1))$ \cite{sharpe}. It is
therefore natural to extend our discussion of equivariant cohomology
to this case.

The equivariant cohomology groups $H^3_G(X,U(1))$ are computable.  
As explained in appendix \ref{app:equiv}, 
in the case of a $T^4/\Z_2$ orbifold we find
\begin{equation}
H^3_{\Z_2}(T^4,U(1)) \cong (\Z_2)^{15}.
\end{equation}
In terms of the previously introduced generators $\th^i$ and $\xi$, a
general element can be written as
\begin{equation}
C = \sum_{i<j<k} c_{ijk}\th^i\th^j\th^k + 
\sum_{i<j} c_{ij} \th^i \th^j\xi + \sum_i c_i\th^i\xi^2 + c_0\xi^3,
\label{C}
\end{equation}
with
\begin{equation}
c_{ijk},c_{ij},c_i,c_0 \in \Z_2.
\end{equation}

Note that we now have contributions of various type. First of all
there are obvious 3-forms that come from invariant fields on
$T^4$. The flat 3-form field on $T^4$ are described by $H^3(T^4,U(1))$
and this group is (up to a determinant) canonically isomorphic with
$T^4$ itself.  The $\Z_2$ group acts as $C \to -C$ on the 3-form
field, so the invariant configurations are
\begin{equation}
C = \sum_{i<j<k} \frac{1}{8\pi^2} C_{ijk} dx^i dx^j dx^k, \qquad 
C_{ijk}=0,1.
\end{equation}
As we explained, there is a well-defined map of the equivariant forms
into the invariant forms, and clearly a class of the form (\ref{C})
gets mapped to the invariant 3-form $C_{ijk}=c_{ijk}$. So the 16
classes $c_{ijk}\th^i\th^j\th^k$ modulo forms of lower degree in the
$\th^i$ can be given this geometric interpretation. 

The class $\xi^3$ is the image of the group
cohomology class in $H^3(\Z_2,U(1))$. It describes a trivial 2-gerbe
with a non-trivial group action. This is for example the class that
describes a non-zero $C$-field in the neighbourhood of the singularity
$\C^2/\Z_2$.  It would be an analogue of discrete torsion for the NS
$B$-field. The other classes of type $c_{ij}\th^i\th^j\xi$ and $c_i
\th^i\xi^2$ describe mixed cases that are partially geometric and
partly group theoretic.

\subsubsection{Three-form holonomies}
\label{threeformhol}

The local computation of the consistent 3-form fluxes is 
much easier in this case. 
If we consider the manifold with boundary $X_0$ obtained
by cutting the 16 singularities out of the $K3$ manifold $T^4/\Z_2$,
we have again $2^{16}$ possibilities, because at each singularity we
have the non-trivial class $\xi^3$ coming from $H^3(\Z_2,U(1))$. Now
the global topology gives only one condition. Because the boundary $Y
=\partial X_0$ bounds a four-manifold, the sum of all classes should add
up to zero. This reduces the $(\Z_2)^{16}$ to $(\Z_2)^{15}$.

In the
same way that flat connections have well-defined holonomies along
closed curves of a given homology class, flat $2$-gerbes have
$U(1)$-valued holonomies along $3$-cycles.
The holonomy of the 3-form $C$ around a fixed point $p$ is determined
as follows. The boundary of a neighbourhood of $p$ is given by a
3-cycle $\Sigma_p \cong \R\P^3$ and the holonomy is defined locally as
\begin{equation}
\exp i\int_{\Sigma_p} C = \exp i \pi \v(C;p).
\end{equation}
We propose the following formula for the holonomy of a class $C$ as
represented as in (\ref{C}) around the fixed point $p\in (\Z_2)^4$,
\begin{equation}
\v(C;p) = \sum_{i<j<k} c_{ijk}p^ip^jp^k + 
\sum_{i<j} c_{ij} p^i p^j
+ \sum_i c_ip^i + c_0,
\label{Chol}
\end{equation}
This formula follows again trivially from the definition of the
holonomy as the restriction to the fixed point
\begin{equation}
i^*_p(C) = \v(C;p) \xi^3
\end{equation}
combined with the localisation formula (\ref{loc}) that $i^*_p(\th^i)
= p^i \xi$.

Note that these formulas have a nice algebraic interpretation. Because
our generators $\th^i$ satisfy $(\th^i)^2=\th^i \xi$ we have for the
variables $p^i$ the relation
\begin{equation}
(p^i)^2 = p^i.
\end{equation}
This means that we can consider the holonomy as a function of the
variables $p^i$ over the field $\Z_2$ (with $p^i=0,1$). 

So quite generally we have the following algebraic interpretation of
the holonomy of an equivariant $k$-cocycle $C$ of the torus
$T^n$. Consider the affine space $A^n=(\Z_2)^n$ with coordinates
$p^i$, and let $F(A^n)=\Z_2[p^1,\ldots,p^n]$ be the space of functions
\begin{equation}
\v:\ A^n \to \Z_2.
\end{equation}
Then there is a filtration, where $F_k(A^n)$ denotes the functions of
at most degree $k$. According to the above localisation formula, we can
think of the holonomy $\v(C;p)$ as an element of $F_k(A^n)$. We can
identify the quotients $F_k(A^n)/F_{k-1}(A^n)$ with the exterior
products $\Lambda^k(\Z_2)^n$. Clearly the affine group $(\Z_2)^n
\ltimes SL(n,\Z_2)$ acts on $F(A^n)$ and the orbits form the different 
types of equivariant classes.

There is an interesting check of the holonomy formula. We can pick a
3-cycle that is represented by a 3-plane in $T^4$. This will give a
3-cycle on the orbifold $T^4/\Z_2$ that will divide the 16 fixed
points in 2 groups of 8. Clearly the holonomy of $C$ around that plane
should be equal to the sum of the holonomies around the fixed points
``to the right'' or ``to the left'' of the cycle. It is not difficult
to check that this property is indeed reflected in the above
formula. For example, if we pick the plane say spanned by
$x^1,x^2,x^3$ the holonomy will be $c_{123}$ and this equals the sum
\begin{equation}
c_{123} = \sum_{p,\ p^4=0} \v(C;p) = \sum_{p,\ p^4=1} \v(C;p).
\end{equation}

In order to compute the holonomy directly in terms of the $C$-field,
we should make use of the formalism of gerbes. Let us consider in all
generality an $n$-gerbe defined on a manifold $M$. A connection on
this $n$-gerbe is an $(n+1)$-form $C_i$, defined over the patches
$U_i$ of the manifold. The forms over patches $U_i$ and $U_j$ are
related on the non-empty intersection $U_{ij} = U_i \cap U_j$ by $C_i
= C_j + d C_{ij}$. The $C_{ij}$ can be regarded as connections on
$(n-1)$ gerbes defined at the overlap patches $U_{ij}$. The $C_{ij}$
are only defined up to closed forms (with integer periods). On triple
intersections $U_i \cap U_j \cap U_k $ there is the consistency
condition $d(C_{ij} + C_{jk} + C_{ki}) = 0$ which is compatible with
the ambiguity of adding closed forms. On the triple overlap, $C_{ij} +
C_{jk} + C_{ki}$ and $C_{ij} + C_{jk} + C_{ki} + dC_{ijk}$ define the
same cocycle. The $C_{ijk}$ can be thought of as connections on an
$(n-2)$-gerbe, defined over the $U_{ijk} = U_i \cap U_j \cap
U_k$. This extends all the way until we arrive at zero-forms
(functions), which can be thought of as transition functions for a
0-gerbe (that is, a line bundle).

For a concrete holonomy formula, we do not consider open patches with
finite overlap, but reduce the overlaps to infinitesimal size; that
is, we cut up $M$ in pieces $M_i$ by a partition of unity. If the
pieces $M_i$ and $M_j$ have a common boundary we denote this $M_{ij}$;
a common boundary between $M_i$, $M_j$ and $M_k$ is written as
$M_{ijk}$ etc. Define $(n-1)$-forms $C_i$ over the pieces $M_i$; we
think about the $M_{ij}$ as the places where we ``jump'' from one
patch to another. The change in connection is given by $C_i = C_j + d
C_{ij}$. $C_{ij}$ itself is defined over the $M_{ij}$, and has an
intrinsic ambiguity by a closed form. One regards $C_{ij}$ as a
connection on an $(n-1)$-gerbe over $M_{ij}$ with a $U(1)$ gauge
invariance etc.

The concrete holonomy formula is now (note the alternating sign):
\begin{eqnarray} \label{hol}
\int_M C = \sum_i \int_{M_i} C_i - \sum_{ij} \int_{M_{ij}} C_{ij} +
\sum_{ijk} \int_{M_{ijk}} C_{ijk} - \ldots.
\end{eqnarray}
This is invariant (for $M$ without boundary) under
\begin{eqnarray}
C_i  & \rightarrow & C_i + dL_i \nonumber \\
C_{ij} & \rightarrow & C_{ij} + L_i + L_j + d L_{ij} \nonumber \\
C_{ijk} & \rightarrow & C_{ijk} + L_{ij} + L_{jk} + L_{ki} + d L_{ijkl}
\\
 & \ldots & \nonumber 
\end{eqnarray}
Two extreme cases of this formula are for a globally well-defined form
$C$, in which case the sum on the r.h.s.\ reduces to a single term, and
the case when only the last sum of integrals in the expression
contributes. These are analogues of the bundle case where physicists
are used to either using well-defined connections over large patches, or
to  ``putting the holonomy in the transition functions.'' For the case of
connections on gerbes, there is a much larger freedom to ``put'' the
holonomy somewhere. 

As an example, let us return to the 4-torus $T^4$ mod the 
$\Z_2$ reflecting all coordinates, where for
convenience we now choose coordinates $x$ in $(\R/ 2\Z)^4$. 
We label the 16 fixed points
of the $\Z_2$ by $x^i=p^i$, $p\in (\Z_2)^4$.

Pick a fixed point $p$ and define a ``hyperbox'' around this point by
the coordinate ranges 
$$ p^1 < x^1 < p^1 + \e,  \quad p^i - \e < x^i < p^i + \e $$ 
for $i=2,3,4$.  The hypersurface of this hyperbox is a
3-surface; we will compute a 3-form holonomy over it.

Assume the existence of a constant 3-form field $\sum_{i<j<k}c_{ijk}
dx^i dx^j dx^k/8$ in the bulk. From now on we will always assume that
the summed over
indices are ordered. The first integral in the
holonomy formula is over the 2-faces of the cube; there are 7 of
these, of which 6 pair up as opposite faces. The opposite faces do not
contribute to the integral, because their contributions cancel. The
only contribution comes from the face at $x^1 = p^1 + \e $.  This
gives
\begin{equation} 
\label{c1} 
\sum_i \int_{M_i} C_i = c_{234} \e ^3. 
\end{equation}

We next consider the transition functions: these are defined at the
edges of $x^1 =  p^1$, $  p^2 < x^2 <   p^2 - \e $, $  p^i -
\e < x^i <   p^i + \e $ for $i=3,4$. The constant 3-form jumps upon
traversing the plane at $x^1 = p^1$ by an amount $c_{ijk} d x^i dx^j
dx^k/4$, and we should write this as $d$ of something. This is
inherently ambiguous, so we choose some conventions and write
\begin{equation}
c_{ijk} d x^i dx^j dx^k = d \left( \sum_{i<j<k} c_{ijk} x^i dx^j
dx^k + c_{ij} dx^i dx^j \right).
\end{equation}
Here the second term is a closed 2-form that does not contribute to the
transition function, but will contribute to the holonomy.

We have to integrate this 2-form over the edge of $x^1 = p^1$, $ p^2 <
x^2 < p^2 + \e $, $ p^i - \e < x^i < p^i + \e $ for $i=3,4$.  Again
only one face contributes (the one with $x^1 = p_1$, $x^2 = p^2 + \e
$), resulting in
\begin{equation} 
\label{c2}
\left(c_{134} p_1 \e  + c_{234} (p_2 + \e ) + c_{34}\right) \e ^2.
\end{equation}
The next transition functions are defined at the edges of the surface
$x^1 = p^1$, $x^2 = p^2 $, $ p^i < x^3 < p^3 + \e $, $ p^4 - \e < x^4
< p^4 + \e $. They are
\begin{equation}
\left(\sum_{i<j<k} c_{ijk} x^i x^j dx^k + \sum_{i<j} c_{ij} x^i dx^j+ c_i 
dx_i \right)/2.
\end{equation}
The integrals result in
\begin{equation} \label{c3}
\left( c_{124} p^1 p^2 + c_{134} p^1 (p^3 + \e ) + c_{234} p^2 (p^3 +
\e ) + c_{14} p^1 + c_{24} p^2 + c_{34} (p^3 + \e ) + c_4\right)
\e.  
\end{equation}
The last contribution comes from the point $x^1 = p^1$, $x^2 = p^2 $,
$x^3 = p^3$, $x^4 = p^4 + \e $. The transition function is
\begin{eqnarray}
\left(\sum_{i<j<k} c_{ijk} x^i x^j x^k + \sum_{i<j} c_{ij} x^i x^j + 
c_i x^i + c_0 \right)/2.
\end{eqnarray}
Inserting values for the coordinates gives
\begin{eqnarray} 
c_{123} p^1 p^2 p^3 + c_{124} p^1 p^2 (p^4 + \e ) + c_{134} p^1
p^3 (p^4 + \e ) + c_{234} p^2 p^3 (p^4 + \e ) + \nonumber \\
c_{12} p^1 p^2 + c_{13} p^1 p^3 + c_{14} p^1 (p^4 + \e ) + c_{23} p^2
p^3 + c_{24} p^2 (p^4 + \e ) + c_{34} p^3 (p^4 + \e ) + \nonumber \\
c_1 p^1 + c_2 p^2 + c_3 p^3 + c_4 (p^4 + \e ) + c_0. \label{c4}
\end{eqnarray}
Adding the contributions (\ref{c1}), (\ref{c2}), (\ref{c3}) and (\ref{c4})
gives formula (\ref{Chol}).

\subsubsection{An M theory dual of the CHL string?}
\label{dualCHL}

In this section, let us continue to work under the premise that the 
3-form can be treated 
as a connection on a 2-gerbe. We still expect that there will
be further conditions, from equations of motion or anomalies, 
constraining us to a subset of the choices predicted by equivariant
cohomology. If this is the right approach---and this is far from clear
to us---then it is interesting to study 3-forms on the global 
orbifold $T^4/\widehat{\cal D}_4$. By $T^4/\widehat{\cal D}_4$, 
we mean the particular global orbifold specified in appendix 
\ref{app:generalquotient} with singularities,   
$$D_4^2 \oplus A_3^3 \oplus A_1^2. $$
If there is a 3-form on $T^4/\widehat{\cal D}_4$ with local
holonomies at the $D_4$ singularities only, it would provide 
a possible candidate M theory dual to the 7-dimensional CHL 
string. 

There is a simple expression for $H^3_G(T^4,U(1))$ that suggests
that this is indeed possible. Suppose that $T^4/G$ has singularities
$\oplus G_i$, i.e., $ADE$-singularities of type $G_i$. Then
\begin{equation} \label{jj8}
H^3_G(T^4,U(1)) \cong \frac{\oplus_i H^3(G_i,U(1))}{H^3(G,U(1))} .
\end{equation}
Here we have used $H^3(G_i,U(1))\cong \Z_{|G_i|}$. 
Therefore, there is a natural inclusion
\begin{equation}
H^3(G_i,U(1)) \to H^3(G,U(1)) .
\end{equation}
Equation (\ref{jj8}) has a simple interpretation in terms of  holonomies
of  2-gerbes around $S^3/G_i$. It simply states that the set of
equivariant 3-forms is the same as the set of all inequivalent local
holonomies, subject only to one overall flux cancellation condition. 
Thus, all possible holonomies can appear, as  long as their sum, 
viewed as an element of $H^3(G,U(1))$, vanishes. 
In particular, there is a case of 
$T^4/\widehat{\cal D}_4$  with equal and opposite holonomies
at the $D_4$ singularities, and no holonomies at any of the other
singularities. This is precisely what we want for an M theory
dual of the CHL string.

In an attempt to make this more precise, 
we give a tentative expression for the holo\-nomies 
associated to 3-forms. Points on
$T^4$ that are invariant under a non-trivial group  element
lie  either on ${\mathbb F}_2^4$ or on ${\mathbb F}_3^2$, where 
${\mathbb F}_2$ and ${\mathbb F}_3$ represent the fields of two and three
elements. The group $G$ still acts on ${\mathbb F}_2^4$ 
and/or ${\mathbb F}_3^2$. We propose that the holonomy around points
on ${\mathbb F}_2^4$ is given by $G$-invariant cubic polynomials
in the four variables $p_i\in\{0,1\}$, $i=1,2,3,4$, with 
coefficients in $\Z_{|G|}$. In the monomials, the highest
power of $p_i$ that can appear is one. To compute the
holonomy, we still have to sum over all preimages
of a given point in $T^4/G$. 
This can be  compared to our discussion of gerbe holonomies 
for $G=\Z_2$ in section~\ref{threeformhol}.
Similarly, the holonomy around points on ${\mathbb F}_3^2$ is given
by $G$-invariant cubic polynomials in the two variables
$p_i\in\{0,1,2\}$, $i=1,2$, with coefficients in $\Z_{|G|}$.
The highest power or $p_i$ appearing in monomials is two.
Again a sum over preimages is required. 
The final answer is to be viewed as an element of 
$\Z_{|G|}$ in which all local holonomies can
embedded.

For example, for $\widehat{\cal D}_4$, the holonomy is
given by by a polynomial,
\bea
H & = & c_{123} p^1 p^2 p^3 + c_{124} p^1 p^2 p^4 + 
 c_{134} p^1 p^3 p^4 + c_{234} p^2 p^3 p^4 \nonumber \\
& & + b_{12} p^1 p^2 + b_{13} p^1 p^3 + b_{14} p^1 p^4 +
 b_{23} p^2 p^3 + b_{24} p^2 p^4 + b_{34} p^3 p^4 \nonumber \\
& & + a_1 p^1 + a_2 p^2 + a_3 p^3 + a_4 p^4 + a_0,
\eea
with all coefficients in $\Z_8$, and with the entire expression
viewed as an element of $\Z_8$. In other words, the
holonomy will be $\exp(2 \pi i H /8)$.
The group $\widehat{\cal D}_4$ acts on the $p^i$. The generator
$\beta$ sends $p^1 \leftrightarrow p^2$ and $p^3 \leftrightarrow p^4$,
whereas the generator $\alpha$ sends $p^1 \leftrightarrow p^3$ 
and $p^2 \leftrightarrow p^4$. Imposing $\widehat{\cal D}_4$-invariance
on $H$ restricts it to the form
\bea
H & = & c(p^1 p^2 p^3 + p^1 p^2 p^4 + 
 p^1 p^3 p^4 + p^2 p^3 p^4) \nonumber \\
& & + b_1 (p^1 p^2 + p^3 p^4) + 
b_2 (p^1 p^3 + p^2 p^4) + 
b_3 (p^1 p^4 + p^2 p^3) \nonumber \\
& & + a(p^1 + p^2 + p^3 + p^4) + a_0 .
\eea
Our prescription for the holonomy states that we should sum over
preimages. In other words, if $\pi$ denotes the projection
$\pi: T^4 \rightarrow T^4/\widehat{\cal D}_4$, the holonomy around
a singularity $x\in  T^4/\widehat{\cal D}_4$ is given by
\begin{equation}
 H(x) \equiv \sum_{p \in \pi^{-1}(x)} H(p) .
\end{equation}
One can verify that when $c$ is divisble by two, it will
only contribute a multiple of $8$ to
the holonomy, and therefore $c$ is naturally restricted to
$c\in \Z_2$. Similarly we find that $b_i \in \Z_4$, $a\in \Z_2$, 
and $a_0 \in \Z_8$. These values are in precise one-to-one
correspondence with the groups $H^p(\widehat{\cal D}_4, 
H^{3-p}(T^4,U(1)))$ we compute in (\ref{d4e2pq}). 
We can now evaluate the holonomy around each of the singularities
in the quotient $T^4/\widehat{\cal D}_4$. The result is
given in table~\ref{table:holod}. If we take $a_0=4$ and all other
coefficients equal to zero, we find holonomy $-1$ around
the two $D_4$ singularities, and vanishing holonomy around all
other singularities. 
We shall explain in section \ref{threeformgeom}\ why we believe that
this particular holonomy is realized, while other possibilities, like 
choices
with holonomy around $A$-type singularities, are unlikely to be 
consistent.
This then appears to be a concrete proposal
for the singular $K3$ with 2 frozen $D_4$ singularities that 
appears in the 7-dimensional M theory description of the CHL string.

\begin{table}
\begin{center}
\renewcommand{\arraystretch}{1.5}
\begin{tabular}{|c|c|} \hline
Singularity     & Holonomy $\in\Z_8$ \\ \hline
 $D_4^{(1)} $ & $a_0  $ \\ \hline
 $D_4^{(2)} $ & $4 c + 2(b_1 + b_2 + b_3) + 4 a + a_0 $ \\ \hline
 $A_3^{(1)} $ & $ 2 b_2 + 4 a + 2 a_0 $ \\ \hline
 $A_3^{(2)} $ & $ 2 b_3 + 4 a + 2 a_0 $ \\ \hline
 $A_3^{(3)} $ & $ 2 b_1 + 4 a + 2 a_0 $ \\ \hline
 $A_1^{(1)} $ & $ 4 c + 4(b_1 + b_2 + b_3) + 4 a + 4 a_0 $ \\ \hline
 $A_1^{(2)} $ & $ 4 a + 4 a_0 $ \\ \hline
\end{tabular}
\renewcommand{\arraystretch}{1.0}
\caption{Three-form holonomies of $T^4/\widehat{\cal D}_4$}
\label{table:holod}
\end{center}
\end{table}

\subsubsection{Some comments on type IIA versus M theory}
\label{compareiia}

One of the puzzles in the duality between type IIA and M theory is
that RR fields in type IIA appear to be classified by K-theory, while
the M theory 3-form requires a quite different framework---perhaps
cohomology.  A detailed analysis showed that for partition functions
computed on large smooth compact spaces, this is not a
contradiction~\cite{moore2}. However, a sum over fluxes in M theory is
needed to reproduce the partition function of type IIA string theory.
One may wonder whether our analysis sheds any light on the relation
between M theory and type IIA. Specifically, let us consider type IIA
on $\C^2/G$, with RR 1 and 3-form flux. We expect a relation between,
$$ K(\C^2/G) \leftrightarrow H^{\ast}((\C^2 \times S^1)/G), $$
or more precisely, a relation between
\begin{equation} \label{dual3}
 K^0_G(pt)/( K^0_G(pt) )_5 \leftrightarrow
 H^3((\C^2 \times S^1)/G, U(1)).   
\end{equation}
The action of $G$ on the right is determined by the 1-form
flux on the left, i.e., by an element 
of $ K^0_G(pt)/( K^0_G(pt) )_3=H^1(G,U(1))$.
At first sight, the relation (\ref{dual3}) appears very problematic.
When the 1-form flux is such that $G$ acts faithfully on the 
$S^1$ factor in $(\C^2 \times S^1)/G$, the latter quotient is
smooth and does not have any non-trivial 3-cohomology.
Therefore, we cannot detect any phases using a non-trivial
Euclidean M2-brane. On the other hand, in type IIA we can consider
Euclidean D2-branes, and according to our previous calculations, these
pick up phases $\frac{n}{4}\,\, {\rm mod}\,\, \Z$. 

This discrepancy can be traced back to the fact that not all 
D2-branes in IIA can be lifted to M2-branes in M theory, at least
not when there is non-trivial RR 1-form flux in IIA. For example,
consider again the case where $G=\Z_2$ with non-trivial
1-form flux. The M theory description is 
on $(\C^2 \times S^1)/\Z_2$. It is easy to see that there
is no M2-brane in this space that reduces to a single D2-brane
wrapping $S^3/\Z_2$ in type IIA. At most we can get two D2-branes
wrapping $S^3/\Z_2$, and these only see twice the 3-form flux
in type IIA. 

This mismatch between M theory and type IIA is removed once we take
the Freed--Witten anomaly for D-branes into account. The only
non-anomalous brane configurations are those for which the NS-NS 2-form
field strength $H$ is cohomologically trivial when
restricted to the brane world-volume.
Applying S- and T-dualities in a cavalier manner 
implies that D2-branes are only consistent if the RR 1-form
field strength is cohomologically trivial on the D2-brane world-volume.
It would be nice to derive this anomaly from a world-volume approach.

In particular, the D2-brane that wraps $S^3/\Z_2$ has a non-trivial
RR 1-form field strength living on its world-volume, and is therefore
inconsistent. The 1-form field strength is the non-trivial element
of $H^2(S^3/\Z_2,\Z)=\Z_2$ so we can wrap two D2-branes.
Therefore, anomaly cancellation removes those D2-branes from the spectrum
that do not appear in M theory. It is pleasing to see that anomaly
cancellation for D2-branes has such a simple interpretation in
M theory. 

The same argument applies more generally: take any D2-brane wrapping
some non-trivial three-manifold $B$, and assume it has RR 1-form flux
$[F^{RR}]$.  Then there is a circle bundle $E$ over $B$ which is the M
theory lift of $B$. This circle bundle has second Chern class
$[F^{RR}]$. However, taking any M2-brane in M theory and projecting it
down to $E$ only gives D2-branes if the pullback of the cohomology
class of $[F^{RR}]$ is zero.  This follows from the fact that
$[\pi^{\ast} F^{RR}]$ is cohomologically trivial on $E$, where
$\pi:E\to B$ is the projection. This in turn can be seen from the
Gysin exact sequence for sphere fibrations; see, for example,
\cite{botttu}. M theory therefore implements the condition for anomaly
cancellation quite generally.

Ultimately, we would like to generalize all this to K-theory, which is
the proper setting both for RR fields and for D-branes at small string
coupling.  We will very briefly discuss a possible generalization of
the Freed--Witten anomaly to K-theory. In the spirit of speculative
exploration, we shall use S- and T-dualities without worrying about
the deep unresolved incompatibilities between K-theory and string
dualities.  Our hope is to find a conjecture about which
configurations are consistent which is natural in the framework of
K-theory.

The Freed--Witten anomaly arises in the string path-integral from open
Riemann surfaces that end on the D-brane. In the bosonic case, there
is a contribution to the path-integral of the form
$$ \exp(i\oint_{\partial\Sigma} A)\exp{i \int_{\Sigma} B}. $$ In
\cite{Freed:1999vc}, this is interpretated in the following way: the
first term provides a trivialization for the second term, viewed as a
section of a line bundle over the space of Riemann surfaces that end
on the D-brane. Such a trivialization only exists if $H=dB$ is
cohomologically trivial when restricted to the D-brane.

Using S- and T-dualities, we find a corresponding statement for D$(p-2)$-branes
ending on D$p$-branes. Let $M$ denote the space on which the D$(p-2)$-brane 
is wrapped which ends on a manifold $Q$ on which the D$p$-brane is wrapped. 
In the path-integral for the D$(p-2)$-branes, there is a phase
factor 
$$ \exp(i\oint_{\partial M}\tilde{A}) 
\exp{(i \int_M C^{RR} e^{F+B} \sqrt{\hat{A}}) }
$$
where $\tilde{A}$ is the dual $(p-2)$-form of the gauge field on the
D$p$-brane, and the second term is the Wess--Zumino term of the
D$(p-2)$-brane action.  By analogy with the argument in
\cite{Freed:1999vc}, we interpret the first term as providing a
trivialization of the second term, viewed as a section of a line
bundle over the space of D$(p-2)$ branes ending on $Q$. Such a
trivialization only exists if the cohomology class of
$$ d(C^{RR} e^{F+B} \sqrt{\hat{A}}) $$ restricted to $Q$
vanishes. Taking $B=\sqrt{\hat{A}}=0$ for simplicity, the class $
d(C^{RR} e^{F})=G^{RR} e^{F}$ restricted to $Q$ should be trivial. In
principle we have to sum over all possible topological sectors
described by $F$. These will include, for example, bound states of
D$(p-2)$ and D$(p-4)$-branes ending on $Q$. For each of these
separately, the restriction of $G^{RR} e^{F}$ has to vanish, when
restricted to $Q$. This will typically imply that $G^{RR}$ itself,
restricted to $Q$ has to vanish. Rewriting all this in K-theory
language gives a generalization of the Freed--Witten anomaly
cancellation condition which we now describe.

Suppose that (in type IIB) the RR field strength $G$ is represented by
an element of $K^1(Y)$, where $Y$ denotes space-time, and consider a
brane represented by an element $Q\in K^0(Y)$. Recall that K-theory is
a graded ring, and in particular we can form the product $Q.G \in
K^1(Y)$.  We propose the following generalization of the Freed--Witten
anomaly cancellation condition,
\begin{equation}
\label{fw}
Q.G=0 \,\, {\rm in}\,\, K^1(Y).
\end{equation}
A similar result applies to type IIA. The product $Q.G$ represents,
roughly speaking, the restriction of $G$ to $Q$, which is how it arose
in the preceeding discussion.

For Euclidean branes, we have to be quite careful when applying this.
First, Euclidean and Minkowski D-branes are apparently classified by
different K-theories. If one is classified by $K^0$ then the other is
classified by $K^1$. Secondly, we have to throw out the form of
highest degree in $G^{RR}$, since the argument does not apply to that
form. This can be accomplished by projecting the appropriate K-theory
classes on those that vanish on the $(p+1)$-skeleton for Euclidean
D$p$-branes.

For IIA on $\C^2/\Z_2$, this implies the following. The product in
(\ref{fw}) becomes the product of two elements of $\Z_4$, and the
projection to forms that vanish on the $3$-skeleton is multiplication
by two on $\Z_4$.  If the RR flux is $1/4$, as discussed in section
\ref{secKtheory}, we pick up a phase of $1/4$ around $S^3/\Z_2$. Once
we project this phase down to $\Z_2$, it remains non-zero, and the
corresponding Euclidean brane is inconsistent.  We can however take
twice $S^3/\Z_2$, which has phase $1/2$ in $\Z_4$, and this vanishes
after projection. What is the interpretation of this object in M
theory? It has exactly the same phase as seen by a Euclidean D0-brane
wrapping the non-trivial cycle in $S^3/\Z_2$, and in M theory it is
therefore described by a ``half-momentum'' graviton.
The worldline of this graviton is the obvious lift of the
non-trivial one-cycle in $S^3/\Z_2$ to $(S^3 \times S^1)/\Z_2$.
For RR flux $1/2$,
there is no RR 1-form flux, and indeed there is never an anomaly. We
can therefore directly match the M theory membranes on $\C^2/\Z_2
\times S^1$ with D2-branes on $\C^2/\Z_2$.  Our proposal (\ref{fw}),
albeit preliminary, is therefore consistent with M/IIA duality. It
would be interesting to develop it in detail.

\subsubsection{The geometry of the three-form} 
\label{threeformgeom}

The nature of the geometry of the M theory $3$-form is a
deep question which is unlikely to have a simple classical answer. 
It is therefore worth
stopping at this point and pondering what we can learn about the
$3$-form from our results. Our `experimental' data follows from
the classification of allowed discrete fluxes in $ADE$ singularities
given in table \ref{table:fluta}. 

There are two natural questions to ask: first, there is a purely local
question. Given an $ADE$-singularity of the form $\C^2/G$, what are
the possible discrete choices of $3$-form flux localized at the
singularity? The second question is a global one. Given a compact
manifold with assorted singularities, are there any relations between
the fluxes at different singularities? In all the examples we have
considered, the answer to the latter question is that the total flux
has to vanish.  This is exactly as we would na\"{\i}vely expect for a
compact manifold.

The answer to the first question is much more difficult.  One natural
candidate is the equivariant 3-cohomology of $\C^2/G$ with values in
$U(1)$, which classifies the flat orbifold $2$-gerbes. This cohomology
group is equal to $H^3(G,U(1))=Z_{|G|}$ with $|G|$ the number of
elements of $G$. This certainly contains all the fluxes appearing in
table \ref{table:fluta}, but it is much too large. Nevertheless, it is
still possible the physical fluxes are a subset of the choices given
by $H^3(G,U(1))$. For this to be the case, there must exist additional
anomaly constraints or non-linear equations of motion. Non-linear
equations might provide an explanation for the lack of any group
structure in table \ref{table:fluta}. The simplest example of a
consistency condition that we know is the shifted quantization of the
$3$-form field strength on curved spaces \cite{fluxquant}.

Let us consider a related possibility, again in the spirit of
speculative exploration. We shall find a picture that seems to
reproduce our ``data.'' Although many questions and puzzles remain, it
seems worth describing.  The key appears to be the study of wrapped
branes. Let us consider a 7-dimensional M theory dual of one our
heterotic orbifolds. This is M theory on a singular $K3$ with flux
localized at some combination of the singularities.  The heterotic
string is constructed in M theory by wrapping an M5-brane on the
$K3$. However, we cannot wrap a single M5-brane because of a
generalization of the Freed--Witten anomaly---computed for a single
M5-brane in \cite{Witten:2000vg}. Let us assume for the moment that
equivariant cohomology classifies the possible fluxes, as in the
preceeding discussion. If the flux is $N$ torsion, we should wrap $N$
M5-branes.\footnote{This claim involves an extrapolation of known
results, but seems quite reasonable.} Is there a bound state of $N$
wrapped M5-branes? If not, then this configuration of singularities
and fluxes cannot give rise to a heterotic string, and so cannot have
a perturbative heterotic dual.

The appearance of bound states of five-branes in terms of irreducible
(or connected) covers also appears in the computation of the partition
function on $K3\times T^2$ \cite{Minahan:1998vr}. Here the bound state
of $N$ M5-branes that produces the $U(N)$ ${\cal N}=4$ Yang--Mills
theory on $K3$, is obtained by wrapping the five-branes by an
irreducible cover of $T^2$. These five-brane configurations are in
complete analogy with the long string discussion of
\cite{Dijkgraaf:1997vv}. So we expect a bound state if the $K3$ is
locally a $G$-orbifold and $G$ has an $N$-dimensional irreducible
representation.

Turning to the local singularity $\C^2/G$, we note that the $A_n$
series have only 1-dimensional irreducible representations (irreps),
while the $D_{n+4}$ series has $n+5$ inequivalent irreps: 4 of
dimension 1 and $(n+1)$ of dimension 2. For $E_6$, there are 3 irreps
of dimension 1, 3 irreps of dimension 2, and 1 of dimension 3. For
$E_7$, there are 2 irreps of dimension 1, 3 of dimension 2, 2 of
dimension 3, and 1 of dimension 4. Finally for $E_8$, there is 1 irrep
of dimension 1, 2 irreps of dimension 2, 2 irreps of dimension 3, 2
irreps of dimension 4, 1 irrep of dimension 5, and 1 irrep of
dimension 6.  The easiest way to determine the list of irreps is from
the Dynkin labels of the extended Dynkin diagram for $G$. It is also
worth noting (and hardly coincidental!)  that these labels are also
the starting point for the analysis of triples appearing in \cite{bfm,
kacsmil}.

Let us now consider the possibilities: take a singularity with a flux
of order $k$ (perhaps glued in a $K3$).  The above arguments suggest
that a wrapped M5-brane is anomalous unless $G$ has an irrep of
dimension $k$. Since all irreps for $A_n$ are 1-dimensional, we cannot
wrap an M5-brane at all on a singularity with non-zero flux.
For $D_{n+4}$, there are 2 possibilities. Of all the possible fluxes
in the group of equivariant 3-form fluxes $\Z_{4n+8}$, only flux of
order 1 or order 2 permit the wrapping of an M5-brane. The group of
fluxes for $E_6$ is $\Z_{24}$. However, there are irreps of dimension
$1,2$ and $3$ only.  Therefore, only fluxes of order $1,2$ and $3$ are
allowed. Note that as we desire, this set does not form a group! For
$E_7$, we have a priori $\Z_{48}$ as the group of choices.  Again,
only fluxes of order $1,2,3$ and $4$ are allowed. Finally, for $E_8$,
we have the group $\Z_{120}$. However, only fluxes of order
$1,2,3,4,5$ and $6$ are possible.  Note also that we can deform an
$E_6$ singularity, for example, with order $2$ flux to a $D_4$
singularity with order 2 flux, because this subgroup also has an irrep
of dimension 2. In this way, we completely reproduce the data of table
\ref{table:fluta}.\footnote{What we seem to be requiring for an
anomaly-free theory can be interpreted via the McKay
correspondence. It appears to be the existence of a globally defined
reflexive sheaf of rank $k$ for flux of order $k$. For $\C^2/G$, the
existence of such a sheaf requires that $G$ have an irrep of dimension
$k$. Away from the singularity, the sheaf is constructed by taking the
trivial $\C^k$ bundle over $\C^2$ and quotienting by $G$ using the
irreducible representation.}
 
Let us check which combinations of singularities with flux satisfy our
global constraint that the total flux vanish, and which can be
embedded in $\G_{3,19}$. The sum of the ranks of the singularities
must then be less than $20$.  To find the minimal list, we discard any
deformable singularities. We therefore need not consider $A_n$
singularities.  With only $D_4$, it is clear that we need an even
number for flux conservation.  The only possibilities are $2$ or $4$
and both cases are realized in M theory.  Analogously, with only $E_6$
singularities, we have 2 possibilities: 2 $E_6$ singularities with
opposite $\Z_3$ flux, or 3 with $\Z_3$ flux. With only $E_7$
singularities, there is a single possibility: 2 $E_7$ singularities
with opposite $\Z_4$ flux. With only $E_8$ singularities, there are 2
possibilities: 2 $E_8$ singularities with opposite $\Z_5$ or $\Z_6$
flux.

We can now consider the mixed cases: for $D_4$ with $\Z_2$ flux and
$E_6$ with $\Z_3$ flux, we note that a brane must wrap $6$ times. To
cancel the flux implies one possibility; namely, adding a single $E_8$
with $\Z_6$ flux. For $D_4$ and $E_7$, a similar argument requires an
additional $E_7$. it is not too hard to check that there are no other
consistent combinations. This list reproduces table
\ref{table:Mduals}.

This argument is natural in the global context where we can see that
the resulting M theory fails to have a perturbative heterotic dual.
What seems critical to show is why models which do not permit the
wrapping of M5-branes---both local and global models---are
inconsistent.

\subsubsection{F theory compactifications with flux}

On the topic of F theory compactifications with flux, we find no new
models. However for completeness, we recall the known
results. Starting with M theory on $K3$ with 2 frozen $D_4$
singularities, we can see from the lattice or from its various dual
realizations that there is a non-trivial 8-dimensional limit. This is
an F theory model in which the 2 $D_4$ singularities are replaced by a
frozen $D_8$ singularity. This gives an 8-dimensional dual of the CHL
string. The second case starts with M theory on a $K3$ with 4 frozen
$D_4$ singularities. This is the model dual to the compactification of
the 9-dimensional $(+,-)$ orientifold. Taking its F theory limit
results in a compactification with 2 frozen $D_8$ singularities. Both
cases were first considered in~\cite{wittoroid}.  These models have
been further studied in \cite{berglund, bershad}.

Can there be any new F theory models with flux? The 7-dimensional M
theory duals of our heterotic asymmetric orbifolds only lift
non-trivially to 8 dimensions for the $\Z_2$-triple.  This is the
first case recalled above. The 7-dimensional M theory duals of our
7-dimensional F theory models again only lift non-trivially in 8
dimensions for the case of $\Z_2$. However, the M theory dual of the
$\Z_2$ F theory compactification is the second case recalled above
with 4 frozen $D_4$ singularities. The only other place we might
expect to find a new model is from our 6-dimensional $\Z_2\times\Z_2$
F theory vacuum.  This model may well have a 6-dimensional M theory
dual with frozen singularities (a precise determination of its lattice
is needed to find a candidate model.). However, since the F theory
model without fluxes has no new 7-dimensional limit, we cannot obtain
a new F theory model with flux this way.

\section{Acknowledgements}

We would like to thank P.~Aspinwall, P.~van~Baal, D.~Freed, R.~Friedman,
A.~Mikhailov, R.~Plesser, N.~Saulina, C.~Schweigert, G.~Segal and E.~Witten
for helpful conversations.  The work of A.~K. was
partly supported by EU contract HPRN-CT-2000-00122.   
K.~H. was supported in part by NSF-DMS 9709694.
J.~M. was supported by NSF grants DMS-9704507 and DMS-0074126.
D.~R.~M. was supported by the Institute for Advanced Study and by NSF
grants DMS-9401447 and DMS-0074072.  S.~S. was
supported in part by the William Keck Foundation and NSF grant
PHY-9513835.  This work was also supported in part by NSF grant PHY-9907949
at ITP, Santa Barbara.

R.~D., J.~M., D.~R.~M. and S.~S. would like to thank the organizers of
the 1999 I.H.E.S. Summer Workshop for their hospitality, while
J.~B. and S.~S. would also like to thank the organizers of the 1999
Amsterdam workshop for their support. In addition, J.~B., K.~H.,
D.~R.~M. and S.~S. would like to thank the Aspen Center for Physics
for hospitality both in 1999 and 2000 as this work matured. R.~D. and
D.~R.~M. would also like to thank the Institute for Advanced Study for
hospitality at various stages of this project.  J.~B., K.~H. and
S.~S. are also happy to thank the Institute for Theoretical Physics at
UCSB for hospitality during the penultimate phase of this project.
Last but not least, S.~S. would like to thank the Rutgers theory group
for their hospitality during the final stage of this project.

\appendix

\section{Lattice Conventions and Some Useful Definitions} \label{app:lattice}

In this appendix, we summarize some useful information about lattices
and our conventions. A {\it lattice}\/ is a free $\mathbb Z$-module
$L$ equipped with a $\mathbb Z$-valued bilinear form $b:L\times
L\to\mathbb Z$.  The lattice is {\it even}\/ if $b(x,x)\in2\mathbb Z$
for all $x\in L$.  In the even case, there is an associated {\it
integer-valued quadratic form}\/ $q:L\to\mathbb Z$ defined by
\[q(x)=\frac12b(x,x).\]

We use lattice conventions which agree with those of Lie algebraists
rather than those of algebraic geometers.  In particular, $A_n$,
$D_n$, and $E_n$ are {\it positive definite}\/ even lattices, the root
lattices of the corresponding complex semisimple Lie algebras.  We use
$\Gamma_{k,k}$ to denote the even lattice of rank $2k$ whose bilinear
form has matrix
\[\begin{pmatrix}0&I_k\\I_k&0\end{pmatrix}.\]
We occasionally describe a lattice by directly writing the matrix of the bilinear form
in some basis.

If $L$ is a lattice, then $L(n)$ denotes the same lattice with the
bilinear form multiplied by $n$ (this is sometimes denoted by
$\sqrt{n}L$).  Thus, $A_n(-1)$, $D_n(-1)$ and $E_n(-1)$ are the
negative-definite lattices which appear in algebraic geometry, and
$\Gamma_{k,k}(n)$ is the even lattice of rank $2k$ whose bilinear form
has matrix
\[\begin{pmatrix}&&&n&&\\&&&&\ddots&\\&&&&&n\\
n&&&&&\\&\ddots&&&&\\&&n&&&\end{pmatrix}.\]
If $x_1,\dots,x_k\in L$, then $\langle x_1,\dots, x_k\rangle$ denotes the
sublattice of $L$ spanned by $x_1$, \dots, $x_k$.

Finally, let us review the definition of the discriminant group and
discriminant form given
by Nikulin \cite{Nikulin80}. Given any vector $l$ of an integer
lattice $L$, we can construct an element of $Hom(L, \Z) = L^*$ which
acts in the following way: given an $x\in L$, we obtain an integer by
evaluating $b(x, l)$.  This gives us an injective map from $L
\rightarrow L^*$.  With this canonical embedding of $L$ into $L^*$, we
can consider the quotient $L^*/L$ which is a finite abelian group known as
the discriminant group. We
can equip this group with a bilinear form $b_L$ known as the
discriminant form which takes values in $\Q /\Z$,
\begin{equation}
b_L(x_1 + L, x_2 +L) \, = \, b(x_1, x_2) + \Z, 
\end{equation}
where $x_1, x_2 \in L^*.$ Under suitable restrictions, this form can
determine the isomorphism type of the lattice.

\section{Heterotic-Heterotic Duality} \label{app:hetdual}

\subsection{Duality on $S^1$}

We will review the argument of \cite{arjan} and correct a little
inaccuracy in that derivation. Consider either of the heterotic string
theories compactified on a circle. We write the standard deformation
of the heterotic string under the inclusion of a Wilson line
$\mathbf{a}$ as follows:
\begin{equation}
W({\mathbf{a}})\left( \begin{array}{c} \mathbf{q} \\
\sqrt{\frac{\alpha'}{2}} (\frac{n}{r} + \frac{wr}{\alpha'}) \\ \sqrt{
\frac{\alpha'}{2}} (\frac{n}{r} - \frac{wr}{\alpha'}) \end{array}
\right) = \left( \begin{array}{c} {\mathbf{q}} + w{\mathbf{a}} \\
\sqrt{\frac{\alpha'}{2}} (\frac{n-{\mathbf{q}}\cdot {\mathbf{a}}- w
{\mathbf{a}}^2/2}{r} + \frac{wr}{\alpha'}) \\ \sqrt{\frac{\alpha'}{2}}
(\frac{n-{\mathbf{q}}\cdot {\mathbf{a}}- w {\mathbf{a}}^2/2}{r} -
\frac{wr}{\alpha'}) \end{array} \right).
\end{equation}
The matrix $W(\mathbf{a})$ represents an $SO(17,1)$ rotation acting on
the lattice $\Gamma_{17,1}$. To connect the lattice $\Gamma_{17,1}$ to
a heterotic string theory, we have to single out a $\Gamma_{1,1}$
sublattice that will be interpreted as the lattice of the spatial
momenta $n$ and winding numbers $w$. The complement of $\Gamma_{1,1}$
in $\Gamma_{17,1}$ is then the lattice of vectors $\mathbf{q}$ of the
gauge group, which can be $D^{0,s}_{16}$---the composition of the
root and spin lattice of $Spin(32)$---for a $Spin(32)/\Z_2$ theory or
$E_8 \oplus E_8$ for the $E_8 \times E_8$ string. Which of these
occurs depends only on how we choose the $\Gamma_{1,1}$ lattice. Put
another way, it depends only on how we choose coordinates for
$\Gamma_{17,1}$.

The states of the two heterotic theories are thus connected by a
simple coordinate transformation. Consider the vectors $W(\mathbf{a})
\psi$ with $\psi \in \Gamma_{17,1}$, corresponding to the states of
the heterotic theory with gauge group $G$ on a circle of radius $R$
with holonomy $\exp 2 \pi i \mathbf{a}$. Also consider the vectors
$W(\mathbf{a}') \psi'$ with $\psi' \in \Gamma_{17,1}$ corresponding to
the states of the heterotic theory with gauge group $G' \neq G$ on a
circle of radius $R'$ with holonomy $\exp 2 \pi i \mathbf{a}'$. We
have argued that $W(\mathbf{a}) \psi = U W(\mathbf{a}') \psi'$ with
$U$ a coordinate transformation. The mass spectrum of the theories
should be preserved by $U$, which therefore should be an element of
$O(17) \times O(1)$. The transformations $W(\mathbf{a})$ and
$W(\mathbf{a}')$ allow us to vary all possible continuous parameters
of the theories: namely, the holonomies $\mathbf{a}$, $\mathbf{a}'$,
and the radii $R$, $R'$. Since the two heterotic theories cannot
possibly be continuously connected, the matrix $U$ must correspond to
a discrete symmetry. An inversion of all coordinates $n,w$ and
$\mathbf{q}_i$ corresponds to a parity transformation on the spatial
circle, combined with an element of the Weyl group acting on the group
lattice. Therefore $U$ is not connected to either the product of the
identities of $O(17) \times O(1)$, or to the product of minus the
identities. Fixing $U = \textrm{diag}(\mathbf{1}_{16},1,-1)$, fixes
most of the possible choices for $\mathbf{a}$ and $\mathbf{a}'$.

A single ansatz fixes most of the remaining freedom. We want to move
from a theory with one gauge group to one with another gauge group.
This is possible if one of the Kaluza--Klein bosons exchanges roles with
one of the $10$-dimensional gauge bosons. Explicitly, let a
state $\psi$ of the theory with gauge group $G$ with $\mathbf{q} =0$ and
non-zero $n,w$ in one theory correspond to a state $\psi'$ with non-zero
$\mathbf{q}'$ and $n'=w'=0$ in the other theory with gauge group $G'$.
{}From the transformation found previously, we can then immediately
deduce that $\mathbf{a}^2 = 2$ and that $\mathbf{q}' = \mathbf{a}$:
therefore $\mathbf{a}$ is a root of $G'$. Furthermore, we find that
$\mathbf{a} \cdot \mathbf{a}' = -RR'/\alpha'$. Exchanging the roles of
$G$ and $G'$ in the ansatz leads to the conclusion that $\mathbf{a}'$ is
a root of $G$.

Studying the image of the state $\mathbf{q}=0$, $n=1$, $w=0$ tells us
that $\alpha'/RR'$ is an integer, say $k$. Studying the image of
the states with $n=w=0$ reveals the consistency condition that $k \mathbf{q}
\cdot \mathbf{a}$ be an integer for every $\mathbf{q}$. This implies
that $k\mathbf{a}$ is a (co)weight of $G$. In a similar way, we find
that $k\mathbf{a}'$ is a (co)weight of $G'$.

There is no simple way to fix $k$ as there are in fact more solutions.
It is clear that $k=0$ will lead to nonsense. Setting $k = \pm 1$ is a
possibility, but does not solve our requirements. Both for $Spin(32)$
and $E_8 \times E_8$, the requirement that $\mathbf{a}$ is a coweight
with length $\sqrt{2}$ implies that $\mathbf{a}$ is a root, and
therefore inevitably implies that $G=G'$, which is not what we desire.
It does lead to a duality transformation teaching us that the
heterotic theory with group $G$ on a circle of radius $R$ with a
trivial holonomy is dual to the same theory on a
circle of radius $R' = \alpha'/R$ with a trivial holonomy, as was
already noted in \cite{ginsparg}.

The next possibility is $k= \pm 2$. This leads to the cases, 
\begin{itemize} 
\item $\mathbf{a}$ is a root of $G'$, and $2\mathbf{a}$ is a coweight
of $G$;   
\item $\mathbf{a}'$ is a root of $G$, and $2\mathbf{a}'$ is a
coweight of $G'$; 
\item ${\mathbf{a}\cdot\mathbf{a}}'= -\hlf$, and
$RR'=\alpha'/2$. 
\end{itemize}
Solutions to these equations can be found. Given $\mathbf{a}$ and
$\mathbf{a}'$, the duality transformation is completely fixed and does
provide a map from the compactification of one heterotic theory to a
compactification of the other.
We have not attempted to evaluate the equations for higher $k$.

\subsection{Duality on $T^n$}

Below $9$ dimensions, explicit duality transformations are in general harder 
to give. In the situations that we will consider, we always
choose $g_{ij} = \delta_{ij}$. We will not yet restrict the NS-NS
two-form $B_{ij}$.

The inclusion of a non-zero $B$-field modifies the momenta
(\ref{hmom1}) and (\ref{hmom2}) in the following way: 
\begin{eqnarray*} 
{\mathbf{k}} & = & ({\mathbf{q}} + \sum_{i} w_i
{\mathbf{a}}_i)\sqrt{\frac{2}{\alpha'}},  \\  k_{iL,R} & =
& \frac{n_i - {\mathbf{q} \cdot \mathbf{a}}_i - \sum_j \frac{w^j}{2}
(\mathbf{a}_i \cdot \mathbf{a}_j+2B_{ij})}{R_i} \pm \frac{w_i
R_i}{\alpha'}.   \end{eqnarray*} 
We will use the following shorthand notation,
\begin{equation}
\left( \begin{array}{c}
\mathbf{k} \\ k_{iL} \\ k_{iR} \end{array} \right) =
\sqrt{\frac{2}{\alpha'}} \, W( \lbrace \mathbf{a}_i \rbrace ; B_{ij})
\left( \begin{array}{c} \mathbf{q} \\ \sqrt{\frac{\alpha'}{2}}
(\frac{n_i}{R_i} + \frac{w_i R_i}{\alpha'}) \\ \sqrt{
\frac{\alpha'}{2}} (\frac{n_i}{R_i} - \frac{w_i R_i}{\alpha'})
\end{array} \right).  \end{equation}  
The explicit expression for $W( \lbrace \mathbf{a}_i \rbrace ; B_{ij})$
can be easily deduced or found in \cite{ginsparg}. In light of the
discussion of the previous subsection, we would like to factorise $W(
\lbrace \mathbf{a}_i \rbrace ; B_{ij})$ into a number of factors
representing contributions that can be traced back to the various
directions labelled by $i$. In a gauge theory, this would be
possible because of commutativity of the holonomies. In string
theory, this seems difficult because the mixing of contributions from
various directions by the term quadratic in the $\mathbf{a}_i$ and the
$B$-field. 

However, a little calculation shows that
\begin{equation}
W( \lbrace \mathbf{a}_i \rbrace ; B_{ij}) W( \lbrace
\tilde{\mathbf{a}}_i \rbrace ; \tilde{B}_{ij}) =  W( \lbrace
\mathbf{a}_i + \tilde{\mathbf{a}}_i \rbrace ; B_{ij}+\tilde{B}_{ij} -
\Delta_{ij}), 
\end{equation}
with $\Delta_{ij}$ given by, 
\begin{equation}
\Delta_{ij}= \hlf \left( \mathbf{a}_i \cdot
\tilde{\mathbf{a}}_j -\mathbf{a}_j \cdot
\tilde{\mathbf{a}}_i \right).
\end{equation}
Note that the $\Delta_{ij}$ can be appropriately thought of as components of a
two-form. These equations are the key to our problem.

Suppose we have a theory with a set of $\mathbf{a}_i$, where for one of
these, say $\mathbf{a}_1$, there exists an $\mathbf{a}'_1$ such that
the pair $\mathbf{a}_1$, $\mathbf{a}'_1$ has the properties mentioned
in the previous subsection. We now want to dualize in the direction
labelled by $1$. In the theories under consideration, the $B$-field is a
modulus and we can set it to an appropriate value. Set 
\begin{displaymath} B_{1j} = - B_{j1} = \hlf \left( \mathbf{a}_1 \cdot
\mathbf{a}_j \right), \qquad B_{ij} = 0, \quad i,j \neq 1. 
\end{displaymath}
Then the momenta of the theory are given by $W( \lbrace \mathbf{a}_i
\rbrace ; B_{ij})$ acting on the vectors of the theory with
$\mathbf{a}_i = B_{ij} = 0$. Now with this specific value of the
$B$-field, we can write
\begin{equation}
W( \lbrace \mathbf{a}_i \rbrace ; B_{ij}) = W( \lbrace  0,
\mathbf{a}_{i \neq 1} \rbrace; 0)W( \lbrace \mathbf{a}_1, 0 \rbrace; 0).
\end{equation} 
Duality now amounts to replacing $W( \lbrace \mathbf{a}_1, 0 \rbrace;
0)$ by $ U_1 W( \lbrace \mathbf{a}'_1, 0 \rbrace; 0)$. The matrix
$U_1$ acts as the identity on all coordinates except for $k_{1R}$,
which is replaced by $-k_{1R}$. However, $U_1$ commutes with $W( \lbrace 0,
\mathbf{a}_{i \neq 1} \rbrace; 0)$. The holonomies and $B$-field in
the dual theory are then summarized as follows, 
\begin{equation} U_1 W( \lbrace
\mathbf{a}'_1, \mathbf{a}_{i \neq 1} \rbrace; B'_{ij}) \quad \textrm{
with } \quad B'_{1j} = - B'_{j1} = \hlf \left(
\mathbf{a}'_1 \cdot \mathbf{a}_j \right), \qquad \ B'_{ij} = 0, \quad i,j \neq 1.
 \end{equation}
The most convenient situation arises when it is possible to
choose holonomies so that, 
$$\left( \mathbf{a}_1 \cdot \mathbf{a}_j
\right) = \left( \mathbf{a}'_1 \cdot \mathbf{a}_j
\right) = 0. $$
Although this is not always the case, we will do this whenever possible.

\section{Classifying Orientifold Configurations}
\label{app:orient}
\newcommand{\mip}{-^{\prime}}

In this appendix, we classify all possible $T^n/\Z_2$
orientifolds involving only $-$ and $-'$ orientifold planes,
for the case $n\leq 5$.

Let $x_1,\ldots,x_n$ be periodic coordinates for $T^n$
with period $2\pi$ on which the $\Z_2$ orientifold group
acts by inversion $\{x_{\mu}\} \mapsto \{-x_{\mu}\}$.
The $2^n$ $\Z_2$-fixed points are located at
$\{x_{\mu}\}=\{a_{\mu}\pi\}$ where $a_{\mu}$ are mod 2 integers.
We denote this fixed point set by $A_{T^n}$ (or $A_n$ or just
$A$ if there is no chance of confusion).
We represent the distribution of O$^{\mip}$ planes by a function on
$A$ with values in the integers mod 2,
$$D:A\to\{0,1\}.$$ 
The function simply counts the (mod 2) number of D-branes at the fixed points.
Namely,
$D(a)=0$ if $a$ is O$^-$ and
$D(a)=1$ if $a$ is an O$^{\mip}$ plane.
The consistency conditions on the distribution
are stated easily in a T-dual picture,
where the locations of 32 D-branes are represented by
the $O(32)$ Wilson lines on the dual torus $\widehat{T}^n$.
Let us denote the coordinates of the dual torus by
$\widehat{x}^1,\ldots,\widehat{x}^n$.
A single D$(9-n)$-brane at the fixed point $\{x_{\mu}\}=\{a_{\mu}\pi\}$ 
is represented by the flat real line bundle
$\xi(a)$ on $\widehat{T}^n$ which has 
holonomy $(-1)^{a_{\mu}}$ in the $\widehat{x}^{\mu}$-direction.
Therefore, a distribution $D$ of D-branes is represented by the flat bundle
\begin{equation}
\xi(D)=\bigoplus_{a\in A}\xi(a)^{\oplus D(a)}.
\end{equation}
Since there are exactly 32 D$(9-n)$-branes,
the number of fixed points with $D=1$ must be even and also
must be less than or
equal to $32$. This latter condition is vacuous as long as $n\leq 5$.
There is a further condition on the distribution $D$ that
\begin{equation}
w_1(\xi(D))=w_2(\xi(D))=CS(\xi(D))=0.
\label{cond}
\end{equation}
The vanishing of $w_1$ and $w_2$ is required since the $O(32)$ bundle
must lift to a $Spin(32)$ bundle.
The condition $CS(\xi(D))=0$ means the vanishing of the Chern--Simons
invariant of the flat $Spin(32)$ bundle.
This is required from heterotic world-sheet anomaly cancellation as 
discussed in section \ref{anomaly}.

\newcommand{\bD}{\overline{D}}

To determine the allowed distributions, it is useful to introduce
the notion of {\it reduction of a distribution $D$ along circles}.
Let $T^1\subset T^n$ be a circle subgroup
where $T^n$ is regarded as an abelian group defined by addition in
$\{x_{\mu}\}$. Then $T^1$ is invariant under $\Z_2$ and
contains exactly two $\Z_2$ fixed points;
the origin $0$ and the midpoint $a_{T^1}$.
The $\Z_2$ action on $T^n$ descends to a $\Z_2$ action on
$T^n/T^1\cong T^{n-1}$.
The fixed point set $A_{T^n/T^1}$ is the quotient of
$A_{T^n}$ by shifts by $a_{T^1}$.
Now let us define a function $\bD_{T^1}:A_{T^n/T^1}\to\{0,1\}$
by the average
\begin{equation}
\bD_{T^1}([a])=D(a)+D(a+a_{T^1}).
\end{equation}
We call this the {\it reduction of $D$ along $T^1$}.
We note that $T^n/T^1$ is dual to the subtorus
$\widehat{T}^{n-1}\subset\widehat{T}^n$ orthogonal to $T^1$.
It is easy to see that the bundle corresponding to
$\bD_{T^1}$ is equal to the restriction of $\xi(D)$ on this subtorus,
\begin{equation}
\xi(\bD_{T^1})=\xi(D)|_{\widehat{T}^{n-1}}.
\end{equation}
Namely, reduction of $D$ along $T^1$ corresponds to a restriction
of $\xi(D)$ on the subtorus $\widehat{T}^{n-1}$ orthogonal to
$T^1$.
For a sufficiently large $n$,
a distribution $D$ satisfies the conditions (\ref{cond})
if and only if its reduction on an arbitrary subgroup
$T^1\subset T^n$ satisfies (\ref{cond}).
To be precise, this is true when $w_1=0$ for $n\geq 2$,
when $w_2=0$ for $n\geq 3$, and when $CS=0$ for $n\geq 4$.
In what follows,
we make use of this fact and inductively
determine the allowed distributions starting from small $n$.

\subsection{Configurations on $T^2$}

We first determine the allowed distributions on $T^2$
where there are four $\Z_2$ fixed points
$(a_1,a_2)=(0,0),(0,1),(1,0),$ and $(1,1)$.
The condition $CS=0$ is vacuous in this case
and we only need require the topological conditions
$w_1=w_2=0$.
It is useful to note that the total Stiefel--Whitney class of $\xi(a)$
is $w(\xi(a))=1+a_1\theta_1+a_2\theta_2$ where
$\theta_1$ and $\theta_2$
are generators of $H^1(T^2,\Z_2)\cong \Z_2\oplus \Z_2$.
Using the product formula of the total Stiefel--Whitney class \cite{Milnor}\
we find $w(\xi(D))=\prod_{a\in A_2}(1+a_1\theta_1+a_2\theta_2)^{D(a)}$.
{}From this, it is easy to see that only $D\equiv 0$
is allowed.
For example, if $D\equiv 1$
(all four are O$^{\mip}$),
we have $w=1\cdot (1+\theta_1)(1+\theta_2)(1+\theta_1+\theta_2)=
1+\theta_1\theta_2$ and we find $w_2\ne 0$.
If $D$ is not constant $D\not\equiv 0,1$ (mixture of 
O$^-$ and O$^{\mip}$), we do not even have $w_1=0$.
Thus, $D\equiv 0$ (all 4 are O$^-$) is the only allowed distribution.

\subsection{Configurations on $T^3$}

We move on to $T^3$ where there are 8 fixed points.
The reduction along a circle must be an allowed
distribution in $T^2$, which is identically zero as we have seen above.
Thus, $\bD_{T^1}\equiv 0$ for any
$T^1\subset T^3$.
By taking $T^1$ as the circle in the first entry,
we obtain $D(a_1,a_2,a_3)+D(a_1+1,a_2,a_3)=0$ mod 2.
Likewise we obtain similar conditions along the shift
in the second and the third entries.
We therefore have
$$D(a_1+1,a_2,a_3)=D(a_1,a_2+1,a_3)=D(a_1,a_2,a_3+1)=D(a_1,a_2,a_3)\quad
{\rm mod \, 2}.$$
Namely, $D$ is a constant function on $A_3$.
There are two possibilities:
$D\equiv 0$  and so all 8 are O$^-$, 
or $D\equiv 1$ and all 8 are O$^{\mip}$.
Since $n=3<4$, we must still impose the condition $CS=0$.
It is obvious that $CS=0$ for $D\equiv 0$ but, as noted before,
for $D\equiv 1$ we have $CS\ne 0$. So only
$D\equiv 0$ (all 8 are O$^-$) is an allowed distribution.

\subsection{Configurations on $T^4$}

We next consider $T^4$ where there are 16 fixed points.
Since the allowed distribution in the $T^3$ case is identically zero
(all 8 are O$^-$) as we have just seen,
the reduction of $D$ along any circle subgroup must be
identically zero.
As in $T^3$ case, this means that $D$ must be a constant function on
$A_4$.
Since $n=4$, this is sufficient for $w_1=w_2=CS=0$ to be obeyed.
Thus, $D\equiv 0$ (all 16 are $-$)
and $D\equiv 1$ (all 16 are $\mip$) are the allowed distributions.

\subsection{Configurations on $T^5$}

We finally consider $T^5$ where there are 32 fixed points.
$D\equiv 0$ (all 32 are O$^-$) and $D\equiv 1$ (all 32 are O$^{\mip}$)
are allowed dsitributions since, for these cases,
 the reduction along any circle is
identically zero which is an allowed distribution in $T^4$.
In addition
\begin{equation}
\left\{
\begin{array}{l}
D(0,a_{(4)})=0\\[0.1cm]
D(1,a_{(4)})=1
\end{array}\right.
~~\forall a_{(4)}\in A_4,
~~~~~\mbox{and}~~~~~
\left\{
\begin{array}{l}
D(0,a_{(4)})=1\\[0.1cm]
D(1,a_{(4)})=0
\end{array}\right.
~~\forall a_{(4)}\in A_4
\label{type}
\end{equation}
are also allowed distributions.  To see this, let $T^1$ be a circle
subgroup of $T^5$ and let $a_{T^1}$ be the $\Z_2$ invariant midpoint
of $T^1$.  Then, it is easy to see that $\bD_{T^1}\equiv 0$ if
$a_{T^1}=(0,\ldots)$ and $\bD_{T^1}\equiv 1$ if $a_{T^1}=(1,\ldots)$.
Thus, the reduction along an arbitrary cricle is an allowed
distribution in $T^3$ and therefore (\ref{type}) is indeed allowed.
We show below that an allowed distribution, if not identically $0$ or
$1$, must be of this type up to a coordinate transformation.

Suppose $D$ is an allowed distribution which is not identically $0$ or
$1$. This means that the reduction along some $x_{\mu}$ is identically
$1$. We may assume that this is true in the $x_5$ direction.  By a
change of coordinate, if necessary, we may also assume that the
reduction along $x_4$ is identically $0$. The fixed point set of $T^3$
then has a disjoint partition into two components $\tilde{A}_3$ and
$\tilde{\tilde{A}}_3$, where $A_3=\tilde{A}_3\cup\tilde{\tilde{A}}_3$
and $\tilde{A}_3\cap\tilde{\tilde{A}}_3=\emptyset$.  Then the
O$^{\mip}$ planes, which are located where $D=1$, are found at
$(\tilde{a}_{(3)},0,0)$, $(\tilde{a}_{(3)},1,0)$,
$(\tilde{\tilde{a}}_{(3)},0,1)$, and
$(\tilde{\tilde{a}}_{(3)},1,\pi)$. We let $\tilde{a}_{(3)}$ and
$\tilde{\tilde{a}}_{(3)}$ run over the disjoint partition
$\tilde{A}_3$ and $\tilde{\tilde{A}}_3$, respectively.

Now let us redefine coordinates so that
$x^{\prime}_{1,2,3}=x_{1,2,3}+x_4$ and $x^{\prime}_{4,5}=x_{4,5}$. In
these new coordinates, the O$^{\mip}$ planes are at
$(\tilde{a}_{(3)},0,0)$, $(\tilde{a}_{(3)}+1^3,1,0)$,
$(\tilde{\tilde{a}}_{(3)},0,1)$, and
$(\tilde{\tilde{a}}_{(3)}+1^3,1,1)$, where $1^3=(1,1,1)$.  Let $T^1$
be the circle in the $x^{\prime}_4$ direction.  The reduction
$\bD_{T^1}$ must be identically $0$ or $1$.  It is easy to see that
$\tilde{A}_3+1^3=\tilde{A}_3$ and
$\tilde{\tilde{A}}_3+1^3=\tilde{\tilde{A}}_3$ if $\bD_{T^1}\equiv 0$,
whereas $\tilde{A}_3+1^3=\tilde{\tilde{A}}_3$ if $\bD_{T^1}\equiv 1$.
We discuss these cases separately.
\begin{itemize}
\item
For $\bD_{T^1}\equiv 0$, both $\tilde{A}_3$ and $\tilde{\tilde{A}}_3$
are invariant under shift by $1^3$. There are five possibilities.
$\#\widetilde{A}_3=0,2,4,6$, or $8$.  If $\#\tilde{A}_3=0$, all 16
O$^{\mip}$ have $x^5=1$ and $D$ is of type (\ref{type}). Similarly for
the case $\#\tilde{A}_3=8$.  If $\#\widetilde{A}_3=2$ or $6$, the
reduction of $D$ along $x^1$ is not a constant function and therefore
$D$ is not allowed.  Finally let us consider the case
$\#\tilde{A}_3=4$.  The set $\tilde{A}_3$ must be of the type
$\tilde{A}_3=\{(0,a,b),(1,a+1,b+1),(0,c,d),(1,c+1,d+1)\}$ with
$(c,d)\ne (a,b)$.  If $(c,d)=(a+1,b)$, $x^1+x^3+x^5=b$ at all 16
O$^{\mip}$.  If $(c,d)=(a,b+1)$, $x^1+x^2+x^5=a$ at all 16 O$^{\mip}$.
If $(c,d)=(a+1,b+1)$, then $x^2+x^3+x^5=a+b$ at all 16 O$^{\mip}$.
Therefore for every case, the distribution $D$ is of type (\ref{type})
for a suitable choice of coordinates.
\item
For $\bD_{T^1}\equiv 1$, $\tilde{A}_3$ and $\tilde{A}_3+1^3$ determine
a partition of $A_3$ and therefore $\tilde{A}_3$ consists of four
points.  In the coordinate system
$x^{\prime\prime}_{1,2,3}=x_{1,2,3}+x_5$,
$x^{\prime\prime}_{4,5}=x_{4,5}$, the O$^{\mip}$ planes are at
$(\tilde{a}_{(3)},0,0)$, $(\tilde{a}_{(3)},1,0)$,
$(\tilde{a}_{(3)},0,1)$, and $(\tilde{a}_{(3)},1,1)$.  If the
reduction in the $x^{\prime\prime}_1$ direction is identically zero,
$\widetilde{A}_3$ consists of four points $(0,a,b)$, $(1,a,b)$,
$(0,c,d)$ and $(1,c,d)$ where $(c,d)=(a,b+1)$ or $(a+1,b)$.  In the
former case, $x^{\prime\prime}_2=a$ for all four points (and therefore
at all 16 O$^{\mip}$ in $T^5$) while $x^{\prime\prime}_3=b$ at all 16
O$^{\mip}$ in the latter case.  Thus, in both cases $D$ is of type
(\ref{type}).  If the reduction in the $x^{\prime\prime}_1$ direction
is identically $1$, $\widetilde{A}_3$ consists of $(a,0,0)$,
$(a,1,1)$, $(b,0,1)$ and $(b,1,0)$ where $a=b$ or $a=b+1$.  In the
case $a=b$, $x^{\prime\prime}_1=a$ at all 16 O$^{\mip}$ while
$x^{\prime\prime}_1+x^{\prime\prime}_2+x^{\prime\prime}_3=a$ at all 16
O$^{\mip}$ in the case $a=b+1$.  Therefore in both cases $D$ is of
type (\ref{type}).
\end{itemize}
In summary, we have shown that an allowed distribution $D$ with
$D\not\equiv 0,1$ must be of type (\ref{type}). The results obtained
in this appendix are compiled in table \ref{table:distributions} which
lists the allowed distributions for $T^n$ with $2\leq n\leq 5$.

\begin{table}
\begin{center}
\renewcommand{\arraystretch}{1.5}
\begin{tabular}{|c|l|} \hline
Dimension              & Allowed Configurations \\ \hline
 $T^2$ & 4 O7$^-$ \\ \hline
 $T^3$ & 8 O6$^-$ \\ \hline
 $T^4$ & 16 O5$^{-}$ \\
       & 16 O5$^{-'}$\\ \hline
 $T^5$ & 32 O4$^{-}$ \\
       & 32 O4$^{-'}$ \\
       & 16 O4$^{-}$ + 16 O4$^{-'}$ \{{\rm as in (\ref{type})}\}\\\hline
\end{tabular}
\renewcommand{\arraystretch}{1.0}
\caption{Allowed configurations of $Op^-$ and $Op^{-'}$ planes.} \label{table:distributions}
\end{center}
\end{table}

\section{Equivariant Bundles and Equivariant Cohomology}
\label{app:equiv}
\subsection{Equivariant cohomology via spectral sequences}
\label{app:spectral}

Vector bundles over an orbifold $X/G$ are best described in terms of
equivariant bundles over the cover $X$. Equivariant bundles are
bundles $E \to X$ where the action of the group $G$ on the base is
given a lift to the fibers. Geometrically an equivariant bundle can be
seen as a bundle over the `homotopy quotient' $X_G$. This is a smooth
space that is homotopy equivalent to $X/G$. It is a
fiber bundle over the classifying space $BG$ with fiber $X$. By
definition, there is a principal $G$-bundle $EG$ over the classifying
space whose total space is contractible, and the homotopy quotient
$X_G$ is defined as the associated bundle
\bea
X_G = (EG \times X)/G.
\eea
Both $BG$ and $X_G$ are in general infinite-dimensional spaces.  The
equivariant cohomology $H^*_G(X)$ is by definition the cohomology of
the space $X_G$
\bea
H^*_G(X) = H^*(X_G).
\eea
Note that as a special case when $X$ is a point (or when $X$ is
contractible) we obtain the group cohomology of $G$, which (in the
discrete topology) is defined as the cohomology of the classifying
space
\bea
H^*(G) = H^*(BG) = H^*_G (pt).
\eea

Since we have a bundle $X_G \to BG$ with fiber $X$, the cohomology of
$X_G$ can be computed by spectral sequence techniques from the
cohomology of $X$ and $BG$. The fibration also gives us some useful
maps. In particular, there is a map
\bea
H^*(BG) \to H^*_G(X)
\eea
that maps the group cohomology into the equivariant cohomology. This
makes the equivariant cohomology a module for $H^*(BG)$.  There is
also an inclusion of the fiber $X$ in the space $X_G$ and this gives a
map $H^*(X_G) \to H^*(X)$. Since the image is obviously $G$-invariant,
we have in general a map of the equivariant cohomology of $X$ onto the
invariant cohomology of $X$
\bea
H^*_G(X) \to H^*(X)^G.
\eea
So, there is a canonical way to associate an invariant form to an
equivariant form, but in general there is no opposite map. That is, if
we are given an invariant form there is no unique equivariant
representative.

We will give two examples of group cohomologies that will be relevant
for this paper. First, for $G=\Z_2$ the classifying space $B\Z_2$ can
be chosen to be the infinite real projective space $\R\P^\infty$. This
has the non-vanishing cohomology groups
\bea
H^{2k}(\R\P^\infty,\Z) \cong \Z_2,\qquad k>0.
\eea
We can write this succinctly as
\bea
H^*(\R\P^\infty,\Z) \cong \Z[y]/(2y)
\eea
where $y$ is a generator of degree two that satisfies $2y=0$. We will
also need below the cohomology with twisted coefficients $\widetilde
\Z$, where we twist the module $\Z$ with the non-trivial
representation of $\Z_2$. Then we find
\bea
H^{2k-1}(\R\P^\infty,\widetilde\Z) \cong \Z_2,\qquad k>0.
\eea
or, with $\xi$ a generator of degree 1,
\bea
H^*(\R\P^\infty,\widetilde\Z) \cong \xi \Z[y]/(2y,2\xi)
\eea
With $\Z_2$ coefficients we get a particular simple result
\bea
H^*(\R\P^\infty,\Z_2) \cong \Z_2[\xi]
\eea
where now $\xi$ is a generator of degree one. This results generalizes
to cyclic groups.

Another important generalization of relevance to this paper is the case
where $G$ is a finite subgroup of $SU(2)$. These groups are of
$ADE$-type. In that case we find
\bea
H^{2+4k}(BG,\Z) \cong \Z_p,\qquad k \geq 0,
\eea
where $\Z_p=G/[G,G]$ is the abelianization of $G$, i.e., for the binary dihedral
group of order $4n$ denoted $\widehat{\cal D}_{n}$, there are two cases: 
$\Z_2\times \Z_2$ for even $n$ and $\Z_4$ for odd $n$.  The remaining cases 
are $\Z_3$ for the binary tetrahedral subgroup $\mathbb T$ which 
gives an $E_6$ singularity, $\Z_2$ for the binary octahedral subgroup $\mathbb O$
which gives an $E_7$ singularity, and trivial for the icosahedral subgroup $\mathbb I$
which gives an $E_8$ singularity. 
The other non-vanishing cohomology groups are
\bea
H^{4k}(BG,\Z) \cong \Z_{|G|},\qquad k > 0.
\eea

As an example we will compute the equivariant cohomology of $T^4$ with
the $\Z_2$ action $x \to -x$ (thanks to D.~Freed). The cohomology of
$T^4$ is generated by the classes $\h^i$ of degree 1 that satisfy
$(y^i)^2=0$. We can think of them as the images of the 1-forms $dx^i$
in integer cohomology. So we find
\bea
H^*(T^4,\Z) \cong \Z[\h^1,\h^2,\h^3,\h^4]/((\h^i)^2).
\eea
Now the $E_2$ tem in the spectral sequence is
\bea
E_2^{p,q} = H^p(\R\P^\infty,H^q(T^4,\Z))
\eea
Now we have to decompose the $\Z_2$ module $H^q(T^4)$ in irreducible
representations. This is quite simple because we get the trivial
representation $\Z$ for $q$ even, and the non-trivial representation
$\widetilde\Z$ for $q$ odd. This gives the following generators for
$E_2^{p,q}$ (all free indices are ordered $i<j<k$ etc) 
$$
\renewcommand{\arraystretch}{1.5}
\begin{tabular}{|c|cccccc|}
\hline
4 & $\h^1\h^2\h^3\h^4$ & 0  \ \ \ \ \ \ \ \ &     \ \ \ \ \ \ \ \ \  & 
  \ \ \ \ \ \ \ \ \  &  \ \ \ \ \ \ \ \ \ & \ \ \ \ \ \ \ \ \ \\
3 & 0                 & $\xi \h^i\h^j\h^k$  & 0          &    & &\\
2 & $\h^i\h^j$        & 0                   & $\h^i\h^j y$ & 0  & &\\
1 & 0                 & $\xi \h^i$          & 0          & $\xi \h^i y$ & 0 &\\
0 & $\Z$              & 0                   & $y$        & 0  & $y^2$  & 0\\
\hline
$q/p$  & 0 & 1 & 2 & 3 & 4 & 5  \\
\hline
\end{tabular}
\renewcommand{\arraystretch}{1.0}
$$ Since there is no cohomology in odd degree the differential is
trivial and $E_2$ equals the $E_\infty$ term. There is a further
extension problem that has to be solved to find the actual cohomology
groups, but turns out to be trivial too. The result is that the
non-trivial groups are
\bea
H^2_{\Z_2}(T^4,\Z) \cong \Z^6 \oplus (\Z_2)^5
\eea
and
\bea
H^4_{\Z_2}(T^4,\Z) \cong \Z \oplus (\Z_2)^{15}.
\eea
Since the odd cohomology vanishes we find that
\bea
H^1_{\Z_2}(T^4,U(1)) \cong (\Z_2)^5,\quad {\rm and} \quad
H^3_{\Z_2}(T^4,U(1)) \cong (\Z_2)^{15}.
\eea

\subsection{The case of $T^4/G$ for more general $G$}
\label{app:generalquotient}

We now briefly describe the generalization of our results
for $T^4/\Z_2$ 
to certain other quotients of four-tori. The possible groups
which can act on $T^4$ giving a Calabi--Yau are 
$G=\Z_3,\Z_4,\Z_6,\widehat{\cal D}_4,\widehat{\cal D}_5, \mathbb{T}$
\cite{fujiki}.
For $\widehat{\cal D}_4$ and $\mathbb{T}$ (the binary tetrahedral group), 
there are actually multiple
possible actions which result in different singularities for $T^4/G$.
We list these possibilities in table \ref{table:singquot}. 

\begin{table}
\begin{center}
\renewcommand{\arraystretch}{1.5}
\begin{tabular}{|c|c|} \hline
Space     & Singularities \\ \hline
 $T^4/\Z_2 $ & $A_1^{16}$ \\ \hline
 $T^4/\Z_3 $ & $A_2^{9}$ \\ \hline
 $T^4/\Z_4 $ & $A_3^{4}\oplus A_1^6$ \\ \hline
 $T^4/\Z_6 $ & $A_5 \oplus A_2^4 \oplus A_1^5$ \\ \hline
 $T^4/\widehat{\cal D}_4 $ & $D_4^2 \oplus A_3^3 \oplus A_1^2$ \\ \hline
 $T^4/\widehat{\cal D}_4 $ & $D_4^4 \oplus A_1^3$ \\ \hline
 $T^4/\widehat{\cal D}_4 $ & $A_3^6 \oplus A_1$ \\ \hline
 $T^4/\widehat{\cal D}_5 $ & $D_5 \oplus A_3^3 \oplus A_2^2 \oplus A_1
 $ \\ \hline
 $T^4/\mathbb{T} $ & $E_6 \oplus D_4 \oplus A_2^4 \oplus A_1$ \\ \hline
 $T^4/\mathbb{T} $ & $A_5 \oplus A_3^2 \oplus A_2^4$ \\ \hline
\end{tabular}
\renewcommand{\arraystretch}{1.0}
\caption{Singularities of torus quotients.} \label{table:singquot}
\end{center}
\end{table}

We shall restrict to the specific cases of
$G=\Z_3,\Z_4,\Z_6,\widehat{\cal D}_4, \widehat{\cal D}_5$. The
particular $\widehat{\cal D}_4$ quotient to be described below results
in a $T^4/G$ with singularities,
$$D_4^2 \oplus A_3^3 \oplus A_1^2. $$
If we
use two complex coordinates $(z_1,z_2)$
to describe the four-torus, the group $\Z_m$ is generated by
\begin{equation} \beta: (z_1,z_2) \to 
(\zeta z_1, \zeta^{-1} z_2) \end{equation}
with $\zeta=e^{2\pi i/m}$
an $m^{th}$ root of unity. The binary dihedral groups 
$\widehat{\cal D}_m$ have an $\Z_{2(m-2)}$ subgroup which acts just as
$\Z_m$ described above, and in addition they have a generator $\alpha$
that we choose to act by
\begin{equation}
\alpha: (z_1,z_2) \to (z_2,-z_1)  .
\end{equation}
The generators $\alpha$ and $\beta$ of $\widehat{\cal D}_m$
satisfy the relations
\begin{equation} \label{reldhat}
\alpha^2 = \beta^{m-2},\qquad  \alpha^4=\beta^{2m-4}=1,\qquad
\beta \alpha \beta = \alpha. \end{equation}
Notice that we use the rank of the singularity to label the binary
dihedral group $\widehat{\cal D}_m$; it is a group of dimension
$4(m-2)$, and in the mathematics literature sometimes denoted by
$D^{\ast}_{4(m-2)}$. 

We first calculate the equivariant cohomology groups 
$H^{\ast}_G(T^4,U(1))$ that will give us the flat equivariant
$n$-forms. We can use the same spectral sequence described in
section~\ref{app:spectral}, which is the Leray spectral sequence associated
to the fibration $X\to X_G \to BG$. In the present case, the $E_2^{p,q}$
terms of this spectral sequence are given by
\begin{equation} 
\label{e2pq}
E_2^{p,q} = H^p(G,H^q(T^4,U(1))) .\end{equation}
The right hand side is defined as follows. The action of $G$ on $T^4$
makes $H^q(T^4,U(1))$ into a module for the finite group
$G$. Then $E_2^{p,q}$ is defined as the cohomology of the finite
group $G$ with values in this module.

Let us briefly review the 
standard definition of the cohomology of a finite group 
with values in a module.
Given any $G$-module $M$, $n$-cochains with values in $M$
are maps 
$$c: \quad G^n\to M, $$
i.e., $M$-valued functions $c(g_1,\ldots,g_n)$
of $n$ group elements. On these cochains, we define a coboundary
operator
\bea
\delta c (g_1,\ldots,g_{n+1}) & = & c(g_2,\ldots,g_n) - 
(-1)^n g_1 c(g_2,\ldots, g_{n+1}) \nonumber
\\ & & + \sum_{i=1}^n (-1)^i
 c(g_1,\ldots,g_{i-1},g_i g_{i+1},g_{i+2},\ldots,
g_{n+1}) .
\eea
The finite group cohomology
$H^{\ast}(G,M)$ is given by ${\rm Ker}(\delta)/{\rm Im}(\delta)$.

Although this definition is quite simple, it is not very convenient
for practical calculations. A more efficient way to compute finite
group cohomology is to use a suitable resolution of the finite group
$G$. More precisely, we need a right projective resolution of $\Z$
by $\Z G$-modules (see e.g. \cite{robinson,thomas}). 
Such a resolution is an exact sequence
\begin{equation} \label{exseq}
\cdots \longrightarrow P_k \longrightarrow P_{k-1} \longrightarrow
\cdots P_0 \longrightarrow \Z \longrightarrow 0
\end{equation}
where all $P_k$ are projective modules over the group algebra $\Z G $,
the maps are homomorphisms of $\Z G$-modules, and $\Z$ is viewed
as the trivial $\Z G$-module. Projective modules are modules
which are the direct summand of a free module. 
An exact sequence like (\ref{exseq}) induces a sequence
\begin{equation}
\label{exseq2}
M^G \longrightarrow {\rm Hom}_G (P_0,M)
 \longrightarrow {\rm Hom}_G (P_1,M)
 \longrightarrow {\rm Hom}_G (P_2,M) \cdots 
\end{equation}
for any $G$-module $M$. Here, $M^G$ denotes the $G$-invariant 
elements of $M$, and ${\rm Hom}_G(P,M)$ denotes the homomorphisms
of $P$ to $M$ that commute with the action of $G$. 
It is a standard fact that (\ref{exseq}) is no longer exact,
but is still a complex. The cohomology of that complex
is precisely the finite group cohomology $H^{\ast}(G,M)$. 

A convenient resolution of finite groups is the Gruenberg
resolution (see section~11.3 in \cite{robinson}), which
is associated to a presentation of the finite group (i.e.,
a set generators and relations). For $\Z_m$, we can use
as generator $\beta$ subject to $\beta^m=1$, and for
$ \widehat{\cal D}_k$ we can use the presentation with
generators $\alpha,\beta$ and relations (\ref{reldhat}).
The respective resolutions are given on page 6 and 35
of \cite{thomas}. Using (\ref{exseq2}) they yield a simple
algorithm to compute the finite group cohomologies $H^{\ast}(G,M)$.
Here we merely summarize the algorithm for $\Z_m$ and our particular
$ \widehat{\cal D}_k$ actions.

For $G=\Z_m$, the cohomology groups $H^i(\Z_m,M)$ are given
by the cohomology of the complex
\begin{equation} \label{reso1}
M \stackrel{d_0}{\longrightarrow} M 
 \stackrel{d_1}{\longrightarrow} M 
 \stackrel{d_0}{\longrightarrow} M 
 \stackrel{d_1}{\longrightarrow} \cdots
\end{equation}
where 
\begin{equation}
d_0(m) = (\beta-1)m, \qquad d_1(m)=\sum_{g \in \Z_m} g m .
\end{equation}
Recall that $\beta$ was defined as the generator of $\Z_m$. In particular, 
we see that the cohomology in degrees above zero will be periodic
with period 2. As an example, take $M=U(1)$ with the trivial
action of $\Z_m$. Then $d_0=0$, $d_1=m$, and we get
that $H^0(\Z_m,U(1))=U(1)$, $H^{2i+1}(\Z_m,U(1))=\Z_m$, and
$H^{2i}(\Z_m,U(1))=0$ for $i>0$.

For $G=\widehat{\cal D}_k$, the cohomology groups are given
by the cohomology of the complex
\begin{equation} \label{reso2}
M \stackrel{d_0}{\longrightarrow} M \oplus M
 \stackrel{d_1}{\longrightarrow} M \oplus M
 \stackrel{d_2}{\longrightarrow} M
 \stackrel{d_3}{\longrightarrow} M 
\stackrel {d_0}{\longrightarrow} M \oplus M
\stackrel{d_1}{\longrightarrow} \cdots
\end{equation}
where
\bea
d_0 (m) & = & ((\alpha-1)m,(\beta-1)m) \nonumber, \\
d_1 (m_1,m_2) & = & ((\beta-1)m_1 + (\beta\alpha + 1)m_2,
 (-1-\alpha)m_1 + (1 + \beta + \ldots + \beta^{k-3}) m_2) \nonumber , \\
d_2(m_1,m_2) & = & (1-\beta\alpha)m_1 + (\beta-1)m_2 \nonumber , \\
d_3(m) & = & \sum_{g \in \widehat{\cal D}_k } g m. 
\eea
Thus, this cohomology will be periodic with period four.

We are now ready to compute the $E_2^{p,q}$ terms defined in
(\ref{e2pq}). First, we determine the $G$-module structure of 
$H^q(T^4,U(1))$. We then insert this module in either (\ref{reso1}) or
(\ref{reso2}), and work out the cohomology of the corresponding
complex. In practice, the maps $d_i$ can always be written
as matrices with integer coefficients with respect to an
integral basis of $H^q(T^4,U(1))$. Using suitable changes
of basis by acting with $SL(p,\Z)$, these matrices can be
brought to a form with only diagonal elements. If the $k^{th}$
differential $d_k$ has diagonal non-zero entries $d_1,\ldots,d_r$
then $H^k$ will be equal to $\Z_{d_1} \oplus \ldots
\oplus \Z_{d_r}$. 

The results of the calculation of $E_2^{p,q}$ are given below.
\begin{equation}
\renewcommand{\arraystretch}{1.5}
G=\Z_2 :
\begin{tabular}{|c|cccccc|}
\hline
$ 4 $&$ U(1) $&$ \Z_2 $&$ 0 $&$ \Z_2 $&$ 0 $&$ \Z_2 $ \\
$ 3 $&$ \Z_2^4 $&$ 0 $&$ \Z_2^4 $&$ 0 $&$ \Z_2^4 $&$ 0 $ \\
$ 2 $&$ U(1)^6 $&$ \Z_2^6 $&$ 0 $&$ \Z_2^6 $&$ 0 $&$ \Z_2^6$  \\
$ 1 $&$ \Z_2^4 $&$ 0 $&$ \Z_2^4  $&$ 0 $&$ \Z_2^4 $&$ 0$  \\
$ 0 $&$ U(1) $&$ \Z_2 $&$ 0 $&$ \Z_2 $&$ 0 $&$ \Z_2$  \\
\hline
$q/p  $&$ 0 $&$ 1 $&$ 2 $&$ 3 $&$ 4 $&$ 5 $  \\
\hline
\end{tabular}
\renewcommand{\arraystretch}{1.0}
\end{equation}
\begin{equation}
\renewcommand{\arraystretch}{1.5}
G=\Z_3 :
\begin{tabular}{|c|cccccc|}
\hline
$ 4 $&$ U(1) $&$ \Z_3 $&$ 0 $&$ \Z_3 $&$ 0 $&$ \Z_3$  \\
$ 3 $&$ \Z_3^2 $&$ 0 $&$ \Z_3^2 $&$ 0$&$  \Z_3^2 $&$ 0 $ \\
$ 2 $&$ U(1)^4 $&$ \Z_3^3 $&$ 0 $&$ \Z_3^3 $&$ 0 $&$ \Z_3^3 $ \\
$ 1 $&$ \Z_3^2 $&$ 0 $&$ \Z_3^2  $&$ 0$&$  \Z_3^2 $&$ 0 $ \\
$ 0 $&$ U(1) $&$ \Z_3 $&$ 0 $&$ \Z_3 $&$ 0 $&$ \Z_3$  \\
\hline
$q/p  $&$ 0 $&$ 1 $&$ 2 $&$ 3 $&$ 4 $&$ 5 $  \\
\hline
\end{tabular}
\renewcommand{\arraystretch}{1.0}
\end{equation}
\begin{equation}
\renewcommand{\arraystretch}{1.5}
G=\Z_4 :
\begin{tabular}{|c|cccccc|}
\hline
$ 4 $&$ U(1) $&$ \Z_4 $&$ 0 $&$ \Z_4 $&$ 0 $&$ \Z_4$  \\
$ 3 $&$ \Z_2^2 $&$ 0 $&$ \Z_2^2 $&$ 0$&$  \Z_2^2 $&$ 0$  \\
$ 2 $&$ U(1)^4 $&$ \Z_4^2 \oplus \Z_2^2 $&$ 0 $&$ \Z_4^2 
\oplus \Z_2^2 $&$ 0 $&$ 
\Z_4^2 \oplus \Z_2^2$  \\
$ 1 $&$ \Z_2^2 $&$ 0 $&$ \Z_2^2  $&$ 0 $&$ \Z_2^2 $&$ 0$  \\
$ 0 $&$ U(1) $&$ \Z_4 $&$ 0 $&$ \Z_4 $&$ 0 $&$ \Z_4$  \\
\hline
$q/p  $&$ 0 $&$ 1 $&$ 2 $&$ 3 $&$ 4 $&$ 5 $  \\
\hline
\end{tabular}
\renewcommand{\arraystretch}{1.0}
\end{equation}
\begin{equation}
\renewcommand{\arraystretch}{1.5}
G=\Z_6 :
\begin{tabular}{|c|cccccc|}
\hline
$ 4 $&$ U(1) $&$ \Z_6 $&$ 0 $&$ \Z_6 $&$ 0 $&$ \Z_6 $ \\
$ 3 $&$ 0 $&$ 0 $&$ 0 $&$ 0$&$ 0 $&$ 0 $ \\
$ 2 $&$ U(1)^4 $&$ \Z_6^3\oplus \Z_2 $&$ 0 $&$ \Z_6^3\oplus \Z_2 $&$ 0 $&$ 
\Z_6^3\oplus \Z_2 $ \\
$ 1 $&$ 0 $&$ 0 $&$ 0  $&$ 0$&$ 0 $&$ 0 $ \\
$ 0 $&$ U(1) $&$ \Z_6 $&$ 0 $&$ \Z_6 $&$ 0 $&$ \Z_6 $ \\
\hline
$q/p  $&$ 0 $&$ 1 $&$ 2 $&$ 3 $&$ 4 $&$ 5  $ \\
\hline
\end{tabular}
\renewcommand{\arraystretch}{1.0}
\end{equation}
\begin{equation} \label{d4e2pq}
\renewcommand{\arraystretch}{1.5}
G=\widehat{\cal D}_4 :
\begin{tabular}{|c|cccccc|}
\hline
$ 4 $&$ U(1) $&$ \Z_2^2 $&$ 0 $&$ \Z_8 $&$ 0 $&$ \Z_2^2$  \\
$ 3 $&$ \Z_2 $&$ 0 $&$ \Z_2 $&$ 0$&$  \Z_2 $&$ 0 $ \\
$ 2 $&$ U(1)^3 $&$ \Z_4^3 $&$ 0 $&$ \Z_4^3 $&$ 0 $&$ \Z_4^3$  \\
$ 1 $&$ \Z_2 $&$ 0 $&$ \Z_2  $&$ 0$&$  \Z_2 $&$ 0$  \\
$ 0 $&$ U(1) $&$ \Z_2^2 $&$ 0 $&$ \Z_8 $&$ 0 $&$ \Z_2^2$  \\
\hline
$q/p  $&$ 0 $&$ 1 $&$ 2 $&$ 3 $&$ 4 $&$ 5 $  \\
\hline
\end{tabular}
\renewcommand{\arraystretch}{1.0}
\end{equation}
\begin{equation}
\renewcommand{\arraystretch}{1.5}
G=\widehat{\cal D}_5 :
\begin{tabular}{|c|cccccc|}
\hline
$ 4 $&$ U(1) $&$ \Z_4 $&$ 0 $&$ \Z_{12} $&$ 0 $&$ \Z_4$  \\
$ 3 $&$ 0 $&$ 0 $&$ 0 $&$ 0$&$ 0 $&$ 0$  \\
$ 2 $&$ U(1)^3 $&$ \Z_4^2 \oplus \Z_6 $&$ 0 $&$ 
\Z_4 \oplus \Z_6 \oplus \Z_{12} $&$ 0 $&$ \Z_4^2 \oplus \Z_6$  \\
$ 1 $&$ 0   $&$ 0 $&$ 0 $&$ 0$&$ 0 $&$ 0 $ \\
$ 0 $&$ U(1) $&$ \Z_4 $&$ 0 $&$ \Z_{12} $&$ 0 $&$ \Z_4 $ \\
\hline
$q/p  $&$ 0 $&$ 1 $&$ 2 $&$ 3 $&$ 4 $&$ 5 $  \\
\hline
\end{tabular}
\renewcommand{\arraystretch}{1.0}
\end{equation}

For $G=\Z_m$, the rows in these tables are periodic with
period two, whereas for $G=\widehat{\cal D}_k$ they are periodic
with period four.  In all cases, the spectral sequence collapses 
at the $E_2^{p,q}$ term. 
In addition, by redoing the calculation with coeffients in
$\Z_r$ rather than $U(1)$, we find that 
the extension problem is trivial.
Thus, the equivariant cohomologies $H^n_G(T^4,U(1))$ are 
isomorphic to $\oplus_i E_2^{i,n-i}$.

As a simple test, we notice that all cohomology above degree four
has to be supported purely by the singularities of $T^4/G$.
This can be seen, for example, from a Mayer--Vietoris argument.
Using Table~\ref{table:singquot} (see also Table~\ref{table:torus}), 
one readily verifies that this is indeed the case. 
Notice that $H^{4i+1}(\widehat{\cal D}_{2k})=\Z_2 \oplus \Z_2$,
$H^{4i+1}(\widehat{\cal D}_{2k+1}) = \Z_4$, and 
$H^{4i+3}(\widehat{\cal D}_{k}) = \Z_{4(k-2)}$.

\noindent
\underline{One-forms}\\
The equivariant 1-forms receive contributions from two
different sources. One is from
$H^0(G,H^1(T^4,U(1)))$, which are the $G$-invariant 1-forms
on $T^4$. They are in one-to-one correspondence with the 
$G$-fixed points on $T^4$. The second contribution is from
$$H^1(G,H^0(T^4,U(1)))=H^1(G,U(1)),$$ 
whose elements correspond to  
the one-dimensional representations of $G$. Both have a clear
interpretation in terms of line bundles over $T^4$: the
$G$-invariant 1-forms represent a flat connection on $T^4$ whereas
$H^1(G,U(1))$ represents possible actions of $G$ on the
fiber of the line bundle. Also note that $H^1_G(T^4,U(1))$
matches with the group $(E^{\perp})^{\perp}/E$ given in
table~\ref{table:torus}.

What is the M theory description of these theories? 
Let $A\cdot dx \in H^0(G,H^1(T^4,U(1)))$ denote an invariant 1-form,
and $\rho \in H^1(G,U(1))$ a 1-dimensional representation
of $G$. The pair $(A,\rho)$ labels a general element
of $H^1_G(T^4,U(1))$.
The 1-form $A$ defines a five-torus $T^5$, which is
not of the product form $T^4 \times S^1$. Instead, the
extra circle is fibered non-trivially 
over $T^4$. The five-torus has a metric of the
form 
\begin{equation} \label{jj1}
ds^2 = ds^2_{T^4} + R_{5}^2 (d\phi +A\cdot dx)^2
\end{equation}
where $\phi$ denotes the coordinate along the extra circle,
whose radius we denoted by $R_{5}$.
The group $G$ acts on this five-torus by
\begin{equation} \label{jj2}
(x,\phi) \to (g(x),\phi+\rho(g) + A \cdot (x-g(x))) .
\end{equation}
The M theory description of these theories is a compactification
on $T^5/G$, with the five-torus and group action given in (\ref{jj1})
and (\ref{jj2}).

To relate this to compactifications on $(M \times S^1)/G$, we need
to find a description where the five-torus is a direct product of
a four-torus and a circle. 
As in section~\ref{RRone}, this can be accomplished by taking a suitable
cover of $T^5$. Suppose that the four-torus is described by
$\R^4/\Lambda$. Then $A\in \R^4/\Lambda^{\ast}$ which is the dual 
four-torus. Let $p$ be the smallest positive integer such that
$p A=0$ in $\R^4/\Lambda^{\ast}$. We define a $p$-fold
cover $\hat{T}^4$ of $T^4$ as $\R^4/\Lambda_A$, where
$\Lambda_A$ is the lattice 
$$\Lambda_A = \{ x\in \Lambda | A \cdot x \in \Z\}.$$
The group action on $T^4$ lifts to an action on $\hat{T}^4$. To prove
this, we need to show that $g(\Lambda_A)\subset \Lambda_A$.
Thus for $x\in \Lambda_A$, then we need to show that
$g(x) \cdot A \in \Z$. The statement that $A$ is $G$ invariant
implies that $g(x)\cdot A - x \cdot A = x \cdot B$, with $B \in \Lambda^{\ast}$.
Applying this to $x\in \Lambda_A$, we immediately get that 
$g(x) \cdot A \in\Z$. Therefore, $G$ lifts to an action
on $\hat{T}^4$. 

Clearly, $T^4=\hat{T}^4/G_p$, where $G_p$ is finite group of 
translations in $\hat{T}^4$. It is not difficult to see that $G_p=\Z_p$,
because we can always use $SL(4,\Z)$ to make only one of
the components of $A$ non-zero. In particular, $\Lambda/\Lambda_A
=\Z_p$, and we can choose an element $x_0\in \Lambda$ that
generates  $\Lambda/\Lambda_A$. Now on $\hat{T}^4 \times S^1$,
we have both an action of $\Z_p$ and of $G$. They act as
\begin{equation}
G: (x,\phi)\to(g(x), \phi+\rho(g)),\qquad
\Z_p: (x,\phi) \to (x+x_0,\phi+ A \cdot x_0) ,
\end{equation}
and one can easily verify that these two group actions commute.
The M theory description of IIA string theory on $T^4/G$
with 1-form flux $(A,\rho)$ is given by 
M theory on
\begin{equation}
\label{mdes}
(T^5)/G \equiv (\hat{T}^4 \times S^1)/(G\times \Z_p) .
\end{equation}

We argue in section~\ref{geometry} that the only possible M theory
compactifications that preserve the right amount of
supersymmetry are of the form $(M \times S^1)/\Z_r$, 
with $M=T^4$ or $M=K3$. Indeed, the space (\ref{mdes}) is 
always of this form. To find $M$, we need to study the 
action of $G\times \Z_p$ on the $S^1$ factor. 

The image  of $\rho$ in $U(1)$ defines a subgroup $\Z_{|\rho|}$
of $U(1)$. We also have a 1-dimensional representation of
$G \times \Z_p$, acting on the $S^1$ factor in $\hat{T}^4 \times S^1$.
This representation has as image in $U(1)$ the group 
$\Z_{ {\rm lcm}(p,|\rho|) }$, where ${\rm lcm}$ denotes
the least common multiple. Thus there is an exact sequence
\begin{equation}
\hat{G} \to G\times  \Z_p \to \Z_{ {\rm lcm}(p,|\rho|)} ,
\end{equation}
where $\hat{G}$ is a normal subgroup of $G$.
Therefore, M theory compactified on (\ref{mdes}) is the
same as M theory compactified on 
\begin{equation}
((\hat{T}^4/\hat{G}) \times S^1)/ \Z_{ {\rm lcm}(p,|\rho|)} .
\end{equation}
This is always of the form $(M \times S^1)/\Z_r$, with
$M=T^4$ of $M=K3$, as expected. We obtain $M=T^4$ only
if $G$ is cyclic, $\rho$ is a faithful representation and
$A=0$. In all other cases, $M=K3$. 

As an example, let us take our favorite non-abelian 
action $G=\widehat{\cal D}_4$. 
We showed that $E_2^{1,0}=\Z_2^2$ and
that $E_2^{0,1}=\Z_2$. There is one case (the trivial
1-form) where $|\rho|=1$  and $p=1$. This corresponds
to M theory on $(T^4/\widehat{\cal D}_4)\times S^1$, as
expected. There are three cases with $|\rho|=2$ and
$p=1$. These correspond to M theory on
$((\hat{T}^4/\Z_4)\times S^1)/\Z_2$. Furthermore, there
is one case with $|\rho|=1$ and $p=2$. This corresponds to
M theory on $((\hat{T}^4/\widehat{\cal D}_4)\times S^1)/\Z_2$.
Finally, there are three cases with $|\rho|=2$ and $p=1$.
These correspond to M theory on 
$((\hat{T}^4/\Z_4)\times S^1)/\Z_2$.

What about the holonomies of these 1-forms? Consider
a point $p\in \R^4$ that is fixed under a part $G_p$ of the group
$G$ when viewed as an element of
$\R^4/\Lambda$. Given a group element $g$ that
fixes $p$, we need to determine the holonomy of the 1-form
around the loop associated to $g$ in $S^3/G_p$. Here, we view
$g$ as an element of $\pi_1(S^3/G_p) \cong G_p$. Generalizing
the discussion at the end of section~\ref{RRone}, we find for the holonomy
\begin{equation} 
\exp \left[ 2 \pi i \left(
(p-g(p))\cdot  A + \rho(g)\right) \right] . \end{equation}

\subsection{Explicit generators for $H^2_{\Z_2}(T^n,\Z_2)$}

To conclude our discussion of equivariant cohomology, we shall explicitly
construct generators for  $H^2_{\Z_2}(T^n,\Z_2)$ where $n=1,2,3$. This makes
the localization of the $B$-field concrete in the orientifold constructions
discussed in section \ref{orient}.  
We use the cell decomposition of the space $T^n_{\Z_2}=
S^N\times_{\Z_2}T^n$ which is obtained from
a $\Z_2$-equivariant cell-decomposition of
$S^N\times T^n$. We are interested in the second cohomology so it 
is sufficient to take $N=3$ here.
\begin{figure}[htb]
\begin{center}
\epsfxsize=3in\leavevmode\epsfbox{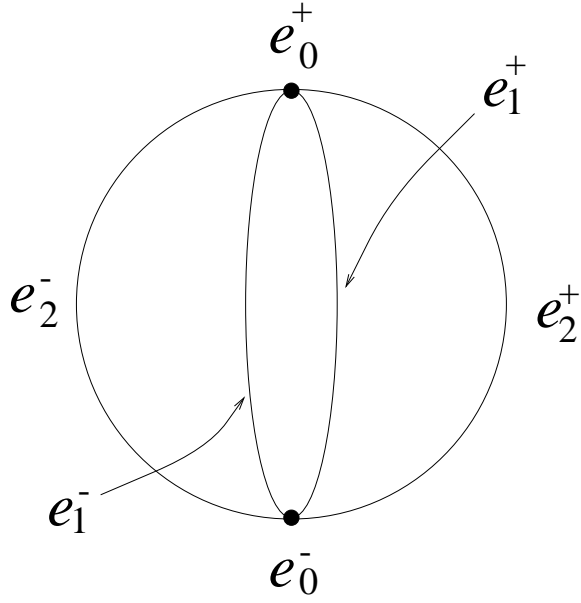}
\end{center}
\caption{Cell decomposition of $S^2$.}
\label{fig:S2}
\end{figure}  
Throughout our discussion, we will use the standard cell decomposition
of $S^N$,
\bea
S^N=e_0^+\cup e_0^-\cup e_1^+\cup e_1^-\cup e_2^+\cup e_2^-\cup\cdots,
\eea
where the $\Z_2$ exchanges $e_m^+$ and $e_m^-$, 
\bea
e_m^{\pm}\to e_m^{\mp}.
\eea
This decomposition is illustrated in figure \ref{fig:S2}.
The boundary operator is given by
\bea
&&\partial e_0^{\pm}=0,\nonumber\\
&&\partial e_1^{\pm}=e_0^++e_0^-,\nonumber\\
&&\partial e_2^{\pm}=e_1^++e_1^-,\nonumber\\
&&\quad \vdots \nonumber
\eea
We note that we do not care about the sign since we are considering
the $\Z_2$ coefficient, $e_m^{\pm}=-e_m^{\pm}$.

\bigskip

\begin{figure}[htb]
\begin{center}
\epsfxsize=2in\leavevmode\epsfbox{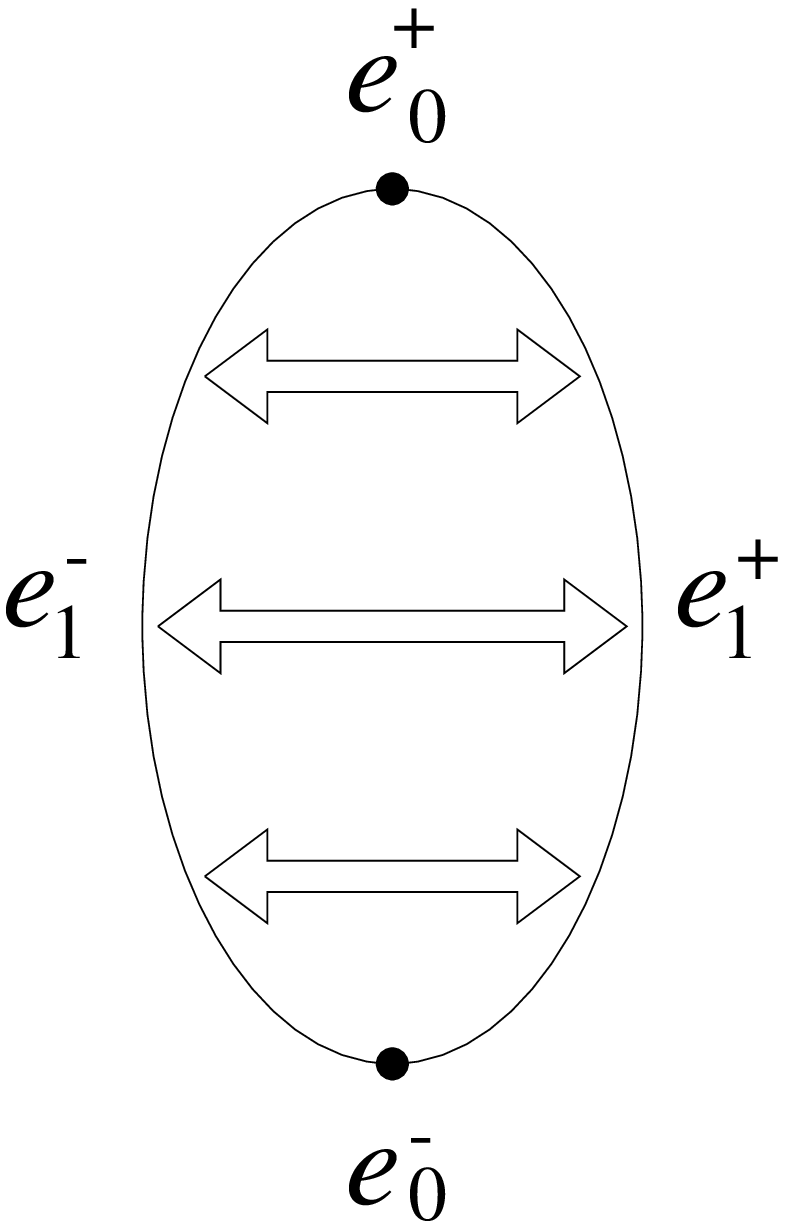}
\end{center}
\caption{Cell decomposition of $T^1$.}
\label{fig:S1}
\end{figure} 

\noindent
\underline{$n=1$}

We consider the cell decomposition of $T^1$ as depicted in figure \ref{fig:S1}.
The $\Z_2$ inversion fixes the $0$-chains but exchanges the two
$1$-chains,
\bea
e_0^{\pm}\to e_0^{\pm},~~~
e_1^{\pm}\to e_1^{\mp}.
\eea
The boundary operator is just
$\partial e_0^{\pm}=0$ and $\partial e_1^{\pm}=e_0^++e_0^-$.

The cell decomposition of the space $T^1_{\Z_2}=(S^N\times T^1)/\Z_2$
is given as follows:
\bea
&&~~e_{00}^{\pm}:=e_0^+\times e_0^{\pm}\equiv e_0^-\times e_0^{\pm},
\nonumber\\
&&\left.
\begin{array}{l}e_{\ell\, 0}^{\pm}:=e_{\ell}^{+}\times e_0^{\pm}
\equiv e_{\ell}^-\times e_0^{\pm},\\
e_{(\ell-1)1}^{\pm}:=e_{\ell-1}^{+}\times e_1^{\pm}
\equiv e_{\ell-1}^-\times e_1^{\mp},
\end{array}\right\}~~~\ell=1,2,3,\ldots.
\eea
It is then easy to see that the boundary operator is given by,
\bea
&&\partial e_{\ell 0}^{\pm}=0,~~\ell=0,1,2,\ldots\nonumber\\
&&\partial e_{01}^{\pm}=e_{00}^++e_{00}^-,\nonumber\\
&&\partial e_{\ell 1}^{\pm}=e_{(\ell-1)\,1}^++e_{(\ell-1)\,1}^-
+e_{\ell\, 0}^++e_{\ell\, 0}^-,~~\ell=1,2,3,\ldots
\eea
By dualization we find the following coboundary operator, where 
the notation for the cochain should be self-evident:
\bea
\delta c_{\ell \alpha}^{\pm}=c_{(\ell+\alpha)1}^+
+c_{(\ell+\alpha)1}^-,~~~(\ell=0,1,2,\ldots;~~\alpha=0,1).
\eea
{}From this expression, we see that $B^2$ (the space of coboundaries 
of dimension $2$)
is spanned by $(c_{11}^++c_{11}^-)$ while
$Z^2$ (the space of $2$-cocycles) is spanned by
$(c_{20}^++c_{11}^+)$, $(c_{20}^-+c_{11}^+)$, and $(c_{11}^++c_{11}^-)$.
The cohomology group $H^2=Z^2/B^2$ is therefore given by
\bea
\Z_2[c_{20}^++c_{11}^+]\oplus \Z_2[c_{20}^-+c_{11}^+].
\eea
The first generator has value $1$ on $e_{20}^+$ 
i.e., $\RP^2$ at the $\Z_2$ fixed point $e_0^+$, 
while it vanishes on $e_{20}^-$,
i.e., $\RP^2$ at the other $\Z_2$ fixed point $e_0^-$. 
The second generator vanishes on $e_{20}^+$ ($\RP^2$ at $e_0^+$)
while it has value $1$ on $e_{20}^-$ ($\RP^2$ at $e_0^-$).

\bigskip

\begin{figure}[htb]
\begin{center}
\epsfxsize=2.7in\leavevmode\epsfbox{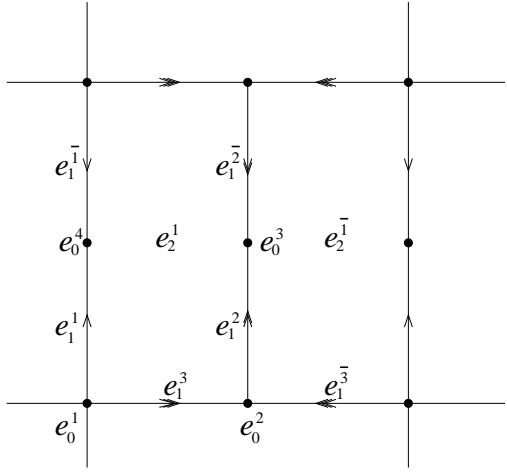}
\end{center}
\caption{Cell decomposition of $T^2$.}
\label{fig:T2}
\end{figure} 

\noindent
\underline{$n=2$}

We consider the $\Z_2$-equivariant
cell decomposition of $T^2$ depicted in figure \ref{fig:T2}.
There are four $0$-chains $e_0^i$ where $i=1,2,3,4$. In addition,
there are
six $1$-chains $e_1^{\mu}$ with $\mu=1,\bar 1,2,\bar 2,3,\bar 3$
and two $2$-chains $e_2^I$ with $I=1,\bar 1$.
The $\Z_2$ acts on them in the following way:
\bea
e_0^i\to e_0^i,~~
e_1^{\mu}\to e_1^{\bar \mu},~~
e_2^I\to e_2^{\bar I}.
\eea
In this expression, $\bar{\bar 2}=2$ etc.
The cell decomposition of $T^2_{\Z_2}$ is given by
\bea
&&~~e_{00}^i:=e_0^+\times e_0^i\equiv e_0^-\times e_0^i,
\nonumber\\
&&\left.\begin{array}{l}
e_{10}^i:=e_1^+\times e_0^i\equiv e_1^-\times e_0^i,\\
e_{01}^{\mu}:=e_0^+\times e_1^{\mu}\equiv e_0^-\times e_1^{\bar\mu},
\end{array}\right\}
\nonumber\\
&&\left.\begin{array}{l}
e_{\ell \,0}^i:=e_{\ell}^+\times e_0^i\equiv e_{\ell}^-\times e_0^i,\\
e_{(\ell-1)1}^{\mu}:=e_{\ell-1}^+\times e_1^{\mu}
\equiv e_{\ell-1}^-\times e_1^{\bar\mu},\\
e_{(\ell-2)2}^I:=e_{\ell-2}^+\times e_2^{I}
\equiv e_{\ell-2}^-\times e_2^{\bar I}
\end{array}\right\}~~\ell=2,3,\ldots
\eea
Here, we do not show all the details of the computation
but just present the result.
The $2^{nd}$ coboundary group $B^2$ is spanned by $c_{02}^1+c_{02}^{\bar 1}$ and
$c_{11}^{\mu}+c_{11}^{\bar \mu}$ with $\mu=1,2,3$.
The cohomology is of rank $4$,
$H^2_{\Z_2}(T^2,\Z_2)=(\Z_2)^{  4}$,
 and the four generators are represented by
\bea
c_{20}^1+c_{11}^1+c_{11}^3+c_{02}^1,~~
c_{20}^2+c_{11}^2+c_{11}^3+c_{02}^1,~~
c_{20}^3+c_{11}^2+c_{02}^1,~~
c_{20}^4+c_{11}^1+c_{02}^1.
\eea
Note that the $i^{th}$ generator has a non-trivial value on the $\RP^2$
at the fixed point $e_0^i$ (i.e., on the cycle $e_{20}^i$),
and is vanishing at the other fixed points $e_0^j$, $j\ne i$.

\bigskip

\begin{figure}[htb]
\begin{center}
\epsfxsize=4.5in\leavevmode\epsfbox{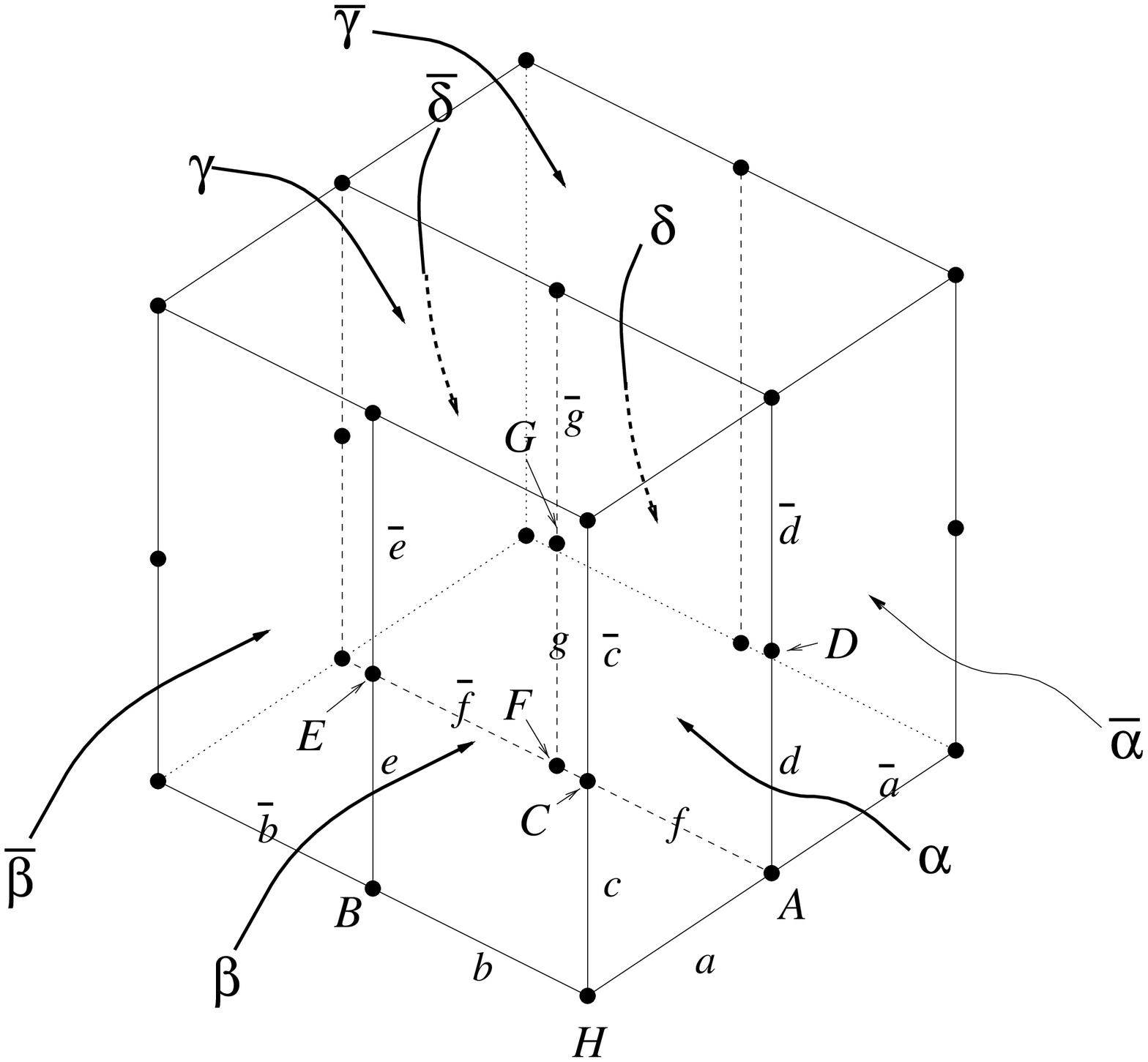}
\end{center}
\caption{Cell decomposition of $T^3$.}
\label{fig:T3}
\end{figure}

\noindent
\underline{$n=3$}

We consider the $\Z_2$-equivariant
cell decomposition of $T^3$ depicted in figure \ref{fig:T3}.
There are eight $0$-chains $e_0^i$ ($i=A,B,C,\ldots,H$);
fourteen $1$-chains $e_1^{\mu}$ ($\mu=a,\bar a,b,\bar b,\ldots, g,\bar g$);
eight $2$-chains $e_2^I$ ($I=\alpha,\bar \alpha,\ldots,\delta,\bar\delta$);
and two $3$-chains $e_3^{\lambda}$ ($\lambda=1,\bar 1$).
The $\Z_2$ acts on them as follows,
\begin{equation}
e_0^i\to e_0^i,~~
e_1^{\mu}\to e_1^{\bar \mu},~~
e_2^I\to e_2^{\bar I},~~
e_3^{\lambda}\to e_3^{\bar\lambda}.
\end{equation}
The cell decomposition of $T^3_{\Z_2}$ is given by
\bea
&&~~e_{00}^i:=e_0^+\times e_0^i\equiv e_0^-\times e_0^i,
\nonumber\\
&&\left.\begin{array}{l}
e_{10}^i:=e_1^+\times e_0^i\equiv e_1^-\times e_0^i,\\
e_{01}^{\mu}:=e_0^+\times e_1^{\mu}\equiv e_0^-\times e_1^{\bar\mu},
\end{array}\right\}
\nonumber\\
&&\left.\begin{array}{l}
e_{2\,0}^i:=e_{2}^+\times e_0^i\equiv e_{2}^-\times e_0^i,\\
e_{11}^{\mu}:=e_{1}^+\times e_1^{\mu}
\equiv e_{1}^-\times e_1^{\bar\mu},\\
e_{02}^I:=e_{0}^+\times e_2^{I}
\equiv e_{0}^-\times e_2^{\bar I}
\end{array}\right\}
\nonumber\\
&&\left.\begin{array}{l}
e_{\ell \,0}^i:=e_{\ell}^+\times e_0^i\equiv e_{\ell}^-\times e_0^i,\\
e_{(\ell-1)1}^{\mu}:=e_{\ell-1}^+\times e_1^{\mu}
\equiv e_{\ell-1}^-\times e_1^{\bar\mu},\\
e_{(\ell-2)2}^I:=e_{\ell-2}^+\times e_2^{I}
\equiv e_{\ell-2}^-\times e_2^{\bar I},\\
e_{(\ell-3)3}^{\lambda}:=e_{\ell-3}^+\times e_3^{\lambda}
\equiv e_{\ell-3}^-\times e_3^{\bar\lambda},
\end{array}\right\}~~\ell=3,4,\ldots
\eea
As in the previous case,
we only present the result.
The $2^{nd}$ coboundary group $B^2$ is spanned by $c_{02}^{I}+c_{02}^{\bar I}$
for all $I$ and
$c_{11}^{\mu}+c_{11}^{\bar \mu}$ for all $\mu$.
The cohomology is of rank $7$,
$H^2_{\Z_2}(T^3,\Z_2)=(\Z_2)^{ 7}$,
 and the seven generators are represented by
\bea
&&c_{20}^A+c_{20}^H+c_{11}^b+c_{11}^c+c_{11}^d+c_{11}^f+c_{02}^{\beta}
+c_{02}^{\delta},\nonumber\\
&&c_{20}^B+c_{20}^H+c_{11}^a+c_{11}^c+c_{11}^d+c_{02}^{\beta}
+c_{02}^{\delta},\nonumber\\
&&c_{20}^C+c_{20}^H+c_{11}^a+c_{11}^b
+c_{02}^{\gamma},\nonumber\\
&&c_{20}^D+c_{20}^H+c_{11}^a+c_{11}^b+c_{11}^c+c_{11}^d+c_{02}^{\beta}
+c_{02}^{\gamma}+c_{02}^{\delta},\nonumber\\
&&c_{20}^E+c_{20}^H+c_{11}^a+c_{11}^b+c_{11}^c+c_{11}^e+c_{02}^{\alpha}
+c_{02}^{\gamma},\nonumber\\
&&c_{20}^F+c_{20}^H+c_{11}^a+c_{11}^b+c_{11}^c+c_{11}^f
+c_{11}^g+c_{02}^{\alpha}
+c_{02}^{\beta}+c_{02}^{\delta},\nonumber\\
&&c_{20}^G+c_{20}^H+c_{11}^a+c_{11}^b+c_{11}^c+c_{11}^g
+c_{02}^{\alpha}
+c_{02}^{\beta}+c_{02}^{\gamma}+c_{02}^{\delta}.\nonumber
\eea
Note that the $i^{th}$ generator has a non-trivial value on the $\RP^2$
at the fixed point $e_0^i$ (i.e., on the cycle $e_{20}^i$)
and also at the point $e_0^H$ (i.e., on the cycle $e_{20}^H$),
and is vanishing at the other fixed points $e_0^j$, $j\ne i,H$.

%

\end{document}